%% file: ms.tex
\documentclass{emulateapj}

\begin{document}

\input mymac.tex

\title{DEMOGRAPHICS AND PHYSICAL PROPERTIES OF GAS OUT/INFLOWS  AT $0.4 < z < 1.4$}

\author{
  Crystal L. Martin\altaffilmark{1}
  Alice E. Shapley\altaffilmark{2,8}
  Alison L. Coil\altaffilmark{3,9}
  Katherine A. Kornei\altaffilmark{2}
  Kevin Bundy\altaffilmark{4}
  Benjamin J. Weiner\altaffilmark{5}
  Kai G. Noeske\altaffilmark{6}
  David Schiminovich\altaffilmark{7}
}

\altaffiltext{1}{Department of Physics, University of California, 
Santa Barbara, CA, 93106, cmartin@physics.ucsb.edu}

\altaffiltext{2}{Department of Physics and Astronomy ,University of California, 
Los Angeles, CA, 90025}

\altaffiltext{3}{Center for Astrophysics and Space Sciences,
Department of Physics, University of California, San Diego, CA 92093}

\altaffiltext{4}{Kavli Institute for the Physics and Mathematics of
the Universe, Todai Institutes for Advanced Study, University of
Tokyo, Kashiwa, Japan 277-8583 (Kavli IPMU, WPI)}

\altaffiltext{5}{Steward Observatory, 933 N. Cherry St., University of
Arizona, Tucson, AZ 85721, USA}

\altaffiltext{6}{Space Telescope Science Institute, Baltimore, MD 21218, USA}

\altaffiltext{7}{Department of Astronomy, Columbia University, New York, NY
10027, USA}

\altaffiltext{8}{Packard Fellow}

\altaffiltext{9}{Alfred P. Sloan Fellow}

\slugcomment{Submitted June 22, 2012; Accepted October 12, 2012}

\begin{abstract}
We present Keck/LRIS spectra of over 200 galaxies with well-determined redshifts between 
0.4 and 1.4. We combine new measurements of near-ultraviolet, low-ionization absorption lines 
with previously measured masses, luminosities, colors, and star formation rates to describe 
the demographics and properties of galactic flows.
Among star-forming galaxies with blue colors, we find a net blueshift of the \feII\ absorption 
greater than 200\kms (100\kms) towards 2.5\% (20\%) of the galaxies. The fraction of blueshifted
spectra does not vary significantly with stellar mass, color, or luminosity but does decline
at specific star formation rates less than roughly $0.8$~Gyr$^{-1}$.
The insensitivity of the blueshifted fraction to galaxy properties requires collimated outflows
at these redshifts, while the decline in outflow fraction with increasing blueshift might
reflect the angular dependence of the outflow velocity. The low detection rate of infalling gas, 
3 to 6\% of the spectra, suggests an origin in (enriched) streams favorably aligned with our 
sightline. We find 4 of these 9 infalling streams have projected velocities commensurate with 
the kinematics of an extended disk or satellite galaxy. The strength of the \mgII\ absorption 
increases with stellar mass, B-band luminosity, and U-B color, trends arising from a combination 
of more interstellar absorption at the systemic velocity {\it and} less emission filling in more massive 
galaxies. Our results provides a new quantitative understanding of gas flows between
galaxies and the circumgalactic medium  over a critical period in galaxy evolution.
\end{abstract}

\section{INTRODUCTION} \label{sec:intro}

Key processes in the evolution of galaxies include the flow of cool gas into galaxies,
the conversion of these baryons into stars, and the ejection of gas enriched with
heavy elements.  Determining the factors that govern the circulation of baryons
remains a critical, unsolved problem in cosmology. 

The measured properties of galaxies indicate strong evolution in this baryon cycle
between the peak in star-formation activity at redshift $z \approx 2$ and the
current epoch (Lilly \et 1996; Madau \et 1996; Hopkins \& Beacom 2006).
At any redshift during this period, star-forming 
galaxies populate a fairly well-defined locus in the star formation rate (SFR) -- stellar 
mass plane (Bell \et 2005; Elbaz \et 2007; Noeske \et 2007). However, the observed SFR at a 
fixed stellar mass declines by about a factor of 20 from $z \sim 2$ to the present. 
The Tully-Fisher relation also evolves in the sense that present-day galaxies 
have about 2.5 times more stellar mass than do galaxies at $z \approx 2.2$ with
the same circular velocity (Cresci \et 2009). Since redshift $z \approx 1$ 
when roughly half the mass in present-day galaxies was assembled (Drory \et 2005;
Faber \et 2007; Marchesini \et 2009),
the stellar mass in red sequence galaxies has continued to grow; but
the mass in blue galaxies has remained essentially constant (Faber \et 2007). 
Over this period, some process quenches star formation in massive galaxies, yet
observational efforts to identify this process with feedback from active galactic
nuclei (AGN) have not reached a consensus (cf., Schawinski \et 2007; Aird \et 2012). 

The leading explanation for this drop in cosmic star-formation activity is a
decline in the accretion rate of cool gas onto galaxies. The simplest form of 
this idea, that the mean growth rate of dark matter halos regulates the gas
inflow rate onto galaxies, fails; it yields too much star formation in low mass 
halos at early times (Bouch\'{e} \et 2010) and does not shut down star formation 
in high mass halos at later times (Somerville \et  2008).  Introducing gas 
physics -- heating by virial shocks in massive halos {\it and} outflows in
both low-mass and high-mass halos -- effectively tilts the underlying relationship 
between halo accretion rate and mass (described by the simplest model) into the observed 
SFR -- stellar mass relation and Tully-Fisher relation (Bouch\'{e} \et 2010 ; Dav\'{e}, 
Oppenheimer, \& Finlator \et 2011a; van de Voort \et 2011a, b; Dav\'{e}, Finlator, \& 
Oppenheimer 2012). The balance between gas outflow and inflow then determines
the relationship between the gas-phase metallicity and stellar mass
(Dav\'{e}, Finlator, \& Oppenheimer 2011b). The problem with this scenario
is that this baryon-driven picture of galaxy formation rests on a
rather rudimentary understanding of which galaxies have outflows and inflows and
how the physical properties of flows vary with fundamental galaxy properties.

Gas flows imprint resonance absorption lines on the spectra of their host
galaxies that can often be kinematically distinguished from interstellar gas at
the systemic velocity
(Heckman \et 2000; Schwartz \& Martin 2004; Martin 2005, 2006; Rupke \et 2005;
Schwartz \et 2006; Tremonti \et 2007; Martin \& Bouch\'{e} 2009; Rubin \et 2010a). 
Thus far, studies of gas flows at $0.5 < z  < 2$ have either been based on just 
a few individual galaxies (Sato \et 2009; Coil \et 2011; Rubin \et 2011) or 
on {\it composite spectra}, i.e., the average of many low S/N ratio spectra (Weiner \et 
2009; Rubin \et 2010b). The pioneering study of Weiner \et (2009) concluded that 
blueshifted \mgII\ absorption was ubiquitous in spectra of star-forming galaxies
at $z \sim 1.4$ and demonstrated that the Doppler shift and absorption 
equivalent width of this absorption increase (rather slowly) with both stellar mass 
and SFR.
At lower redshift, however, Rubin \et (2010) did not find blueshifted \mgII\ 
absorption in a composite spectrum of galaxies with stellar masses similar
to the Weiner \et (2009) sample and speculated that the higher specific SFRs
(i.e., SFR per unit stellar mass) characterizing $z \approx\ 1.4$ galaxies
might be required to host such outflows.  These two studies have demonstrated the 
existence of outflows and their possible evolution over a key period in the 
assembly of galaxies but are limited by their reliance on the {\it mean} 
spectrum of a population. Averaging many spectra together in
a composite hides less common features like gas inflows (Sato \et 2009;
Rubin \et 2012) and does not allow de-projection of the line-of-sight
Doppler shift into an outflow velocity.

Here, we present the results of a
survey of near-ultraviolet spectral features in 208 galaxies
with redshifts between  $0.4 < z < 1.4$ to provide an empirical measurement of how 
outflow and inflow properties change with cosmic time and galaxy properties.  The 
galaxies were selected from the DEEP2 (Deep Extragalactic Evolutionary Probe 2) 
survey (Davis \et 2003; Newman \et 2012). This redshift survey provides a relatively un-biased sample for 
investigating the demographics of outflows and inflows in galaxies brighter than $R_{AB} = 24.1$ in 
four fields. More importantly though, selecting from the DEEP2 survey ensures that fundamental
galaxy parameters --  such as stellar mass, $B$-band luminosity, and $U-B$ color  --
have been measured for the sample in a systematic manner (Bundy \et 2006; Willmer \et 2006). The additional 
photometry obtained for the AEGIS (All-Wavelength Extended Groth Strip International Survey) field 
allows us to measure SFRs for 51 of these galaxies; and these SFRs provide important information
about the energy and momentum produced by supernovae, stellar winds, and radiation from massive stars 
and therefore available to drive outflows (Chevalier \& Clegg 1985; Murray \et 2005).

In this paper, we focus on the diagnostics provided by low-ionization,
resonance-absorption lines, reserving the presentation of high-ionization
interstellar absorption  and stellar features, resonance emission, and fluorescent emission 
for future papers.
 Low-ionization, resonance lines in near-ultraviolet 
(rest-frame) spectra provide an empirical bridge between the optical transitions
typically observed for low-redshift galaxies and the far-UV transitions studied
extensively in spectra of high-redshift galaxies (Steidel \et 2010) because the \mgII\ 
$\lambda \lambda 2796$, 2803 doublet is accessible from the ground over the broad redshift range from 
roughly $0.25 < z < 2.5$ (Martin \& Bouch\'{e} 2009; Tremonti \et 2007; Weiner \et 2009; 
Rubin \et 2010; Coil \et 2011).  At the lower end of this redshift range, \mgII\ absorption properties 
can be cross calibrated with rest-frame optical lines such as \naI\ $\lambda \lambda 5890, 96$
 and \ion{Ca}{2} $\lambda 3933, 69$ . 
At $z \sgreat\  1.19$, the \mgII\ absorption properties can be directly compared to 
far-UV transitions in \ion{Si}{2}, \ion{Al}{2}, \ion{C}{2}, and \ion{C}{4}
commonly employed to study outflows in much higher redshift galaxies (Shapley \et 2003;
Steidel \et 2010; Jones \et 2012).

The \mgII\  doublet provides a very sensitive probe of outflows for several reasons: 
singly-ionized magnesium is a dominant ionization state over a broad range of conditions, 
both lines have large oscillator strength, and the cosmic abundance of Mg is fairly high.
Scattered \mgII\ emission, however, partially fills in the intrinsic absorption troughs
in galaxy spectra and complicates the interpretation of the Doppler shift of the absorption
trough. Spectral coverage blueward of the \mgII\ $\lambda \lambda 2796$, 2803 doublet provides
access to a series of strong \feII\ resonance lines which alleviate the concern about emission
filling. 
The \feII\ absorption troughs provide a cleaner view of the intrinsic absorption profile 
because fluorescence (rather than resonance emission) often follows absorption in several 
of these \feII\ transitions (Prochaska \et 2011; Erb \et 2012). Another advantage of \feII\
lines over the \mgII\ doublet arises from the large number of \feII\ transitions in the NUV. 
The oscillator strengths of the NUV \feII\ transitions span a substantial range and therefore
make it possible to place useful bounds on the column density of singly-ionized iron,
thereby constraining the total gas columns for an assumed (i.e., model dependent) metallicity 
and ionization fraction.

The paper is organized as follows. Section~\ref{sec:observations} introduces the sample and describes
our new Keck observations, including a discussion of the systemic velocity determination
and absorption-line sensitivity. After providing a broad overview of the near-UV spectral 
features, Section~\ref{sec:diagnostics} explains the complications caused by emission filling
and how we identify and measure galactic gas flows. 
We then compare the line profiles of the \mgII\ and \feII\ 
absorption troughs in Section~\ref{sec:compare}. In Section~\ref{sec:galaxy_properties},
we calculate the fraction of galaxies with a Doppler shift of 
low-ionization absorption relative to the systemic velocity and use previously measured
galaxy properties to illustrate the demographics of the galaxies showing blueshifts.
We then discuss the physical properties of the outflows and show how the outflow 
properties scale with galaxy properties. In Section~\ref{sec:inflow}, we expound on our 
discovery of net inflows of enriched gas. Section~\ref{sec:conclusions} summarizes our conclusions.

Throughout this paper we assume a cosmological model with $\Omega_m = 0.3$,
$\Omega_{\Lambda} = 0.7$, and $H_0 = 70$\kms\ Mpc$^{-1}$. We adopt the atomic 
oscillator strengths and cosmic abundance ratios given by Morton (2003) as well
as the associated vacuum wavelengths for transitions shortward of 3200\AA. We refer to
optical transitions by their wavelengths in air (for ease of comparison to
previously published work), but we work with their vacuum wavelengths.

\clearpage

\section{DATA}  \label{sec:observations}

We present rest-frame, ultraviolet spectroscopy of $z \sim 1$ galaxies drawn from the Deep 
Extragalactic Evolutionary Probe 2 survey (Davis \et 2003; Newman \et 2012) using the Low 
Resolution Imager and Spectrometer (LRIS) on the Keck~I telescope (Oke \et 1995). Building a 
sample of 208 spectra required a significant observational campaign, custom-designed masks for 
the focal plane, and the remarkable blue sensitivity of Keck/LRIS-B (Steidel \et 2004). We describe 
the sample selection in Section~\ref{sec:sample} and the new observations in Section~\ref{sec:lris_observations}.

\subsection{Sample}  \label{sec:sample}

\begin{figure*}[t]
        \hbox{\hfill \includegraphics[height=8cm,angle=-90,trim= 80 0 0 0]{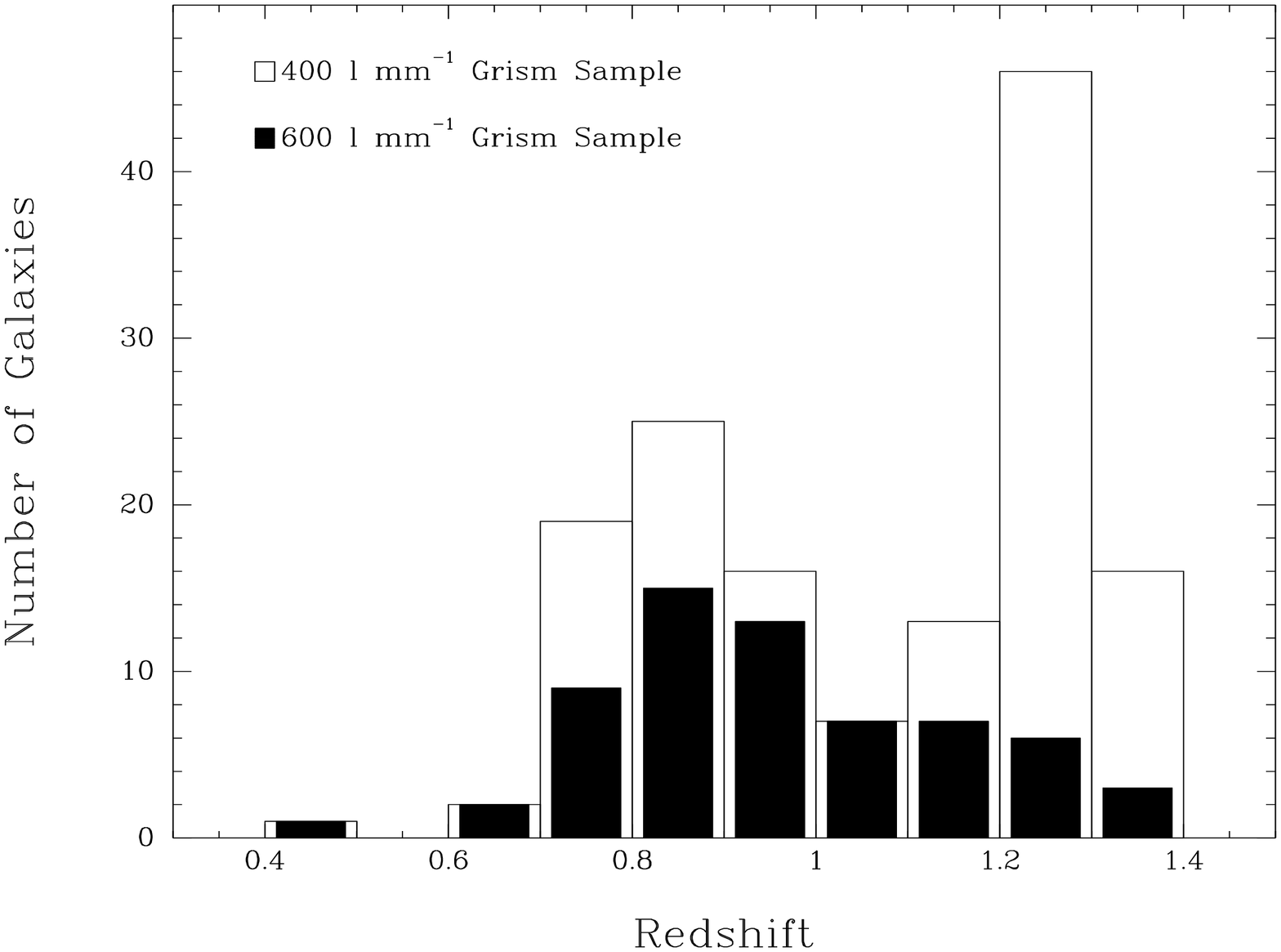}}
         \hbox{\hfill \includegraphics[height=8cm,angle=-90,trim= 60 0 0 0]{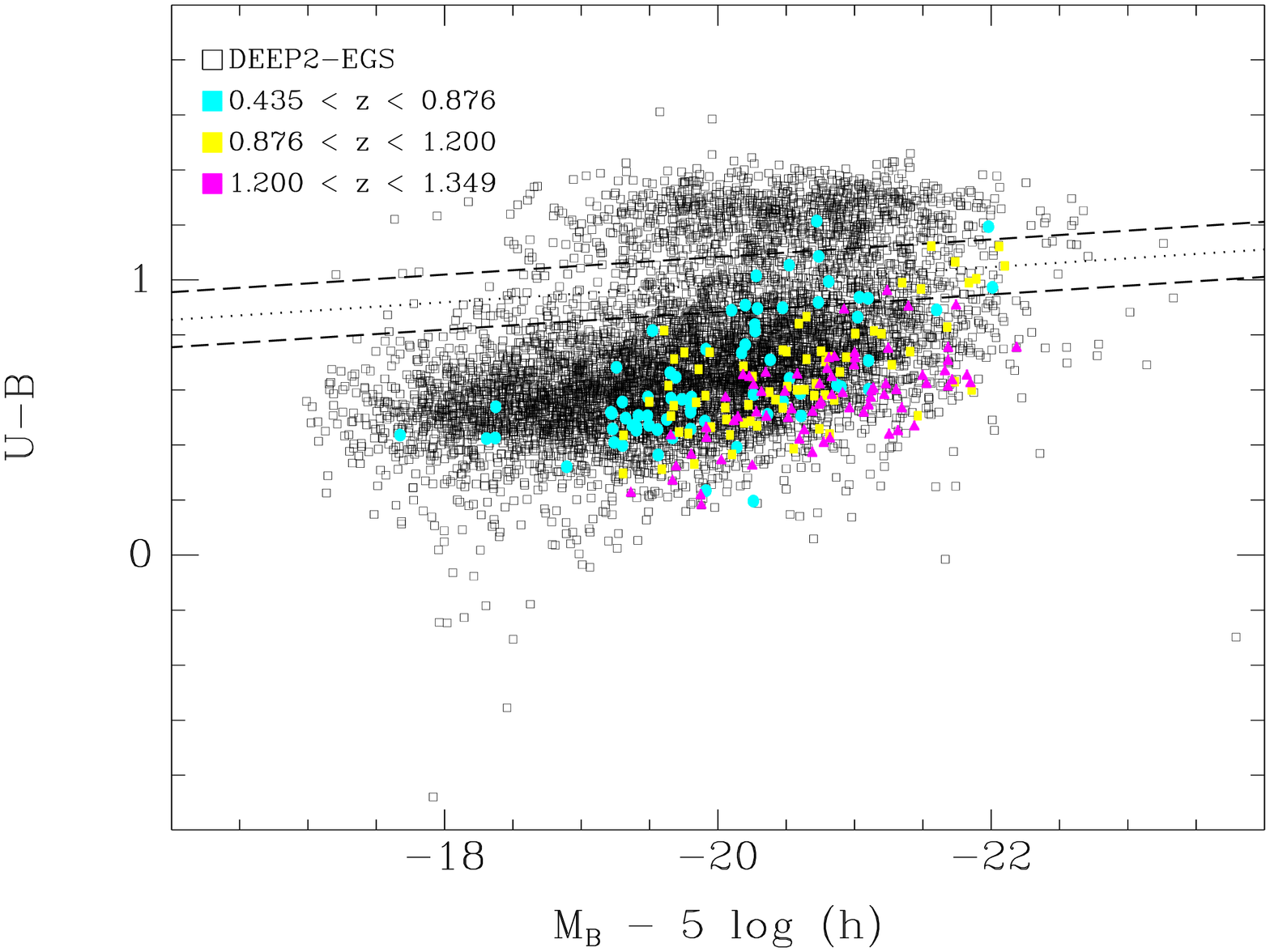}}        
          \hbox{\hfill \includegraphics[height=8cm,angle=-90,trim = 60 0 0 0]{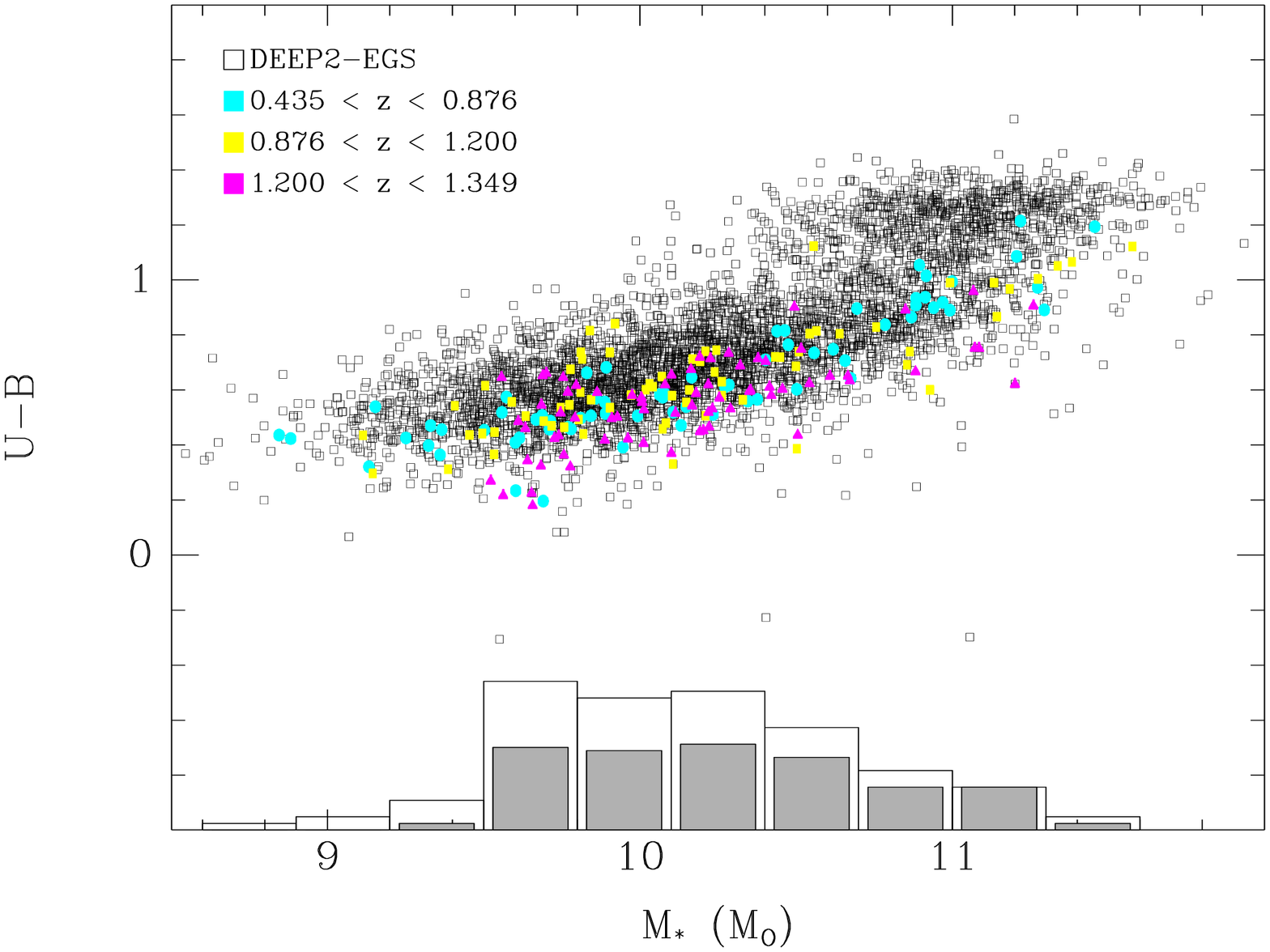}}
                                 \caption{\footnotesize
            Properties of the galaxies observed with LRIS.            
            ({\it Top:})
            Redshift histogram. Higher redshift galaxies were specifically
            targeted for the lower resolution observations in order to obtain spectral coverage of 
            resonance lines in the far-UV spectrum. The median redshifts of the galaxies
            observed with the \dhigh\ (282\kms\ FWHM) and \dlow\ (435\kms\ FWHM)
            configurations are 0.94 and 1.12, respectively. 
            ({\it Middle:}) Color -- magnitude diagram (Willmer \et 2006).  The diagonal, dashed 
            line marks the division between the red sequence and the blue cloud at $z \sim 1$.
            For the LRIS sample, galaxy color is anti-correlated with absolute magnitude,
            7.8 standard deviations from the null hypothesis (i.e., no correlation) with
            Spearman rank correlation coefficient $r_S = -0.54$.  
            The rest-frame optical colors of the 
            galaxies (Willmer \et 2006) become slightly bluer with increasing redshift due to 
            the $R$-band selection of DEEP2 galaxies, and the lowest redshift tertile includes 
            lower luminosity galaxies that would not pass our apparent magnitude cut at
            higher redshift.
            ({\it Bottom:}) Color vs. stellar mass (Bundy \et 2006) showing
            histogram of stellar mass along the bottom. 
            Stellar mass was derived from SED fitting using a Chabrier stellar initial 
            mass function (IMF; Chabrier 2003). The stellar masses would be 0.25 dex larger for 
            a Salpeter IMF. The histogram distinguishes the most secure stellar masses (solid bars), 
            obtained for 127 galaxies detected in the K-band, from the full sample of 208 galaxies
            with LRIS spectroscopy. The median stellar mass of the K-band-detected galaxies in our 
            sample is $1.60 \times 10^{10}$\msun, and the galaxies not detected in the K-band have a 
            median mass of $7.23 \times 10^9$\msun\ as estimated from optical photometry. 
            Color and stellar mass are strongly correlated ($r_S = 0.77$ at $11\sigma$ significance).
            }
            \label{fig:sample} \end{figure*}

The DEEP2 survey obtained redshifts for approximately 38,000 galaxies 
spread over four fields widely separated in right ascension (Newman \et 2012). 
In three of the four fields,
the galaxies were color-selected to have redshifts  $z > 0.7$. The fourth
field, without color selection, is also the target of a deep imaging 
effort from the X-ray through the radio known as the All-Wavelength Extended Groth 
Strip International Survey, AEGIS (Davis \et 2007). The panchromatic data in 
AEGIS provide more accurate estimates of galactic SFRs than are currently
available for the other three fields (e.g., Noeske \et 2007).
Figure~\ref{fig:sample} summarizes the properties of the galaxies observed
with LRIS.

Our primary  sample of galaxies was chosen for spectroscopic follow-up based on redshift,
$1.19 \le\ z \le\ 1.35$,  and apparent magnitude $B < 24.0$. The magnitude cut ensures
adequate continuum S/N ratio for absorption-line spectroscopy. This lower bound on the 
redshift guarantees that the \cIV\ $\lambda \lambda 1548, 1550$ doublet lands longward of the atmospheric cut-off 
in the blue; and requiring $z < 1.35$  ensures that the \mgI\ 2853  transition is
covered by the blue spectrum.
The simultaneous coverage of far-ultraviolet (FUV) and near-ultraviolet (NUV) resonance lines
allows a direct comparison of the transitions used to measure outflow velocities in
$z \ge\ 2$ galaxies (Steidel \et 2010) and $z \sim 1$ galaxies (Weiner \et 2009; Rubin \et 2010).
In total, 68 galaxies meeting these criteria were observed on 6 masks.  Each mask
subtends a 5\farcm5 by 8\farcm0 field of view (Steidel \et 2004). With an average of 11 
primary targets per mask, we filled the masks with the following types of galaxies:
(1) lower redshift ($ z < 1.19$) galaxies with $EW([$ \ion{O}{2} $]) > 10~$\AA\ 
and $B < 24.5$ magnitudes, green valley galaxies, and K+A galaxy candidates (Yan \et 2009).
In total, 145 galaxy spectra were obtained through these masks with the \dlow\
configuration of LRIS described below.

To utilize less than optimal observing conditions, a second set of masks were
designed to obtain NUV spectroscopy of brighter galaxies. Spectra were obtained 
through 3 masks with galaxies prioritized as follows: (1) $B \le\ 23.6$, $ z \ge 0.7$, 
and  $EW([$ \ion{O}{2} $]) > 20 {\rm ~\AA\ } $ (41 objects);
(2) $B \le\ 23.6$, $ z \ge 0.7$, and  $EW([$ \ion{O}{2} $]) < 20 {\rm ~\AA\ } $ (12 objects);
(3) $B=23.6-24.0$ and $z > 0.7$ (7 objects), $B \le\ 23.6$ and $z < 0.7$ ( 1 object),
and 6 foreground galaxies that lie at 
small angular separation from another target.  Among these 67 galaxies, the spectra
of four foreground objects with $z < 0.4$ are not discussed further in this paper due
to their much lower redshifts. Hence, the second sample includes 63 galaxies observed with
the \dhigh\ configuration of LRIS described below.

The primary sample ($z > 1.19$) includes three galaxies with colors that place them between
the red sequence and blue cloud, a region of the color-magnitude diagram known
as the green valley (Bell \et 2003; Mendez \et 2011).  A total of 21 green valley galaxies
were included on the secondary (\dhigh\ backup) and primary (\dlow) 
masks including the filler objects; most
have $EW([OII]) < 10$ \AA. The sample
includes just two K+A (post-starburst) galaxies identified by Yan \et (2009). 
One of these is a green valley galaxy and the other is a K+A galaxy
with colors just redward of the green valley in the color - magnitude diagram.

As illustrated in Figure~\ref{fig:sample}, these targets 
span the redshift range from 0.4 to 1.4, have luminosities between $-17 > M_B > -22$, and sample
the stellar mass function between $8.9 < \log (M/\msun) < 11.6$. With a few exceptions noted above,
the galaxies have blue $U-B$ colors. These spectra therefore probe the properties of low-ionization
gas in and around typical star-forming galaxies 4.3 to 9.0~Gyr ago. The broad range in stellar
mass makes this an appropriate sample to compare gas outflows and inflows over about 3 dex in dark-matter 
halo mass. The clustering of the population studied suggests that the most massive tertile
of our sample resides in halos with $\log (M_h /\msun) > 12$ and that some of the galaxies in the lowest
mass tertile populate halos with $\log (M_h /\msun) < 11$. In the stellar mass range $\log (M_*/\msun) = 
9.0 - 11.5$, abundance matching suggests $\log (M_h/\msun)  = 11.25 - 14.24$ at $z = 1.0$ and 
$\log (M_h/\msun)  = 11.2 - 14.7$ at $z = 0.5$ (Behroozi, Conroy, and Wechsler 2010), consistent
with the halo masses measured from galaxy clustering.

\subsection{LRIS Observations}  \label{sec:lris_observations}

Multislit spectroscopy of the fields listed in Table~\ref{tab:observations} was obtained with LRIS on the 
Keck I telescope between 2007 October and 2009 June. Our primary instrumental configuration used 
the D680 dichroic, the LRISb 400~mm$^{-1}$ grism blazed at $\lambda 3400$\AA, and the
the 831~mm$^{-1}$ grating blazed at 8200\AA\ on the red side. For the backup program, we configured 
LRIS with the D560 dichroic, 600~mm$^{-1}$ grism blazed at 4000\AA, and the 600~mm$^{-1}$ red grating 
blazed at 7500\AA; this setup generally provided continuous spectral coverage between the red and blue spectra. 
In both configurations, the red-channel wavelength coverage usually included rest-frame-optical
emission lines, most commonly [\ion{O}{2}] $\lambda \lambda 3726, 29$. For all observations, 
the Cassegrain Atmospheric Dispersion Corrector (ADC) (Phillips \et 2006) was enabled; and  
the width of the slitlets subtended 1\farcs2 on the sky.
We observed a total of 9 slitmasks, 6 in our primary configuration, and 3 in our
back-up configuration, as described below. Conditions during the observing runs
ranged from excellent (clear with 0\farcs6 seeing) to variable (intermittent clouds
with variable seeing up to 2\farcs0). Details of these observations are presented in Table~\ref{tab:observations}.

The data were primarily reduced using IRAF tasks, with scripts designed 
for cutting up the multi-object slit mask images into individual 
slitlets, flat-fielding using spectra of the twilight sky for the blue 
side and dome flats for the red side, rejecting cosmic rays, subtracting 
the sky background, averaging individual exposures into final stacked 
two-dimensional spectra, extracting to one dimension, wavelength and 
flux calibrating, and shifting into the vacuum frame.\footnote{The
         typical root-mean-square error in the wavelength calibrations are 
         0.30 \AA, 0.15 \AA, 0.05 \AA, and 0.05 \AA, respectively, for the blue
         \dlow, blue \dhigh, red 831~mm$^{-1}$, and red 600~mm$^{-1}$ spectra.}
These procedures 
are described in detail in Steidel \et (2003). We also followed the 
procedures outlined in Shapley \et ( 2006) for background subtraction of 
deep mask spectra. Accordingly, to avoid potential oversubtraction of 
the background, the object continuum location was excluded from the 
estimate of the background fit at each dispersion point. Along with each 
target science spectrum, we extracted an error spectrum from the 
standard deviation of the mean of the individual exposures. We paid 
particularly close attention to deriving accurate wavelength solutions, 
adjusting an initial wavelength calibration derived from arc lamp 
exposures such that sky lines in our spectra appeared at the correct 
wavelengths. Finally, spectra were flux calibrated using observations of 
spectrophotometric standards throughout the run. To calibrate the entire 
observed wavelength range, spectra of these stars were obtained through 
slits at multiple positions in the focal plane because the standard 
longslit spectral coverage is too blue relative to that of the multislit 
coverage.

The spectral resolution realized differed between the two configurations.
In the primary configuration (\dlow\ grism), the atmospheric seeing rather than the slit width 
determined the spectral resolution because the angular sizes of the galaxies (measured along the 
spatial direction of the slit) were slightly smaller than the slit. Scaling the width of arc lamp 
lines by this ratio, the average effective resolution was 435\kms\ FWHM (full width at half
maximum intensity) in the grism spectra and 150 \kms\ in the grating spectra. In the \dhigh\ 
configuration (used to observe lower redshift galaxies in non-optimal conditions), the galaxies 
typically filled the 1\farcs2 aperture; and the slit width set the spectral resolution. We 
measured line widths of 282 \kms\ FWHM in the blue spectra and 220 \kms\ FWHM in the red spectra.

\subsubsection{Determination of the Systemic Velocity} \label{sec:vsys}

Identifying net outflows or inflows requires accurate {\it relative redshifts}
between the galaxy and the low-ionization gas. We derive the galaxy redshifts directly 
from the LRIS spectra whenever possible. The ADC used for the LRIS observations compensates 
for atmospheric differential refraction and eliminates systematic errors caused by the
aperture subtending different regions of a spatially-resolved galaxy at widely
separated wavelengths. In most of our spectra, line emission in the [\ion{O}{2}] doublet drives
the determination of the galaxy redshift. Since dense, photoionized gas near star-forming regions
produce this emission, we expect these redshifts to accurately describe the
{\it systemic velocity}.

We derived LRIS redshifts for 167 galaxies using only the spectrum redward 
of the strong NUV \mgII\ and \mgI\ transitions ($\lambda_r > 3000$\AA). 
The most prominent spectral feature is usually the [\ion{O}{2}] $\lambda
\lambda 3726,29 $ doublet. However, strong [\ion{O}{3}] $\lambda \lambda 4959, 
5007$ and Balmer emission lines
appear in spectra of the lower redshift galaxies; and Balmer absorption and 
the 4000\AA\ break are prominent in some spectra. 
Although we have attempted to avoid strong resonance lines that are
blends of photospheric and interstellar lines, interstellar \ion{Ca}{2} $\lambda 
3933, 69$ absorption could lower some of the LRIS redshifts relative to
the purely stellar/nebular redshift in a few spectra. 
To estimate a redshift, each 
spectrum was cross-correlated with 3 templates - 
an emission line galaxy, a quiescent galaxy, and a K+A galaxy -
using the DEEP2 IDL pipeline;
the best fit to a linear combination of these 3 templates was adopted as
the LRIS redshift. 
How accurately we can find the centroid of spectral lines determines
the statistical error on the redshift (and systemic velocity) for most galaxies. 
For a Gaussian line profile, the standard deviation of the mean, $SDOM 
\approx \sigma / SNR $ describes the centroiding error.
For a single line, the typical uncertainty is $ \delta(v) \approx 19$\kms $(\sigma / 
190$ \kms $) (S/N / 10)^{-1}$, where $\sigma = {\rm FWHM}({\rm km~s}^{-1}) / 2.35$ is the 
spectral resolution.

For the 41 galaxies with no LRIS coverage of [\ion{O}{2}], 
we adopted the DEEP2 redshift and estimated
the typical error in the systemic velocity from the redshift differences
measured among the 167 galaxies. A positive redshift difference, 
$z_{DEEP2} - z_{LRIS} > 0$, means that adopting the DEEP2 redshift
increases the inferred outflow
speed (makes it more negative) by an amount $\Delta v_{sys} = 
c  (z_{DEEP2} - z_{LRIS}) / (1 + z_{LRIS})$. The mean offset computed
from 167 spectra is $\Delta v_{sys} = -14\pm3$\kms,  quite
small compared to typical outflow/inflow speeds. 
The standard deviation of 41\kms\ indicates the size of the
systematic error in the redshift estimate.
For a spatially-resolved source, uncertainties of this magnitude can result 
from an offset between the slit center and the brightest line emission; and
although we use the same galaxy coordinates as the DEEP2 survey, the position
angle of the slitlets generally differ.

\subsubsection{Sensitivity to Absorption Lines}

Prior to making measurements, we deredshifted the spectra to the rest-frame.
We fit a low-order spline to the continuum in each galaxy spectrum avoiding 
regions near spectral lines. We divided the spectra by the fitted continuum
to produce normalized spectra. 
Uncertainty about the continuum level propagates into an error term in 
the equivalent width measurements. We found that the error term from 
the continuum fit could be kept small, compared the term derived from 
the error spectrum, by fitting the NUV continuum (approximately 
$\lambda 2200 - 2900$) independently of the FUV and optical continua.

The resulting sensitivity to absorption lines is described by the minimum
detectable equivalent width. Our observing strategy was designed to reach a rest-frame
equivalent width $W_r(2796) \simeq\ 1~$\AA\ because the distribution 
function of intervening absorbers presents a break here which may signal 
association with galaxy halos (Nestor \et 2006). The weakest line that
we detect depends on the continuum S/N ratio at the observed wavelength
of the transition as well as the intrinsic width of the absorption trough.
In the 2400-2500\AA\ bandpass, the median S/N ratios of the \dlow\ and \dhigh\ spectra
are 6.5 and 5.5 (per pixel), respectively. 
For purposes of illustration, a typical equivalent width limit can be
estimated assuming $S/N \sim 5$ and the FWHM of an unresolved line.
In the near-UV \feII\ series, absorption troughs with rest-frame equivalent
widths stronger than
\begin{eqnarray}
W_{3\sigma} = 0.92 {\rm ~\AA\ } (6.5/SNR)_{\lambda2450} 
\end{eqnarray}
in \dlow\ spectra, or larger than
\begin{eqnarray}
W_{3\sigma} = 0.65 {\rm ~\AA\ } (5.0/SNR)_{\lambda2450}
\end{eqnarray}
in \dhigh\ spectra, are easily detected.
Just blueward of the \mgII\ doublet, the median continuum S/N ratios of the \dlow\ and \dhigh\ 
spectra are, respectively, 4.6 and 6.6. 
The typical rest-frame sensitivities
of the \dlow\ and \dhigh spectra are, respectively, 
\begin{eqnarray}
W_{3\sigma} = 2.02 {\rm ~\AA\ } (5.0/SNR)_{\lambda2790} 
\end{eqnarray}
and 
\begin{eqnarray}
W_{3\sigma} = 0.58 {\rm ~\AA\ } (5.0/SNR)_{\lambda2790}.
\end{eqnarray}

Farther to the red near the \mgII\ doublet, the continuum S/N in a particular 
spectrum is usually a bit lower than it is near \feII\ $\lambda \lambda 2587, 2600 $ due to the 
blue blaze of both grisms, so some of the spectra do not quite reach the
target sensitivity in \mgII. Spectra with S/N ratio adequate to detect
equivalent withs of 1-2~\AA\ in individual galaxies at $z \sim 1$ have never
been previously presented however in such large numbers.

We measure the Doppler shift of resonance absorption relative to 
the systemic velocity.  Since this systemic velocity, as well any
projected gas flow, may vary slightly with slit position, the 
systematic error  in the Doppler shift of the resonance lines could, 
in exceptional geometries, be as large as the systematic error of 
41\kms\ estimated above. Typically, however, the statistical error
in centroiding the absorption lines will characterize the error on
the Doppler shift.
The resolution of the \dlow\ spectra, expressed by the standard deviation $\sigma$, 
is 1.54 times coarser than that of the \dhigh\ spectra; but this lower 
resolution is partly offset by higher average S/N ratio spectra of the \dlow\ subsample.
For emission lines in our best \dlow\ and \dhigh\
spectra, this statistical error is about $\delta V \approx 185 / 25 
\approx  7$\kms\  and $\delta V \approx 120 / 10 \approx  10$\kms, respectively. 
Fitting multiple absorption lines, as described in Section~\ref{sec:measure_fe2},
further reduces the statistical error on our best estimate of Doppler shift $V_1$ 
of the \feII\ series.

\section{DIAGNOSTICS OF OUTFLOWS AND INFLOWS} \label{sec:diagnostics}

In this paper,
we explore the diagnostics in the near-UV absorption-line spectrum longward of
2000 \AA.  The emission line spectra are discussed further in two
companion papers  (Martin \et, in prep; Kornei \et, in prep). We defer the 
discussion of the far-UV spectral features to a future paper.

\begin{figure*}[t]
 \hbox{\hfill \includegraphics[height=19cm,angle=-90,trim=230 100 200 0]{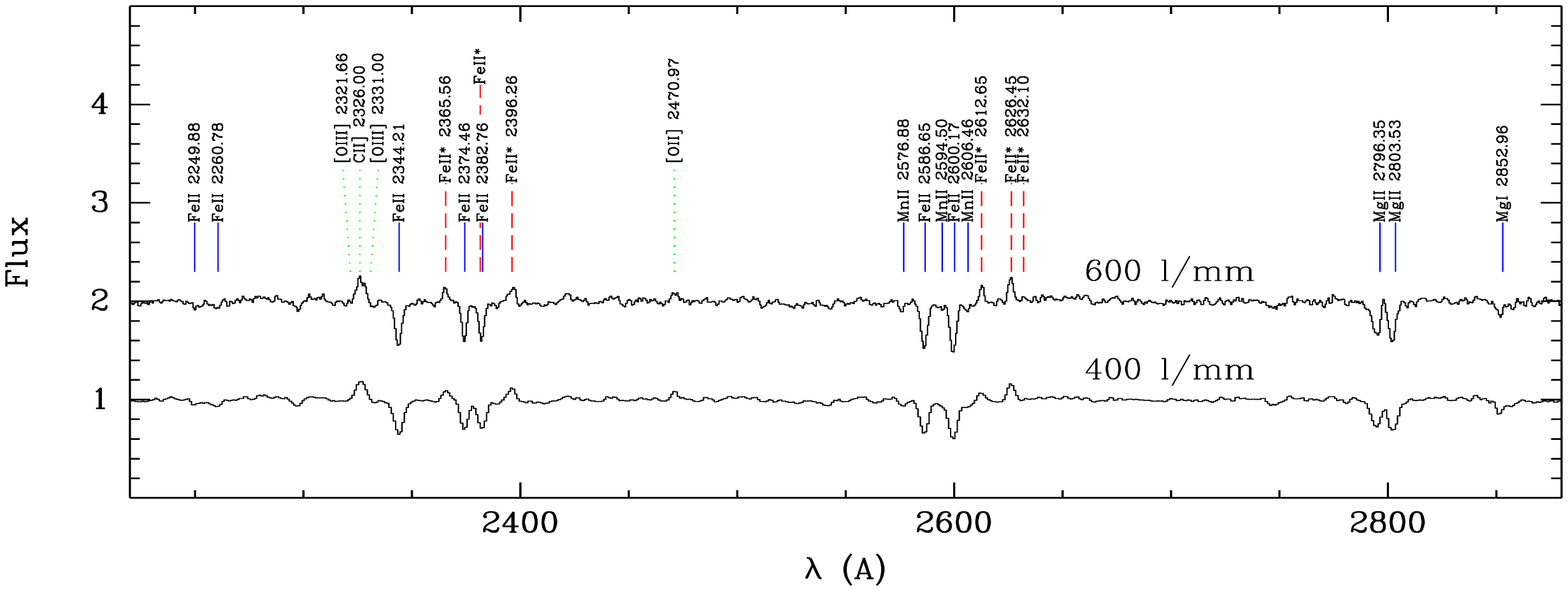} \hfill}
  \vspace{5cm}
          \caption{\footnotesize
            Composite, near-ultraviolet spectra of star-forming galaxies at $z \sim 1$. 
            The continuum level has been normalized to unity.
            The lower spectrum in \fig~\ref{fig:comp_spec} shows the addition of all 
            208 continuum normalized spectra after smoothing them to a common resolution of 
            435\kms\ FWHM. The upper spectrum is the average of 63 spectra observed at a resolution
            of 282\kms\ FWHM. Vertical blue (solid), red (dashed), and green (dotted) lines mark, respectively, 
            resonance absorption lines, fluorescent emission lines, and nebular
            emission lines. The unlabeled \feII$^*$ transition is a weak line at 
            $\lambda 2381.49$; see Martin \et (2012b, in prep.) for a tabular linelist.
}
 \label{fig:comp_spec} \end{figure*}

\subsection{Near-UV Spectral Features} \label{sec:spectral_features}

In Figure~\ref{fig:comp_spec}, we show the spectral region around the following NUV
resonance lines: 
\feII\   $\lambda 2249.88, 2260.78, 2344.21, 2374.46, 2382.77, 2586.65,$ and~ $2600.17$; 
\mgII\   $\lambda \lambda 2796.35$, 2803.53; and \mgI\  $\lambda 2852.96$. 
The absence of strong absorption in some individual spectra will partially fill in 
the absorption troughs in the composite spectra. Inspection of the individual spectra, 
however, shows that other factors prevent the absorption troughs from appearing completely 
black. Foremost, instrumental resolution smooths the spectrum and explains why
the absorption troughs are slightly deeper in the \dhigh\ composite than in the \dlow\ 
composite. In addition, 
the absorbing gas likely only partially covers the continuum source allowing
some continuum light to leak out and fill in the absorption troughs;
this partial covering has been fitted as a function of velocity using 
higher resolution spectra of brighter galaxies (Martin \& Bouch\'{e} 2009).
All three of these factors -- partial covering, instrumental resolution, and
averaging spectra -- imply that saturated absorption troughs need not be black.

Nebular emission lines mark dense gas near star-forming regions
excited by either photoionization or shocks. The LRIS red-side 
spectra typically show strong [\ion{O}{2}] $\lambda \lambda 3726,29$  emission and cover an
increasing number of rest-frame optical lines with decreasing redshift. In the NUV
composite spectra, the [\ion{O}{2}] $\lambda \lambda 2470.97, 2471.09$ lines are detected and blended. We attribute the broader width of the strong emission feature at $\lambda 2326$ 
to a blend of various \ion{C}{2}] transitions; in our picture, the \feII\ $\lambda 2328$ 
feature, which Rubin \et (2010a) suggested contributed to this emission complex, does
not contribute because its upper level cannot be populated by a permitted transition 
from the ground state.

Line blending complicates measurements of the absorption troughs. The individual
transitions of the \mgII\ doublet  blend together in the \dhigh\ composite
in \fig~\ref{fig:comp_spec}. The \feII\ $\lambda \lambda 2587, 2600$ doublet can blend with 
the \mnII\ $\lambda \lambda \lambda
2576.88, 2594.50, {\rm ~and~} 2606.26$ triplet, recognizable in the higher resolution composite in
\fig~\ref{fig:comp_spec}. 
Strong CII] $\lambda 2326$ emission often blends with the blue wing of  $\lambda 2344 $.
In some individual spectra,  the \mnII\ or CII] contamination compromises measurements 
of the maximum blueshift of absorption in the \feII\ $\lambda 2587 $ and \feII\ $\lambda 2344$
line profiles, respectively. In contrast, the \feII$^* \lambda 2365$ emission blueward of \feII\  
$\lambda 2374$ is much weaker than the CII] $\lambda 2326$ emission,
leaving $\lambda 2374$ as the preferred transition for looking for highly blueshifted \feII\
absorption. 

The \feII\ $\lambda 2374$  transition has the lowest oscillator strength
among the NUV \feII\ transitions routinely detected in the individual spectra. 
The oscillator strength of \feII\ $\lambda 2374$ is only a tenth
that of the strongest transition, \feII\ $\lambda2383$. The oscillator strengths of 
\feII\ $\lambda 2383$ and \mgII\ $\lambda 2803$, the weaker line in the \mgII\ doublet,
are nearly equal. The shorter wavelength transition in the \mgII\ doublet
has an oscillator strength twice that of the longer wavelength transition. The
rough equality of the absorption equivalent widths in all these transitions in
\fig~\ref{fig:comp_spec} demonstrates the resonance absorption is typically optically
thick. The weak \feII\ $\lambda 2250$ and $\lambda 2261$ transitions, just barely detected 
in \fig~\ref{fig:comp_spec}, are optically thin and particularly useful for
constraining the column density of single-ionized iron, $N(Fe^+)$.

Among our spectra, we often find suprisingly low \feII\ $\lambda2383$ 
equivalent widths relative to the other \feII\ absorption lines.
We attribute this apparent contradiction of the intrinsic line strengths
to resonance
emission partially filling in the deeper, intrinsic absorption profile.
This {\it emission filling} is more significant in \feII\ $\lambda2383$
than the other strong \feII\ transitions because \feII\ $\lambda 2383$ absorption
does not produce fluorescent emission; i.e., the only permitted transition
from the excited state is decay to the ground state. 
When an absorption trough is partially filled in by emission,
the centroid  will generally
differ from the instrinic absorption profile because absorption samples
only the gas between the observer and the galaxy, while emission may come from 
both the near and far sides of the galaxy.

Many of the individual spectra show prominent \mgII\ emission lines.
In \fig~\ref{fig:comp_spec}, the longer wavelength, lower oscillator
strength transition in the \mgII\ doublet has the larger equivalent width,
an unphysical situation for the intrinsic equivalent width ratio. The
inverted apparent ratio may arise from the troughs being partially filled
in by resonance emission. For example, if the intrinsic absorption
troughs are saturated, then the two troughs will have equal intrinsic
equivalent width; and slightly stronger emission in \mgII\ 2796 will
produce the inverted ratio of the net equivalent widths. Simple,
radiative-transfer models do not produce these inverted ratios
(Prochaska \et 2011), and more detailed modeling is needed to test
this idea. A non-spherical geometry, for example, allows the absorption
and emission to come from physically distinct regions of the galaxy
with differing optical depths. Furthermore, it may be quite important
to consider multiple origins of the \mgII\ line photons, which are
both emitted by HII regions (Erb \et 2012) and generated by continuum absorption
throughout the interstellar medium (ISM) and circumgalactic medium.

\subsection{Emission Filling}

In the normal situation for intervening absorption in quasar spectra,
where the continuum source subtends a small angular size compared to the gas clouds, a
negligible fraction of photons absorbed in the resonance transitions are re-emitted along
our sightline. For extended sources like galaxies, however, the equivalent width of
the emission can be comparable to the absorption equivalent width. 

At low densities, every excitation from the ground state decays as either 
resonance emission or fluorescent emission. In the absence of dust, photon
number is conserved, and the equivalent width of the resonance emission (plus 
any fluorescent emission from the same upper level) will be exactly equal to the 
absorption equivalent width.  The resonance emission will generally not
cancel out the absorption trough due to their velocity offset, generated for example by
radial gas flow. The absorption trough reflects only the kinematics of gas  in front of 
the continuum source while the emission-line profile probes the kinematics of both the near 
and far sides of the galaxy. Departure from a spherically symmetric gas distribution
generates specific viewing angles for which spectra will show stronger absorption 
than emission (and other angles with emission stronger than absorption). In addition to
dust and the geometry of the gas distribution, spectroscopic apertures smaller than
the emission region will also produce unequal emission and absorption equivalent width.
Hence, it is not surprising that the LRIS spectra rarely show equal absorption 
and emission equivalent widths.

The point is that resonance emission should be expected in galaxy spectra, and we
must consider how much it fills in the intrinsic absorption profile, a phenomenon we
call {\it emission filling}.
In spectra of nearby galaxies, the near absence of emission filling in \naI\ $\lambda \lambda 
5890, 96 $ has long been attributed to the high gas densities in regions with a significant 
neutral sodium fraction; the large pathlength of scattered photons produces a high probability 
of absorption by a dust grain (Heckman \et 2000; Martin 2005, 2006; Chen \et 2010). Quite
often, the \mgII\ regions in galaxy spectra exhibit a P~Cygni profile, which clearly
shows resonance emission filling in the part of the intrinsic absorption trough near
the systemic velocity (Martin \& Bouch\'{e} 2009; Weiner \et 2009;  Rubin \et 2011a; 
Coil \et 2011; Erb \et 2012.).

Absorption from the ground state necessarily leads to resonance emission when
the ground state has a single level, such as in the cases of \naI, \mgII, and \cIV\ ions, but
can be followed by fluorescent emission when the ground state has fine structure, such
as for the Fe$^+$ and Si$^+$ ions. 
In the composite spectra in \fig~\ref{fig:comp_spec}, we identify
fluorescent \feII$^*$ lines at $\lambda 2365.56$, 2396.26, 2612.56, and 2626.45; 
the \feII$^*$ $\lambda 2632.10$
line may  be marginally detected in a few individual spectra. Because these \feII$^*$ emission lines all 
have upper levels that are populated by absorption from the ground state, we can 
attribute them to fluorescence.  No absorption from excited levels in \feII\ is detected
consistent with densities well below the critical densities of the \feII\ transitions.

We understand the reduction in \feII\ $\lambda 2383$ equivalent width relative to
\feII\ $\lambda 2374$ in terms of emission filling. The spectroscopic selection rules for
total angular momentum imply that the \feII\ $\lambda 2383$ photons will scatter like the
common transitions (such as \mgII\ $\lambda \lambda 2796, 2803$ ) with a singlet, ground state.
Absorption of \feII\ $\lambda2374$ photons, however, can produce fluorescent \feII$^* ~\lambda2396.26$ photons
that do not fill in the resonance absorption trough. The relative radiative decay rates suggest that 
88\% of \feII\ $\lambda2374$ absorptions will decay as \feII$^* ~\lambda2396.26$ photons.

\begin{figure*}[t]
  \hbox{\hfill   \includegraphics[height=18cm,angle=-90,trim = 40 80 30 0]{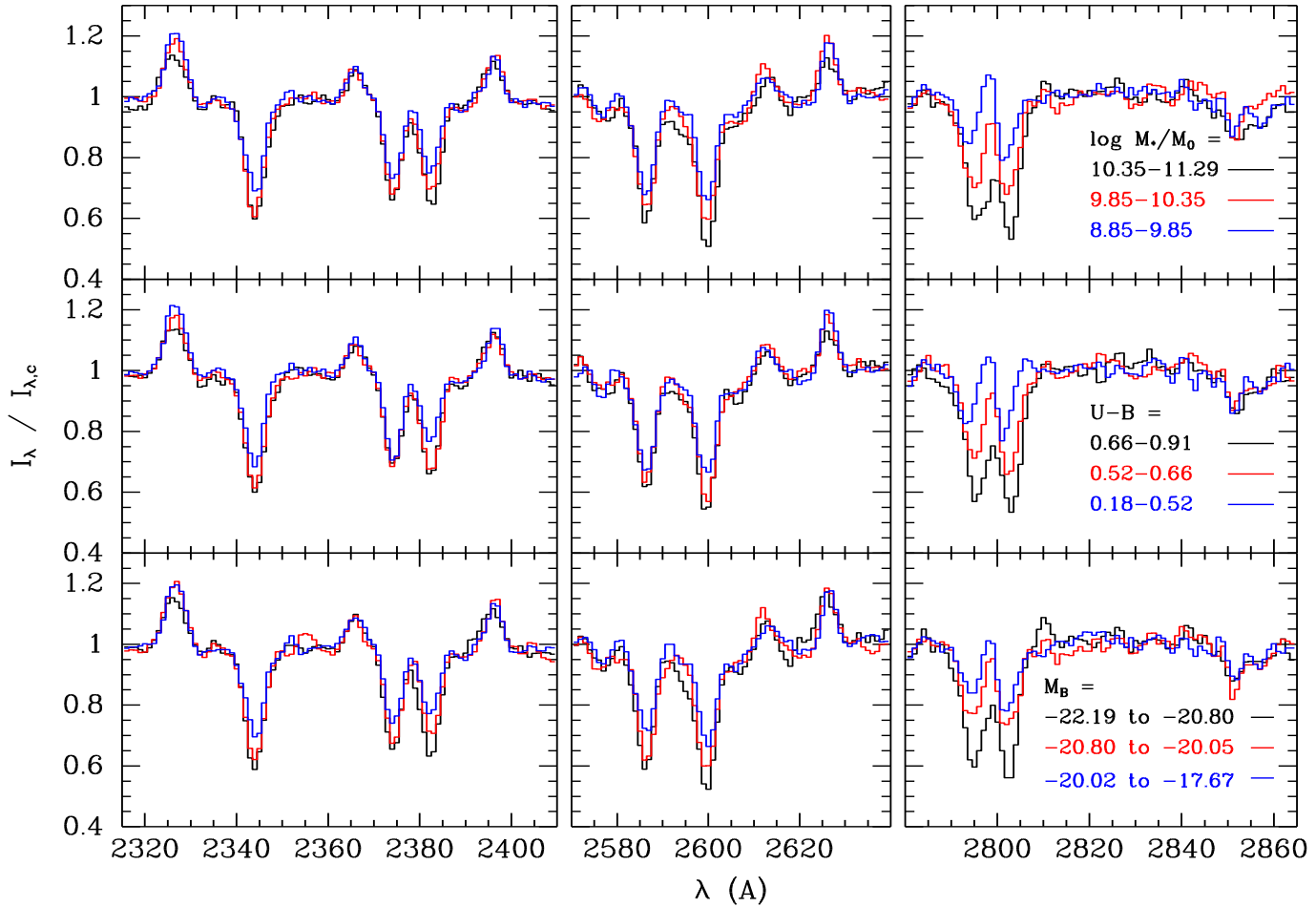}
          \hfill}
          \caption{\footnotesize
Comparison of the \feII\ and \mgII\ line profiles in average spectra of galaxies
with different properties.
{\it Top:}  Towards lower mass galaxies, the \mgII\ absorption troughs become shallower, the
\mgII\ emission becomes stronger, and the \feII$^*$ emission in $\lambda 2612, 2626$
increases. In the highest mass tertile, the \mgII\ doublet ratio is not inverted,
and \feII\ $\lambda 2383$ has a higher equivalent width than \feII\ $\lambda2374$, as it should in
the absence of emission filling.
{\it Middle:}  We see similar trends with color because color is strongly correlated with 
stellar mass (redder galaxies are more massive on average).
{\it Bottom:}  The resonance absorption troughs become deeper in more luminous
galaxies. See the text for a full explanation of why the stronger absorption in
more luminous, redder, more massive galaxies is largely attributable to emission
filling.
}
 \label{fig:mass_composite} \end{figure*}

Fluorescent emission following \feII\ $\lambda 2344$ absorption
in \feII$^*$ $\lambda 2381.49$ will also contribute to filling in the \feII\ $\lambda 2383$ absorption trough, 
but the resonance scattering of the latter will dominate the emission filling of the \feII\
$\lambda 2383$ trough. An analogous situation arises in the 
longer wavelength \feII\ multiplet shown in the middle panel of Figure~\ref{fig:mass_composite}, 
where absorption in the transition with the higher oscillator strength, \feII\ $\lambda 2600 $, usually 
results in emission of a photon at the same wavelength. In contrast, the excited level of the weaker
transition \feII\ $\lambda 2587$ typically decays via a longer wavelength, and therefore fluorescent,
photon. Comparison of the Einstein~A coefficients among all the transitions from the excited
level indicates a decay via \feII$^*$ $\lambda 2612.65$ 45\% of the time, through
\feII$^*$ $\lambda 2632.10$ 23\% of the time, and in the resonance transition $\lambda 2587$ the
remaining 32\% of the time.
These fluorescent emission lines, and the \feII$^* ~\lambda 2626$ fluorecence
that follows the absorption in the \feII\ $\lambda 2600 $\AA\ resonance transition
(13\% of the time), are clearly detected in the spectra shown in Figures~2-5
and will be discussed further in a forthcoming paper (Kornei \et, in prep.; Martin \et, in prep.).

Photons absorbed in the \mgI\ $\lambda 2853$ transition will scatter (rather
than fluorescence) because the ground state is again a singlet. Could emission filling 
also explain the decreasing equivalent width of \mgI\ $\lambda 2853$ 
in the spectra of lower mass galaxies seen in Section~\ref{sec:compare}.
It would be premature to draw this conclusion 
from the equivalent widths of the \mgI\ $\lambda 2853$ feature alone; any of a number
factors -- such as a  decrease
in the gas velocity dispersion, an increase in the \mgII\ ionization fraction, or a 
reduced covering fraction  -- would 
systematically decrease the equivalent width with decreasing stellar mass.

The presence of \mgII\ emission in galaxy spectra (Martin \& Bouch\'{e} 2009; 
Weiner \et 2009) can often be attributed to resonant scattering (Rubin \et 2011; 
Coil \et 2011; Erb \et 2012). Emission that fills in part of the absorption 
trough greatly complicates the interpretation of the Doppler shift and equivalent
width of \mgII\ absorption in galaxy spectra. Radiative transfer models,
however, predict little emission at the highest outflow speeds (Prochaska \et 2011). 
The maximum blueshift of the \mgII\ $\lambda 2796$ absorption is therefore the only
property of the intrinsic absorption profile that we can routinely measure directly. 
The maximum blueshift of the absorbing gas has also been used previously
to discuss whether outflowing gas is bound to the galaxy (e.g., Heckman
\et 2000; Martin 2005; Martin \& Bouch\'{e}; Weiner \et 2009; Coil \et 2011;
Heckman \et 2011).

\subsection{Fitting \feII\ Absorption Troughs}  \label{sec:measure_fe2}

Examination of our spectra showed that emission filling complicates
the direct measurement of intrinsic absorption properties, especially
in the \mgII\ $ \lambda \lambda 2796$, 2803, \feII\ $ \lambda 2383$, and
\feII\ $ \lambda 2600$ transitions. First, the equivalent widths measured
for transitions from a single ion do not always follow the curve of growth (Spitzer 1978).
In the \fig~\ref{fig:comp_spec} spectra, the highest oscillator strength \feII\ 
transition at $\lambda 2383$ has a lower equivalent width than other \feII\ transitions.
Secondly, a partially filled absorption trough has a centroid velocity that 
is bluer than the centroid of the intrinsic absorption profile because 
the resonance line emission is closer to the systemic velocity than
is the absorption. This prevents robust measurement of the Doppler shift
without detailed modeling (Prochaska \et 2011). Figures~7 and 8 in the
recent work of Erb \et (2012) further illustrate these points quantitatively.

We circumvent these problems by adopting the centroid velocity of  \feII\ absorption 
rather than \mgII\ as our primary indicator of the net Doppler shift of the
low-ionization gas.
In our discussion of emission filling, we explained why
we expect the \feII\ $\lambda2344$, 2374, and 2587 absorption troughs to  
reflect the outflow opacity more closely than does the \mgII\ profile.
Absorption in these transitions usually leads to fluorescence (i.e.,
emission of a longer wavelength photon) rather than scattering, so
the intrinsic shape of the absorption trough is much less affected by
emission filling than the \mgII\ or \feII\ $\lambda 2383$ lines. 
We therefore fit only those transitions that have a high probability of 
fluorescence, i.e., decaying via a longer wavelength photon rather than scattering 
and thereby filling in the absorption trough.

We detect resonance absorption lines at high significance in the individual LRIS 
spectra. Since the spectral resolution does not fully resolve the shapes
of most absorption troughs, we can only robustly measure the Doppler shifts
and equivalent widths of these absorption troughs in general. Centroid velocities 
have been widely used to describe galactic outflows in the past, but we caution 
the reader that interstellar absorption near the systemic velocity blends with
the absorption profile from outflowing (inflowing) gas in a low resolution spectrum. 
In Section~\ref{sec:measure_2comp}, we will examine how
interstellar absorption at the systemic velocity biases the estimates
of outflow properties.

\subsubsection{Single-Component Fits Describing the Absorption Centroid Velocity} \label{sec:v1}

\begin{figure*}[t]
 \hbox{\hfill \includegraphics[height=9cm,angle=-90,clip=true]{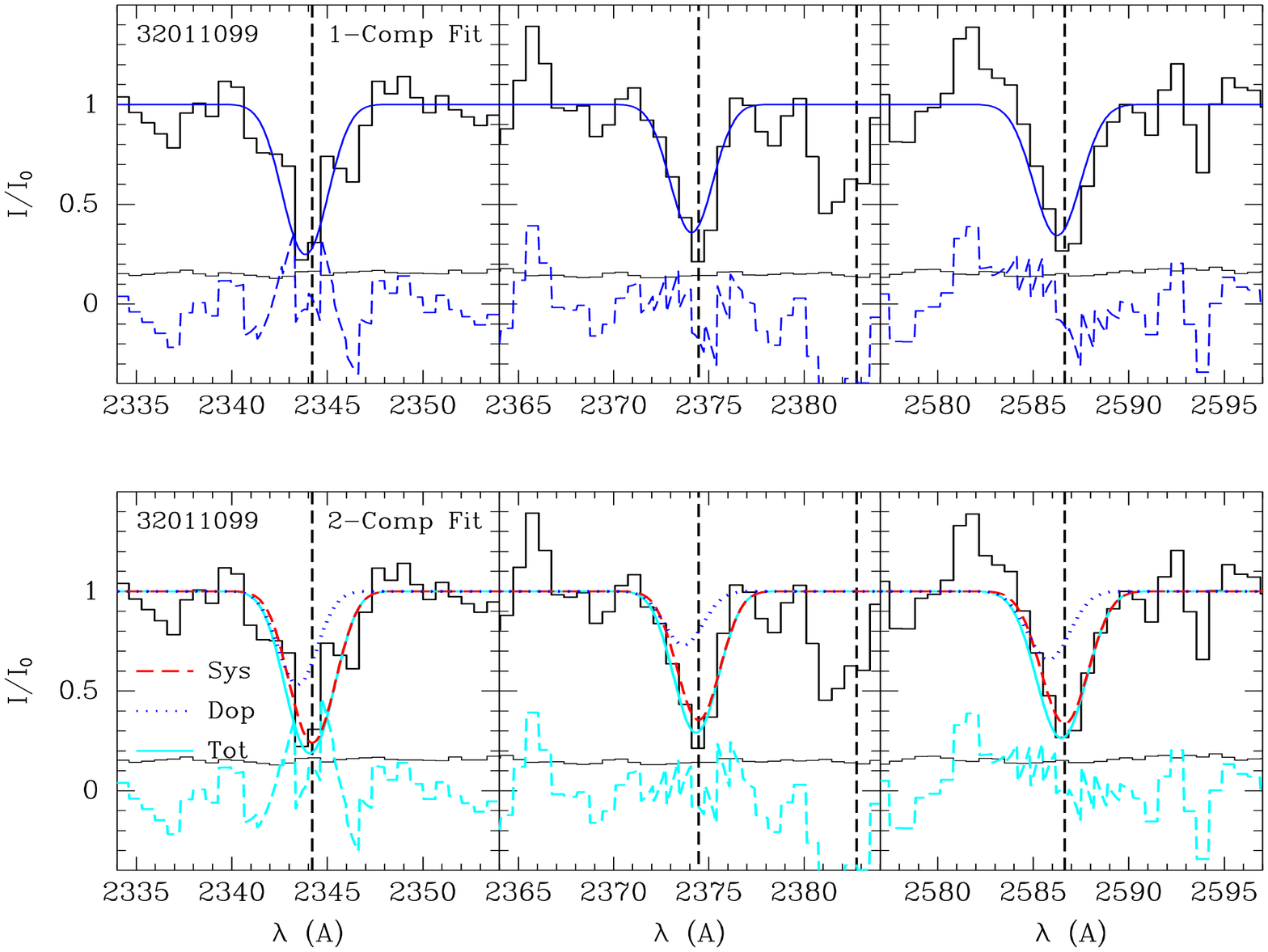}
               \includegraphics[height=9cm,angle=-90,clip=true]{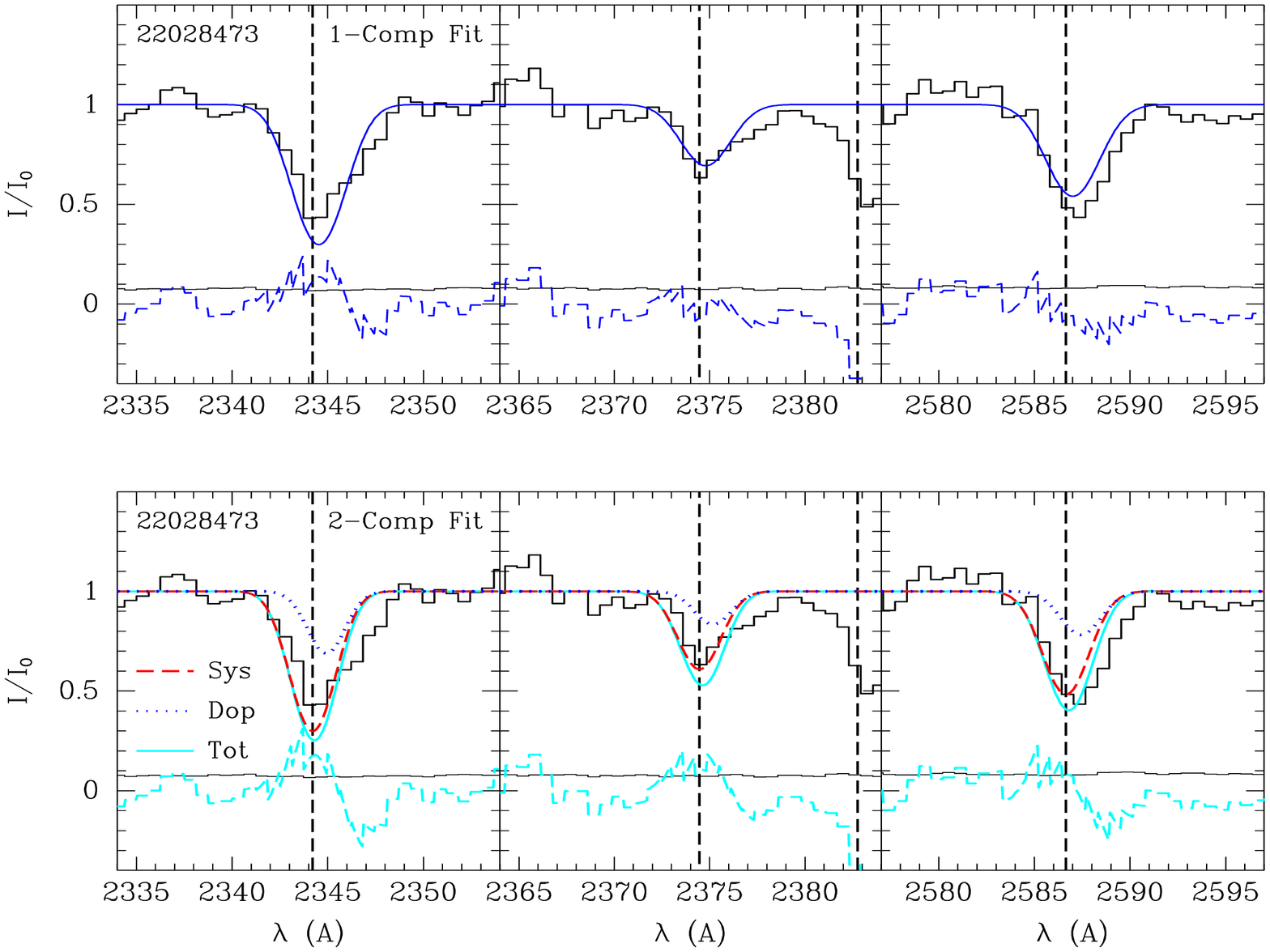}
                \hfill}
 \hbox{\hfill \includegraphics[height=9cm,angle=-90,clip=true]{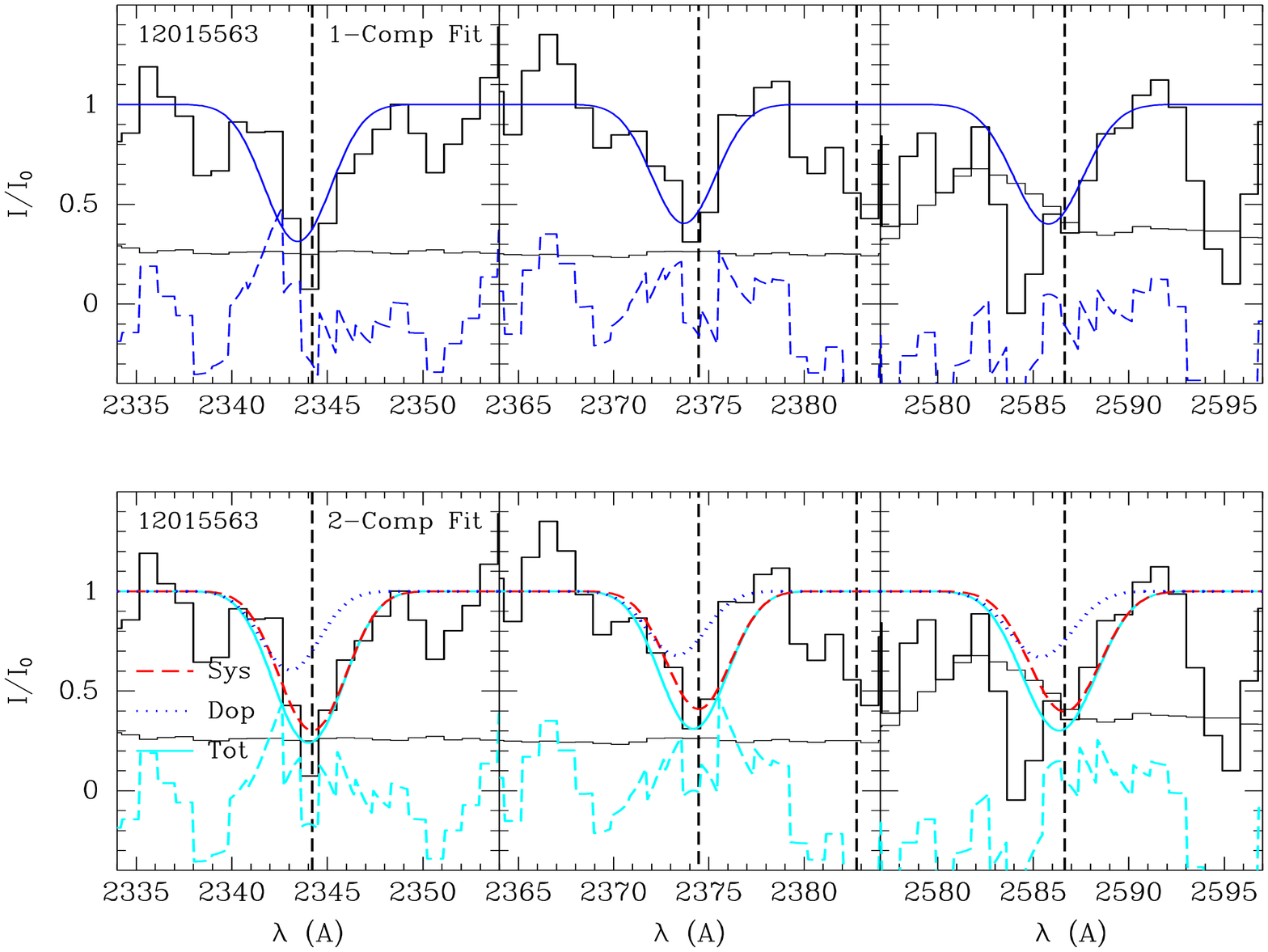}
               \includegraphics[height=9cm,angle=-90,clip=true]{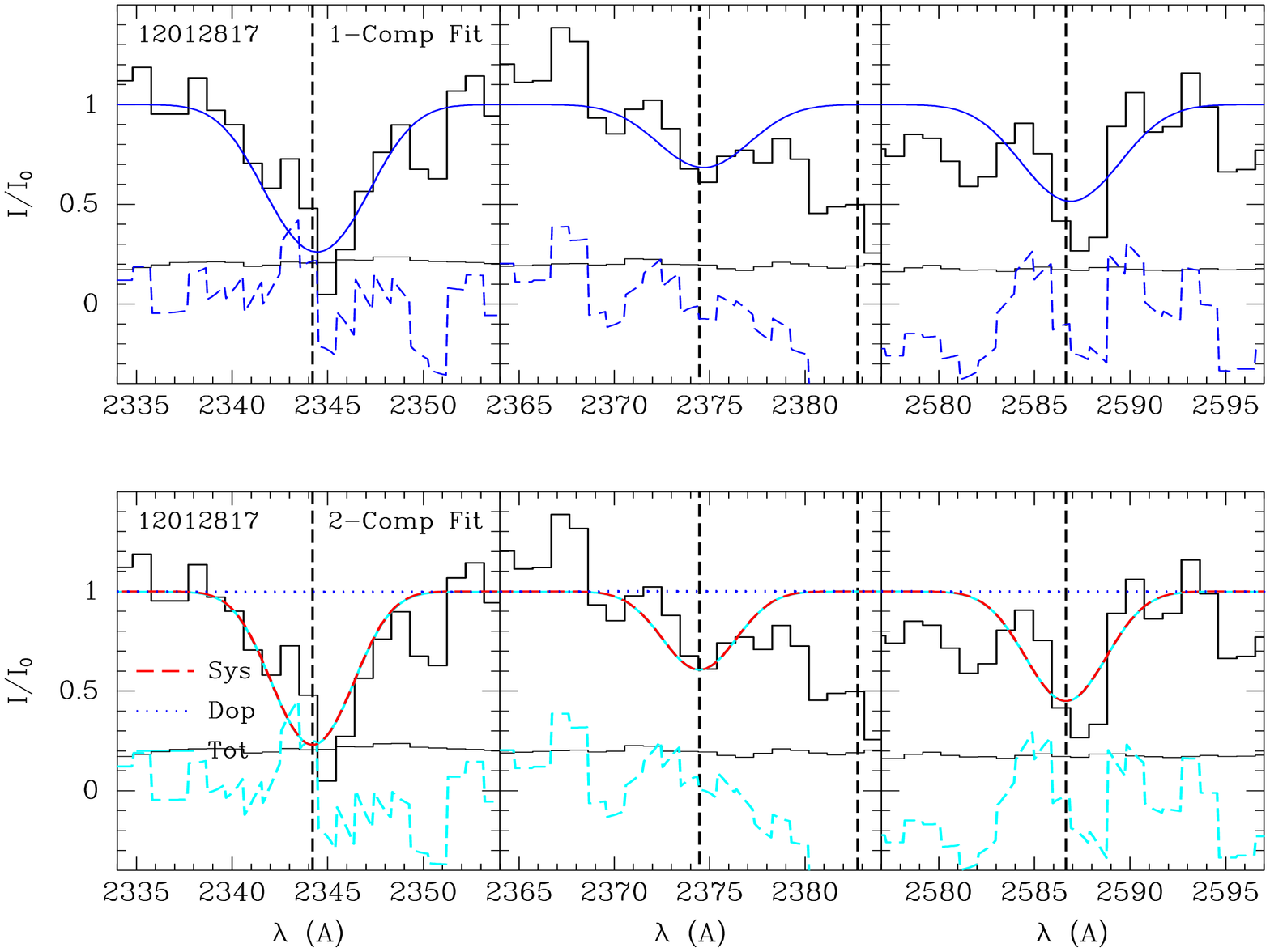}
                \hfill}
          \caption{\footnotesize
Four comparisons of single-component (top panels) and two-component (bottom panels) fits
to the \feII\ series with residuals shown at the bottom. 
{\it Top Left Panels:}
The single-component fit to the \dhigh\ spectrum of 32011099 yields an outflow 
velocity of $-46\pm14$\kms.
In a two-component fit, a systemic component at zero velocity describes much 
of the total absorption equivalent width; but the fit also requires a blueshifted 
Doppler component ($V_{Dop} = -101\pm25$\kms, $W_{Dop} = 1.16$\AA) to model
the net blueshift of the absorption trough. 
{\it Top Right Panels:}
The single-component fit to the \dhigh\ spectrum of 22028473 yields an inflow 
velocity of $42\pm10$\kms.
In a two-component fit, a systemic component at zero velocity describes much 
of the total absorption equivalent width; but the fit also requires a redshifted 
Doppler component ($V_{Dop} = 133\pm47$\kms for the inflow and $W_{Dop} = 0.46$\AA) 
to model the net redshift of the absorption trough. 
{\it Bottom Left Panels:}
The single-component fit to the \dlow\ spectrum of 12015563 yields
an outflow velocity of $-98\pm42$\kms.
In a two-component fit, a systemic component at zero velocity describes much 
of the total absorption equivalent width; but the fit also requires a blueshifted 
Doppler component ($V_{Dop} = -154\pm84$\kms, $W_{Dop} = 1.55$\AA) to model
the net blueshift of the absorption trough. 
{\it Bottom Right Panels:}
The single-component fit to the \dlow\ spectrum of 12012817 yields
 $V_1 = 29\pm44$\kms\ and does not require a net flow.
In a two-component fit, a systemic component at zero velocity describes much 
of the total absorption equivalent width;  we classify the Doppler component as
insignificant because the equivalent width is extremely low and the change in
the fit statistic is tiny.
In general, the two component fits increase the estimated velocity of the flow, 
relative to the single component fit.
}
 \label{fig:vc_2comp} \end{figure*}

In Figure~\ref{fig:vc_2comp}, we show examples of the \feII\ absorption-line
profiles from our LRIS data. The Doppler shift was jointly fit to five \feII\ 
transitions:  \feII\   $\lambda$ 2250, 2261, 2344, 2374, and 2587. 
The \feII\ transitions  at $\lambda2383$ and $\lambda 2600 $\AA\ are excluded from the fit
because they show the strongest emission filling. Even though they significantly constrain 
the column density, the weak \feII\ 2250 and 2261 transitions provide little constraint 
on the Doppler shift.  The fitted model simply
describes the net Doppler shift of the intrinsic \feII\ absorption.

The shapes of the absorption troughs are strongly influenced by the instrumental
response and are therefore often well-described by a Gaussian line profile. We
adopted a more physical description of the line profile in order to properly 
model the relative line strengths in the \feII\ series. For each transition, $j$,
our fitting function takes the form
\begin{eqnarray}
I_j(\lambda) = I_{c,j}(\lambda) e^{-\tau_j(\lambda)}. 
\label{eqn:function} \end{eqnarray}
The optical depth distribution,
\begin{eqnarray}
\tau_j(\lambda) = \tau_{0,j} e^{-(\lambda - \lambda_{0,j})^2 c^2 / (b^2 \lambda_{0,j}^2)},
\label{eqn:tau} \end{eqnarray}
introduces the Doppler parameter $b$ in order to describe the location of the absorbing clouds 
in velocity space relative to line center. The central wavelengths of the five \feII\
transitions, $\lambda_{0,j}$, are tied together to define a single Doppler shift, $V_1$.
The optical depths of the different transitions are tied together by
the ratio of their oscillator strengths (and the rest wavelengths of the transitions), so
one parameter $\tau_0(\lambda 2344)$ is sufficient to define the absorption optical depth at
line center for each transition. Equivalently, the product of the column density and the
covering factor, $N(Fe^+) C_f^{-1}$, determines the optical depth in each transition.
When line profiles from multiple transitions overlap at a given wavelength, e.g. redshifted
\feII\ $\lambda 2374$ and blueshifted \feII\ $\lambda 2383$, we add their 
respective contributions to the total optical depth at that wavelength, $\tau(\lambda)$.
Provided the form of the optical depth distribution
is symmetric about the centroid of the line profile, $\lambda_0$, the shape of this
distribution does not affect the fitted central wavelength $\lambda_0$, which
defines the Doppler shift $V_1$.\footnote{
  In our schematic picture, the angular extent of the stellar population
  producing the continuum emission is much larger than that of a gas cloud. This
  partial covering of the continuum source has often been described by a covering
  factor, $C_f$, which represents the fractional area of the continuum source covered 
  by gas clouds. Since a fraction $(1-C_f)$ of the galaxy continuum is observed 
  directly without any absorption, a widely used, alternative parameterization of the 
  line profile is   $I(\lambda) / I_c(\lambda) = 1 - C_f + C_f e^{-\tau(\lambda)}$ 
  (e.g., Rupke \et 2005).  This parameterization of the absorption trough, however, does 
  not capture the strong velocity dependence of the covering factor found by Martin \& 
  Bouch\'{e} (2009). For this reason, and because velocity-dependent covering factors 
  (Arav \et 2005) introduce structure in the line profile that cannot be observed
  at the resolution of our spectra, we simply fitted the 3-parameter model of
  Eqn.~\ref{eqn:function}. For any symmetric $C_f(v)$ distribution, partial covering 
  will clearly not impact the fitted Doppler shift.
  In our simple parameterization of the line profile, Eqn.~\ref{eqn:function} where $C_f \equiv 1$,
  the width of the absorption trough depends on both $b$ and $\tau_0$.
  When we attempted to included a covering factor in the description of the line profile, we
  found $b$, $C_f$, and $\tau_0$ were all strongly covariant due to the low resolution
  of our data.}

The free parameters $V_1$, $b$, and $\tau_0(\lambda 2344)$ were fitted as follows using
custom software.  Each galaxy spectrum was de-redshifted to the rest frame
and normalized by the continuum. A model
with line profiles centered at zero velocity was created assuming $b = 200$\kms and 
$\tau_0 = 1$. The model was convolved with a Gaussian profile of width
$\sigma = 120~ (185)$\kms, which describes the instrumental response for the 
\dhigh (\dlow) spectra. The residuals between the model 
and the data were minimized iteratively using the Levenberg-Marquardt algorithm 
(Press \et 1992) to make successive parameter estimations; the median reduced
chi-squared value was close to unity.  The resulting
covariance matrix provides a good estimate of the uncertainty in a fitted
parameter when the errors are normally distributed. We tested this assumption by
bootstrap resampling typical spectra. The resulting distributions
of Doppler shift measurements were found  to be normally distributed, so we use the
corresponding element of the covariance matrix to estimate the 68\% confidence interval 
on $V_1$, $\delta V_1 \equiv \sqrt {\sigma_{V1,V1}^2}$.
 The Doppler shifts from the joint fit, $V_{1}$, were also found to be consistent with
the average of the velocities measured in the individual fits to \feII\ 
$\lambda2344$, $\lambda 2374$, {\rm ~and} $\lambda 2587$. 
The error distributions for $\tau_0$ and  $b$, in contrast, are often not normally 
distributed, so we do not use these fitted parameters in our analysis; we choose instead
to discuss measurements of equivalent width and (in a later section) the bluest absorption 
velocity, $V_{max}$.

These joint fits to five \feII\ lines describe the Doppler shift of the entire absorption
trough with one velocity, so we will refer to them as the {\it single-component} fits.
 They avoid transitions with strong emission filling and beat down statistical
error by combining information from multiple transitions.  
Within the LRIS sample of 208 spectra, 36 had no good \feII\ lines to fit
due  sky line residuals (5 cases) or low equivalent widths (31 spectra); and 
the fit failed to converge for an additional 7 spectra. Since the
43 spectra without a fit have low continuum S/N ratio, the absence of
significant absorption does not provide much information; and we drop
these from the analysis of $V_1$ Doppler shifts.
Among 165 single-component fits, we find significant blueshifts, $V_1 / \delta(V_1) \le\ -3 $, 
for 35 spectra and significant redshifts, $V_1 / \delta(V_1) \ge\ 3 $, for 11 spectra.\footnote{As 
      discussed further in Section~\ref{sec:inflow}, two of
      the inflows are not significant due to systematic errors.}
When the significance of the \feII\ Doppler shift is greater than  $3\sigma$, we list
the properties of the galaxy in Table~\ref{tab:properties}.
Table~\ref{tab:outflow} and Table~\ref{tab:inflow}, respectively, list the fitted 
Doppler shifts (column 5) for these outflows and inflows as well as the
\feII\ $\lambda 2374$ equivalent widths (column 2) and upper limits on
the \feII\ $\lambda 2261$ equivalent widths (column 3)

\subsubsection{Two-Component Fits Illustrating ISM Absorption at the Systemic Velocity} \label{sec:measure_2comp}

Having identified bulk outflows and inflows with the single-component fitting, 
we next model the contribution of the interstellar medium (at the systemic velocity)
to obtain an estimate
of the velocity that characterizes the absorption equivalent width produced
by the bulk flow. In spectra of nearby star-forming galaxies, interstellar gas 
is generally the primary source of absorption in resonance lines near
the systemic velocity (Heckman \et 2001; Martin 2005; Martin 2006; 
Rupke \et 2005; Soto \et 2009). Stronger interstellar absorption at the systemic
velocity is observed 
towards low redshift galaxies with higher inclinations (Chen \et 2010). In
these resonance lines, the equivalent width of the interstellar absorption 
only becomes weaker than the stellar photospheric absorption in post-starburst galaxies,
which can be recognized by the presence of absorption lines from excited 
states that are populated in photospheres but not in diffuse gas. For our
purposes here, we will use the term {\it maximum interstellar component} 
in the broadest sense to represent all absorption at the systemic velocity, 
possibly including some contribution from stellar photospheres.

We modeled the \feII\ series, again excluding the $\lambda2383$ and $\lambda 2600 $ lines due to
emission filling, using two velocity components -- a {\it systemic component} with no
Doppler shift and a {\it Doppler component} with a fitted velocity, $V_{Dop}$. Since 
the absorption troughs are not well resolved, we found that five parameter fits 
allow many degenerate solutions.  To produce meaningful Doppler velocities,
we fixed the width of the systemic component. Because the systemic component 
represents interstellar gas, we set its Doppler parameter to be consistent 
with the measured velocity dispersion of nebular gas previously fitted to
the [\ion{O}{2}] doublet.
          These photoionized, emission regions are believed to reside in a gaseous
          disk; and their velocity disperions have been directly measured from the
          [\ion{O}{2}] $\lambda \lambda 3726,29$ doublet. The doublet is resolved in the 831~mm$^{-1}$
          (red channel) spectra, and
          the fitted Gaussian $\sigma$ has been corrected for instrumental broadening.
          We choose $b_{sys} = \sqrt{2} \sigma([OII]) $.
We then fitted the maximum interstellar absorption to the \feII\ series by
varying the optical depth at line center, obtaining the deepest absorption trough 
at $v = 0$ consistent with the data.  With the interstellar component now fully
specified, we next fitted the Doppler shift, Doppler parameter, and central optical 
depth of the Doppler component.

This two-component fitting approach has two critical advantages over simply 
reflecting the redshifted portion of the fitted absorption trough to create a 
model of the systemic absorption, as implemented for example by Weiner \et (2009).
First, it does not require fully resolved troughs because the model is convolved 
with the instrumental response. Second, it can be used to estimate the effect
of the ISM on the inferred velocity of redshifted absorption (gas inflows) as
well as blueshifted absorption (gas outflows). We only fit two components
when the single-component fit indicated a significant outflow or inflow. The
low resolution of our spectra does not constrain the two-component fits well
enough to use them to determine whether the spectrum shows a net flow in the first place.

Figure~\ref{fig:vc_2comp} shows examples of the two-component fits relative
to the single-component fits. The $V_{Dop}$ velocity from the two-component
fit is bluer than the single-component Doppler shift $V_1$, but the difference 
is typically less than the error bar $\delta V_1$. The maximum single-component 
blueshift is $-330$~km~s$^{-1}$; but the blueshift of the Doppler component reaches
$-550$\kms\ in the two-component fits.
This trend is consistent with that shown in Coil \et (2011),
where the impact of the systemic component on the outflow velocity is quantified.
While the one- and two-component fits yield consistent
outflow velocities for many spectra, there is a subset for which accounting for a
symmetric absorption component at the systemic velocity significantly increases
the estimated outflow velocity. It is therefore of interest to understand which
galaxies have this stronger interstellar absorption component.

The equivalent width of the systemic component 
rises strongly with increasing stellar mass. Hence modeling the systemic component
results in fitted outflow velocities about 200 to 300\kms bluer regardless of
whether the systemic absorption is strong or weak.  Only the most massive
galaxies in our sample, however, exhibit exceptionally strong interstellar
absorption with $W_{sys} \sgreat\ 2.3$\AA. 
In this paper, we use the $V_1$  values to determine whether a net flow is
present; and, when it is, we use the values of $V_{Dop}$ and $W_{sys}$ to
illustrate how the interstellar absorption at the systemic velocity may
influence the conclusions drawn from the single-component fits. Columns 6 and 7
of Tables~\ref{tab:outflow} and \ref{tab:inflow} compile the Doppler parameters 
derived for the systemic components and the fitted values of the Doppler shifts, 
$V_{Dop}$.

\subsubsection{Equivalent Widths and Ionic Column Density}  \label{sec:EW}

In general, the {\it intrinsic} equivalent width of each \feII\ line in a spectrum
is determined by a combination of the ionic column density, $N(Fe^+)$, and the
Doppler parameter, $b$, called the curve of growth (Spitzer 1978).
The  equivalent width of optically-thin transitions is sensitive to \feII\ column 
density, but the equivalent width of optically-thick transitions depends primarily  on
the velocity spread of the absorbing gas. 
The measured equivalent widths do not necessarily follow this theoretical curve of growth, 
however, because emission filling reduces the intrinsic area of each absorption trough in
a given spectrum by a different amount. Here we discuss the measured equivalent widths of a 
line not affected much by emission filling, \feII\ $\lambda 2374$, and the upper limit on
the equivalent width of a weak line, \feII\ $\lambda 2261$, detected in composite spectra.
Since the \feII\ 2261 transition has an oscillator strength 12.8 times lower than 
that of \feII\ 2374, the equivalent width measurement and upper limit together
constrain the ionic column denstiy to order-of-magnitude accuracy.

How accurately the ionic column density can be estimated from a single line depends 
on the optical depth of the transition. The column density at which the absorbing gas
becomes optically thick, i.e., $\tau(\lambda) \equiv 1$,  depends on the absorption 
cross section. At line center, the optical depth can be expressed as 
\begin{eqnarray}
\tau_0 = 1.498 \times 10^{-15} \lambda_0 f N C_f b^{-1},
\end{eqnarray}
where $f$ denotes the oscillator strength of the transition, the wavelength
of the transition is in \AA, the column density has units of ${\rm cm}^{-2}$,
and the Doppler parameter is in \kms. To illustrate how partial covering of the 
continuum source affects the inferred ionic column density, we include a covering
factor, $0 < C_f \le\ 1$, which describes the fraction of the continuum light
covering by the absorbing clouds.

The rough equality of the equivalent widths of \feII\ $\lambda 2344$, $\lambda 2374$, and
$\lambda 2587$ in many of the LRIS spectra indicates these transitions are typically 
optically thick. The intrinsic equivalent widths of these transitions therefore provide
lower bounds on the ionic column density. Our measurement of the equivalent width for
the transition with the lowest oscillator strength, \feII\ $\lambda 2374$, yields the
best lower limit on $N(Fe^+)$. In this optically-thin limit, the linear relationship between 
equivalent width and ionic column yields a lower limit 
\begin{eqnarray}
N(Fe^+) \ge\ 6.30 \times 10^{14} {\rm ~cm}^{-2} ~W_{2374}({\rm \AA}) C_{f,1}^{-1},
\end{eqnarray}
where $C_{f,1}$ is the covering factor in units of $C_f = 1$.
The summary of the outflow and inflow properties, Tables~\ref{tab:outflow} and \ref{tab:inflow} 
respectively, provides these lower limits on $N(Fe^+) C_{f,1}$ in column 4 for the total
absorption trough, in column 8 for the interstellar component at the systemic velocity, and
in column 9 for the Doppler component.

The \feII\ $\lambda 2261$ absorption trough is generally much weaker than the \feII\ $\lambda 
2374$ line, so we can be certain that the \feII\ $\lambda 2261$ transition falls on the 
linear part of the curve-of-growth. In individual spectra, we measure an upper limit
on the equivalent width of \feII\ $\lambda 2261$. 
We place an upper bound on the ionic column density, 
\begin{eqnarray}
N(Fe^+) \le\ 9.06 \times 10^{15} {\rm ~cm}^{-2} ~[3 \sigma (W_{2261})] C_{f,1}^{-1},
\end{eqnarray}
where the $3\sigma$ upper limit on the equivalent width has units of \AA.
The summary of the outflow and inflow properties, Tables~\ref{tab:outflow} and \ref{tab:inflow} 
respectively, provides these upper limits on $N(Fe^+) C_{f,1}$ in column 4 for the total
absorption trough, in column 8 for the interstellar component at the systemic velocity, and
in column 9 for the Doppler component.

\subsection{Highest Velocity \mgII\ Absorption} \label{sec:vmax}

Resonance emission is not expected to fill in the high velocity wings of the
the intrinsic \mgII\ absorption profile (e.g., Figure~5 of Prochaska \et 2011).
If the gas column is low at high velocities
we expect to detect \mgII\ absorption even though the \feII\ absorption is too 
weak to detect for several reasons. First, consider equal column densities of Fe$^+$ and Mg$^+$. In the 
optically thin limit, the \feII\  $\lambda 2374$ equivalent width would  be a factor of 20 lower 
than the intrinsic  \mgII\ $\lambda 2796$ equivalent width. Next, this equivalent width
ratio could be higher because the cosmic abundance ratio of Mg to Fe is a factor of 1.2, and 
Fe is more depleted onto grains than is Mg (Savage \& Sembach 1996). Finally,
since the first and second
ionization potentials of iron are similar to those of magnesium, we might expect 
\feII\ and \mgII\ ions to be present in the same phase of the wind.\footnote{
           Their first and second ionization potentials are are
           IP(\mgI) = 7.646 eV, IP(\mgII) = 15.035 eV,
           IP(\feI) = 7.870 eV, and IP(\feII) = 16.18 eV.}
The column density ratios calculated from grids of photoionization models show, however,
that the relative fraction of Fe and Mg in the singly-ionized state is sensitive
to the ionization parameter. Over the range of likely
ionization parameters near galaxies, the \feII\ column density can be 
similar to or significantly less than the \mgII\ column density (Churchill \et 
2003). Hence, the relative optical depths in \feII\ $\lambda 2383$ and
\mgII\ $\lambda 2803$ can be similar, but the \mgII\ optical depth can easily
be much larger for a range of physical reasons.

\subsubsection{Measurement Technique}

The LRIS spectra of 186 of the 208 galaxies cover the \mgII\ doublet.
In 57 spectra, no significant absorption trough was detected at a rest wavelength 
of 2796\AA; some of these spectra
are characterized by strong \mgII\ emission that fills the absorption 
trough. Among the most prominent \feII\ outflows, i.e., the 35 with $3\sigma$ outflows,
33 spectra provide \mgII\ coverage. We measured the properties of the \mgII\ absorption
for composite spectra and 104 individual spectra excluding spectra that had
weak $\lambda 2796$ absorption or sky-subtraction artifacts near the doublet.

Some of the  individual spectra show a \mgII\ profile with a blue wing. To characterize
the minimum (maximum) velocity of the \mgII\ absorption,
we measure the velocity,  $V_{max}$,  blueward (or redward) of the systemic velocity at 
which the spectral intensity is consistent with the continuum level (at the $1\sigma$ level);
specifically, we define $I(\lambda_{vmax}) = 1 - \delta I(\lambda_{vmax})$, where $\delta I(\lambda)$
is the error spectrum. To measure $V_{max}$,
we added a Gaussian random deviate to the original value of the intensity
(with standard deviation matched to the error spectrum at each wavelength), remeasured the 
maximum velocity of the absorption trough in this modified spectrum, and calculated 
the average $V_{max}$ and standard deviation after 1000 iterations. The statistical 
error was taken to be the standard deviation or half a pixel width, whichever was
larger. 

Clearly, this definition is only meaningful when the blue wing of the line profile is well 
resolved. Furthermore, comparisons of $V_{max}$ must be made at similar spectral S/N ratio.
The \mgII\ $\lambda 2796$ absorption trough provides our most sensitive measurement of
$V_{max}$ and is not blended with another transition.  When the \feII\ lines have a 
significant redshift, i.e., the 9 inflow galaxies discussed in Section~\ref{sec:inflow}, we 
measure the the maximum (inflow) velocity from the red wing of 
\mgII\ 2803; but this measurement can be affected by emission filling.

The absence of blue wings on line profiles in our LRIS spectra does not rule
out the presence of an outflow due to sensitivity issues caused by 
instrumental smoothing and the variation in S/N ratio among spectra. The 
$V_{max}$ measurements should {\it not} be compared among the entire sample
to determine which galaxies have outflows but do serve two useful purposes.
First, values of $V_{max}$ blueward
of approximately -435\kms\ in \dlow\ spectra identify a resolved line wing,
as does absorption blueward of roughly -282\kms\ in \dhigh\ spectra.
Second, we can compare the \mgII\ blue wings among galaxies with different properties 
by measuring the $V_{max}$ feature in composite spectra constructed 
to have the same S/N ratio. Previous studies have done this type of analysis at spectral 
resolution higher than our LRIS spectra (Weiner \et 2009; Rubin \et 2010b).

\subsubsection{Contamination from Stellar Absorption}

The analysis of both composite and individual presented in this paper
excludes galaxies with colors placing them in the green valley or red sequence
or K+A spectral classification. The LRIS spectra of many of these 22 galaxies
show significant stellar absorption. The stellar lines become stronger 
with the mass of the intermediate age stellar population, so we recognized
the contamination by comparing absorption features (from excited states rather
than resonance transitions) in spectra of post-starburst galaxies to our spectra.
The excited absorption features are absent or weak in our spectra of blue-cloud galaxies.

This rejection affects the measurement of the maximum absorption blueshift
because the stellar line profiles have broad wings. We measure high $V_{max}$ values for 
several of the 22 rejected galaxies. Because these galaxies are mostly in the high mass 
tertile of our sample, they generate a positive correlation between $V_{max}$ and
stellar mass when we include them. Without them, the maximum blueshift of the 
\mgII\ $\lambda \lambda 2796$, 2803 absorption in Figure~\ref{fig:mass_composite}
does not vary significantly with stellar mass, $B$-band luminosity, or color.

\begin{figure*}[t]
 \hbox{\hfill 
              \includegraphics[height=3.6cm,angle=-90,clip=true]{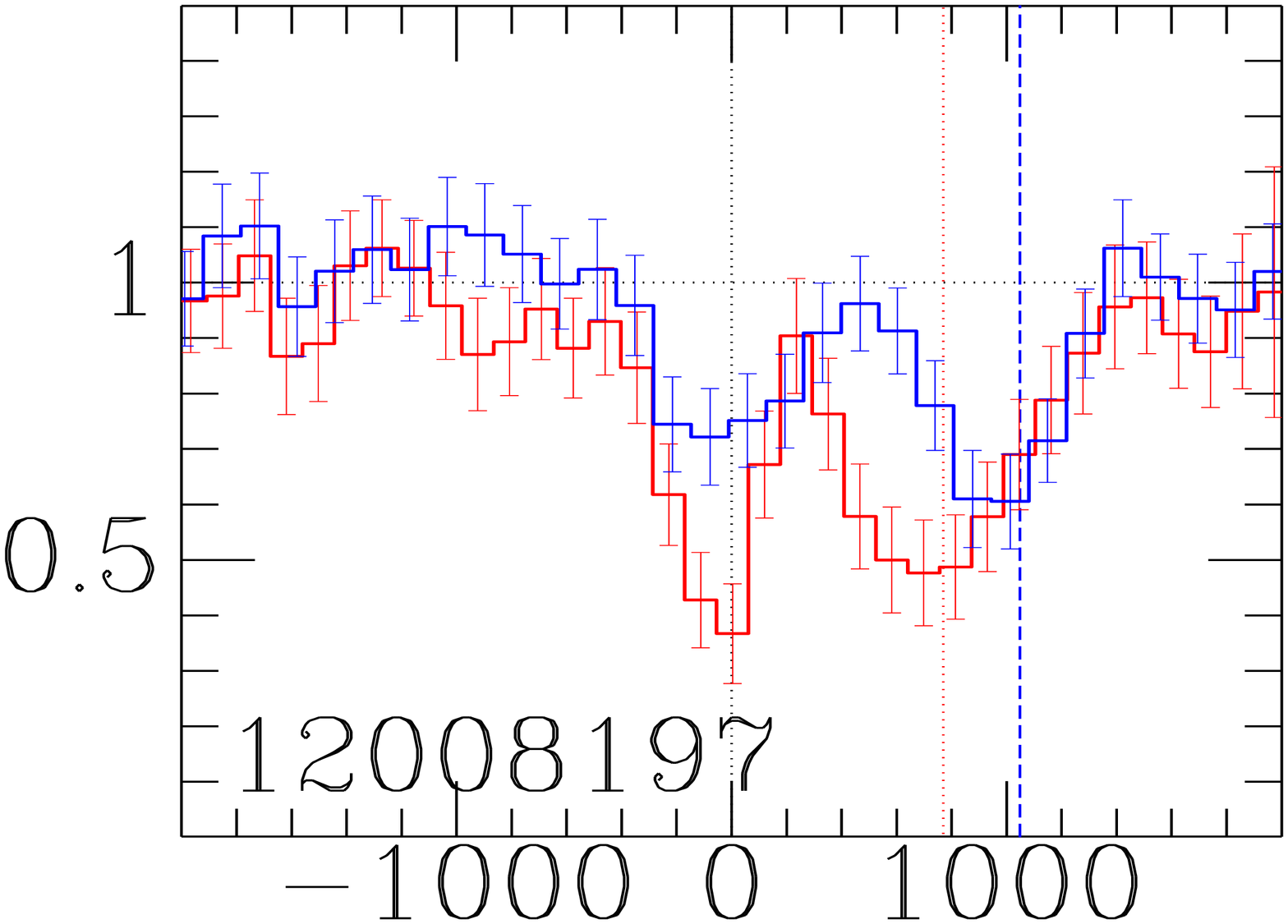}
              \includegraphics[height=3.6cm,angle=-90,clip=true]{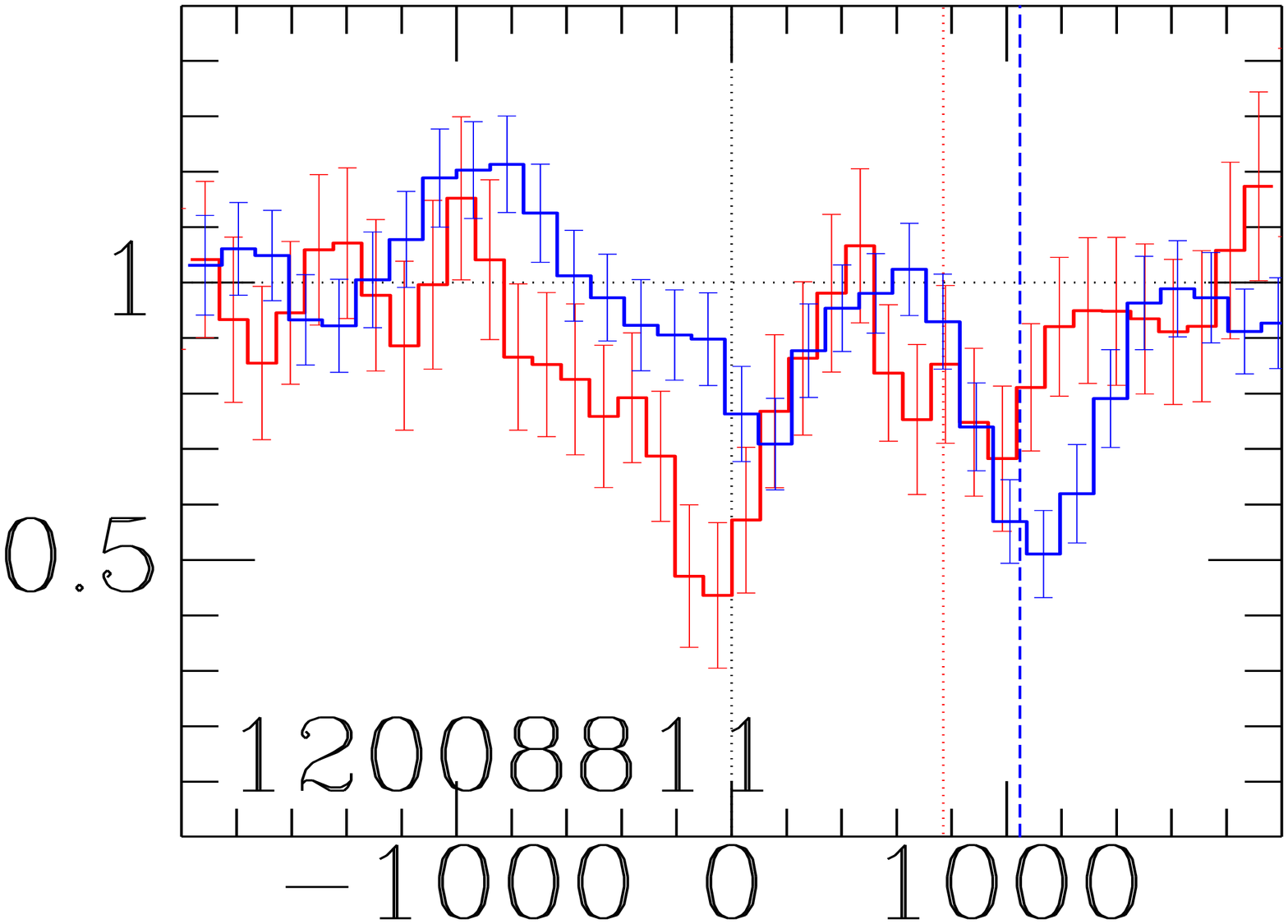}
              \includegraphics[height=3.6cm,angle=-90,clip=true]{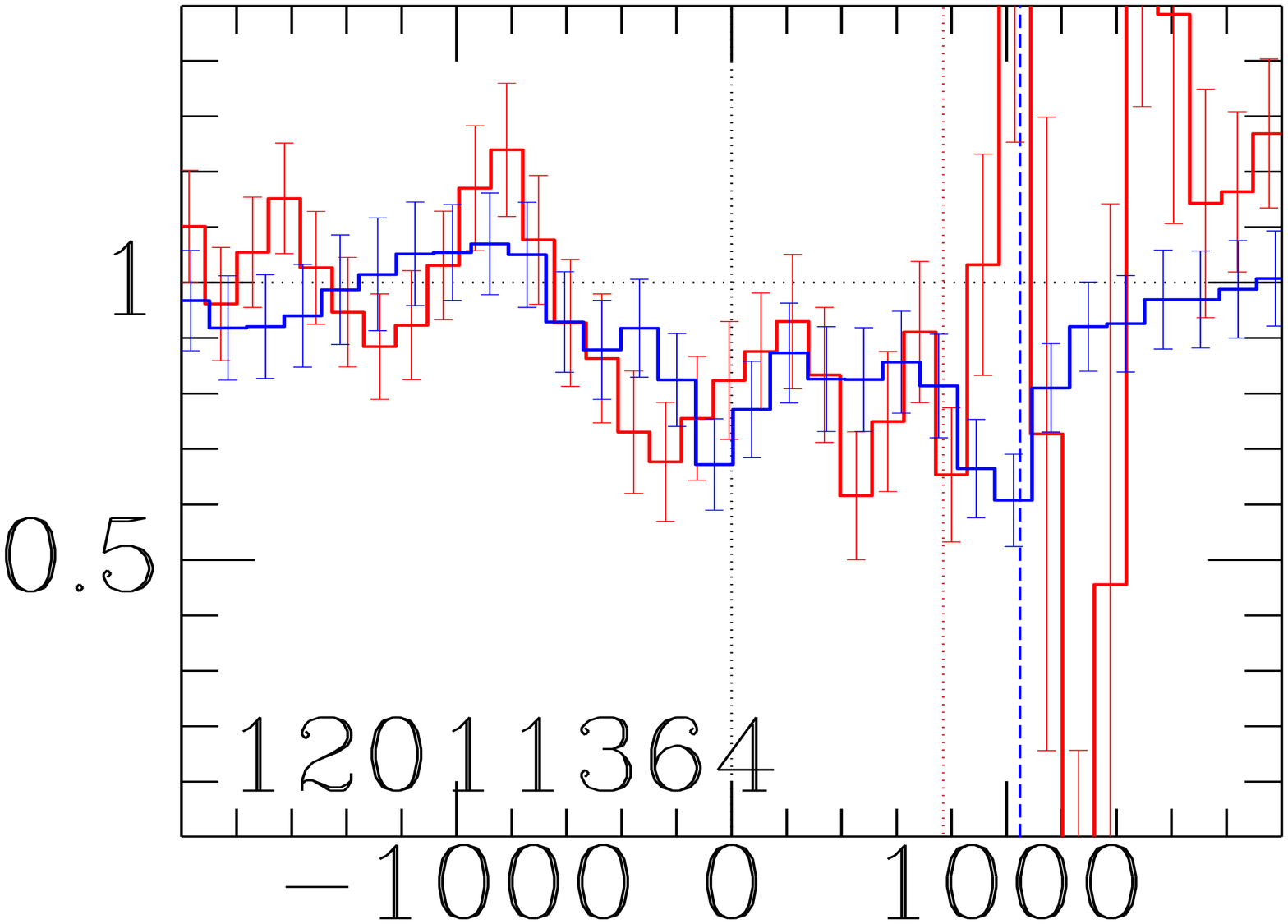}
              \includegraphics[height=3.6cm,angle=-90,clip=true]{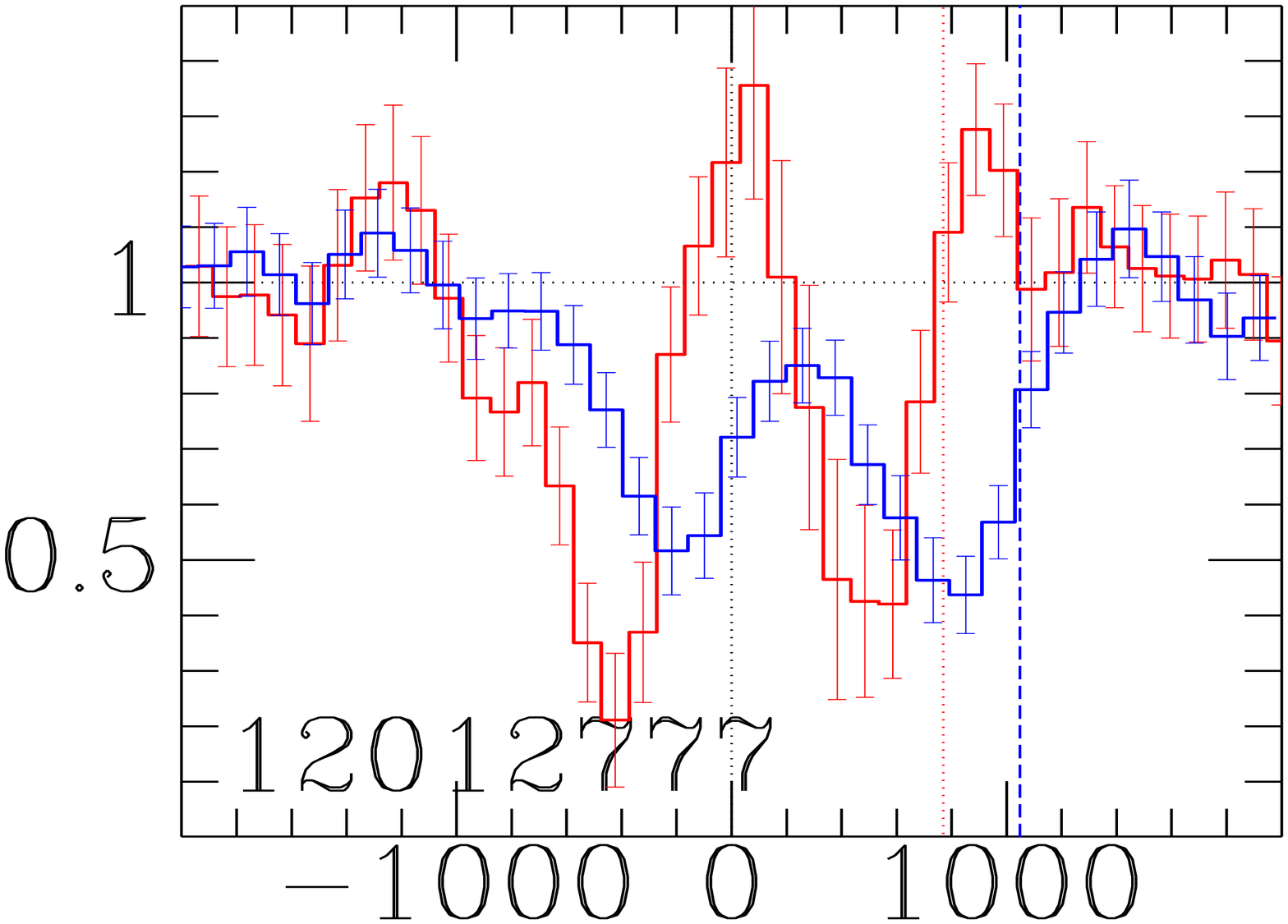}
              \includegraphics[height=3.6cm,angle=-90,clip=true]{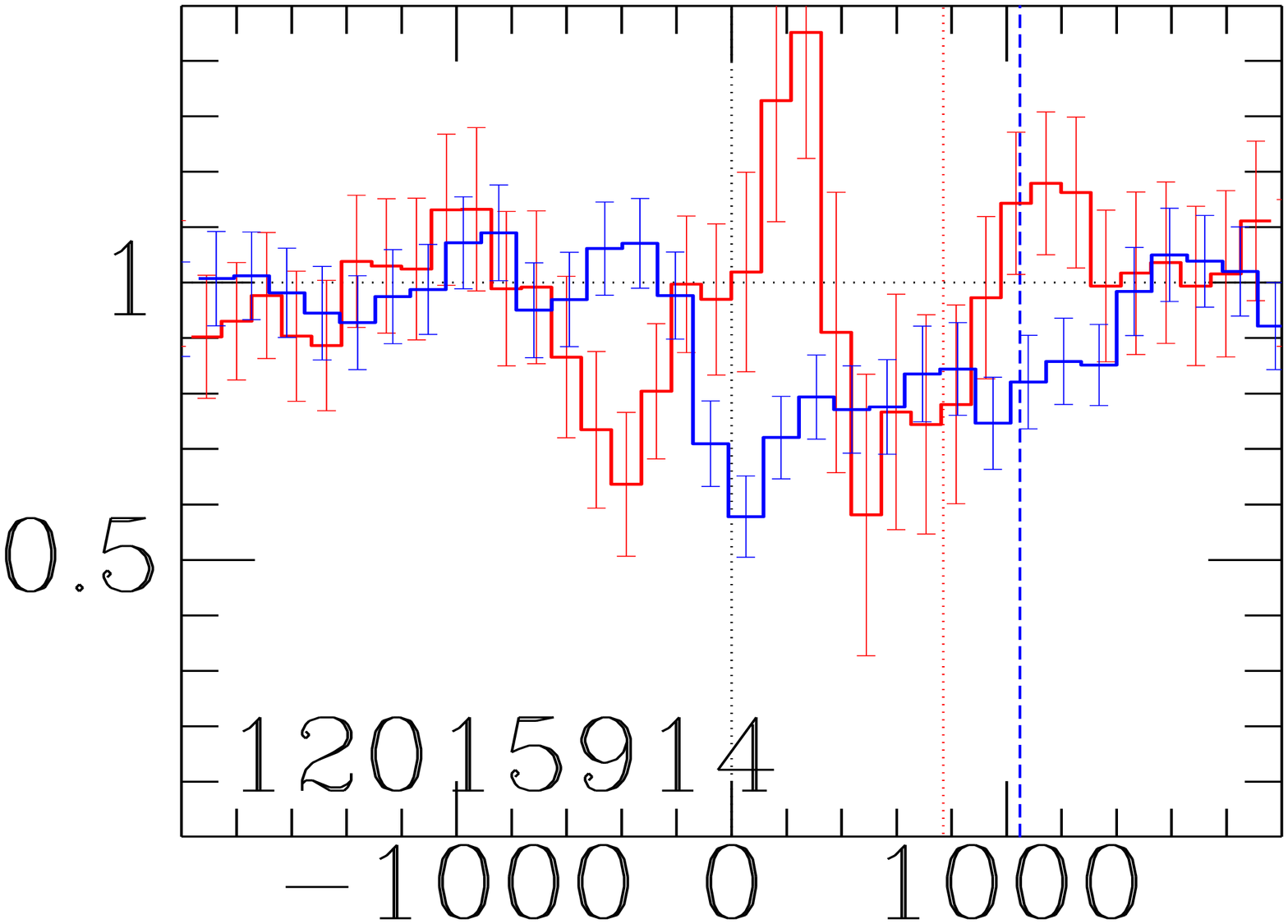}
             }
 \hbox{\hfill 
              \includegraphics[height=3.6cm,angle=-90,clip=true]{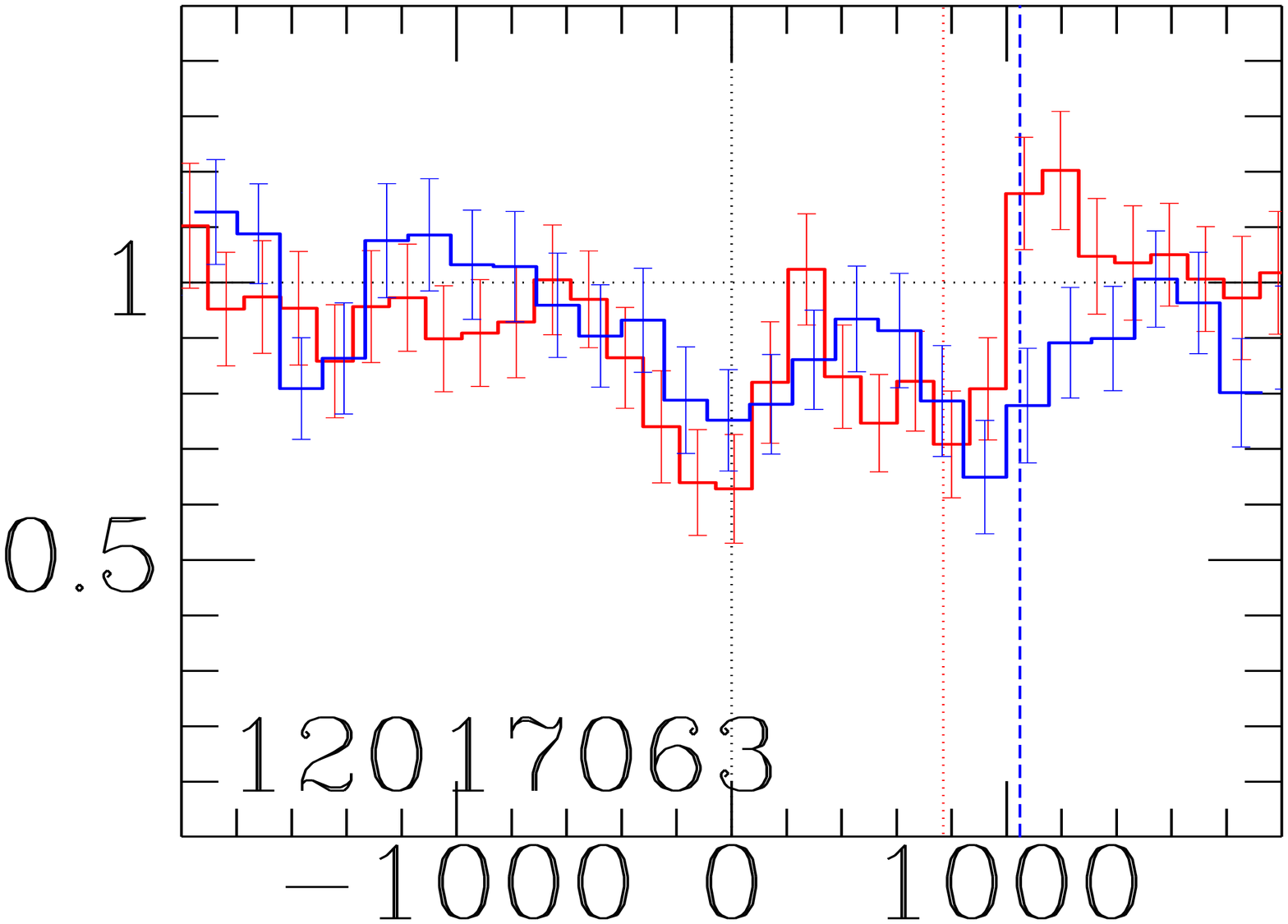}
              \includegraphics[height=3.6cm,angle=-90,clip=true]{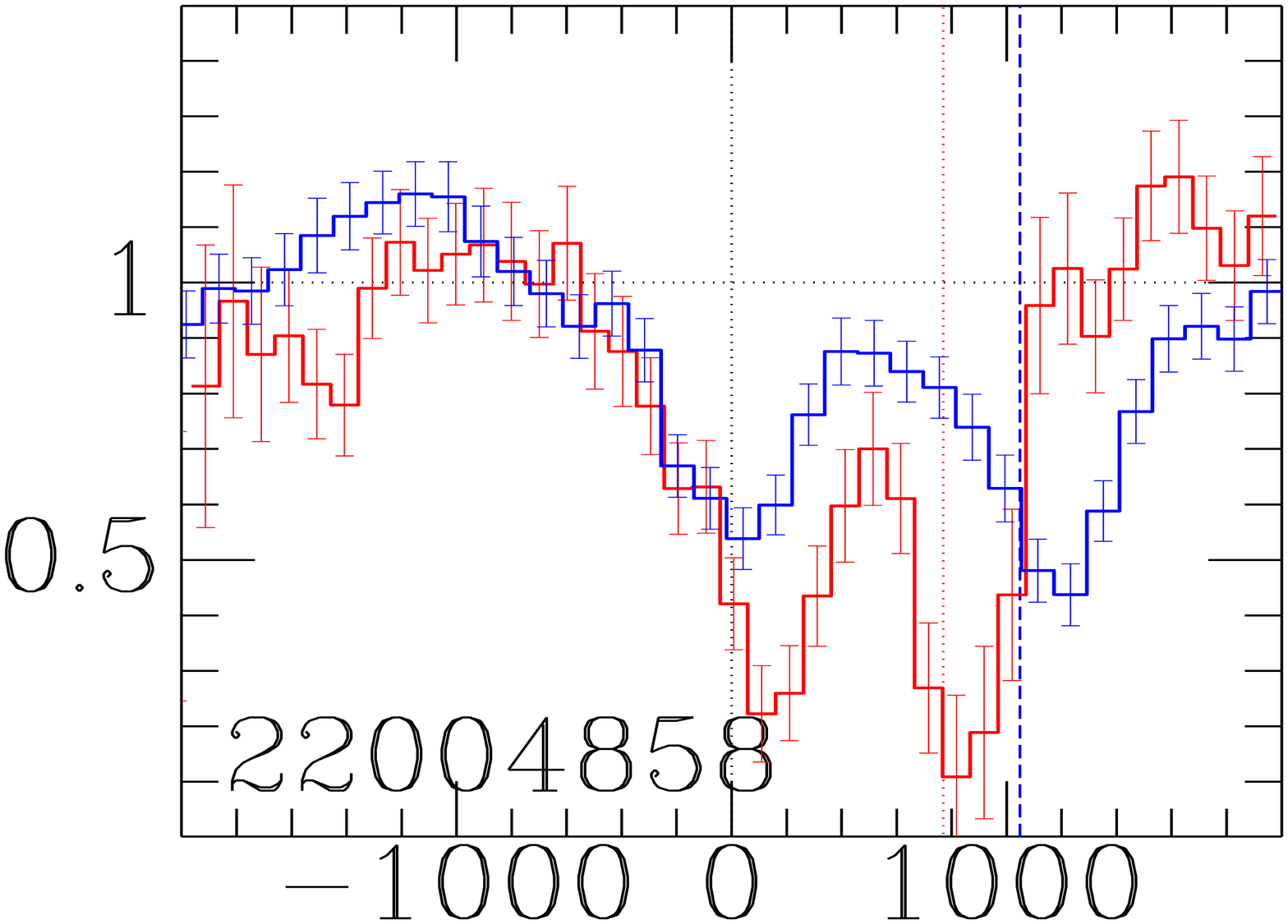}
              \includegraphics[height=3.6cm,angle=-90,clip=true]{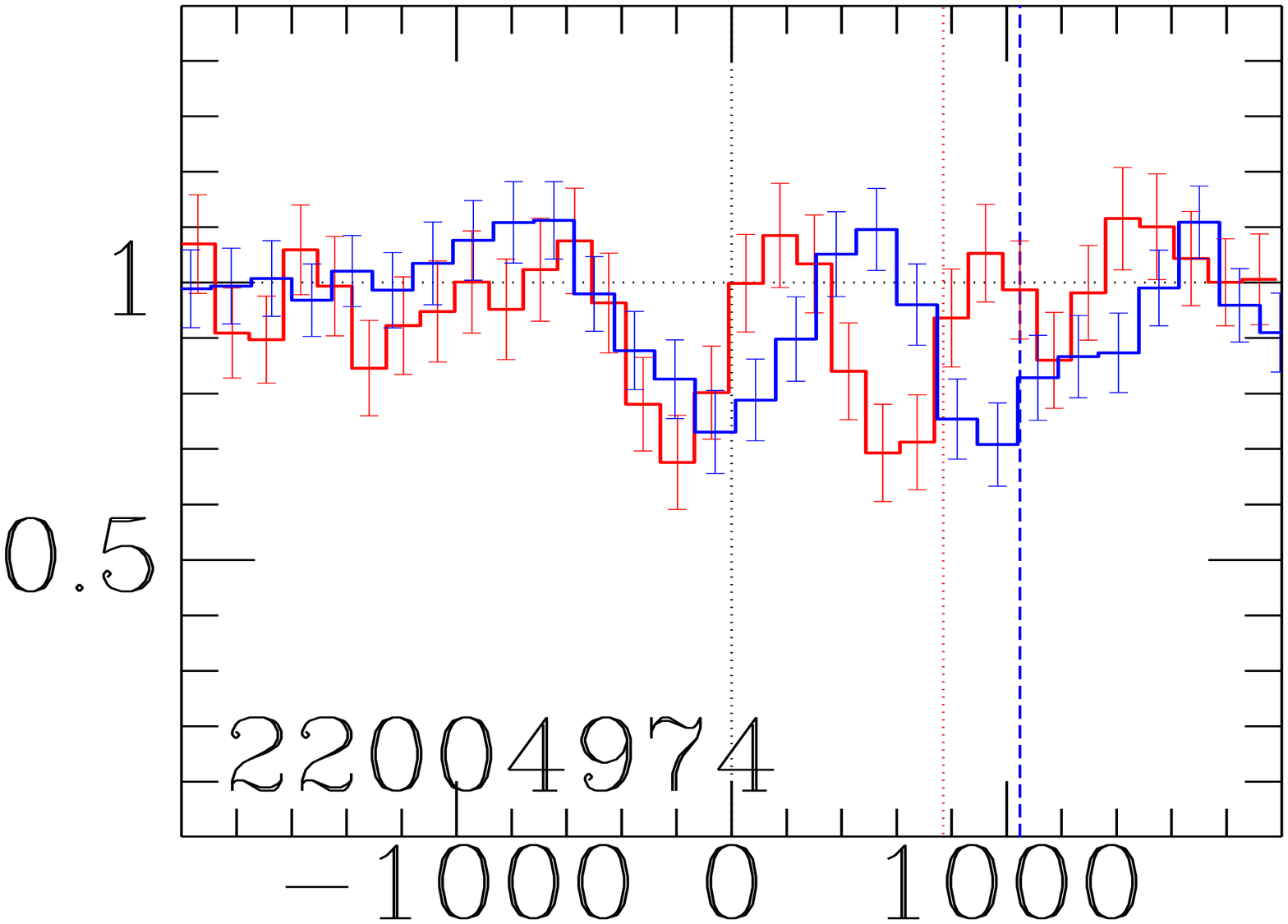}
              \includegraphics[height=3.6cm,angle=-90,clip=true]{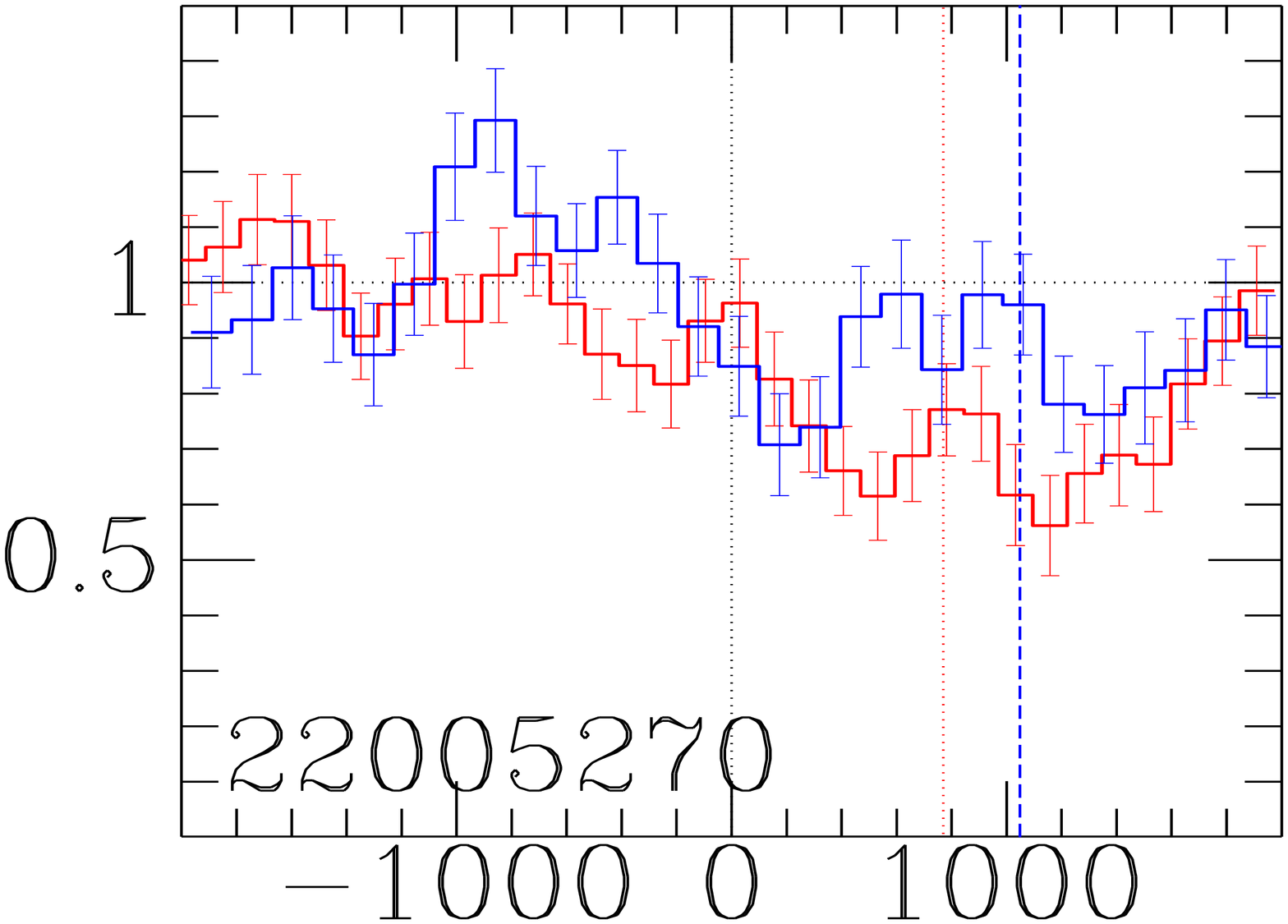}
              \includegraphics[height=3.6cm,angle=-90,clip=true]{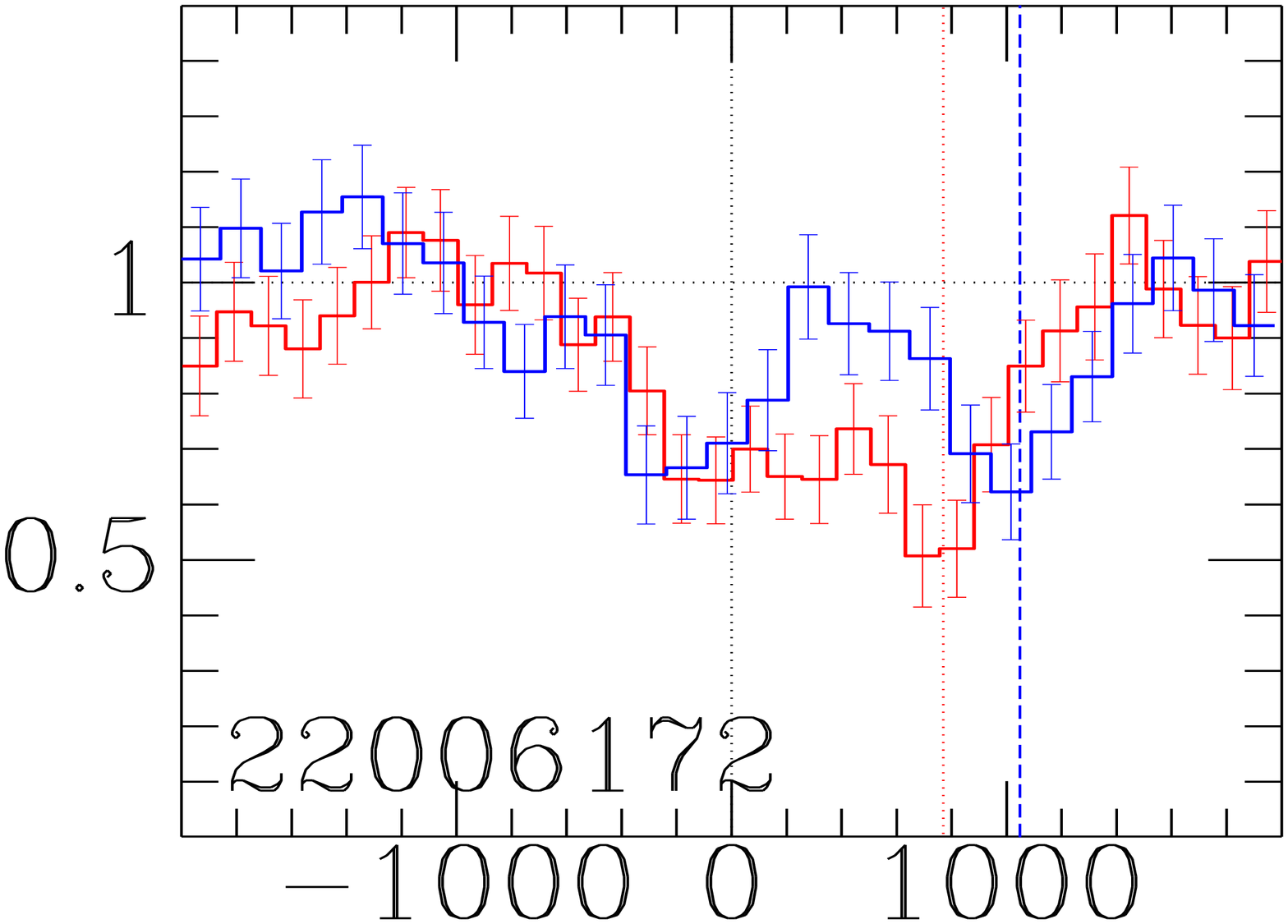}
             }
 \hbox{\hfill 
              \includegraphics[height=3.6cm,angle=-90,clip=true]{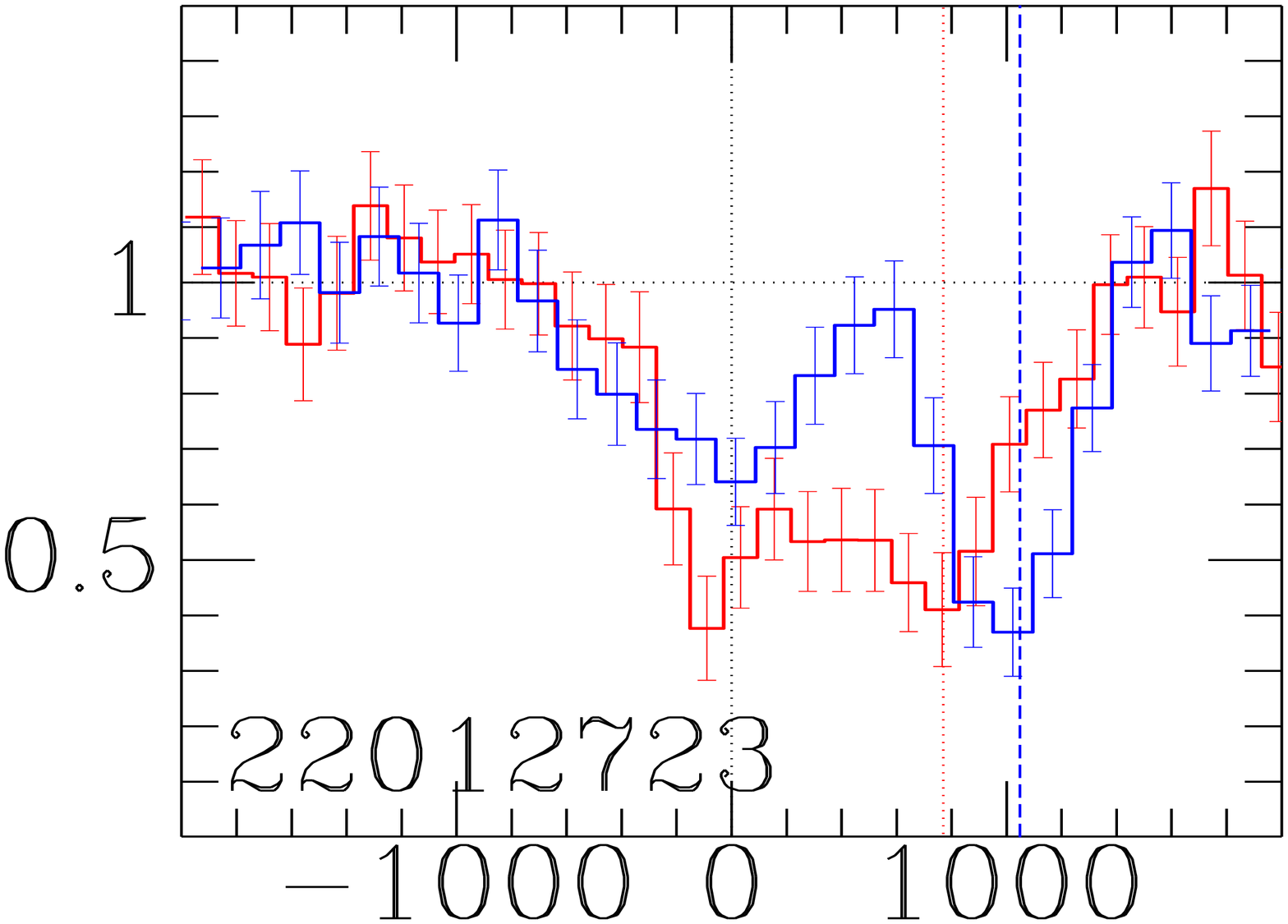}
              \includegraphics[height=3.6cm,angle=-90,clip=true]{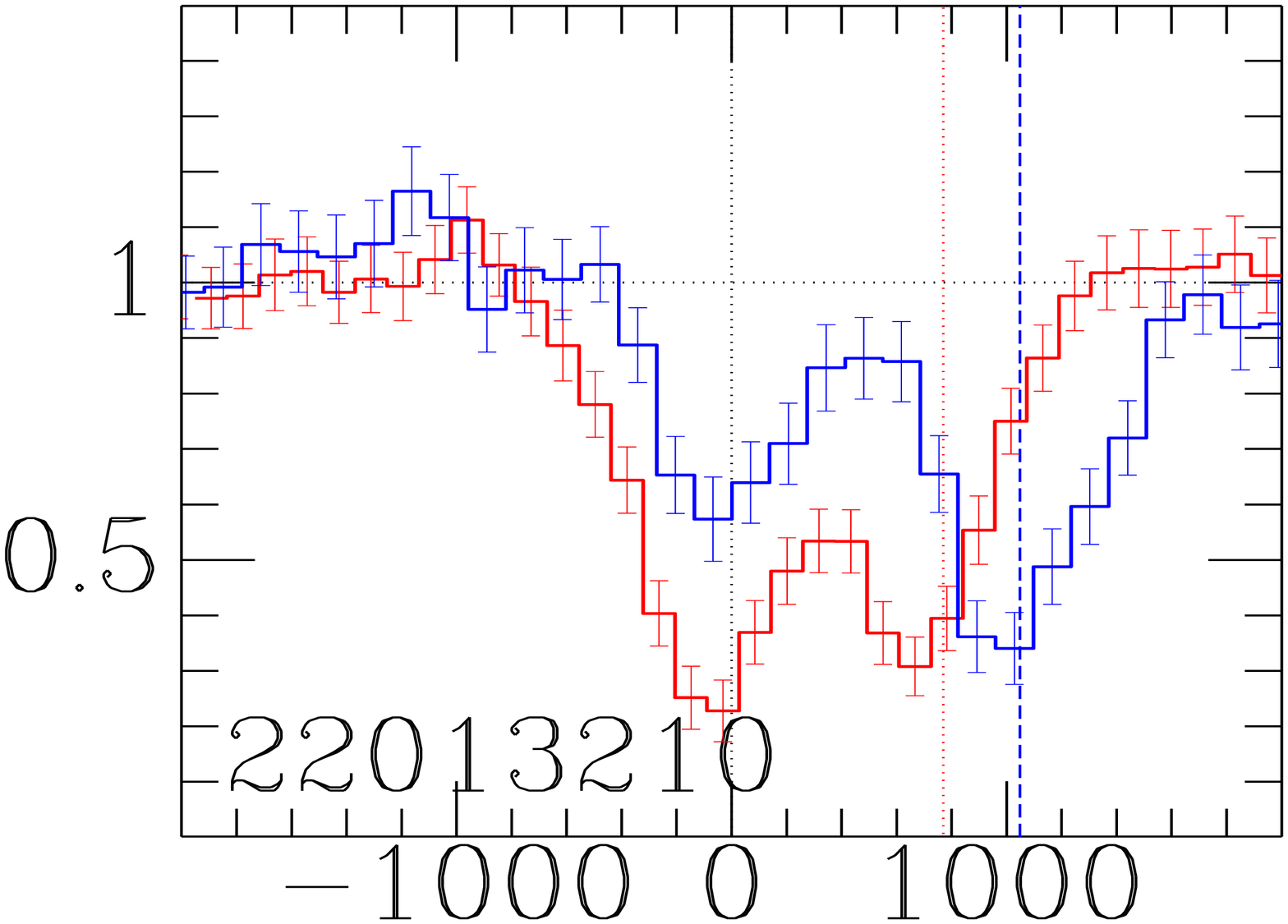}
              \includegraphics[height=3.6cm,angle=-90,clip=true]{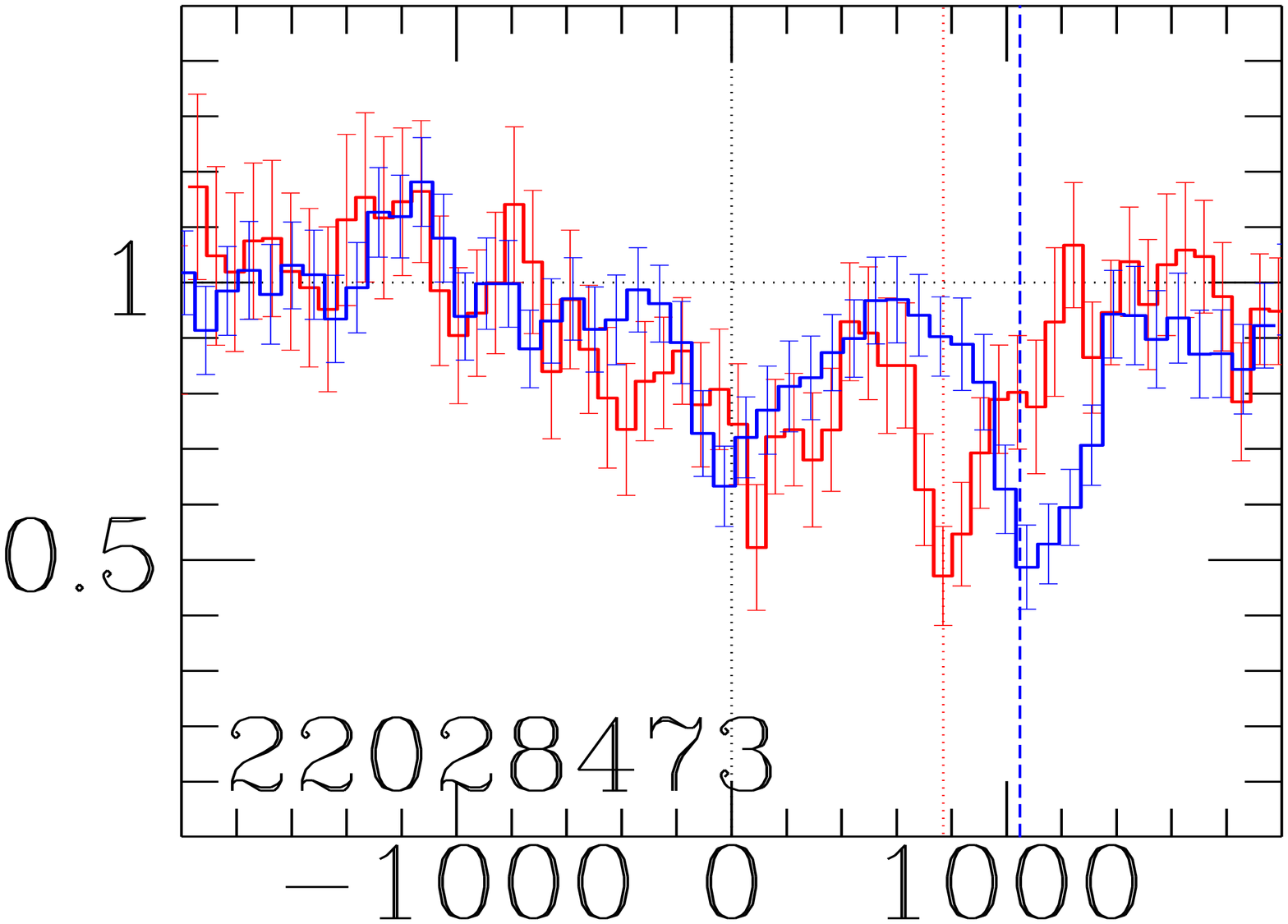}
              \includegraphics[height=3.6cm,angle=-90,clip=true]{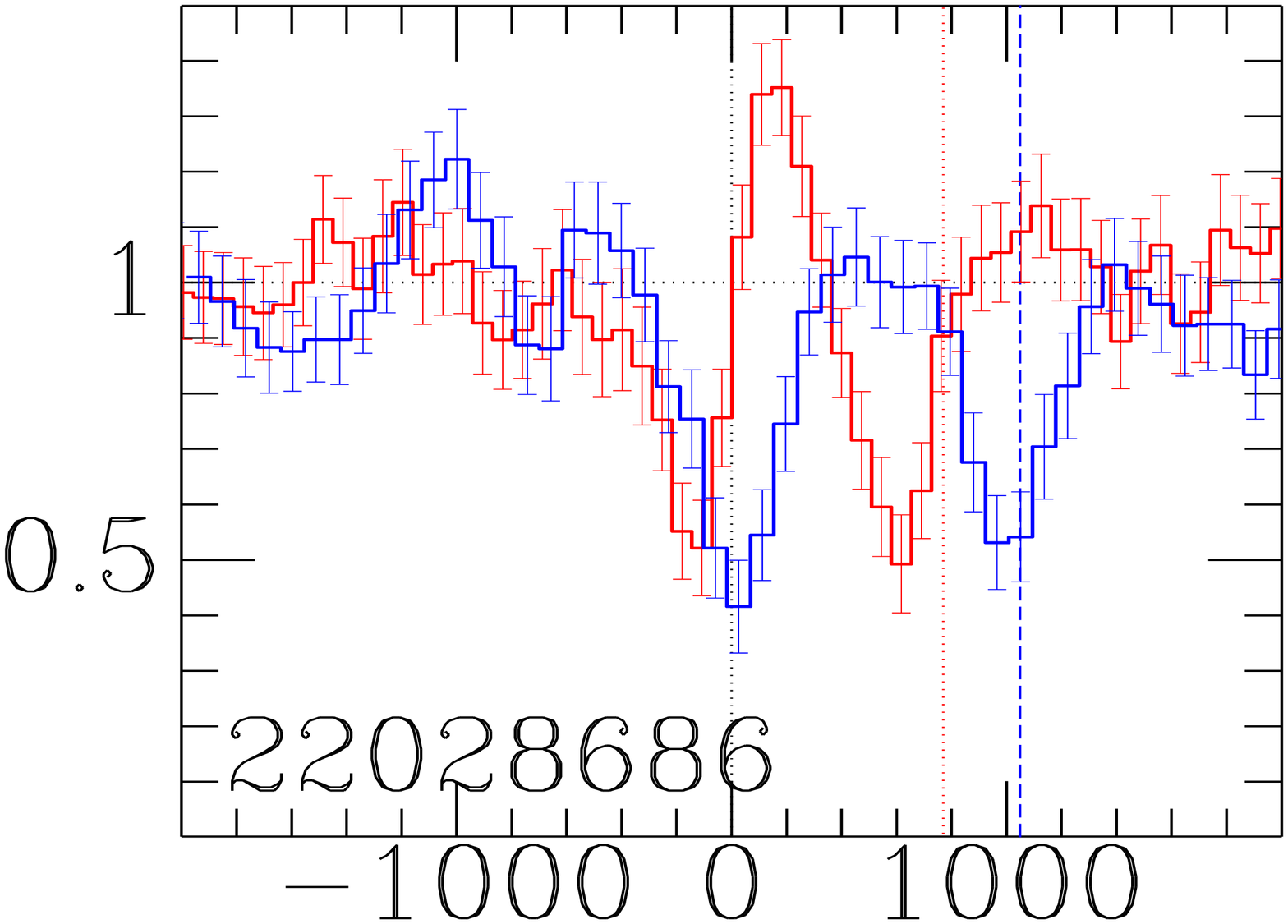}
              \includegraphics[height=3.6cm,angle=-90,clip=true]{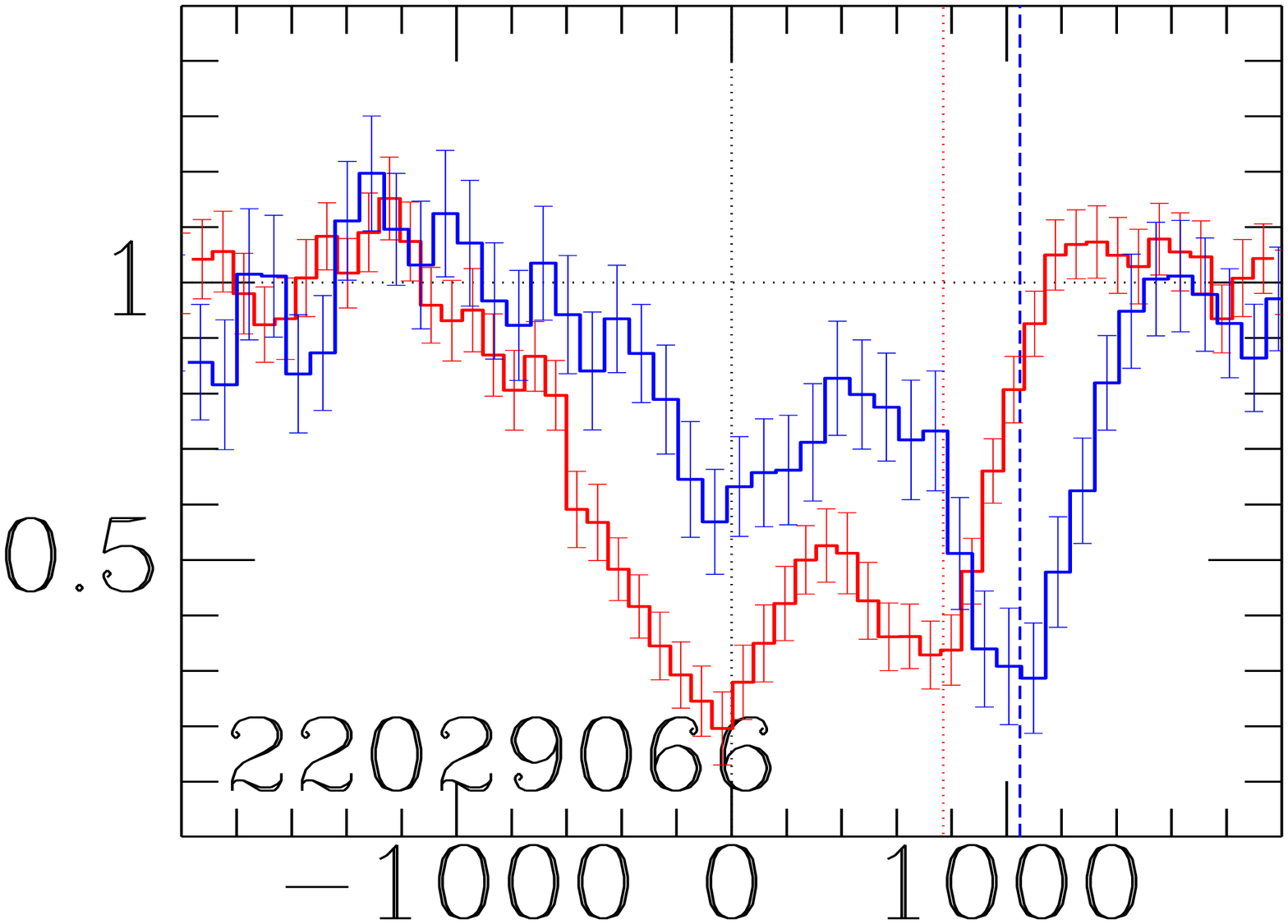}
             }
 \hbox{\hfill 
              \includegraphics[height=3.6cm,angle=-90,clip=true]{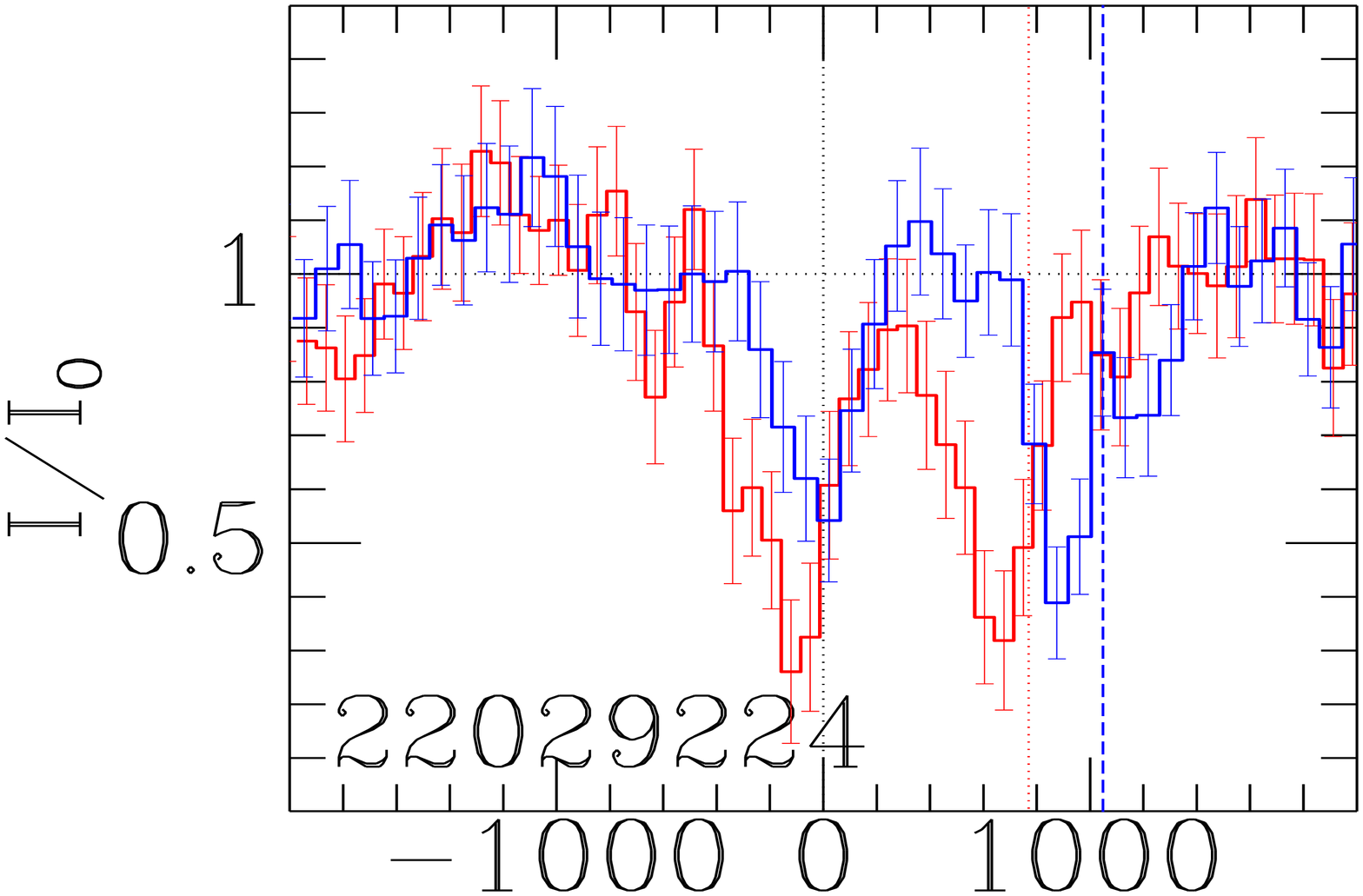}
              \includegraphics[height=3.6cm,angle=-90,clip=true]{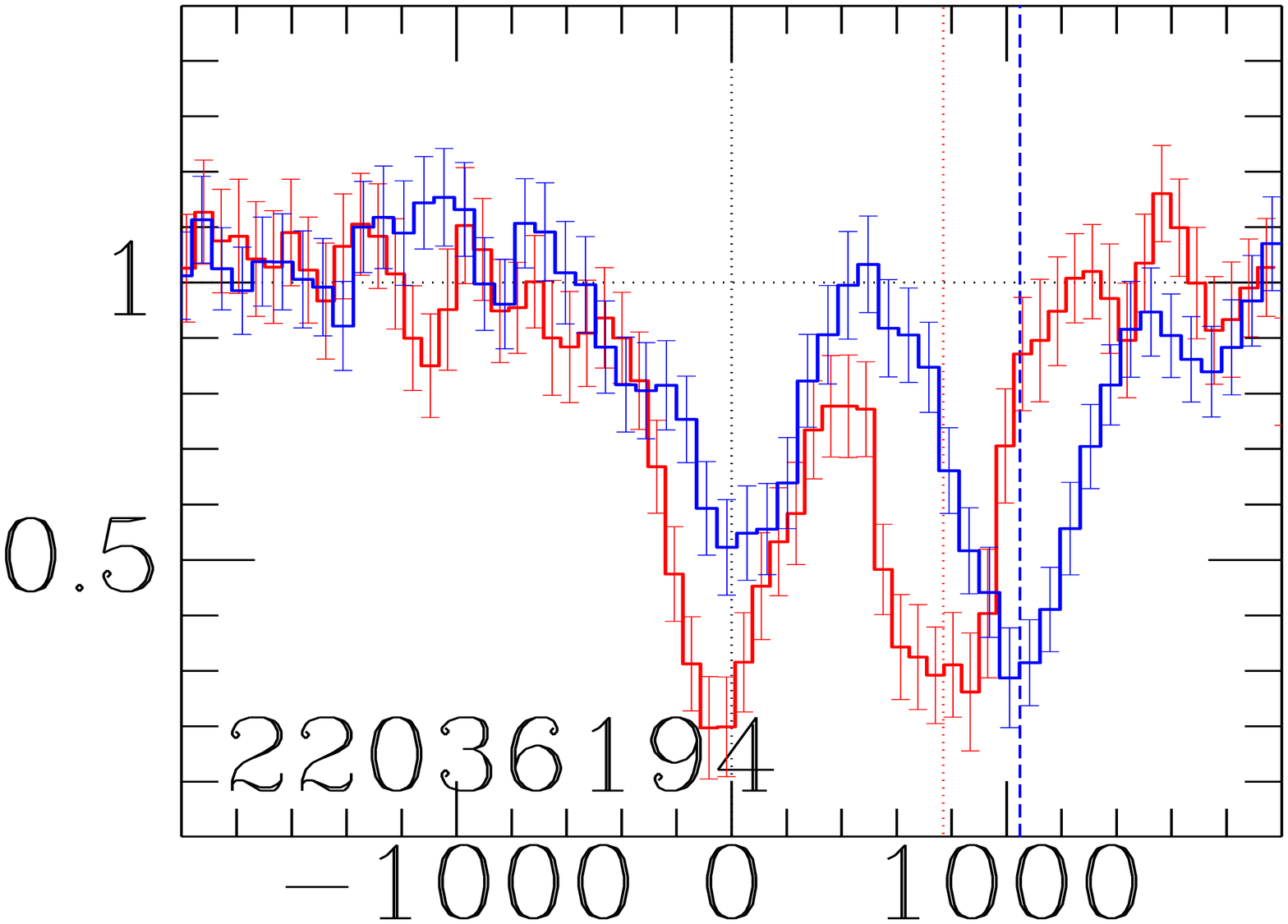}
              \includegraphics[height=3.6cm,angle=-90,clip=true]{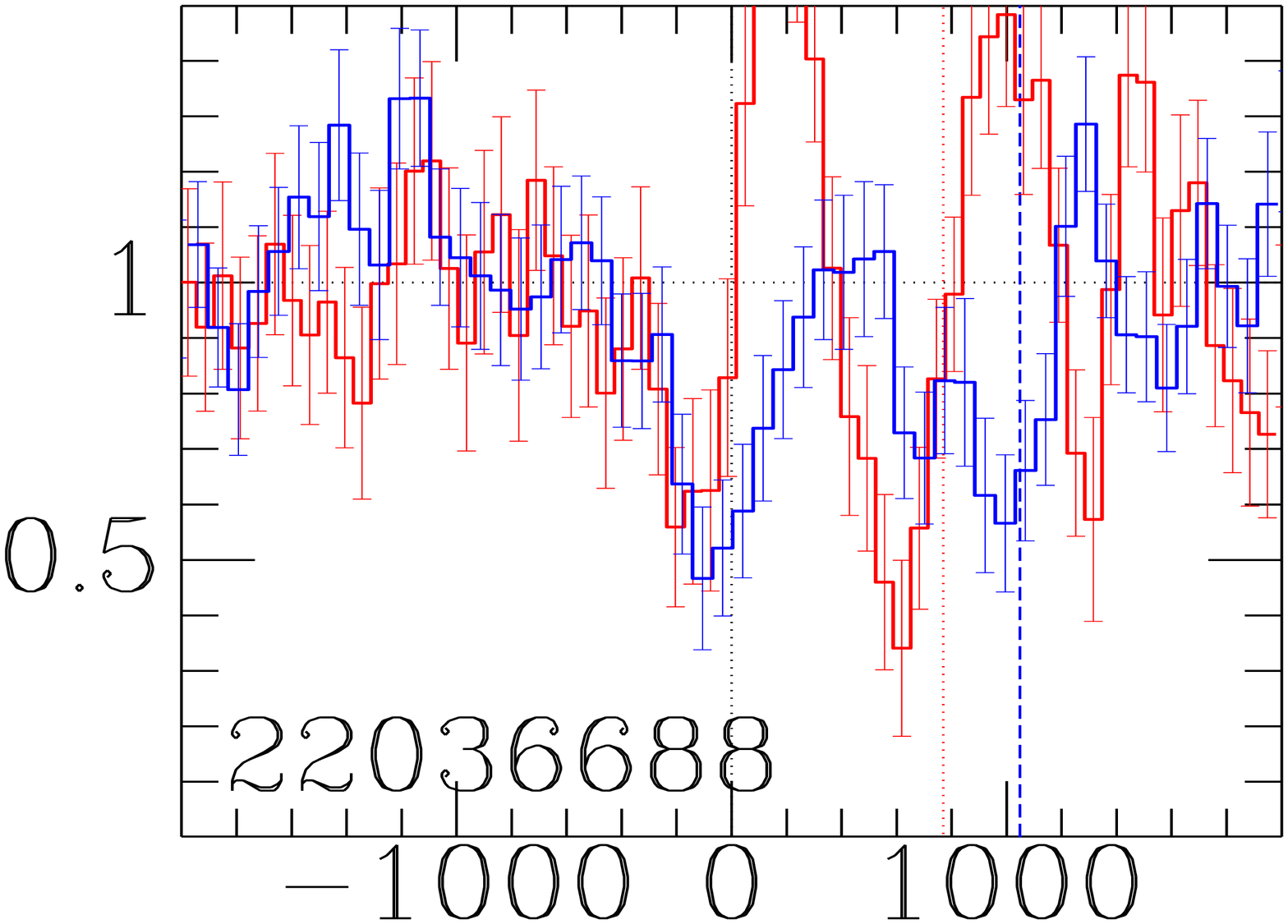}
              \includegraphics[height=3.6cm,angle=-90,clip=true]{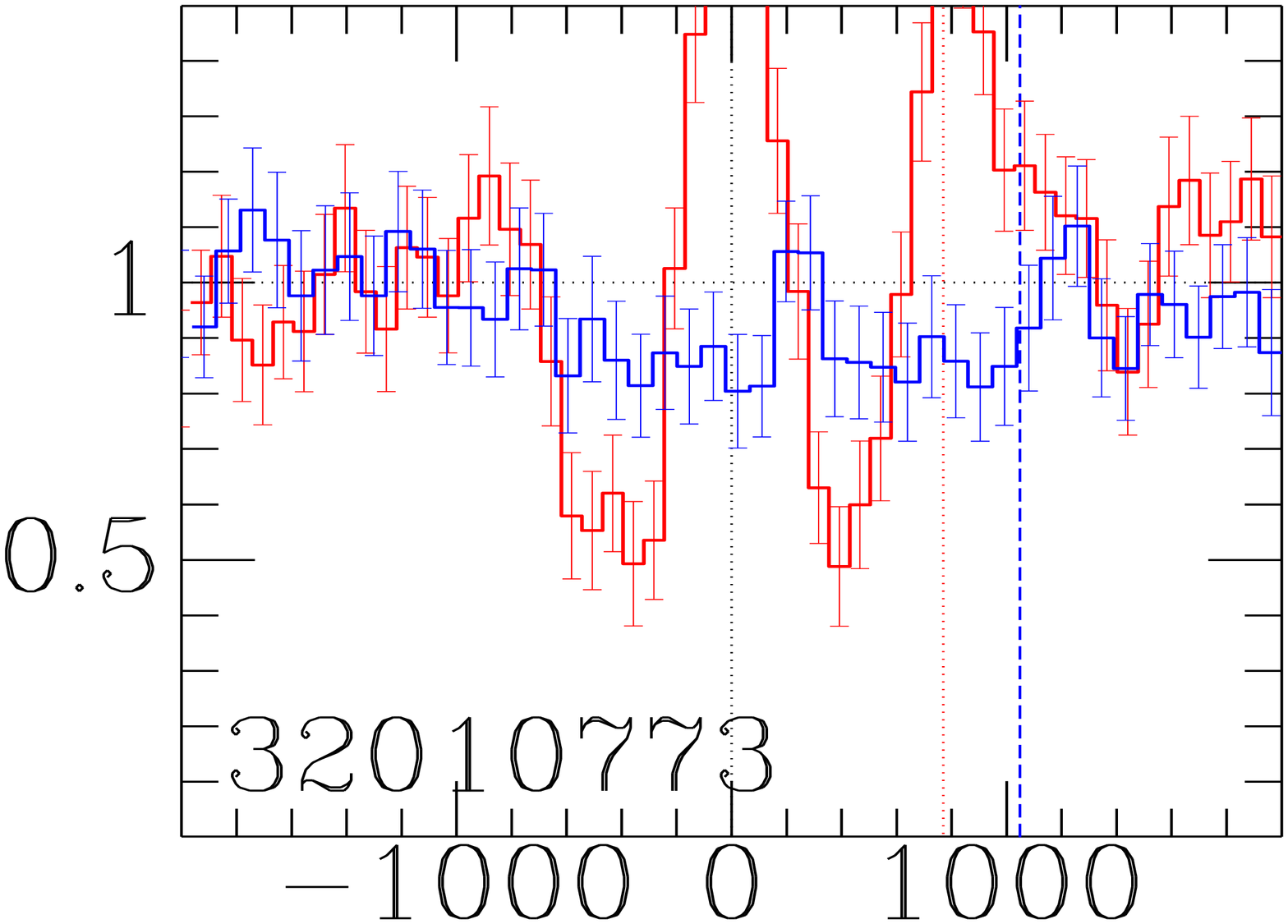}
              \includegraphics[height=3.6cm,angle=-90,clip=true]{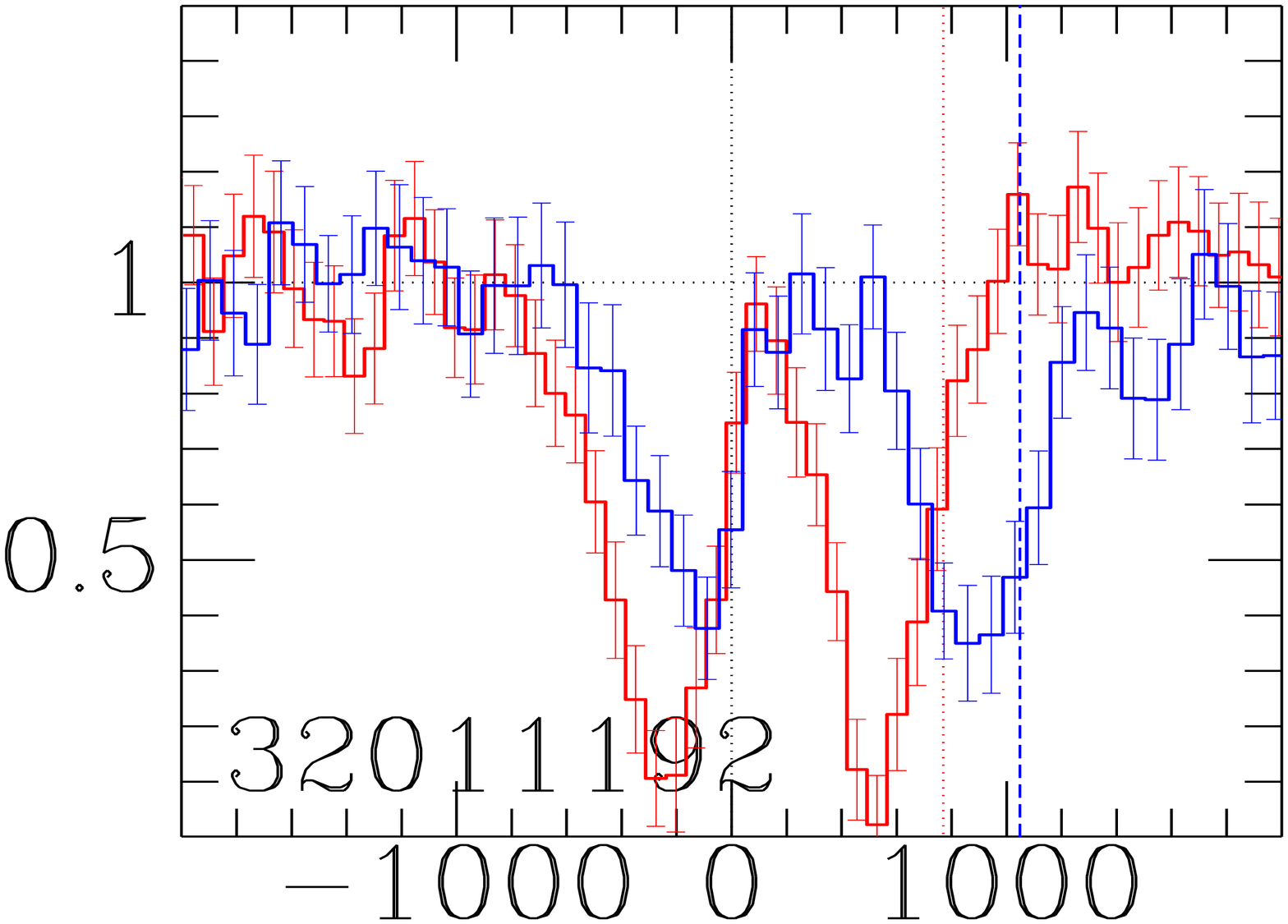}
             }
 \hbox{\hfill 
              \includegraphics[height=3.6cm,angle=-90,clip=true]{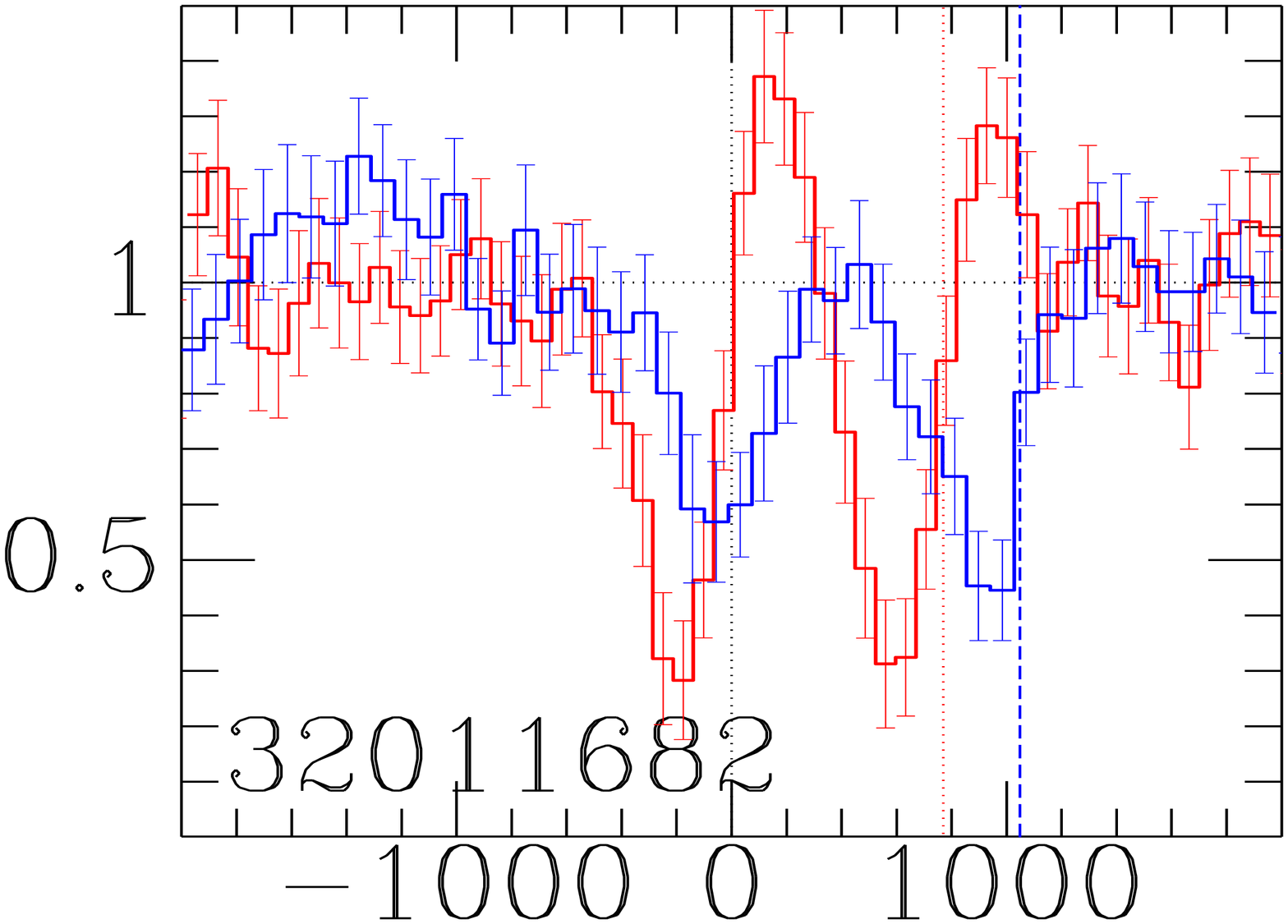}
              \includegraphics[height=3.6cm,angle=-90,clip=true]{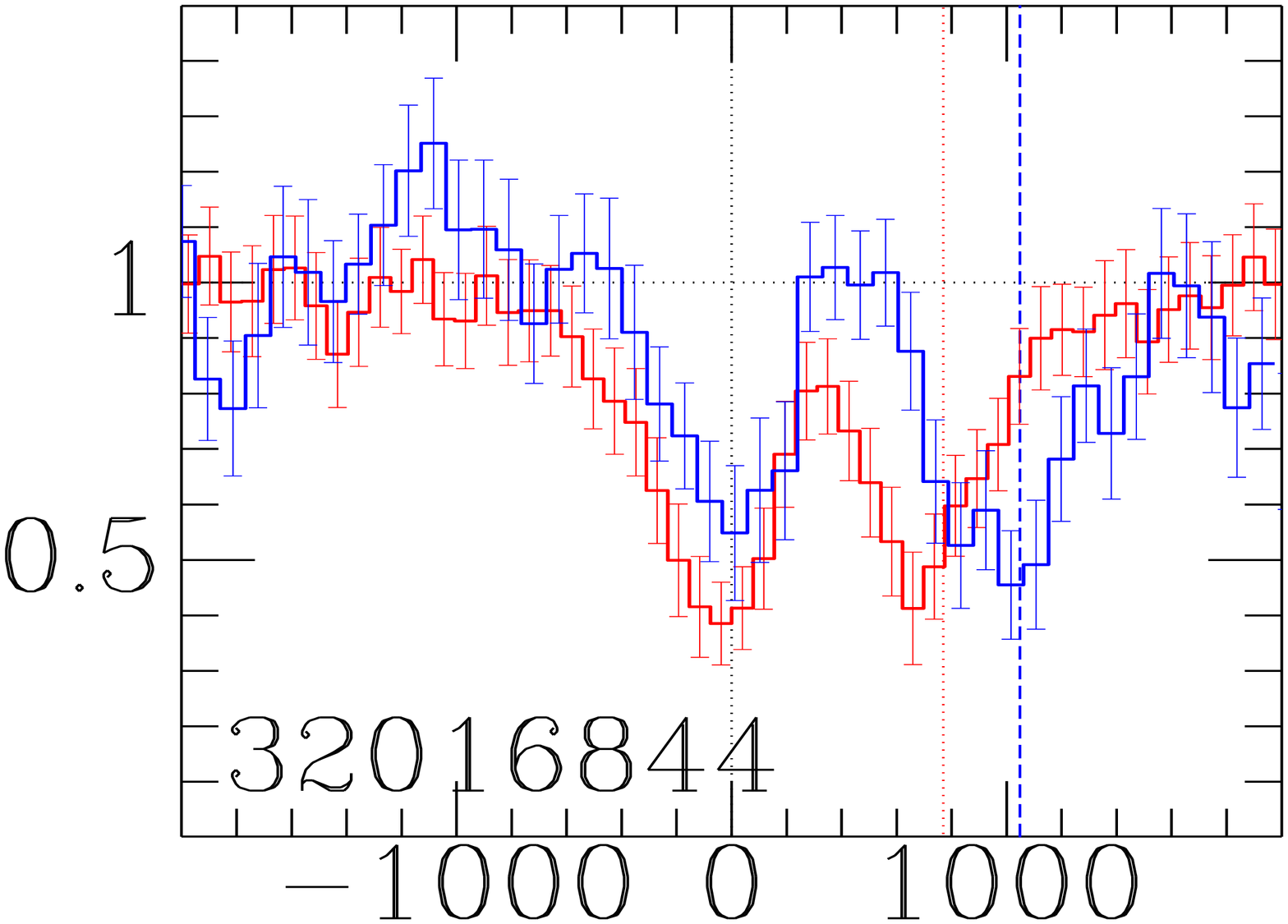}
              \includegraphics[height=3.6cm,angle=-90,clip=true]{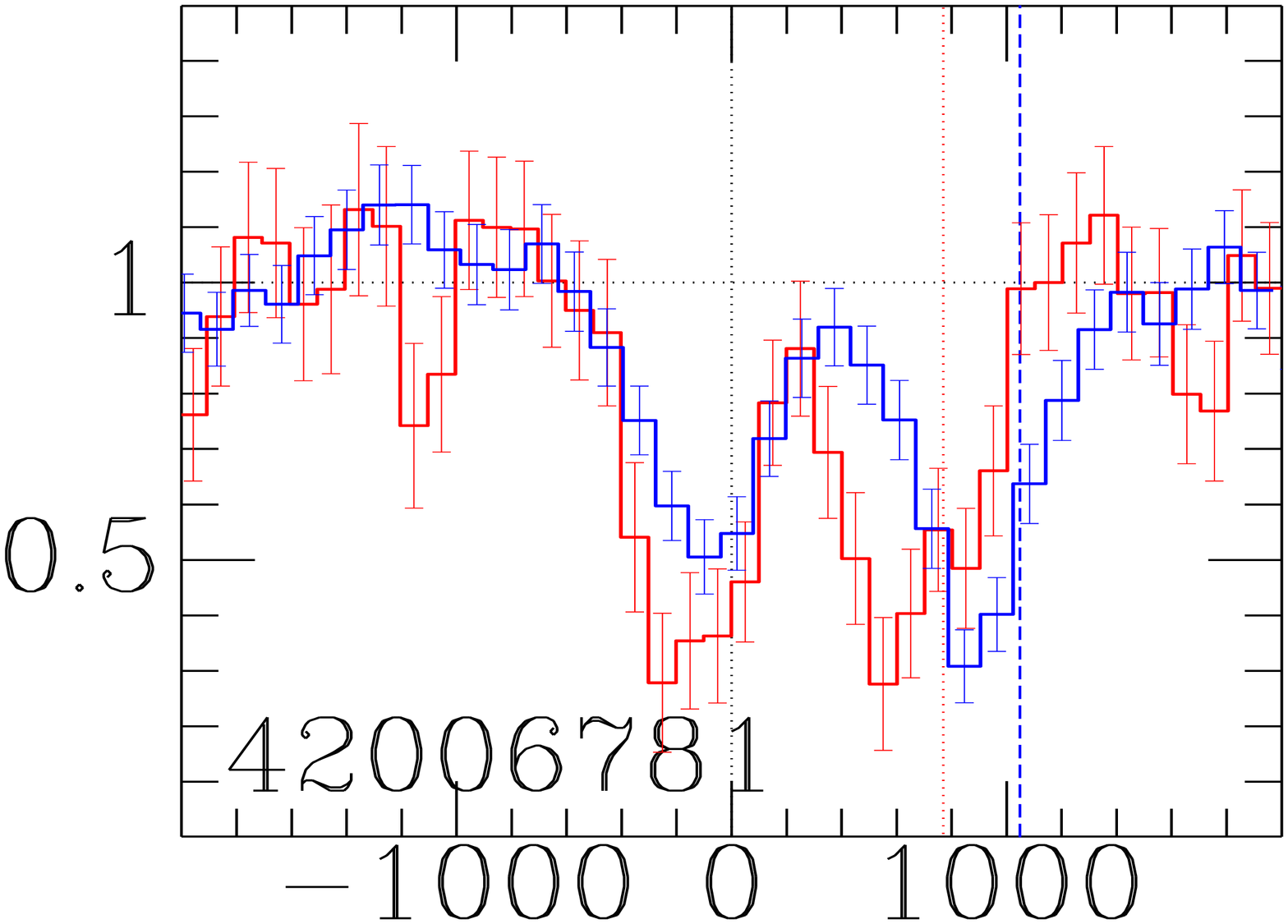}
              \includegraphics[height=3.6cm,angle=-90,clip=true]{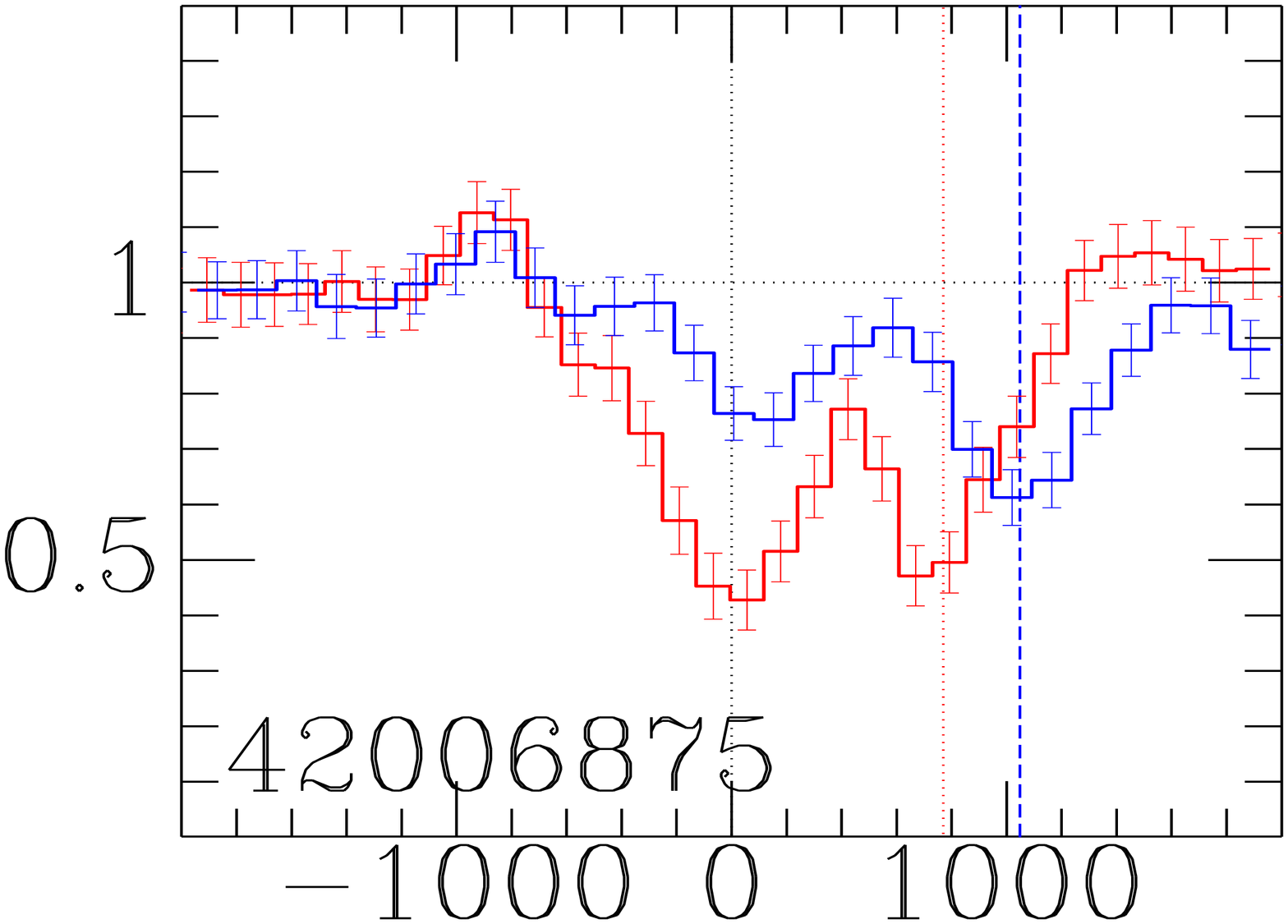}
              \includegraphics[height=3.6cm,angle=-90,clip=true]{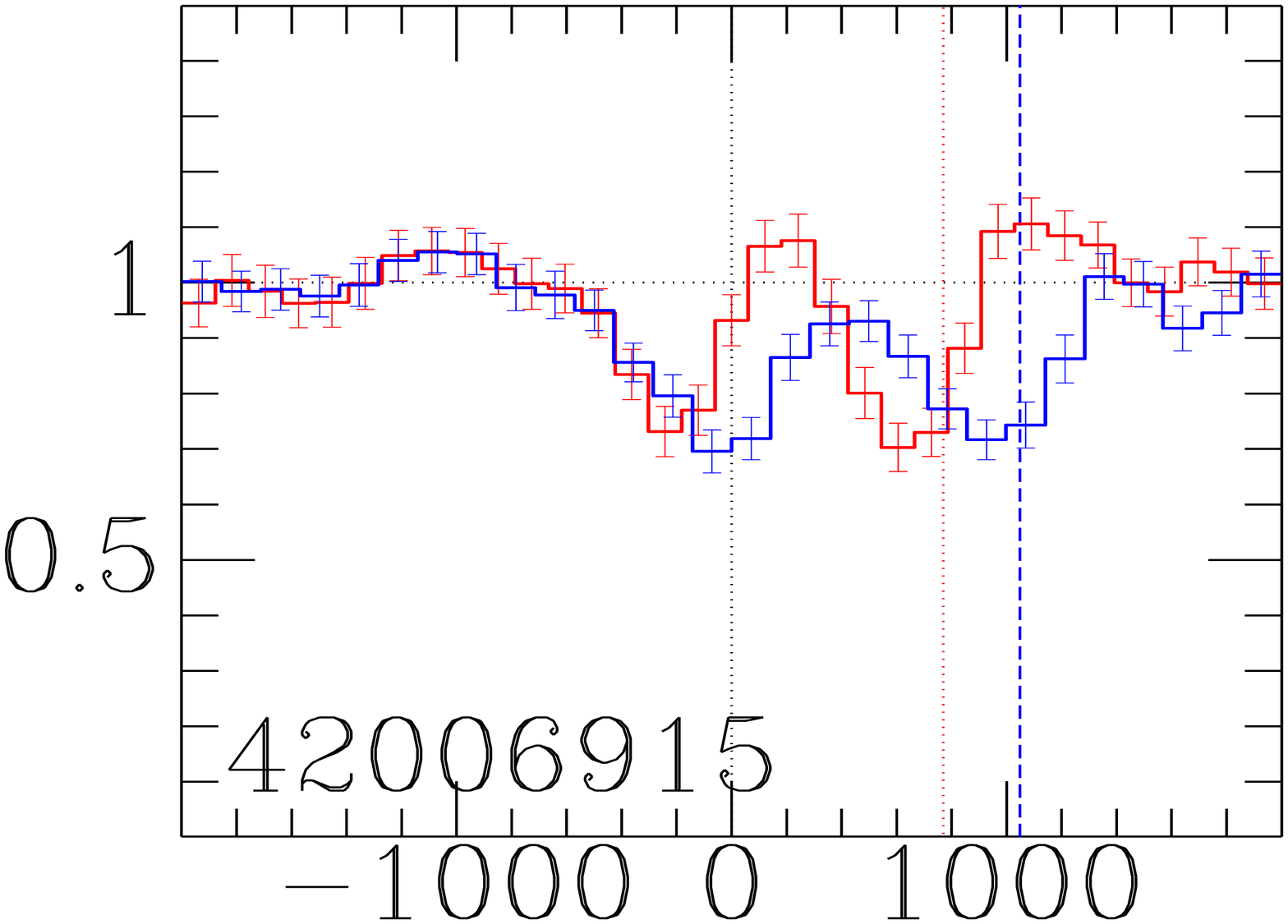}
             }
 \hbox{\hfill 
              \includegraphics[height=3.6cm,angle=-90,clip=true]{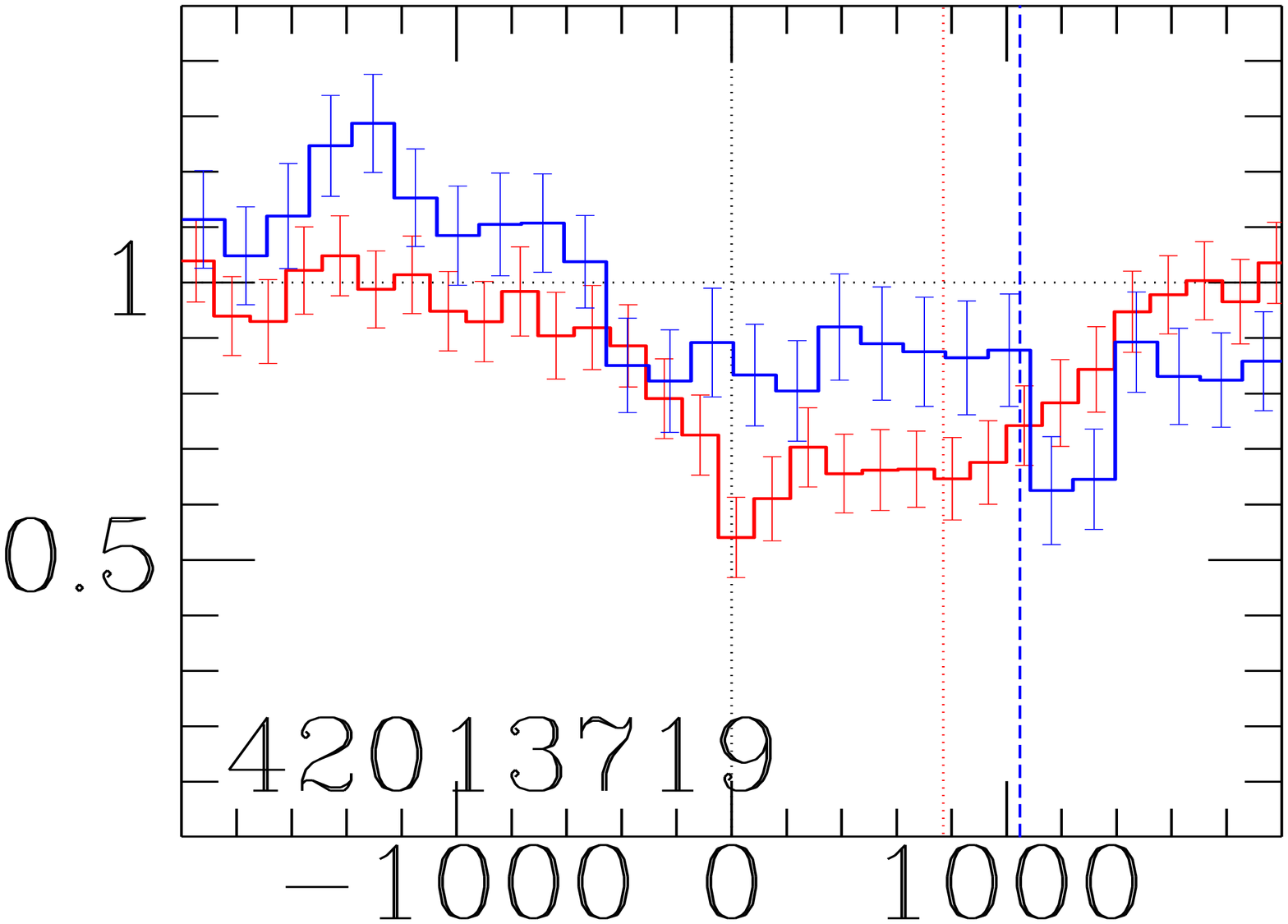}
              \includegraphics[height=3.6cm,angle=-90,clip=true]{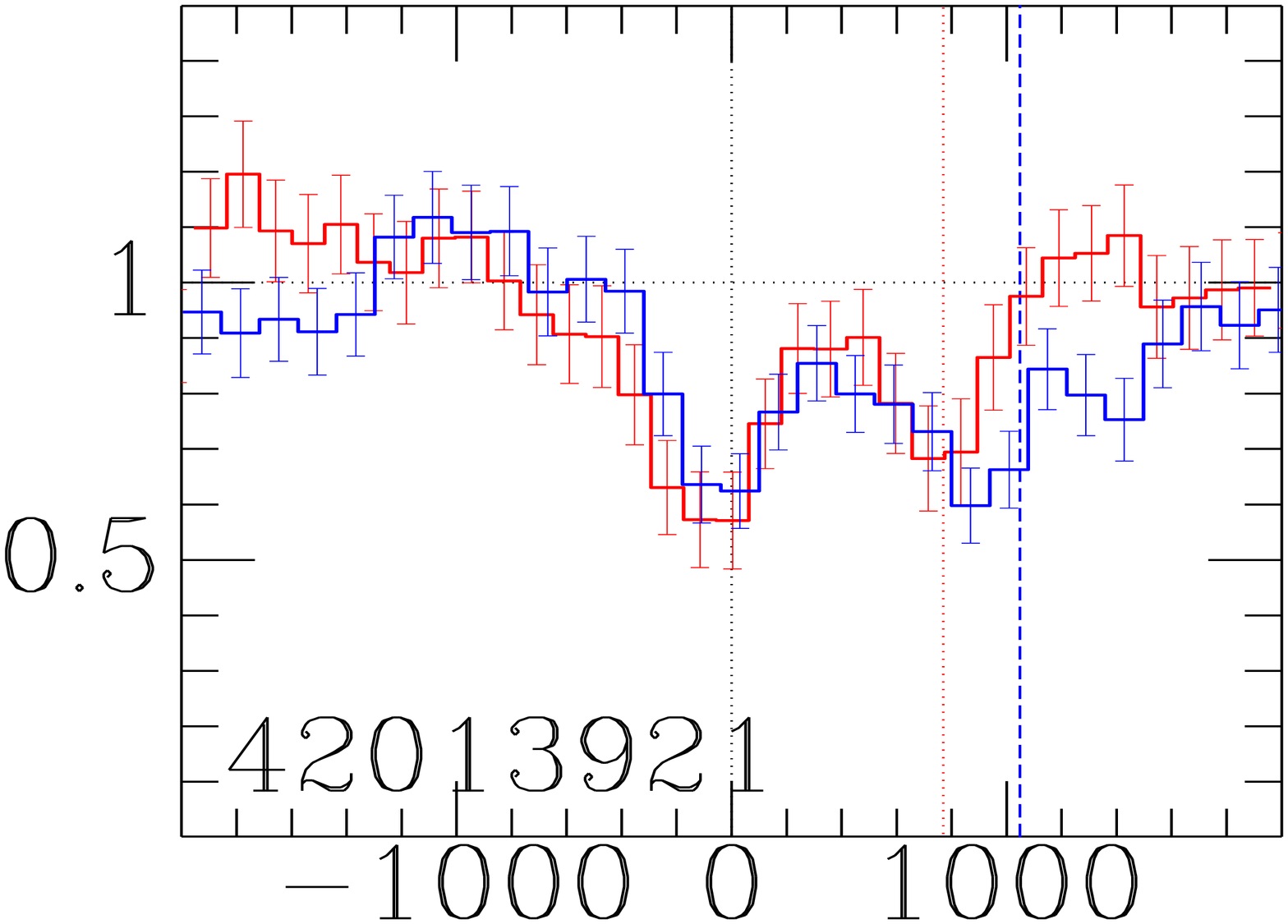}
              \includegraphics[height=3.6cm,angle=-90,clip=true]{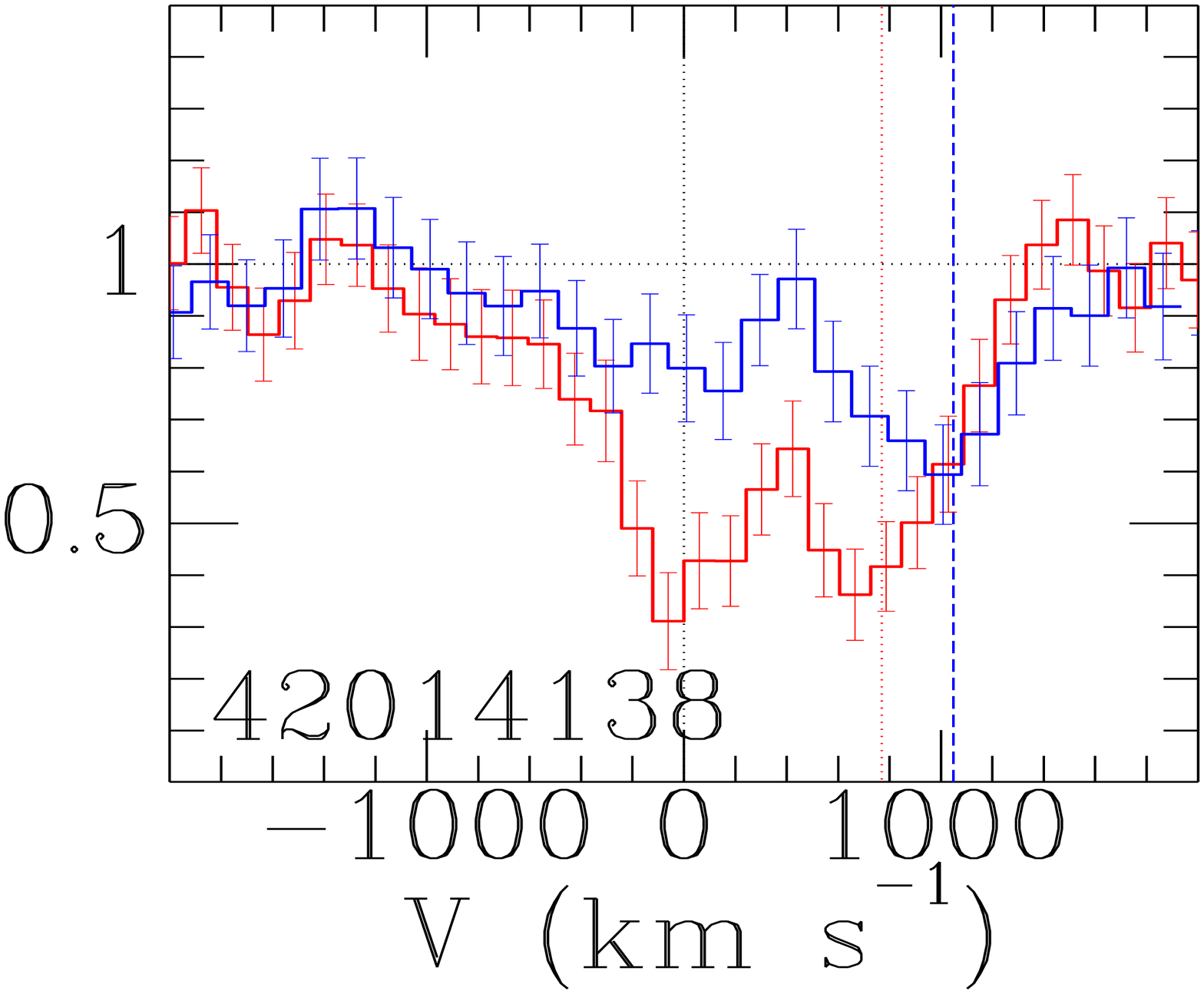}
              \includegraphics[height=3.6cm,angle=-90,clip=true]{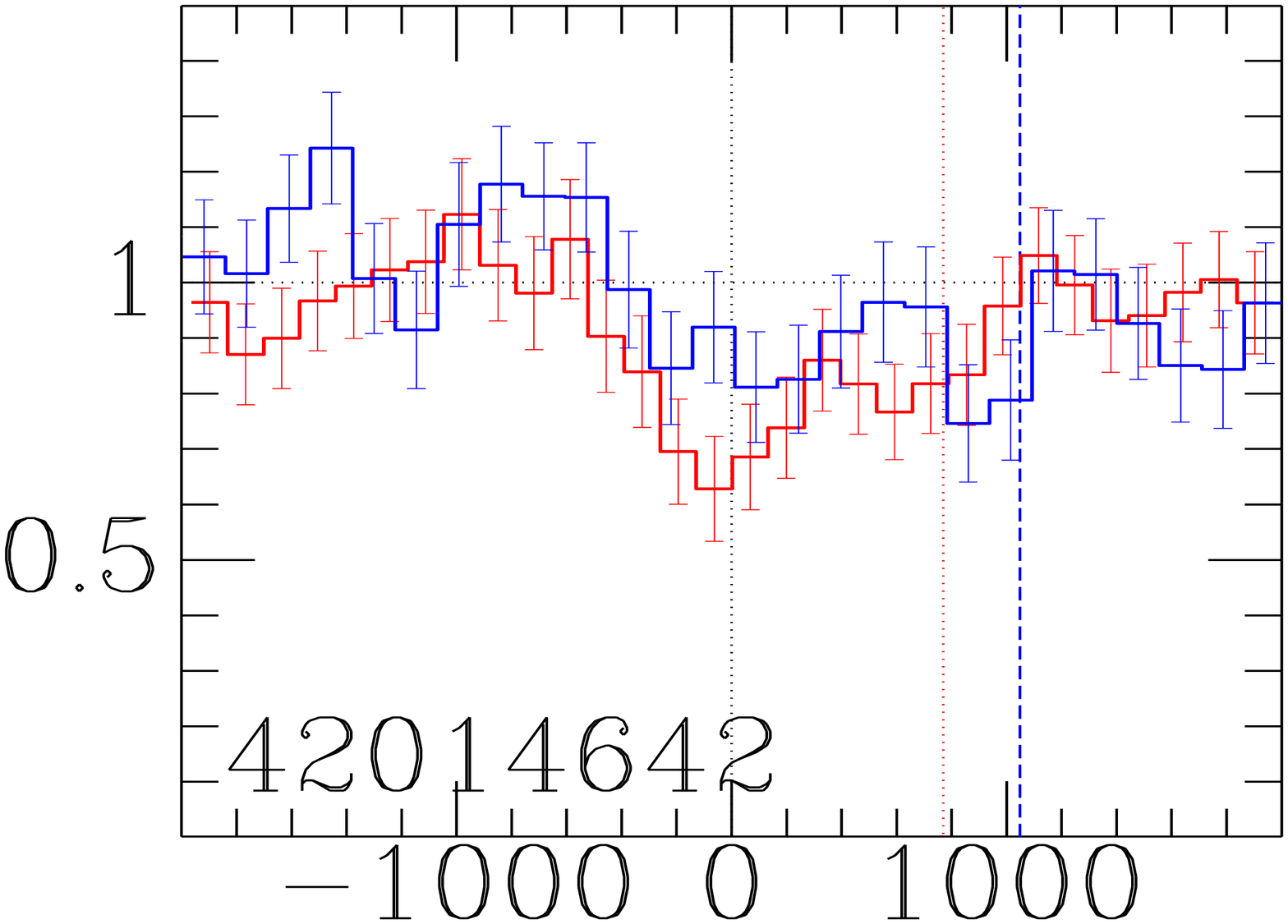}
              \includegraphics[height=3.6cm,angle=-90,clip=true]{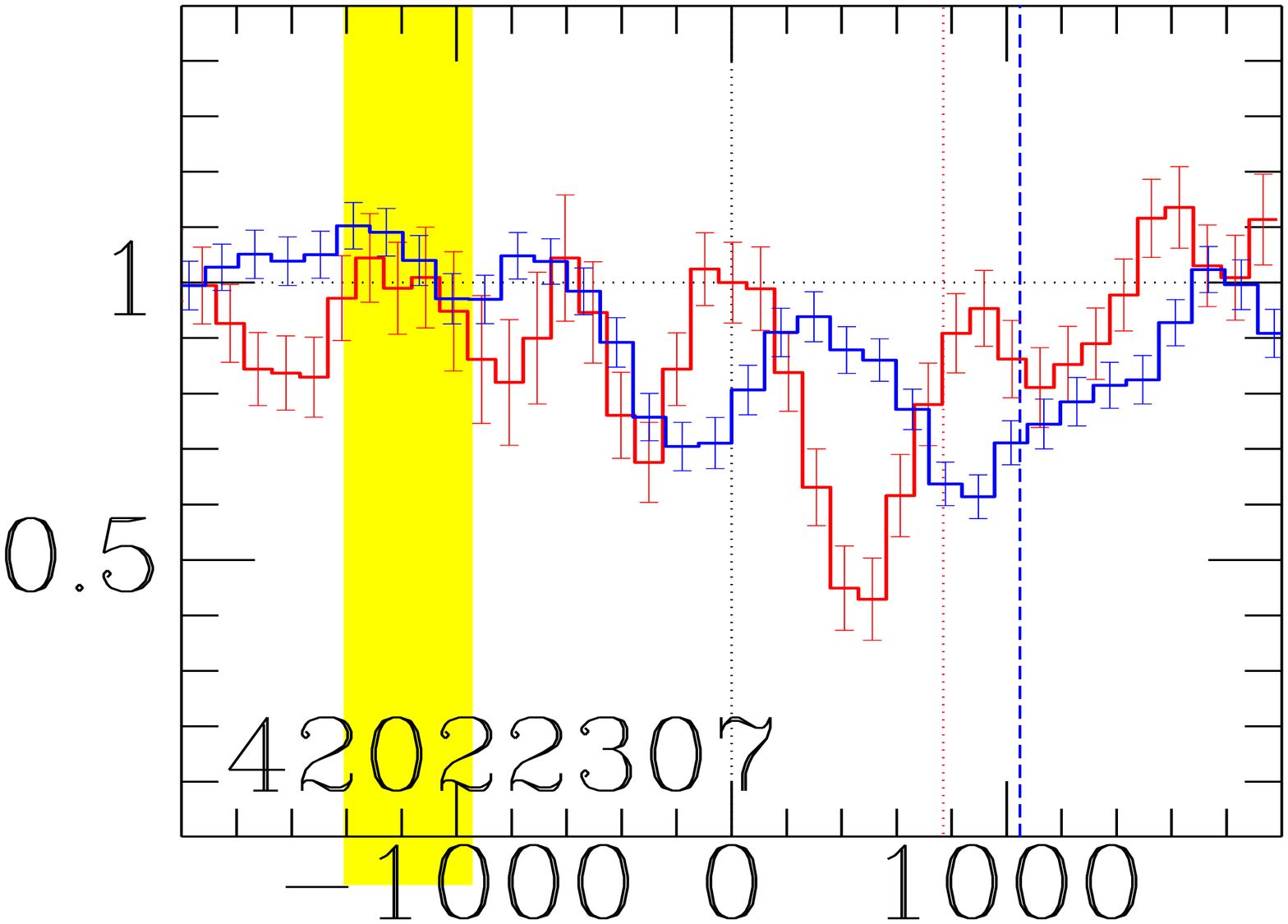}
              }
          \caption{\footnotesize
Comparison of the \mgII\ $\lambda 2796$ (red spectrum) and \feII\ $\lambda 2374$ (blue spectrum) 
line profiles in  spectra with continuum S/N ratio greater than 10 at $\lambda 2450$. The black,
vertical line marks the systemic velocity. Near \feII\ $\lambda2374$, note the location of \feII$^*$ 
2365 at -1124\kms\ (blueward) and \feII\ $\lambda 2383$ at 1048\kms\ (vertical blue line). 
Near \mgII\ $ \lambda 2796$, note the \mgII\ 2803 line at 769\kms\ (vertical red line). 
The location of strong telluric lines that interfere with measurement of the blue wing are noted 
in yellow. In the spectra of 22028686, 22036688, and 42006915, resonance emission prominently fills 
in the \mgII\ absorption troughs near systemic velocity and likely affects the shape of the \mgII\
absorption troughs in many spectra. Despite emission filling, however, the \mgII\ lines generally 
have higher absorption equivalent width than the \feII\ $\lambda 2374$ line, as can be seen from
comparison of the blueshifted portions of their line profiles.  A blue wing on the \mgII\ 
absorption trough is partially resolved in some spectra including 12008811, 12012777, and 42014138.
}
 \label{fig:profiles_hisnr} \end{figure*}

In principle, broad stellar line profiles could be differentiated from outflowing
gas by the symmetry of the former about the systemic velocity. In practice, however,
the \mgII\ $\lambda2803$ line is filled in by resonance emission. In 
Figure~\ref{fig:profiles_hisnr} for example, visual comparison of
the maximum blueshifted \mgII\ $\lambda2374$ and the maximum redshifted \mgII\ 2803 absorption,
shows that the latter is often clearly affected by emission filling (e.g., in the cases 12012777, 22013210, and
32011682) leaving a significant excess of absorption on the blue side. Exceptions to this
rule include the spectra of 22004858, 22005270, and 42014138, where the symmetry might be
used to argue for broad, stellar wings on the line profile. A definitive answer requires 
fitting the spectra with stellar population models to determine the net difference in stellar ages and
is beyond the scope of this paper. However, in general, we think it would be very hard to fill
in the red wing from stellar absorption with resonance emission because the emission is narrow,
unresolved in the LRIS spectra, while stellar absorption lines in single stellar population models
have full widths at half maximum depth of 7-8\AA\ (or more at ages greater than 1.4~Gyr) and broad 
wings on the line profile.
 Hence, it seems clear that high velocity gas is
required to produce the blue wings of many \mgII\ absorption profiles.

\section{COMPARISON OF \mgII\ AND \feII\ ABSORPTION TROUGHS} \label{sec:compare}

Composite spectra allow comparisons between subsets of galaxies at equal 
continuum S/N ratio. Before considering individual spectra, it is
useful to establish the overall trends between absorption properties and
galaxy properties in composite spectra. Table~\ref{tab:composite_spectra}
gives the velocity and equivalent width measurements for the composite spectra
discussed below. To make the comparison between individual spectra more meaningful, 
we focus on \dlow\ (\dhigh) spectra with continuum S/N ratio greater than 10 (7.7) in
this section, excluding the 22 spectra with the strongest stellar absorption
(based on their classification as K+A, red-sequence, or green-valley galaxies). 
Table~\ref{tab:xc_measure} summarizes the correlations between the absorption 
measurements for these individual spectra and the galaxy properties.

\subsection{Absorption Equivalent Widths}

\subsubsection{Composite Spectra} \label{sec:composite_spectra}

In Figure~\ref{fig:mass_composite}, we show composite spectra by tertiles in
stellar mass, $U-B$ color, and rest-frame $B$-band luminosity. 
The equivalent 
width of the \mgII\ absorption increases quite strongly with stellar mass, 
redder color, and higher $B$-band luminosity. 
The lower \mgII\ equivalent width in the less massive, bluer, and intrinsically 
fainter galaxies appears to result largely from increased resonance emission
filling in the absorption troughs.
First, only the composite spectrum of the highest mass quartile of
galaxies has a physical ratio of the $\lambda 2796$ to $\lambda 2803$ equivalent widths; 
composite spectra of the lower mass quartiles have larger equivalent
widths in the transition with the lower oscillator strength. Second,
the composite spectrum of the lowest masses show a P-Cygni emission line 
in the \mgII\ $\lambda 2796$ transition. Third, towards lower mass and bluer color,
the absorption is more suppressed in the red wing of \mgII\ $ \lambda 2803$ than 
it is in the blue wing of \mgII\ $ \lambda2796$. Since models predict the
strongest emission near (and redward of) the systemic velocity, this trend is consistent
with emission filling reducing the equivalent width and maximum absorption velocity
of the red wing.

The \feII\ multiplets in the left and center panels of Figure~\ref{fig:mass_composite} 
also show the impact of increased emission filling at lower mass, bluer color, and fainter 
luminosities. In spectra of lower mass, bluer, and lower luminosity galaxies,
the  equivalent width of \feII\ $\lambda 2383$ decreases much more than that of 
\feII\ $\lambda2374$. In contrast to what
we observed, decreasing the \feII\ column density would reduce the equivalent width of \feII\ $\lambda2374$ faster 
than the \feII\ $\lambda 2383$  equivalent width because the \feII\ $\lambda 2374$
transition, due to its lower oscillator strength,  becomes optically thin when $\tau_0(2383) \approx 10$.

\begin{figure*}[t]
  \hbox{\hfill   \includegraphics[height=18cm,angle=-90,trim = 40 80 30 0]{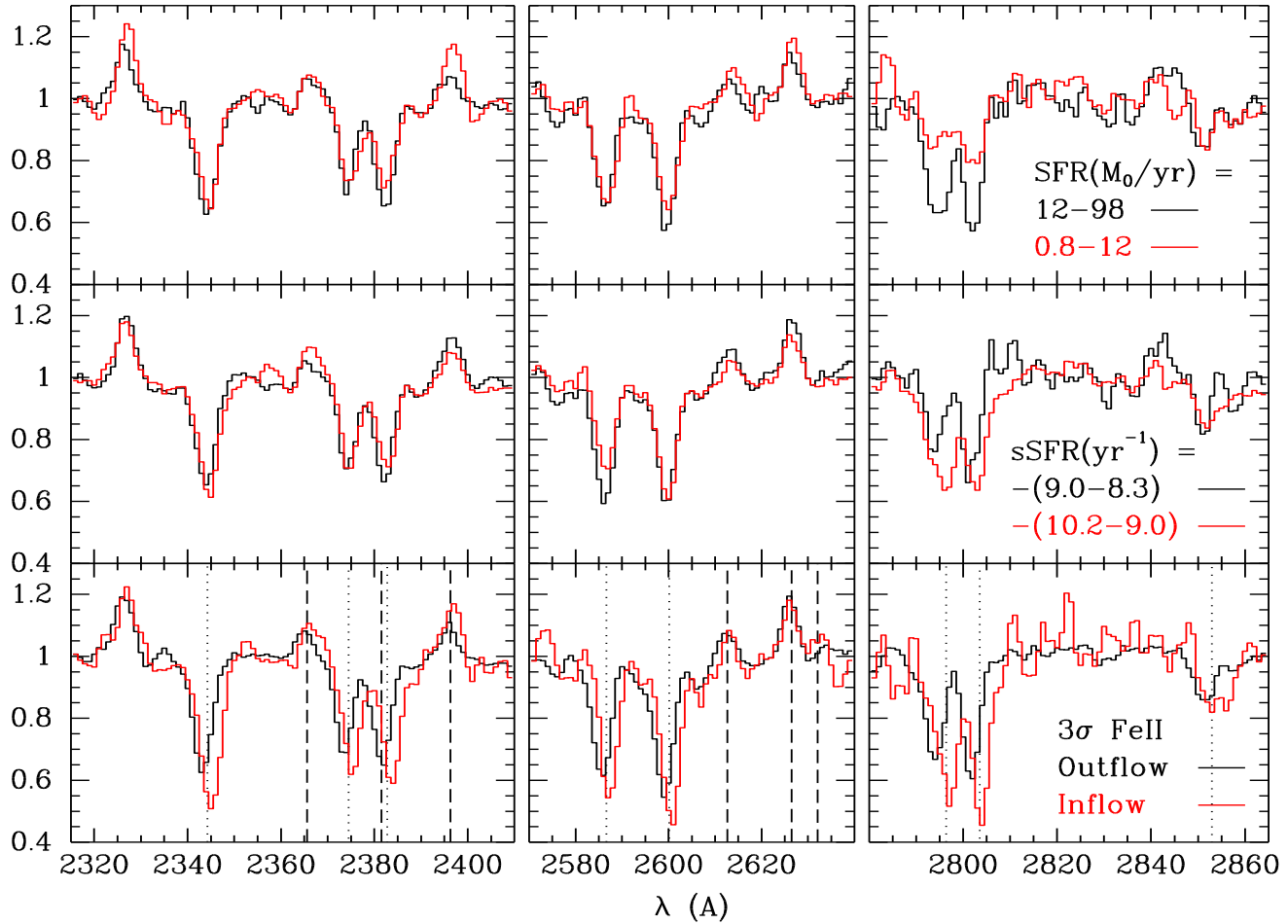}          
         \hfill}
          \caption{\footnotesize
Composite spectra for subsets of galaxies with measured SFR or significant bulk flows.
{\it Top:} The higher SFR galaxies
have higher equivalent widths in \mgII, \feII\ $\lambda 2600 $, and \feII\ $\lambda 2383$.
{\it Middle:} These absorption equivalent widths have a weaker dependence on specific
SFR. The \mgII\ equivalent width is slightly lower in the half of the sub-sample with
the highest SFRs. The inverted equivalent width ratio in \mgII\ clearly indicates this
is due to the emission filling.
{\it Bottom:} The subset of galaxies with the most significant \feII\ 
blueshifts is compared to the subset with significant \feII\ redshifts. The continuum
S/N ratio is lower in the ``inflow''  spectrum (10 galaxies) than the ``outflow'' 
spectrum (35 galaxies). The wavelengths of the resonance absorption lines are
marked with dotted lines for reference; and the \feII$^*$ transtions are marked
with dashed lines.
}
 \label{fig:sfr_composite} \end{figure*}

In \fig~\ref{fig:sfr_composite} and in Kornei \et (2012), 
we show composite spectra for galaxies with lower
and higher SFRs. Robust SFRs have only been measured for the AEGIS subset 
of our sample, so the S/N ratios of these composites are lower than those in \fig~\ref{fig:mass_composite}.
The \mgII\ EW increases towards higher SFR. The equivalent width evidently increases 
more strongly with mass than with SFR, however, because the \mgII\ equivalent width declines 
with increasing specific SFR in the middle panel of \fig~\ref{fig:sfr_composite}.

\subsubsection{Individual Spectra} \label{sec:v_compare}

In Figure~\ref{fig:profiles_hisnr}, we overlay the \feII\ $\lambda 2374$ and \mgII\ 
$\lambda 2796$ absorption profiles.
When the absorption troughs are weak, these line profiles can be indistinguishable
as seen in the 12011364, 12017063, and 42013921 spectra. Among the better resolved 
line profiles, however, the absorption trough of the weaker transition (\feII\ $\lambda2374$) 
is often deeper near the systemic velocity than the \mgII\ $ \lambda 2796$ trough. The prominent
\mgII\ emission in spectra like 12012777, 22028686, 22036688, and 32011682 is clearly the
reason why the \feII\ trough is deeper. Hence, we argue that resonance emission is 
filling in the \mgII\ absorption troughs in spectra like 22004974, 42006915, and 42022307.
To summarize, the most striking difference between the \feII\ and \mgII\ line profiles 
is the paucity of \mgII\ absorption (relative to \feII) near the systemic velocity,
which we attribute in large part to emission filling.

\begin{figure}[h]
  \hbox{\hfill   \includegraphics[height=8.0cm,angle=-90,clip=true]{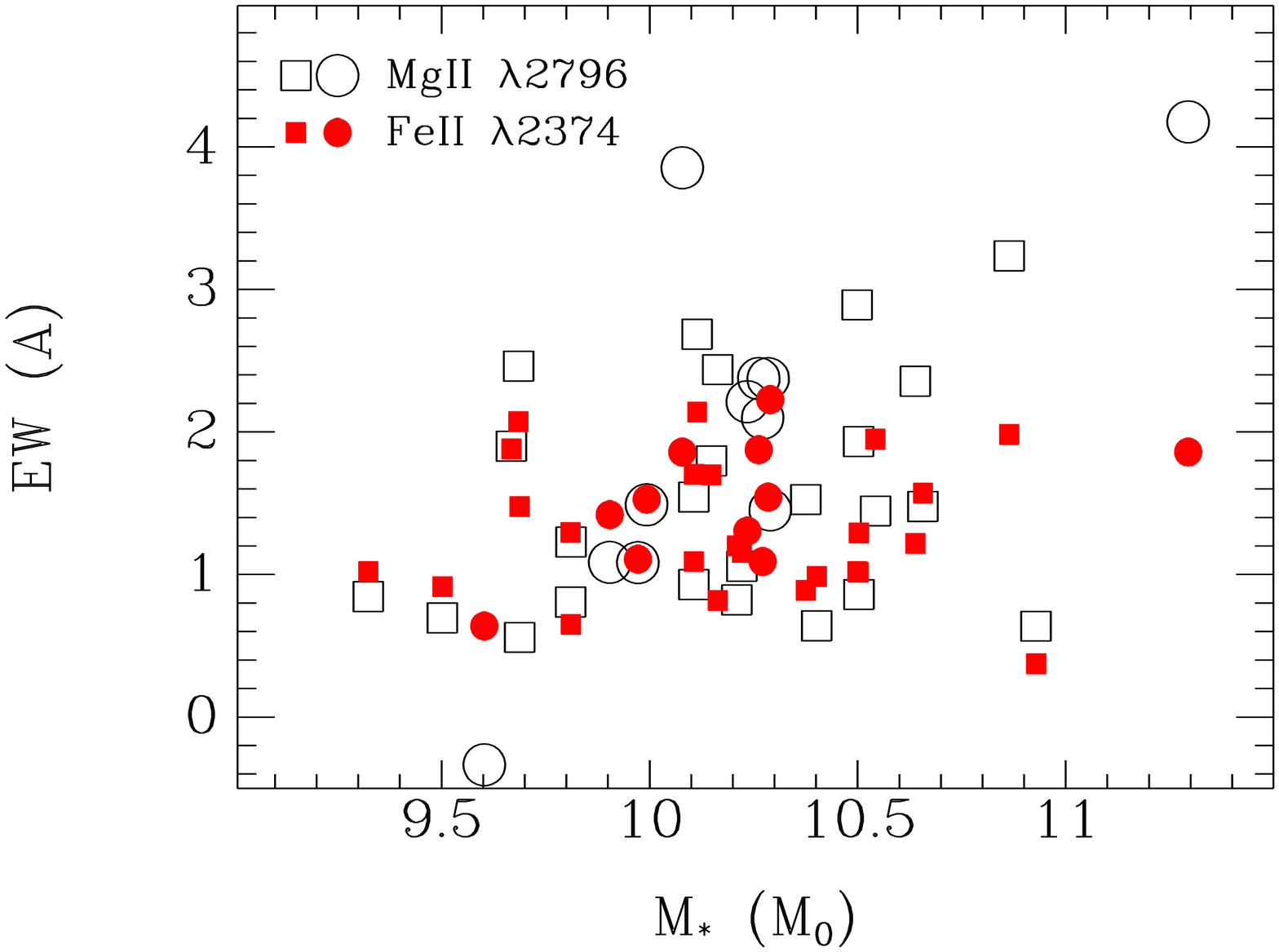}    }      
   \hbox{\hfill   \includegraphics[height=8.0cm,angle=-90,clip=true]{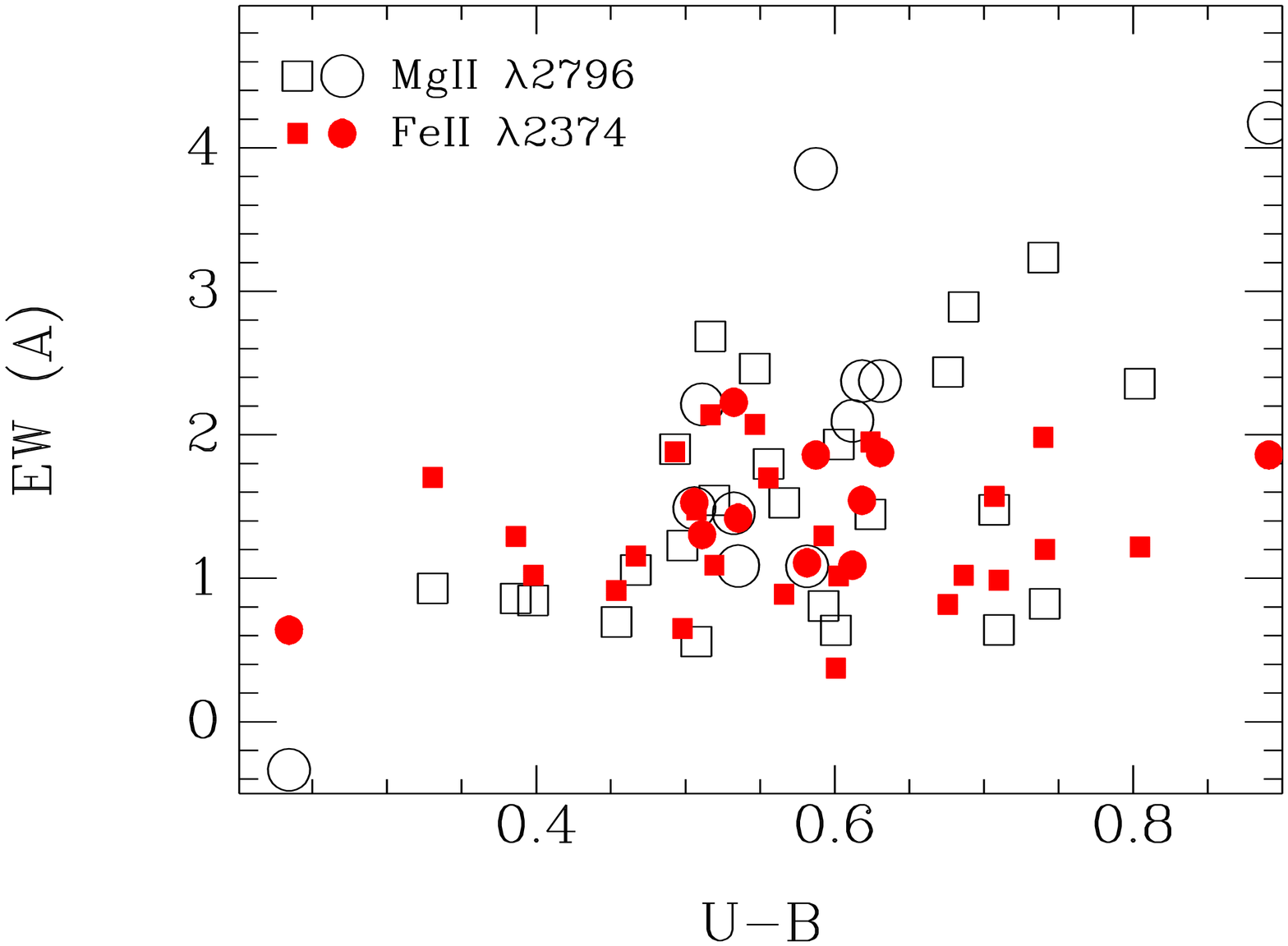}   }       
    \hbox{\hfill   \includegraphics[height=8.0cm,angle=-90,clip=true]{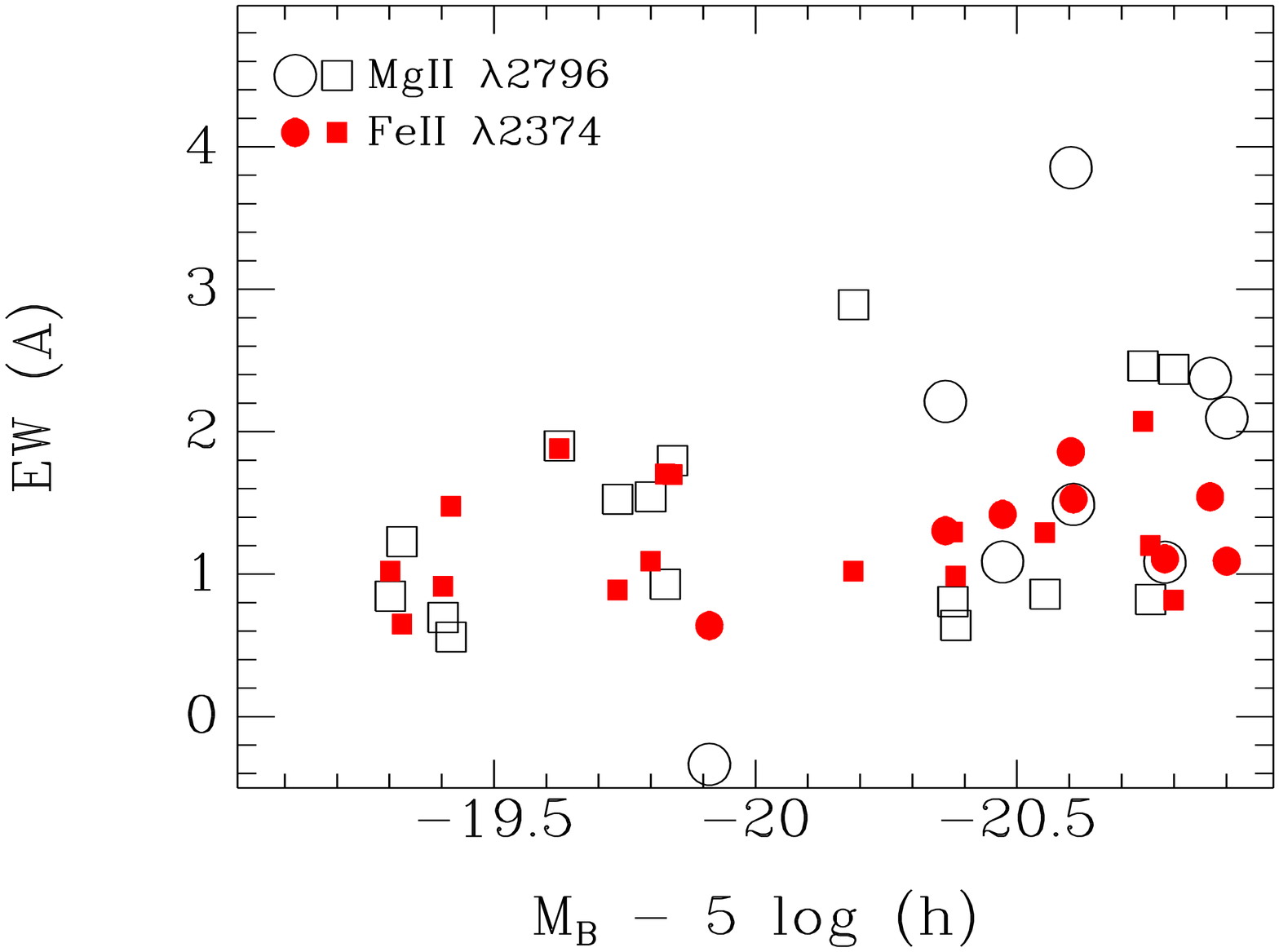}  }        
         \caption{Equivalent widths of the \mgII\ $\lambda 2796$ 
                 and \feII\ $\lambda 2374$  absorption troughs in individual,
                 high S/N ratio spectra. Squares and circles represent the
                 \dlow\ and \dhigh\ spectra, respectively.
                 The \mgII\ equivalent width is positively 
                 correlated with stellar mass, color, and $B$-band luminosity; see
                 Table~\ref{tab:xc_measure}. 
                 (To mitigate the impact of emission filling on
                 the \mgII\ absorption measurement, we have integrated the area of the
                 absorption from -1000 to 0 \kms.)  
                 }
\label{fig:w_mg2} \end{figure}

Figure~\ref{fig:w_mg2} shows the absorption equivalent widths measured in these
individual spectra over the full range in stellar mass, color, and luminosity.
As anticipated from our inspection of the composite spectra, the \mgII\ 
equivalent width measurements for individual spectra are also positively correlated with 
stellar mass, color, and blue luminosity. Among these trends,
the correlation with color is both the most significant (2.58 standard deviations from
the null hypothesis) and strongest (Spearman rank correlation coefficient of
$r_S = 0.44$).  The \mgII\ absorption equivalent width 
is significantly more correlated  with these galaxy properties than is the \feII\ $\lambda2374 $
equivalent width; i.e., comparison of columns~3 and 4 in Table~\ref{tab:xc_measure} 
quantitatively confirms the impression obtained from the composite spectra in this regard.
Even though the \mgII\ equivalent width is correlated with the \feII\ equivalent
width, the ratio $W(2796)/W(2374)$ increases towards redder, more massive galaxies.

In column~2 of Table~\ref{tab:xc_measure}, we show that the Doppler shift of the \feII\ 
absorption has a negative correlation with specific SFR. Since our sign convention 
assigns negative velocities to \feII\ blueshifts,
the \feII\ absorption is more blueshifted in galaxies that are undergoing a
more significant episode of star formation.
This correlation with specific SFR supports the expectation that the blueshifts mark galactic outflows.

We might expect larger blueshifts to generate broader lines and
therefore have larger \feII\ equivalent width; but the measured correlation is in the opposite sense.
The Doppler shift $V_1$ (from single-component fitting)
is positively correlated with the \feII\ $\lambda2374$ equivalent width --
a trend that may  arise, for example, from a significant interstellar absorption 
component at the systemic velocity.
It is certainly more difficult to detect a blueshifted component
if the ISM absorption at the systemic velocity is strong.

\subsubsection{Implications of Emission Filling} \label{sec:interpret_indi}

We have presented evidence for increased emission filling in lower mass, lower
luminosity, bluer galaxies. These trends may be attributed to dust attenuation
because, on average at least, the more massive and redder galaxies are 
dustier than the lower mass and bluer galaxies. Scattered photons absorbed by
dust grains will not emerge from the galaxy, and we observe absorption troughs
which are less affected by resonance emission in spectra of higher mass galaxies. 

Dust is not required, however, to reduce emission filling. Indeed,
we find examples of good agreement between the shape of the \mgII\ 
and \feII\ absorption troughs (i.e., low emission filling) among 
individual spectra from galaxies with a wide range of properties.
Viewing geometry is another way to select for (or against) emission.
The line emission is weaker (and emission filling is reduced) in
more collimated outflows, where the half-opening angle
of the outflow cone is 45\deg\ for example, than a spherical outflow
(Prochaska \et 2011). Adding absorption from the ISM to 
the radiative transfer calculation also reduces the number of resonance-line 
photons in the emergent spectrum. The increased photon trapping in the
ISM gives each continuum photon absorbed in an \feII\ resonance transition more
chances to escape the galaxy as fluorescent emission, thereby reducing the
resonance emission and leaving an observed absorption trough similar to
the intrinsic absorption profile.

\subsection{Identifying Fast Outflows with \mgII\ Absorption} 

In some spectra, such as 22013210, 22029066, and 32011682, the \mgII\ absorption is 
detected to significantly higher blueshifts than is \feII. At the same time, the maximum blueshifts
of the \mgII\ and \feII\ absorption are consistent in many spectra even though
$W(\mgII 2796) >> W(\feII 2374)$, e.g., 12008197, 22004858, 22012723, and 42006781. Overall,
we find no significant correlations between $V_{max}$ and mass ($r_s = -0.31; 1.8\sigma$)
or color ($r_s = -0.21; 1.2\sigma$) among the individual spectra, consistent with
with the lack of any trend in the composite spectra constructed by these galaxy properties.
In addition, the \feII\ Doppler shift is not 
significantly correlated with $V_{max}(\mgII)$; and $V_1(\feII)$ has the stronger
correlation with SFR and specific SFR.
We summarize our observations as follows:
(1) nearly all of the high S/N ratio spectra show a blue wing indicating outflowing
material;
(2) spectra with a blueshifted \feII\ centroid usually show \mgII\ absorption at 
highly blueshifted velocities; but
(3) roughly half of the spectra with \mgII\ absorption at a large 
blueshift do not show a net blueshift of the centroid of the \feII\  absorption.

Among the high S/N ratio spectra, the equivalent widths of the \mgII\ absorption troughs
tend to be larger when the absorption extends to higher blueshifts ($r_s = -0.45$ at $2.64\sigma$ 
in Table~\ref{tab:xc_measure}). We measured the \mgII\ $\lambda 2796$ equivalent width by
integrating the line profile from the systemic velocity blueward 1000~km~s$^{-1}$.
Defined this way,
the equivalent width ranged from 0.25 to 1.0 times the total equivalent width integrated
over the emission and absorption in both transitions. For the purpose of measuring
the outflow equivalent width, and velocity range, this definition mitigates the 
impact of emission filling allowing a more direct comparison among spectra. For
optically thick lines, the positive correlation between equivalent width and $V_{max}$ 
requires either larger spreads in velocity along the sightline (e.g., more clouds or
gas acceleration) or higher gas covering fractions when extremely blueshifted absorption 
is detected.

\fig~\ref{fig:profiles_hisnr} compares the \mgII\ and \feII\ line profiles in spectra
with high continuum S/N ratio. Most galaxies with $3\sigma$ blueshifts of 
the \feII\ lines show a resolved blue wing on the \mgII\ profile. The blue wing  in these 
spectra is typically  detected in \mgII\ to velocities more than twice the fitted single-component 
\feII\ velocity. In spectra where the $V_{max}(\mgII)$ and $V_1(\feII)$ measurements differ, 
the \mgII\ $\lambda 2796$ absorption is detected to larger blueshifts, as expected at low gas 
column densities.
It is difficult to determine, however, whether the \feII\ centroid fitting fails 
to identify some galaxies with outflowing gas. Some spectra, for example, have a well detected blue 
wing of \mgII\ even though the centroids of their \feII\ profiles do not show a significant outlow; 
these include 7 among the nine $3\sigma$-inflow galaxies whose spectra will be discussed in a
Section~\ref{sec:inflow}.
In the absence of broad line wings from stellar absorption, measurements of this maximum absorption 
velocity, $V_{max}$, may be more directly related to the physical speeds in galactic 
outflows than the centroid of the \mgII\ absorption trough.

Among the composite spectra shown in Figure~\ref{fig:mass_composite}, the wavelengths where 
the \feII\ absorption troughs meet the continuum do not vary with stellar mass, $B$-band luminosity, 
color, SFR, or specific SFR. In Table~\ref{tab:composite_spectra}, the most significant change in 
$V_{max}(\mgII)$ is with the Doppler shift of the \feII\ absorption. A composite spectra built from 
individual spectra with significant ($3\sigma$) \feII\ absorption blueshifts show \mgII\ absorption at 
larger blueshifts than does the typical spectrum with redshifted \feII\ absorption. In the bottom 
panel of \fig~\ref{fig:sfr_composite}, we  show the \mgII\ absorption at large blueshifts in
the composite spectrum constructed from the individual spectra with the most significant \feII\ outflows. 
The composite spectrum of the $3\sigma$ outflow galaxies has an \feII\ 
Doppler shift of $-119\pm6$ \kms.  This outflow composite shows \mgII\ absorption out to 
$-901\pm99$~km~s$^{-1}$. The  $V_{max}(\mgII)$ of the outflow composite is higher than that 
of the mass or luminosity composites, so we argue that (at least among star-forming galaxies)
higher velocity outflowing gas is the most likely reason for the blue wings on the \mgII\ profiles.
In Kornei \et (2012), we further show that $V_{max}(\mgII)$ becomes more blueshifted in
composite spectra as the surface density of star formation increases.

\section{GAS OUTFLOWS}  \label{sec:galaxy_properties}

Resonance absorption lines from \mgII\ and \feII\ are the primary near-UV 
spectral diagnostics of galactic outflows.  At spectral resolutions of 
a 285 to 435~km~s$^{-1}$, we can robustly measure the Doppler shifts and 
equivalent widths of the absorption troughs in nearly all spectra. A net
blueshift of the transitions that decay primarily by fluorescence, 
$V_1(\feII)$, best identifies net outflows of low-ionization gas. The
maximum blueshift of the absorption in the strongest line, \mgII\ $\lambda 2796$,
provides a complementary diagnostic of the outflow when spectrally resolved.

To describe the demographics of outflows in Section~\ref{sec:outflow_fraction_estimate},
we introduce a methodology for computing the fraction of spectra showing 
blueshifted \feII\ absorption or blue absorption wing in \mgII\ and examine
how the outflow fraction varies with galaxy properties.  We then discuss
the outflow properties and how they depend on galaxy parameters 
in Section~\ref{sec:outflow_properties}, and Section~\ref{sec:scaling_relations},
respectively.

\subsection{Demographics of Outflow Galaxies} \label{sec:outflow_fraction_estimate}

Since blueshifts mark
gas with a net outflow relative to the stellar system, we want to compute the fraction
of galaxies with \feII\ blueshifts in order to determine which galaxies host outflows.
The challenge is that a blueshifted line detected in a high quality spectrum is
not necessarily detectable in a lower quality spectrum. We illustrate this problem 
Section~\ref{sec:hist} and introduce a statistical method for quantifying the prevalence
of outflows in Appendix~\ref{sec:fout_calc}. We discuss 
the results in  Section~\ref{sec:outflow_fraction}.

\subsubsection{Fraction of Galaxies with Net Flows}  \label{sec:hist}

\begin{figure}[t]
  \hbox{\hfill \includegraphics[height=9cm,angle=-90,clip=true]{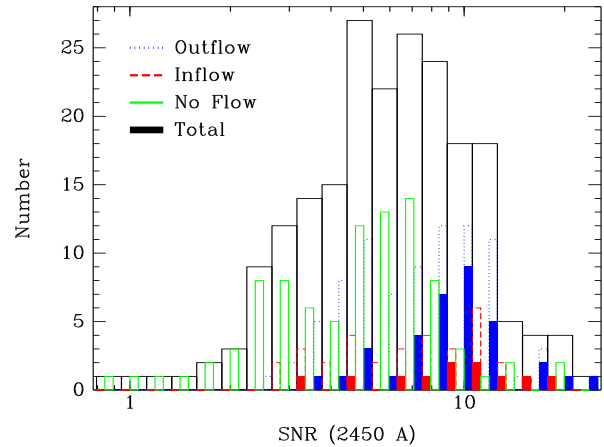} \hfill}
          \caption{\footnotesize
            Histogram of continuum S/N ratio measured between 2400\AA\ and 2500\AA\ 
            (computed from the full sample of 208 galaxies).
            Outflows/inflows are selected by the sign of the Doppler shift, $V_1$, 
            fitted to the \feII\ absorption troughs. The more secure $3\sigma$ blueshifts/redshifts 
            are denotes by the solid histograms, and a more generous $1\sigma$ cut 
            denoted by the extended open histogram of the same color. Among the full 
            sample of 208 spectra, the fraction of \feII\ outflows is very sensitive 
            to the sample's distribution of S/N ratios. Net flows at the $3\sigma$ confidence 
            level are extremely rare in spectra with $SNR < 5$; yet their fraction grows
            to about 58\% at $SNR \approx\ 9$ (36\% with S/N ratio $> 9$). A significant
            fraction of high S/N ratio spectra present no net flow.
}
 \label{fig:c_snr} \end{figure}

Figure~\ref{fig:c_snr} shows a histogram of continuum S/N ratio. As anticipated,
at increased spectral S/N ratio, a higher fraction of net outflow (and inflow) galaxies are
detected. The continuum S/N ratio is not significantly correlated with $B$-band luminosity, 
stellar mass, color, or SFR; hence spectral sensitivity rather than galaxy 
properties shape this growth in the outflow fraction. The continuum S/N ratio is
anti-correlated with the apparent {\it B} magnitude of the galaxy ($4\sigma$ deviation 
from the null hypothesis), but the correlation is not very strong (correlation coefficient
$r_S = -0.32$)
for two reasons: the galaxies have a range of sizes, and
the observations were obtained under a wide range of conditions. Since the
sample selection was by apparent magnitude, the anti-correlation between S/N ratio and 
redshift is not very significant.

Sensitivity to Doppler shifted absorption also depends on spectral resolution 
and the strength of ISM absorption.
At the same S/N ratio (per pixel), a \dhigh\ spectrum is more likely to
yield a net Doppler shift than a \dlow\ spectrum because the uncertainty in
the line centroid is smaller at higher resolution as described in Section~\ref{sec:vsys}.
In addition, high equivalent width absorption at the systemic velocity will
reduce the Doppler shift of the fitted, single-velocity component. In practice, we 
cannot measure the out/inflow fraction at Doppler shifts less than $ |V_1| < 41$\kms\ 
due systematic errors described previously in Section~\ref{sec:vsys}. Focusing
our analysis on these larger Doppler shifts largely removes any dependence on 
spectral resolution.

In Appendix~A, we describe a methodology for calculating outflow fraction
that takes into account the detection 
biases introduced by the variations in S/N ratio and resolution among our spectra. 
Rather than arbitrarily picking a minimum Doppler shift to define an
outflow, we consider {\it threshold velocities} ranging from the largest 
blueshift measured down to the minimum Doppler shift that we can reliably measure, 
roughly  $41$\kms. We assign to each spectrum a probability that the Doppler shift 
of the low-ionization gas is larger than a threshold velocity. Finally, we obtain
the outflow fraction for that threshold velocity by averaging these probabilities
over the entire sample computing the uncetainty directly from the probability
distribution. This definition of outflow fraction is quite general and can
be applied under a broad range of circumstances. For example, the Doppler shift 
may represent the fitted \feII\ centroid velocity $V_1$, the Doppler shift of 
the blue wing of the \mgII\ absorption trough $V_{max}$, or the velocity derived
from a more complex model. This methodology can also be applied to describe the inflow fraction
if we use the net Doppler shift or velocity of the red wing of a line profile.

\begin{figure}[t]
 \hbox{\hfill \includegraphics[height=9cm,angle=-90,clip=true]{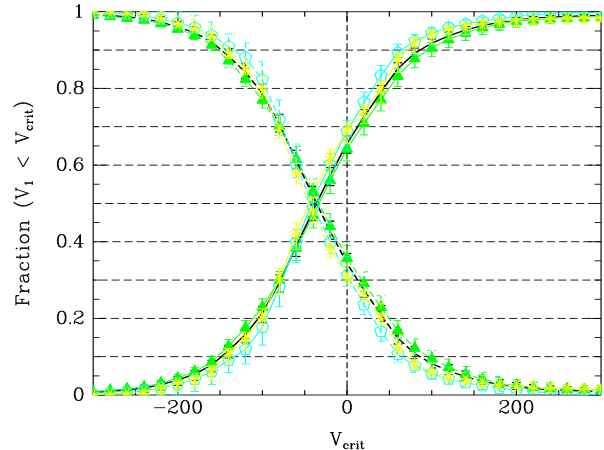}      \hfill}
          \caption{\footnotesize
Fraction of spectra with a net Doppler-shift of the \feII\ absorption lines (relative
to nebular emission lines). The solid lines show the fraction of blueshifted spectra with 
$V_1 < V_{crit}$; and the dashed lines illustrate the complementary quantity, the fraction of 
galaxies with $V_1 > V_{crit}$. The black, green (solid triangles), cyan (open pentagons), 
and yellow (asterisk) curves show the results of 
the calculation defined in Appendix~\ref{sec:fout_calc} for the following four subsamples, 
respectively: all spectra, all \dlow\ spectra, all \dhigh\ spectra, and the highest quality 
spectra defined as $S/N > 8.5 (6.5)$ for \dlow\ (\dhigh). The dominance of net blueshifts (65\%) 
over spectra with redshifted absorption (35\%) is apparent at  $V_{crit} = 0$; and 
the fraction of galaxies with $V_1 < V_{crit}$ (solid line) rises steadily as the threshold speed 
for an outflow decreases. The curves for the four subsamples  agree within the
the error bars showing the 68.27\% confidence interval. This method of calculation
does a good job of correcting for variations in spectral S/N ratio and resolution allowing
us to further examine the influence of galaxy properties on the fraction of spectra with 
Doppler-shifted absorption lines.
}
 \label{fig:fout} \end{figure}

In Figure~\ref{fig:fout}, we show the outflow and inflow fractions computed
for four samples. Roughly 5\%, 20\%, and 45\% of the spectra in our full sample (black
lines) show \feII\ blueshifts faster than -200\kms, -100\kms, and -50\kms, respectively.
In contrast, less than 8\% of the spectra have redshifts higher than 90\kms. Repeating
the analysis with just the highest S/N ratio spectra (yellow lines), we
obtain the same result. Similarly, with the sample restricted to either the lower resolution (\dlow)
spectra or the higher resolution (\dhigh) spectra, the outflow fraction
remains consistent with the results obtained for the full sample. This
test demonstrates that our results for the outflow fraction are insensitive to
S/N ratio and spectral resolution. 
Because many more of the LRIS spectra have blueshifts than redshifts, 
the results for the outflow fraction are better constrained (than for inflow)
and we describe the outflow results in Section~\ref{sec:outflow_fraction}.

\begin{figure*}[t]
 \hbox{\hfill  \includegraphics[height=5.5cm,angle=-90,clip=true]{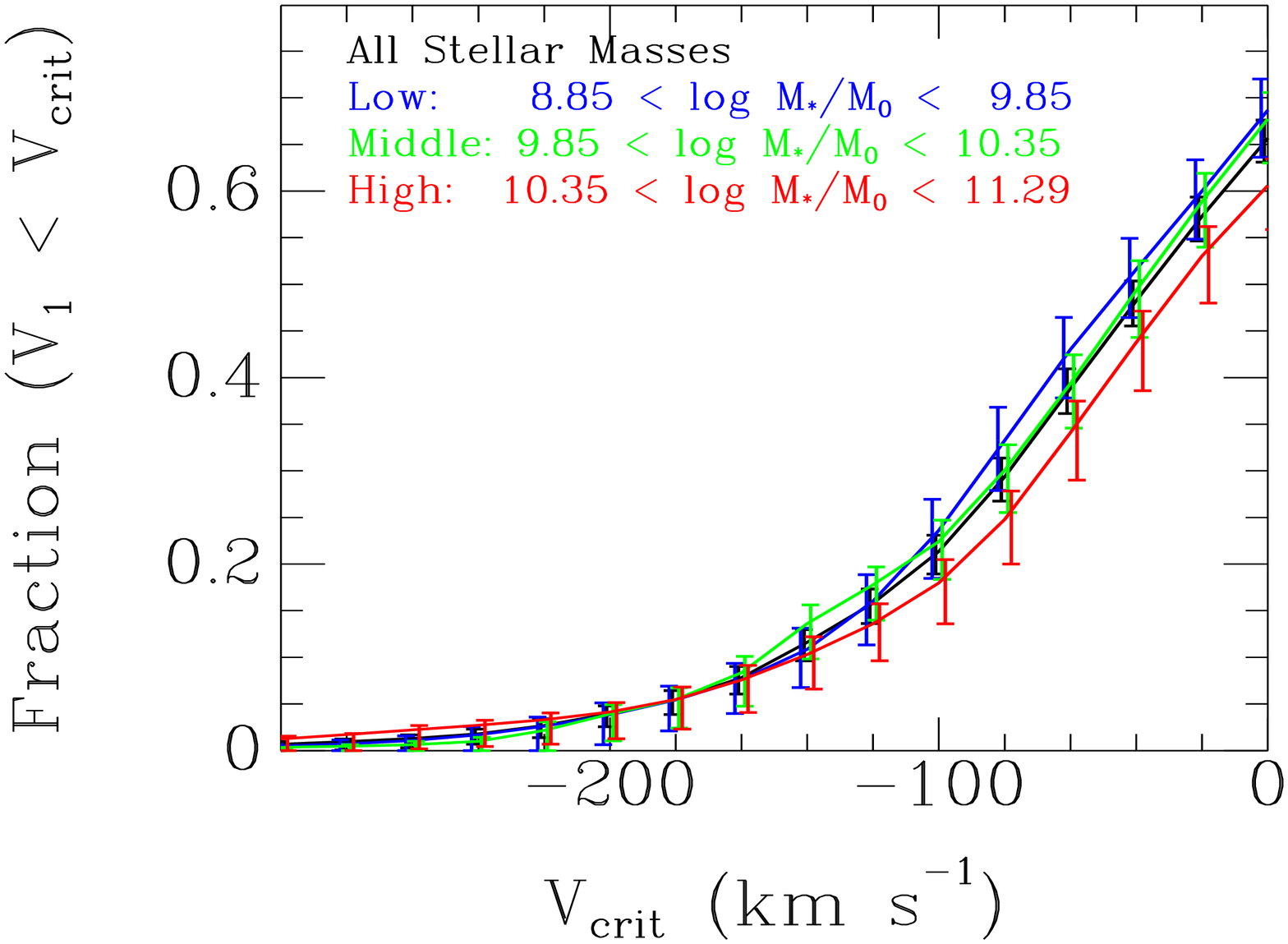} 
                \includegraphics[height=5.5cm,angle=-90,clip=true]{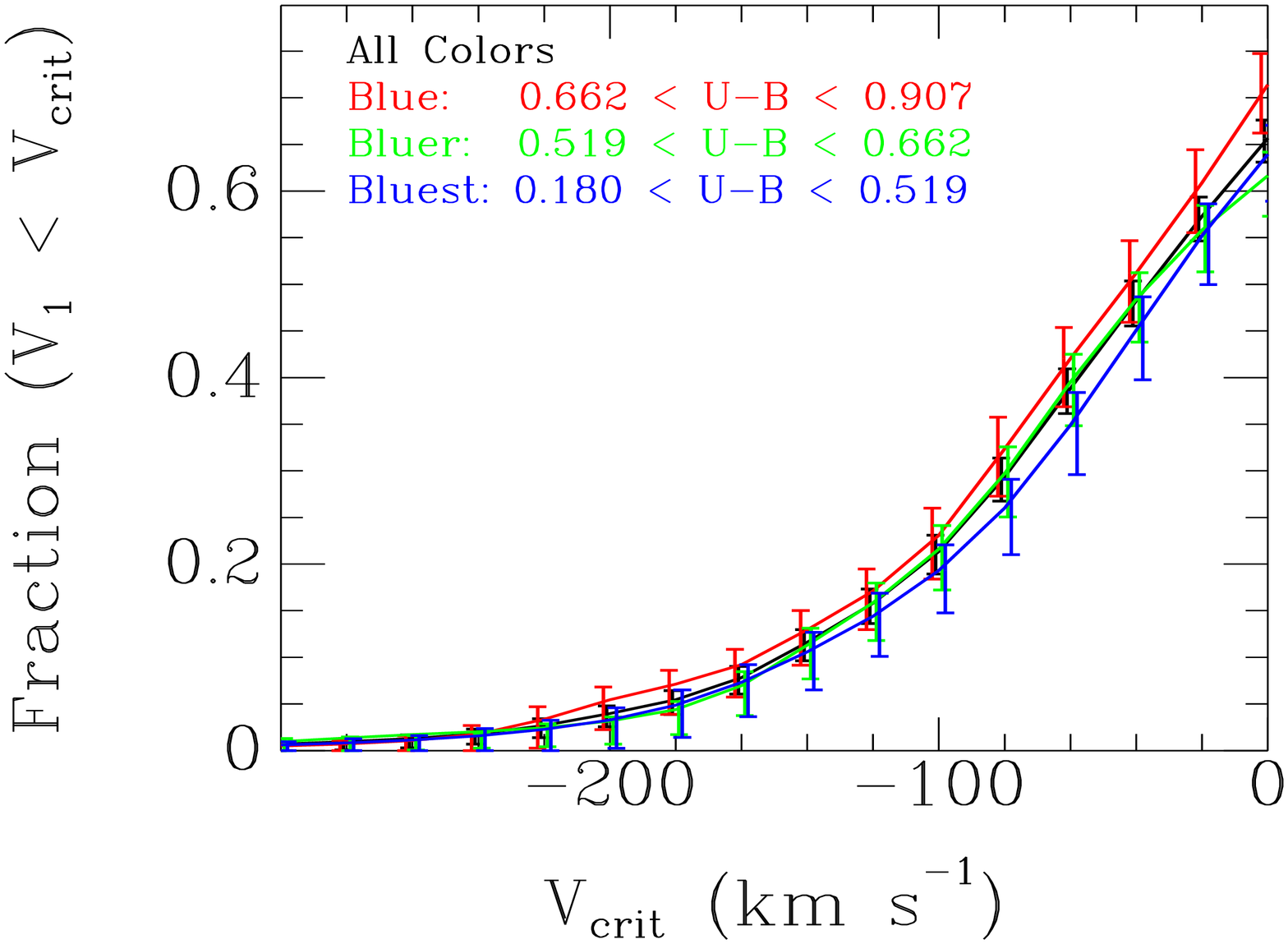} 
                 \includegraphics[height=5.5cm,angle=-90,clip=true]{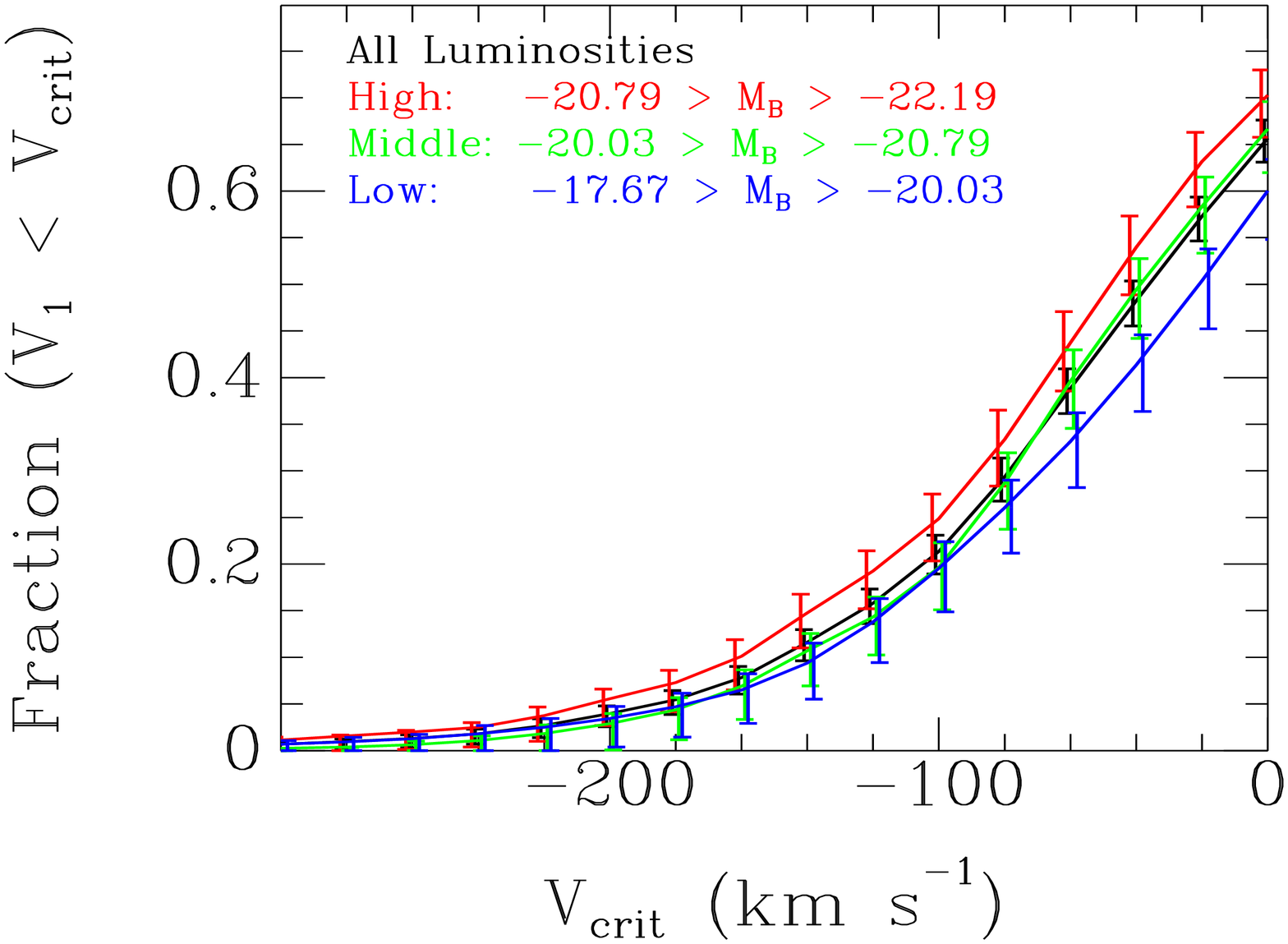}  
                  \hfill}
  \hbox{\hfill \includegraphics[height=5.5cm,angle=-90,clip=true]{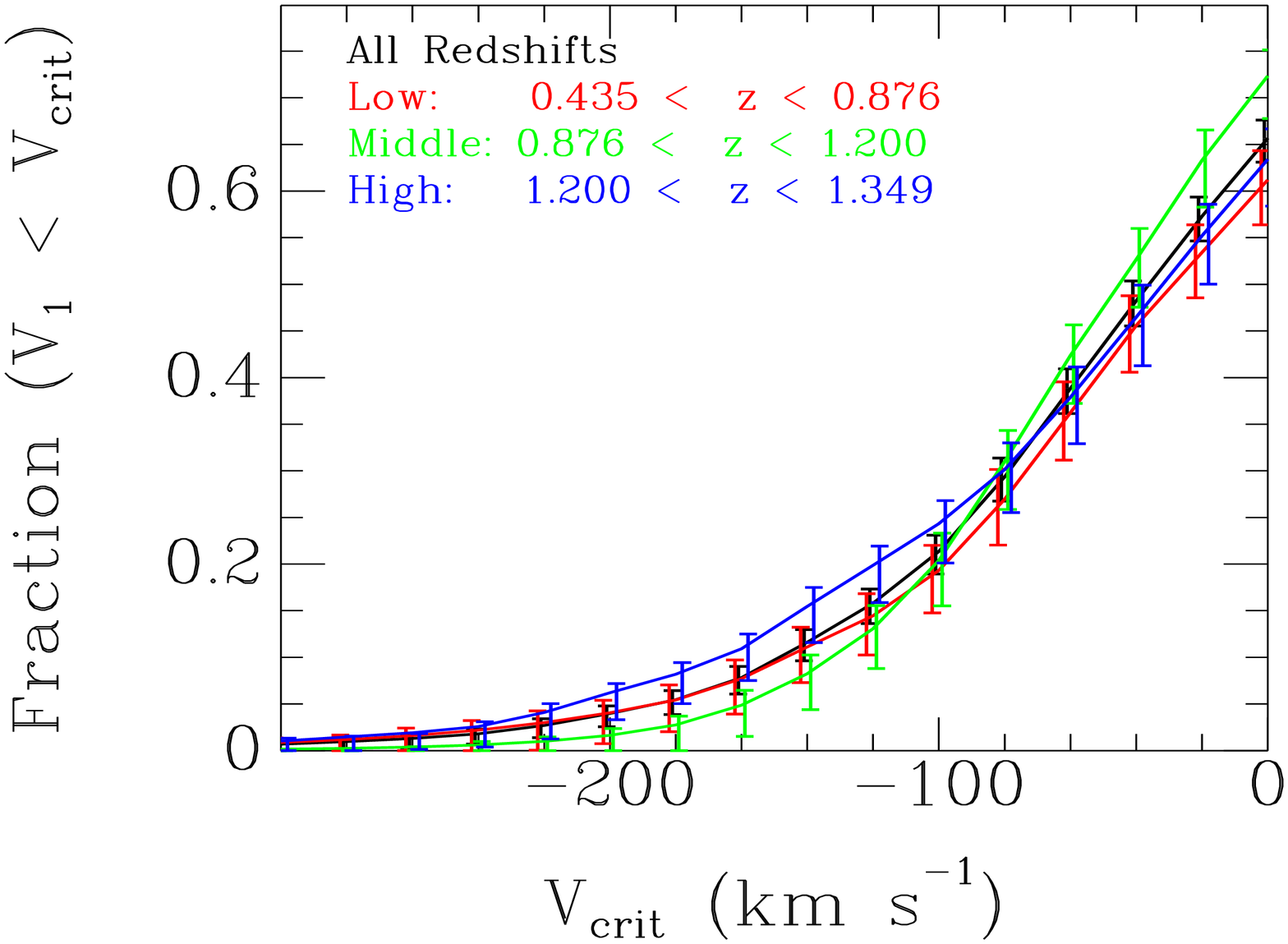} 
                \includegraphics[height=5.5cm,angle=-90,clip=true]{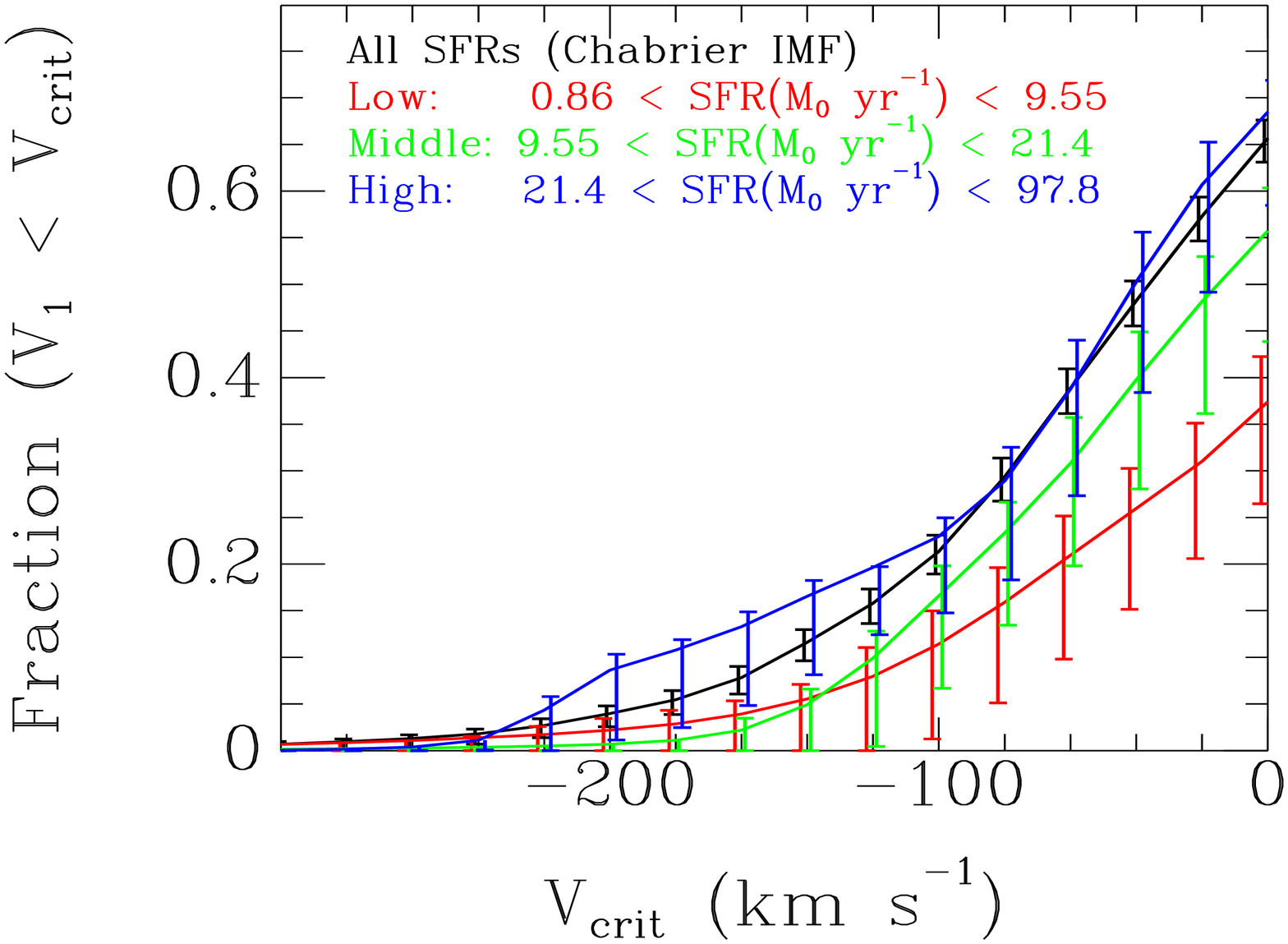} 
                \includegraphics[height=5.5cm,angle=-90,clip=true]{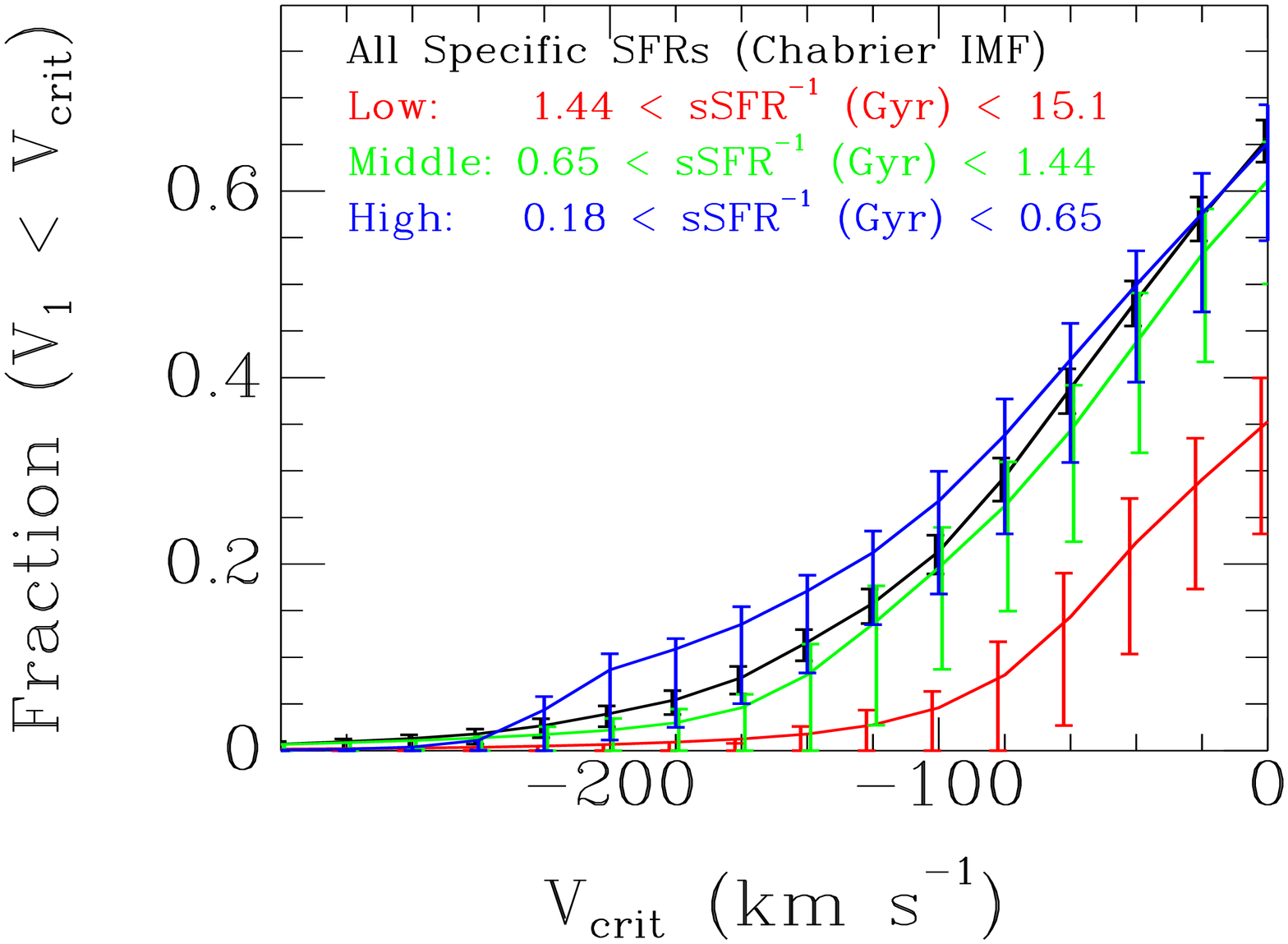} \hfill}
          \caption{\footnotesize
Fraction of galaxies with $V_1$ bluer than $V_{crit}$. This measure of the outflow
fraction shows no dependence on stellar mass (upper left) or color (upper middle).
The outflow fraction is slightly higher among higher luminosity galaxies (middle left) and
higher redshift galaxies (middle right). Over the velocity range $ -180 < V_1 ({\rm km s}^{-1})
< -100$, the outflow fraction among the galaxies in the highest SFR tertile 
($21 < SFR(\msunyr) < 98$) is 
about a factor of four higher than the outflow fraction measured  among the galaxies in the 
lowest SFR tertile ($0.9 < SFR(\msunyr) <  10$). Among the galaxies with the lowest specific star 
formation rates ($-10.2 < \log\ sSFR(yr^{-1}) < -9.16$), outflows appear to be strongly suppressed
at all velocities relative to the average. The outflow fraction is enhanced by a factor
of 1.8 over the average among the tertile of galaxies with the highest sSFRs 
($-8.82 < \log\ sSFR(yr^{-1}) < -8.25$). 
The calculation of the outflow fraction follows the methods outlined in Appendix~\ref{sec:fout_calc}. 
}
 \label{fig:fout_fe2} \end{figure*}

\begin{figure*}[t]
  \hbox{\hfill   \includegraphics[height=18cm,angle=-90,clip=true]{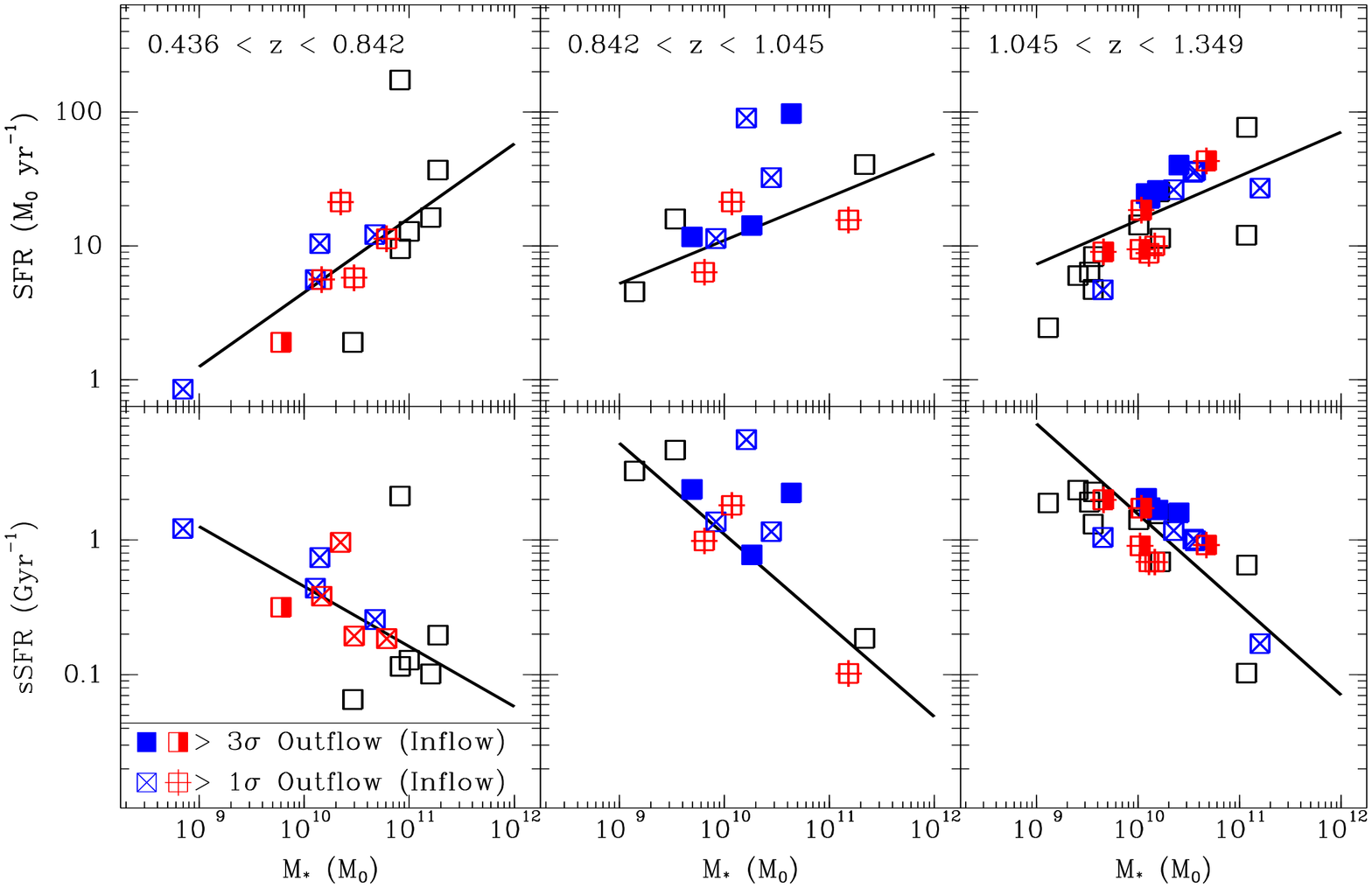}
          \hfill}
          \caption{\footnotesize
Location of galaxies with \feII\ blueshifts in the SFR-\mstar\ and  specific SFR -- stellar
mass diagram. The UV-corrected SFRs have been divided by 1.8 to put both the SFR and 
stellar mass on a Chabrier IMF.  Within each redshift tertile, the bold line
marks the main sequence of star-forming galaxies fitted to the full sample of
DEEP2 galaxies with GALEX photometry; we excluded galaxies with $U-B$ redder 
than the green valley, $U-B > 1.135 - 0.032 (M_B + 21.18)$, from the fit.
The slope of the best fit flattens a bit
from the lowest redshift data ($\log \langle SFR \rangle = 0.5419 \log \langle M_* \rangle - 4.7255$) 
through the middle redshift tertile
($\log \langle SFR\rangle = 0.4020 \log \langle M_* \rangle - 2.9337$)
to the highest redshift spectra
($\log \langle SFR\rangle = 0.2975 \log \langle M_* \rangle - 1.7488$).
The galaxies with the most robust \feII\ blueshifts (solid blue squares) all lie on or
above the main sequence.
}
 \label{fig:mmUV_flow} \end{figure*}

\subsubsection{Dependence of Outflow Fraction on Galaxy Properties} \label{sec:outflow_fraction}

We apply the computational method introduced here to compare the outflow
fraction among galaxies with different properties. 
We divide our sample into tertiles (thirds) by redshift, mass, color, luminosity, and SFR.
To obtain insight about the importance of starbursts, we also normalize the SFR by 
the stellar mass; this specific SFR is the reciprocal of the timescale required to 
assemble the stellar mass at the current SFR.
We present the fraction of galaxies with a blueshifted \feII\ 
centroid velocity and then compare to the fraction of galaxies with 
significant blue absorption wings in \mgII. It is important to keep in mind that 
all the galaxies in our sample are star-forming galaxies; the post-starburst, green 
valley, and red sequence galaxies are excluded from this analysis.

The galaxy properties measured for DEEP2 galaxies were 
a primary motivation for our study. 
Stellar mass is perhaps the most fundamental 
property measured for the entire sample; and relative masses are determined to within 
a factor of two (Bundy \et 2006). Measurements of $B$-band luminosity and $U-B$ color are 
described in Willmer \et (2006) for the full sample. Star formation rates have been measured 
for 51 galaxies in our LRIS sample. In Kornei \et (2012), we show that these extinction-corrected 
SFRs (derived from GALEX imaging of the AEGIS field)
agree with the sum of the SFRs derived from uncorrected UV and 24\um\ photometry.

In \fig~\ref{fig:fout_fe2}, we show the fraction of spectra with $V_1$ bluer than 
$V_{crit}$. The outflow fraction measured this way is clearly independent of stellar
mass. The outflow fraction is enhanced among galaxies in the highest tertile by SFR 
(relative to the lowest tertile). The largest variation with galaxy properties, however,
is seen between the tertiles with the highest and lowest specific SFRs. 
Outflows faster than roughly 100\kms\ are nearly absent at low specific
SFR, $ -10 < \log sSFR ~(\mbox{yr}^{-1}) < -9.15$. Among galaxies with high specific SFR, 
$ -8.8 < \log sSFR ~(\mbox{yr}^{-1}) < -8.2$, the fraction of galaxies with \feII\ Doppler shifts 
(from $-200$ to $-100$\kms) is boosted by up to a factor of 1.8 relative
to the entire subsample with measured SFR. 

In the LRIS sample, $B$-band luminosity is more significantly 
correlated with SFR (6.1 $\sigma$ from the null hypothesis, correlation coefficient
$r_S = -0.87$) than is any 
other galaxy property.\footnote{
       A weaker correlation of somewhat lower significance is found between 
       SFR and stellar mass (4.3 $\sigma$ from the null hypothesis, $r_S = 0.61$).}
Hence, within the full LRIS sample, we might expect to find the outflow fraction
elevated among the more luminous galaxies. In \fig~\ref{fig:fout_fe2}, the outflow
fraction is indeed highest at all $V_{crit}$ for the most luminous galaxies,
but the significance of the distinction is hardly compelling ($\sles\ 1 \sigma$).

The high specific SFR galaxies have bluer colors on average (4.1 $\sigma$ from 
the null hypothesis, $r_S=-0.58$). Since we find a higher outflow fraction among
galaxies with high specific SFR, we might expect an enhanced outflow fraction 
among bluer galaxies in the full sample. From \fig~\ref{fig:fout_fe2}, we simply
note that any enhancement in outflow fraction with bluer color is weaker than the
marginal one seen with $B$-band luminosity. These results are consistent with
outflow fraction depending primarily on specific SFR, and color being strongly
dependent on galaxy luminosity and mass as well as specific SFR.
In light of the increasing outflow fraction with specific SFR,
the absence of any variation in the outflow fraction over the redshift range 
from 0.4 to 1.4 deserves further inspection because
galaxies have larger SFRs and specific SFRs at higher redshift. 

The three redshift tertiles for the LRIS sample with GALEX photometry 
are shown in the SFR -- stellar mass plane in \fig~\ref{fig:mmUV_flow}.
We show the {\it SFR - \mstar\ main sequence} (Noeske \et 2007;
Elbaz \et 2007) fitted to the GALEX-detected 
subsample of 6102 DEEP2 galaxies with stellar mass measurements. 
The specific SFRs of the galaxies with \feII\ blueshifts 
are $1\sigma$  larger than the average sSFR at the same stellar mass.  Furthermore,
the main sequence fitted to the GALEX-detected subsample is entirely consistent 
with our fit to the SFRs and stellar masses estimated for the full sample 
of blue cloud galaxies in DEEP2 (Mostek \et 2011).\footnote{
           Mostek \et (2011) use the AEGIS data to calibrate the results
           of SED fitting to optical photometry. Comparison of this SFR - stellar 
           mass relation to previous work is complicated by differences in both
           redshift range and methods used to derive stellar mass and SFR.
           For example, the Noeske \et (2007) relation was measured at
           lower redshift $0.2 < z < 0.7$, which may explain why our
           fit is about $\sim 0.2$~dex higher with similar slope. 
           For our subsamples at $0.842 < z < 1.349$, the SFR - stellar mass relation 
           in  DEEP2 is remarkably shallower in slope than the Elbaz \et (2007)
           relation fitted at $0.7 \le\ z \le\ 1.2$.  
           }
We can therefore conclude that the galaxies with \feII\ blueshifts have
slightly higher sSFR than the typical blue cloud galaxy.

Among our subsample of galaxies with measured SFRs,
only galaxies with $sSFR > 0.8 $~Gyr$^{-1}$ show secure, $3\sigma$ blueshifts.
Since the slope of the main sequence is less than unity, the typical sSFR 
falls gradually with increasing stellar mass and this may explain why
the \feII\ blueshifts will be detected up to larger stellar masses at higher 
redshift. A sSFR threshold for \feII\ blueshifts around  0.8~Gyr$^{-1}$ would 
explain the paucity of $3\sigma$ blueshifts in the low redshift tertile. 
Inspection of \fig~\ref{fig:mmUV_flow}, however, shows that not all galaxies 
with high sSFRs have blueshifted \feII\ absorption in their spectra. Hence,
it will be important to measure SFRs for the remainder of the LRIS sample and
determine whether the minimum sSFR criterion remains a necessary 
condition for seeing blueshifted \feII\ absorption in galaxy spectra.

\begin{figure*}[t]
  \hbox{\hfill  \includegraphics[height=5.5cm,angle=-90,clip=true]{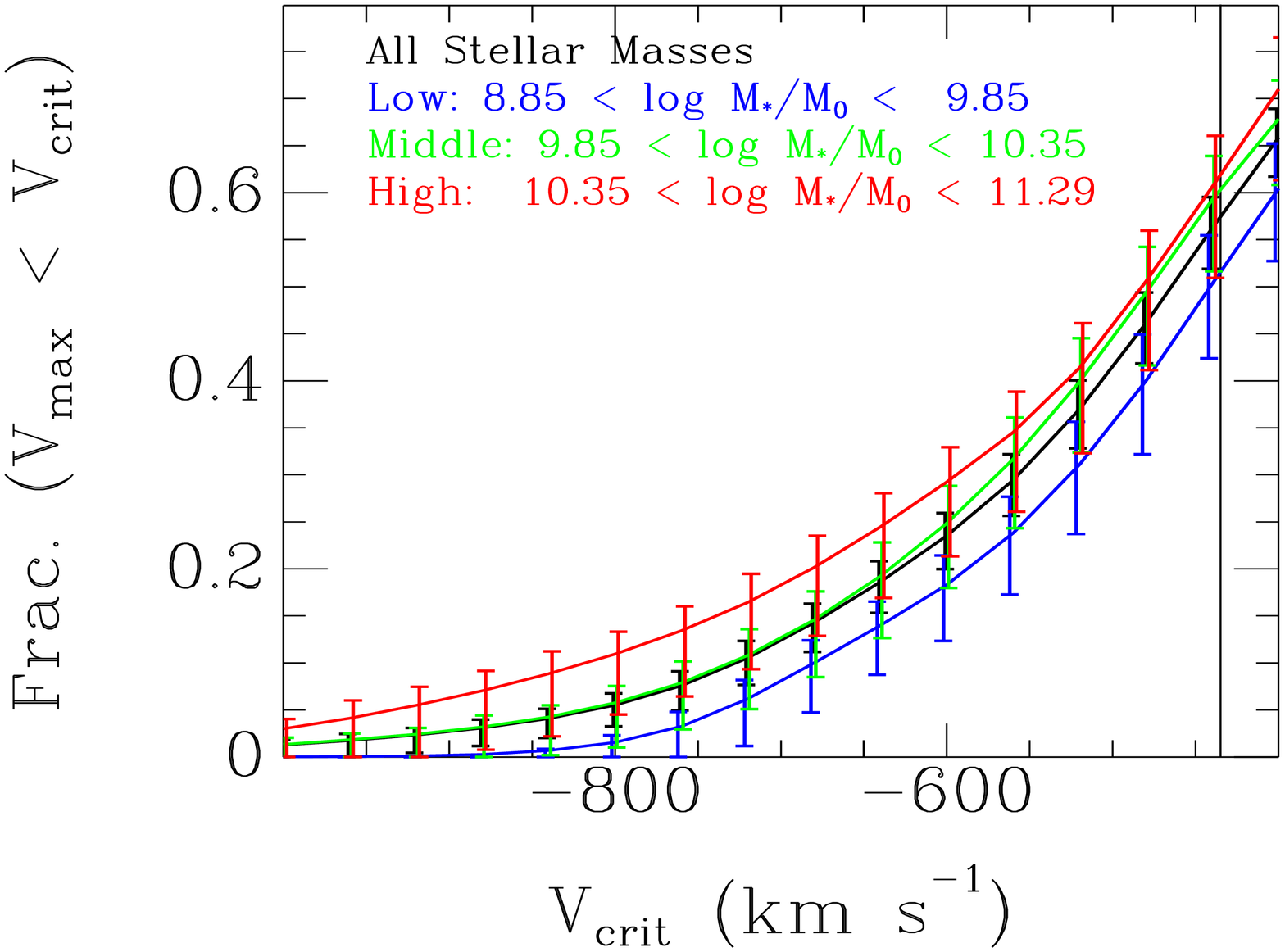} 
                 \includegraphics[height=5.5cm,angle=-90,clip=true]{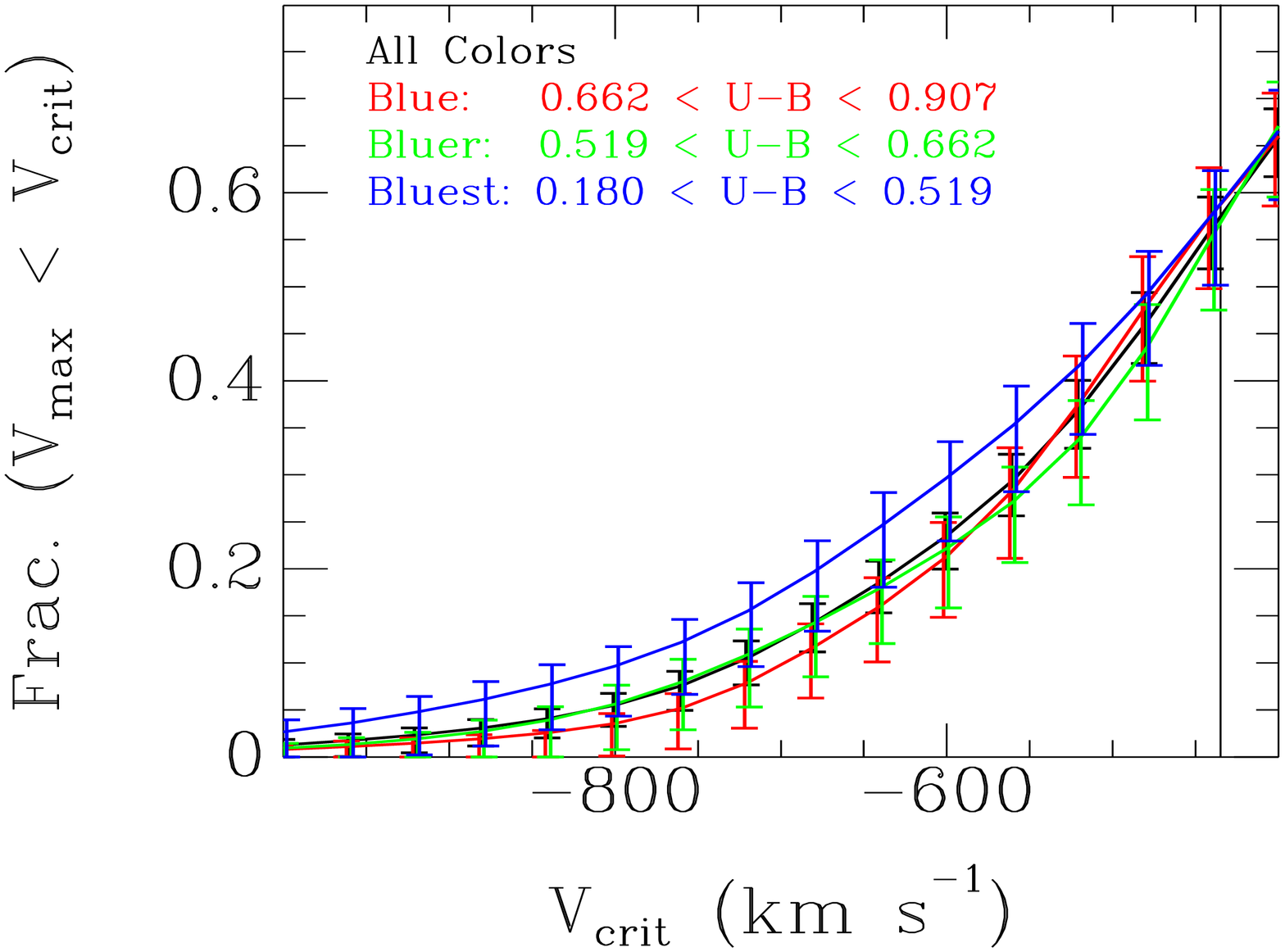} 
                  \includegraphics[height=5.5cm,angle=-90,clip=true]{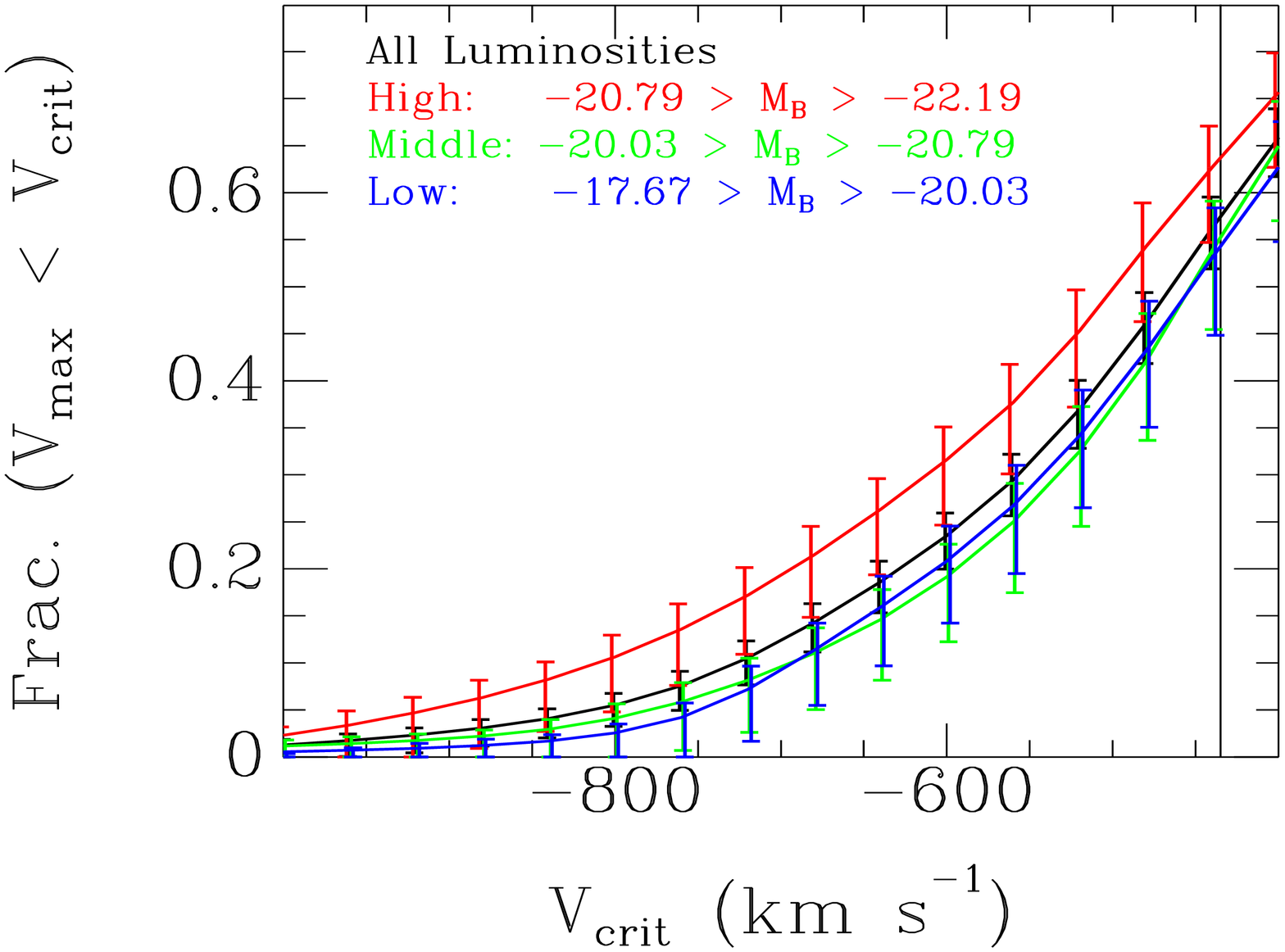} 
                   \hfill}
  \hbox{\hfill \includegraphics[height=5.5cm,angle=-90,clip=true]{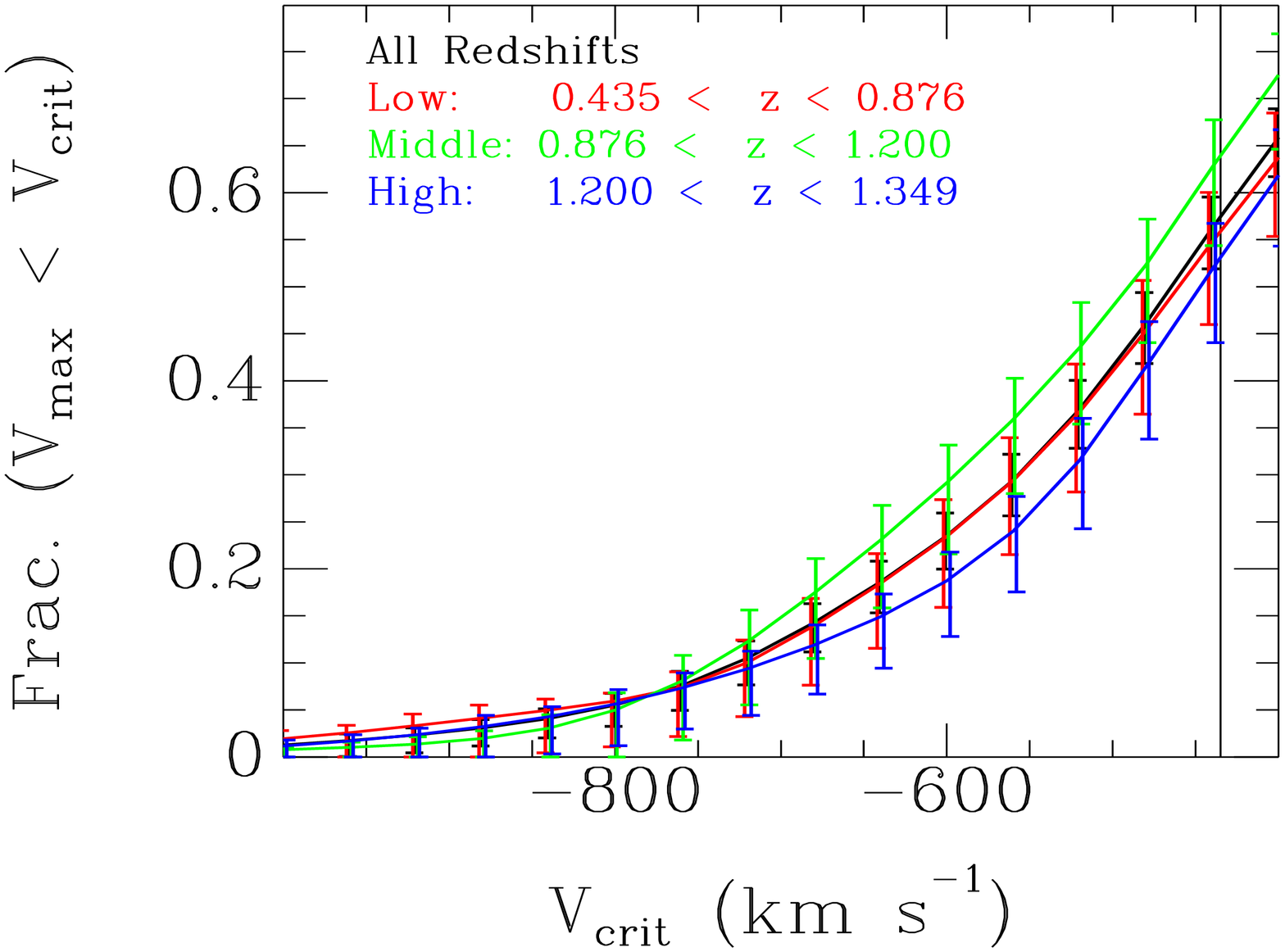}  
                \includegraphics[height=5.5cm,angle=-90,clip=true]{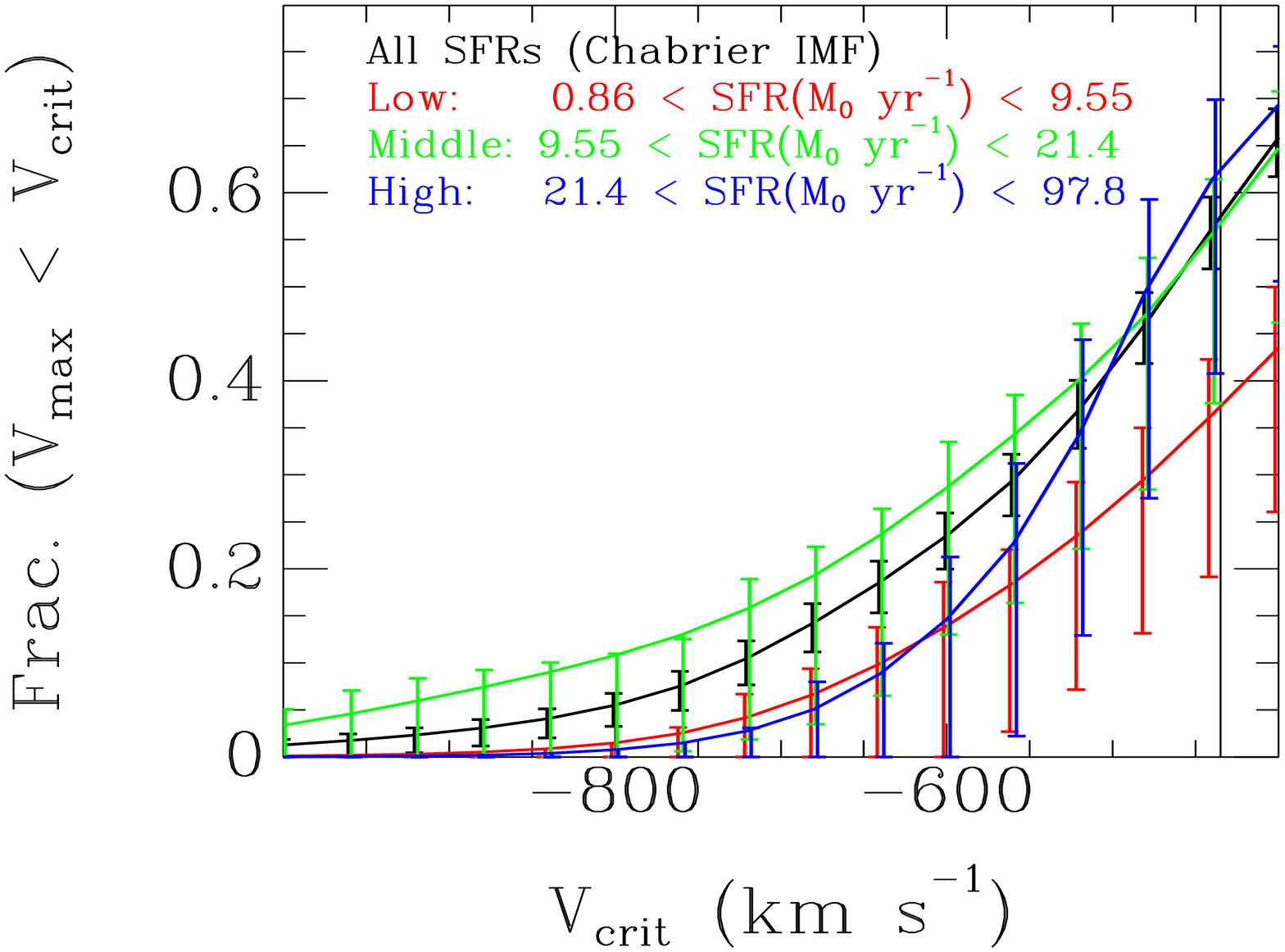} 
                 \includegraphics[height=5.5cm,angle=-90,clip=true]{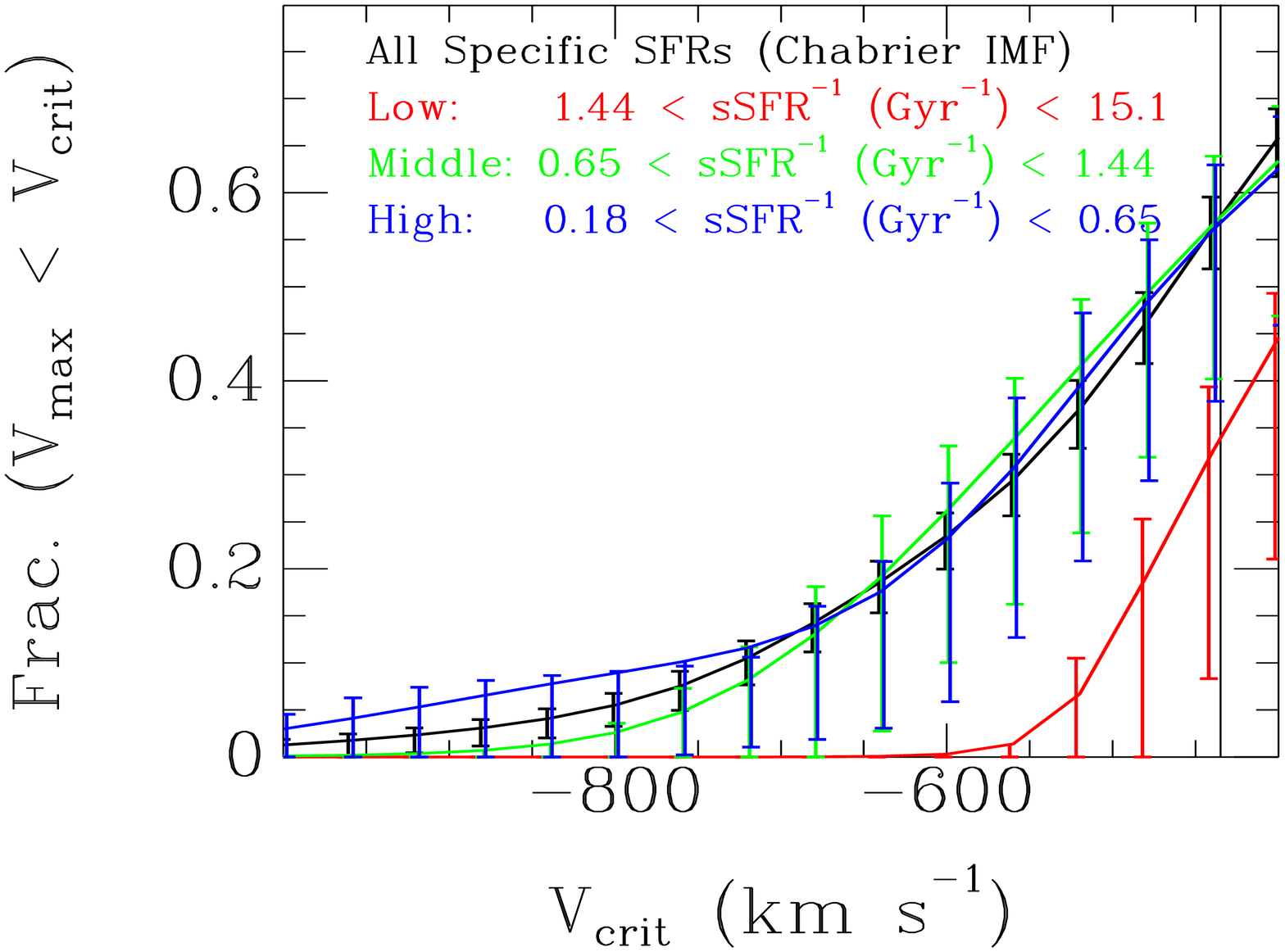} 
                  \hfill}
          \caption{\footnotesize
Fraction of galaxies with \mgII\ absorption detected at velocities larger (bluer)
than a velocity $V_{crit}$. The black curve shows the entire sample with \mgII\ 
measurements; and the colored curves break this sample into thirds by galaxy properties. 
The $V_{max}$ measurements have not been corrected for instrumental smoothing
of the spectra, so the curves are only physically meaningful at blueshifts significantly 
larger than the width of the instrumental profile which is $-435$\kms\ (vertical line in 
each panel) in the \dlow\ spectra. A high velocity blue wing is most prevalent among spectra
of higher mass, higher luminosity, and redder galaxies and does not appear in spectra
of galaxies with the lowest sSFR.
}
 \label{fig:fout_vmax_sfr} \end{figure*}

Given the distinct shapes of the \feII\ and \mgII\ absorption troughs in some
spectra, we compare the demographics of galaxies with blueshifted
\feII\ absorption and those with highly blueshifted \mgII\ absorption.
Figure~\ref{fig:fout_vmax_sfr} shows the fraction of galaxies with \mgII\ 
absorption at blueshifts larger than a threshold velocity, $V_{crit}$.
Once again (as for \feII\ blueshifts), galaxies with high specific SFR appear
more likely to show highly blueshifted \mgII\ absorption in their spectra;
the fraction of spectra with highly blueshifted
\mgII\ absorption (as measured by $V_{max}(\mgII)$) shows no trend 
with redshift.
In contrast to the case for \feII\ blueshifts, however, the outflow fraction
derived from the \mgII\ blue wing shows not even a weak dependence on SFR alone.

While the outflow fraction curves in \fig~\ref{fig:fout_vmax_sfr} are highest
for galaxies with higher masses, higher luminosity, and redder color, the
enhancement with each of these properties is not statistically significant.
The larger error bars in \fig~\ref{fig:fout_vmax_sfr} relative to \fig~\ref{fig:fout_fe2} 
arise from not all spectra showing resolved \mgII\ absorption troughs (and therefore
no $V_{max}$ estimate). Within the measurement uncertainties, however, we conclude
that the fraction of galaxies with outflows as determined by $V_{max}(\mgII)$ appears
to be less dependent on SFR than the outflow fraction determined by $V_1(\feII)$.
Furthermore,  both of these outflow indicators are remarkably insenstive to other fundamental
galaxy parameters including stellar mass, $B$-band luminosity, color, and redshift.

\subsection{Physical Properties of Outflows} \label{sec:outflow_properties}

The properties of winds have generally been varied in cosmological simulations in
order to fit observations of fundamental galaxy properties such as stellar content, 
SFR, and metallicity evolution (Dav\'{e}, Oppenheimer, \& Finlator 2011a;
Dav\'{e}, Finlator, \& Oppenheimer 2011b). Direct measurements of outflow velocity
and mass flux are therefore of great importance for modeling galaxy formation and
evolution. Here we discuss the measured Doppler shifts and the \feII\ column densities.
We will illustrate plausible extrapolations to the wind velocity and mass flux but emphasize
that, in order to accurately model the interplay of the various gas phases, well-resolved
simulations of galactic outflows need to be projected into the quantities we observe.

The escape of metals from low-mass galaxies in the local universe has been
established empirically (Martin 1999; Martin \et 2002; Tremonti \et 2007). This
enrichment of the circumgalactic medium by winds probably started very early in cosmic 
history based on the properties of metal-line systems at high-redshift 
(Meyer \& York 1987; Lu 1991; Songaila \& Cowie 1996; Ellison \et 2000; Schaye
\et 2003; Simcoe \et 2006; Martin \et 2010; Simcoe  2011). The main question is
whether these outflows transport substantial mass. The winds clearly affect the
chemical evolution of galaxies, but their role (if any) in creating the observed
baryon deficit remains unclear (e.g., McGaugh \et 2010).

Since we use low-ionization metal lines to identify the blueshifted absorption, our survey 
does not probe the hot phase of a galactic wind. The low-ionization gas is entrained in the 
hot wind by the breakup of supershells (Fujita \et 2009), the shear between the free wind and
the galactic ISM (Heckman \et 2000), and pre-existing, interstellar clouds
(Cooper \et 2008, 2009). The relationship between this warm gas and hot wind fluid is 
critical to any extrapolation of our measurements to the wind properties. In the recent 
Hopkins \et (2012) simulation of a high-redshift galaxy (which has properties typical of 
our sample) however, the warm, low-ionization gas carries the bulk of the 
outflowing gas mass; hence we may be observing the dominant phase of the wind (by mass). 
The mass-loss rates derived from low-ionization gas should still be viewed as
lower limits on the mass flux in the multiphase wind; for example, inspection of the other
types of galaxies in Figure~4 of Hopkins \et (2012) shows that the hot wind fluid can carry 
most of the mass at some outflow velocities.

\subsubsection{Outflow Solid Angle} \label{sec:opening_angle}

The measurements presented in Section~\ref{sec:outflow_fraction}
provide a statistical characterization of which galaxies have spectra with blueshifted
low-ionization absorption. Among the various velocity measurements discussed,
we have argued that the net Doppler shift of the \feII\ lines provides the 
most robust indication of a net outflow of gas along the sightline. The fraction of 
spectra with blueshifted \feII\ absorption decreases towards larger Doppler shifts.
Possible origins for this trend include variations in outflow velocity with galaxy 
properties, a velocity-dependent covering fraction,
or an angular dependence in outflow velocity generated  by the forces collimating
the outflow.

We can distinguish among these scenarios using the demographics of the sample.
For example, some galaxies with high SFR have spectra which do not show a net
blueshift of the resonance absorption; and we will show in Section~\ref{sec:scaling_relations} 
that the blueshifts do not vary strongly with any galaxy property. Hence, we
find no indication that the small fraction of galaxies with the largest blueshifts
have unique physical properties and conclude that the outflow velocity does not vary
strongly within our sample of blue galaxies. Furthermore, while gas covering
fraction is likely velocity dependent as measured from resolved line profiles of 
lower redshift galaxies (Martin \& Bouch\'{e} 2009), gas covering fraction does
not simply explain why galaxies with essentially the same physical properties 
differ as to whether their spectra show blueshifted absorption.
Over the full range in Doppler shift, the outflow fraction 
varies very little in any measured galaxy parameter; the largest variations in 
outflow fraction are only a factor of two even though the stellar masses of the
observed galaxies span more than 2 dex. We interpret this remarkable result as 
direct evidence for strong collimation of the ouflows at $0.4 < z < 1.4$; and we
predict that the spectra without \feII\ blueshifts select galaxies viewed along a sightline
perpendicular to the outflow direction. 

Support for this interpretation comes
from the inclinations of the AEGIS galaxies. In Figure~20 of Kornei \et (2012), we
show that the five spectra with  $3\sigma$ blueshifts of the \feII\ lines are
found for galaxies viewed  face-on, i.e. $i < 45\deg$. In this scenario, the outflow 
fraction may be interpreted in terms of the solid angle of the outflow.
For example, we consider a spherical outflow geometry.
Let $\theta_B$ be the half-angle of a biconical outflow, then the outflow subtends a solid
angle
\begin{eqnarray}
\Omega = 4\pi ({1 - cos~ \theta_B}),
\end{eqnarray}
and will be detected in a fraction $\Omega / 4 \pi$ of the galaxies we observe. 

We also find a decrease in the outflow fraction as the Doppler shift becomes bluer,
a result deserving further interpretation. We will simply note that 
if the covering factor of low-ionization gas decreases towards higher velocity,
then the absorption troughs develop shallow blue wings as demonstrated by
Martin \& Bouch\'{e} (2009) who had much higher resolution spectra of the \mgII\ 
doublet. The line profiles in the LRIS spectra lack the resolution required
to directly measure the velocity dependence of partial covering. We observe a
distinctly different property, a smaller fraction of the spectra show the 
centroid of the absorption troughs at higher outflow velocities. Since the
troughs are just as deep as the absorption lines detected at lower blueshifts (in 
spectra of other galaxies with similar properties), the velocity-dependence of the 
outflow fraction in \fig~\ref{fig:fout} is not simply explained in terms of the gas 
covering fraction. Instead, we interpret the observation as evidence that outflows 
have a smaller opening angle at higher velocity.  Smaller opening angle at higher 
velocity is qualitatively consistent with dynamical simulations of outflows where 
the opening angle is determined by the scale-height of the ISM (DeYoung \& Heckman 1994).

Our study provides this first evidence for highly collimated outflows at $z \sim 1$.
In \fig~\ref{fig:fout}, we show that the fraction of spectra with blueshifts 
increases as $V_{crit}$ decreases. Specifically, we measure an outflow fraction of 0.025, 
0.20, 0.45, and 0.65  at $V_{crit} = 200, 100, 50, {\rm ~and~} 0$\kms, respectively.  
The half-angle of the outflow cone would be 13\deg, 37\deg, 57\deg, and 70\deg
at velocities of 200, 100, 50, and 0\kms.  Hence our results indicate that
the outflows subtend a larger (smaller) solid angle at slower (faster) speeds.
Also, at all speeds, the outflow solid angle is much less than $4\pi$ steradians. 
While bipolar outflows are well documented at lower redshift (e.g., Heckman \et 1990; 
Chen \et 2010; Bouch\'{e} \et 2012), only recently has evidence for outflow collimation
emerged at redshifts from $0.09 < z < 0.9$ (Bordoloi \et 2011; Kacprzak \et 2012).

The polar angle of $\theta_B \approx 37$\deg\ for the outflow at 100\kms\ derived
from our results appears to be 
consistent with the distribution of outflowing gas in galactic halos. 
Bordoloi \et (2011) mapped the \mgII\ absorption at impact parameters (to 
background galaxies) within 40~kpc of star-forming galaxies at $0.5 < z < 0.9$.
Around blue, star-forming galaxies,  the \mgII\ absorption equivalent 
widths are significantly higher at small polar angles, near the minor axis,
than those measured closer to the major axis. Their results suggest the
outflow cone makes an angle of 45\deg\ or less with the minor axis of the galaxy.

\subsubsection{Mass Outflow Rates}  

We have nearly enough information to measure the mass flux in the low-ionization
outflow. As described in Section~\ref{sec:EW}, 
our measurements bound the \feII\ column density on both sides; this is a significant
improvement over previous studies that detected outflows only in saturated lines, which
yield only lower bounds. We have also shown that the largest Doppler shifts in \feII\ are
roughly 200\kms\ and provided some evidence that slower outflows may be simply 
viewed at higher inclinations (Kornei \et 2012); this result constrains the outflow solid angle,
$\Omega$. The ionization correction for \feII\ is uncertain
but seems likely to lie in the range $\chi(Fe^+) \equiv n(Fe^+)/n(Fe) = 0.1 - 1$
based on photoionization modeling (Churchill \et 2003; Murray \et 2007). In the halo
gas of the Milky Way, the measured depletion of iron onto grains is much lower than its value
in the cold disk; hence we conservatively take $\log d(Fe) \equiv [Fe/H] = -0.69$ from Table~6
of Savage \& Sembach (1996) as our best estimate of the depletion.
For a spherical flow launched at radius $R_0$, the mass flux is 
\begin{eqnarray}
\dot{M} = \Omega v R_0 N(H) \bar{m}, 
\label{eqn:mdot}  \end{eqnarray}
where the average mass per hydrogen atom is $\bar{m} = 1.4$~amu, $\Omega$ represents
the solid angle subtended by an outflow with spherical geometry, $v$ is the radial velocity
of the ouflow, and $N(H)$ the total hydrogen column density. The mass flux is
independent of the covering factor because covering factors less than unity increase
the inferred column density and decrease the solid angle $\Omega$ by the same factor.

The galaxy spectra do not uniquely constrain the location of the gas along the sightline.
In a radially diverging flow, the column density integral ``down the barrel'' will be
dominated by the densest gas near the launch radius; hence, the linear dependence of
the mass flux on the launch radius in Eqn.~\ref{eqn:mdot}. Simple models relate
$R_0$ to the size of the starburst region (Chevalier \& Clegg 1985) and/or 
several pressure scale heights in a gaseous disk (De Young \& Heckman 1994). We
therefore assume that $R_0 \sles\ 1$~kpc in order to illustrate the mass fluxes.
Some of the absorbing gas is clearly at much larger radii based on the detection of
spatially extended \mgII\ emission (Rubin \et 2011a; Erb \et 2012; Martin \et 2012b, in prep)
and intervening \mgII\ absorption at impact parameters $b \le\ 70$~kpc (Bordoloi \et 2011) .

Using the bounds on the \feII\ column from Table~\ref{tab:outflow}, we note
that the total ``ISM plus outflow'' column is less than $\log N(Fe^+) C_f \sles\ 15.8$
in half the outflow galaxies, and the highest upper limit is 16.27. Expressing
Eqn.~\ref{eqn:mdot} in terms of values consistent with the observations yields
\begin{eqnarray}
\footnotesize
\dot{M} = 23 \msunyr \left( \frac{\Omega}{\pi}  \right)
\left( \frac{v}{200 {\rm ~km~s}^{-1}}   \right) \nonumber \\
\times \left( \frac{R_0}{1~{\rm kpc}}   \right)
\left( \frac{N(Fe^+)}{10^{16} {\rm ~cm}^{-2}}   \right)
\left( \frac{3.16 \times 10^{-5}}{n(Fe)/n(H)}   \right) \nonumber \\
\times \left( \frac{0.5}{\chi(Fe^+)}   \right)
\left( \frac{0.20}{d(Fe)} \right),
\label{eqn:mdot_scaled} \end{eqnarray}
independent of the covering factor.
For comparison, the median SFR in the AEGIS subsample is 12.3\msunyr\ for a
Chabrier IMF (or, 21.9\msunyr\ for a Salpeter initial mass function), indicating
a mass loading factor $\eta \equiv \frac{\dot{M}}{SFR} \approx 1.9 (1.1)$.  A lower
ionization fraction would clearly increase our mass flux estimate. Taking  $\chi(Fe^+)
\approx 0.1$, for example, raises the implied mass loading factor by a factor of 5.
We will estimate lower mass fluxes, however, when a correction is made for the
contribution of interstellar gas to the absorption equivalent width. 

In Section~\ref{sec:measure_2comp}, we introduced a model that describes absorption
at the systemic velocity with a maximum ISM component.
We fit the typical properties of this ISM component and those of the Doppler component
to the \feII\ absorption. From Table~\ref{tab:outflow}, the
column density in the Doppler component is typically in the range $14.47 < \log N(Fe^+) C_f
< 15.30$, where the limits are the median values $\log N_{DOP}(Fe^+) C_f$. In the
mass flux estimate, the lower column density of the Doppler component is partially offset 
by its larger blueshift (relative to the single component fit). Deprojecting the line-of-sight
velocity based on the collimation discussed in Section~\ref{sec:opening_angle}, a typical
outflow velocity for the Doppler component is  $V_{Dop} = -476$~km~s$^{-1}$.
From Eqn.~\ref{eqn:mdot_scaled}, we estimate a mass flux in the Doppler component of
$\dot{M} = 2 - 11$\msunyr. For a Chabrier initial mass function, we obtain
$\eta \approx\ 0.2 - 0.9$, where $\eta$ would be 1.8 times lower for a Salpeter initial 
mass function.

Since previous mass-loss rates were
derived from lower limits on gas column density (from saturated lines),
these new results show, for the first time,
that the mass fluxes in the low-ionization gas are not orders of magnitude
higher than the typical SFR in these $z \sim 1$ galaxies. 
Future work should be able to combine our constraint from ``down-the-barrel'' observations,
absorption detections towards background quasars and galaxies, and the extent of
scattered resonance emission to improve the accuracy of the mass flux in the metal
ions.

\subsection{Outflow Scaling Relations} \label{sec:scaling_relations}

Empirical scaling relations for outflows help us to understand which physical
processes shape outflows. For example, for local starbursts, the nearly linear 
increase in the blueshift of \naI\ absorption with increasing rotation velocity 
(Martin 2005) motivated outflow models in which radiation pressure accelerates
the low-ionization gas (Murray \et 2005). Momentum-driven outflows, in general,
also predict an inverse scaling between the mass loading parameter 
and galaxy velocity dispersion, $\eta \equiv \dot{M}/SFR \propto \sigma^{-1}$ 
(Murray \et 2005; Oppenheimer \et 2010).

The evidence presented in this paper for collimated outflows complicates
testing the momentum-drive wind conjecture. The measured Doppler shifts
cannot be deprojected into the outflow velocities without knowledge of
the galaxy inclination. At this time, the best we can do is to look for
trends of the median Doppler shift with galaxy properties. We emphasize, however,
that high-resolution images of the sample would allow us to improve our
interpretation of the spectral measurements.

In contrast to studies of resolved galaxies at much lower redshifts, we have
good statistical constraints on the halo masses of the galaxies in the LRIS sample.
The halo masses are of interest because galaxy evolution models invoke
feedback to suppress star formation in halos less massive than a {\it mass floor} at
 $\log M_h/\msun\ \sles\ 11$ (e.g., Bouch\'{e} \et 2010)
and predict halo gas is mostly virialized (and therefore 
cold accretion and star formation suppressed) when $\log M_h/\msun\ \sgreat\ 12$,
the {\it mass ceiling}.
The observed clustering (of star-forming DEEP2 at $z \sim 1$ ) 
galaxies with $M_B -5 \log(h) < -20$, 
brighter than the faintest tertile of the LRIS sample, indicates a bias of 1.28   (Coil \et 2008).
If 15-20\% of those galaxies are satellites, the minimum halo mass 
is $\log M_h/\msun\ \sgreat\ 11.3$ (Zheng, Coil, \& Zehavi 2007), and the mean halo mass is
$\log M_h/\msun\ \approx 12.0$. Roughly one-third of the galaxies populate halos where
gas accretion should occur primarily by hot-mode accretion. Our survey is sensitive
to changes in outflow properties below the mass floor and above the mass ceiling.

\subsubsection{Outflow Velocity vs. Stellar Mass}

\begin{figure*}[t]
  \hbox{\hfill  \includegraphics[height=5.5cm,angle=-90,clip=true]{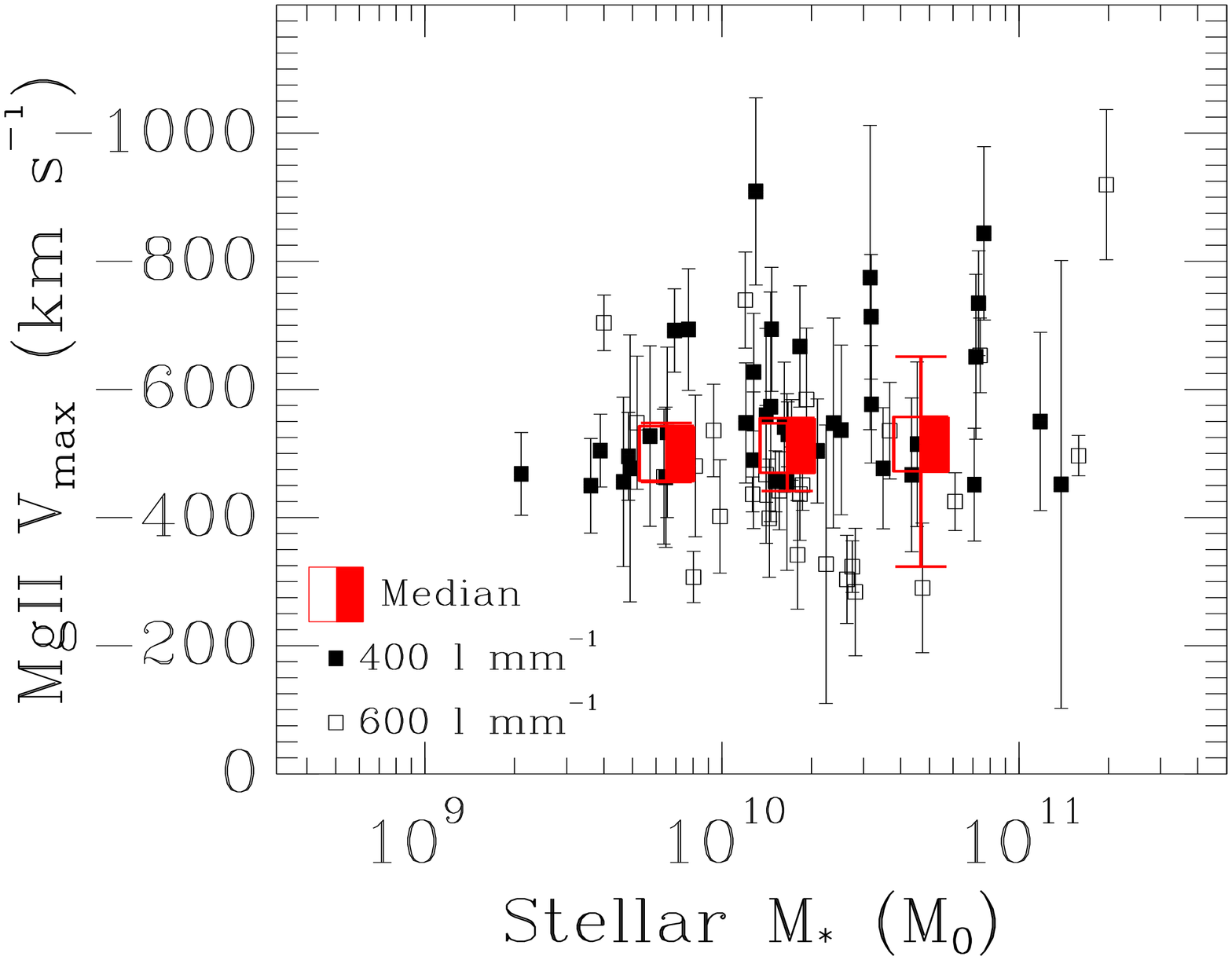} 
                 \includegraphics[height=5.5cm,angle=-90,clip=true]{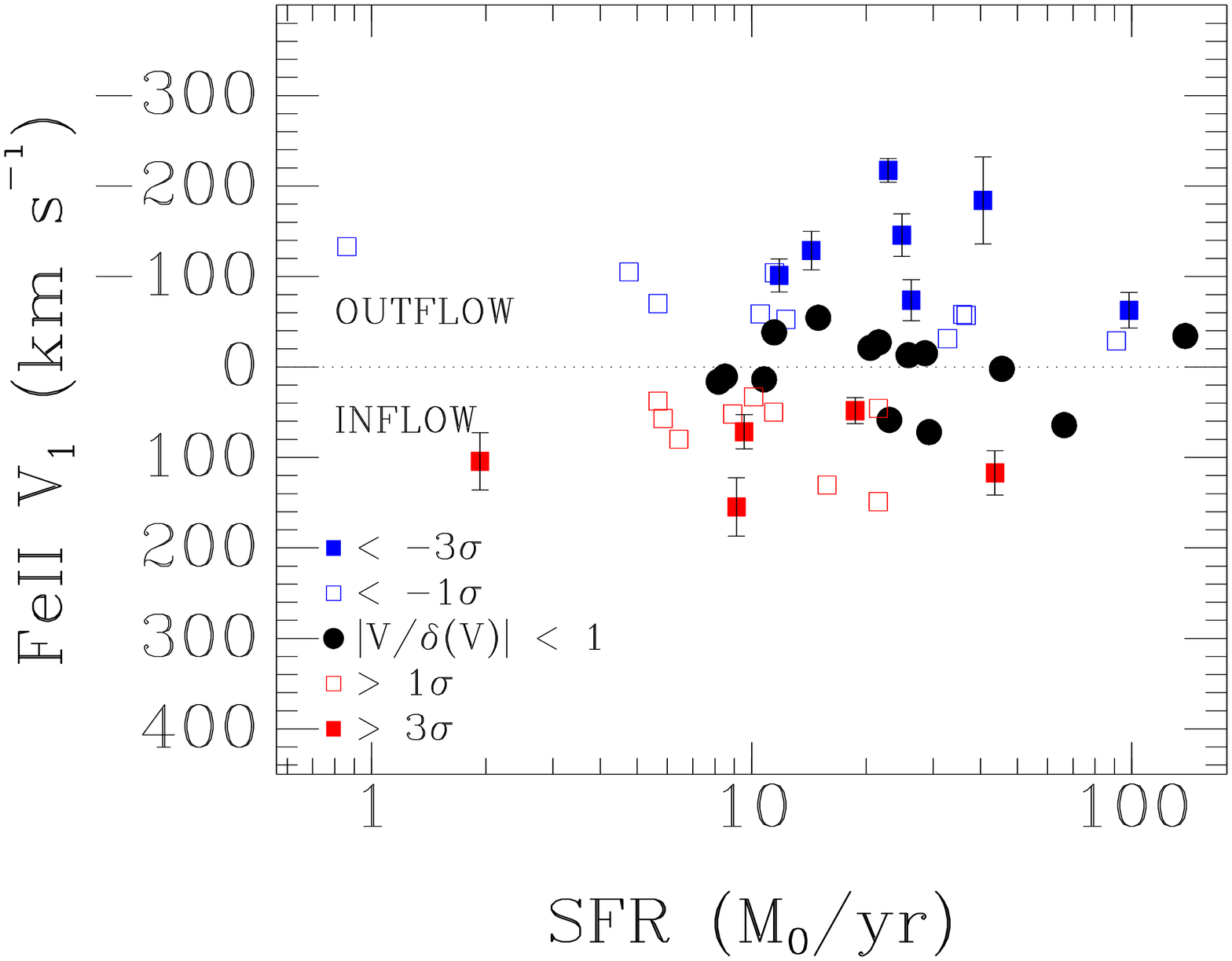} 
                  \includegraphics[height=5.5cm,angle=-90,clip=true]{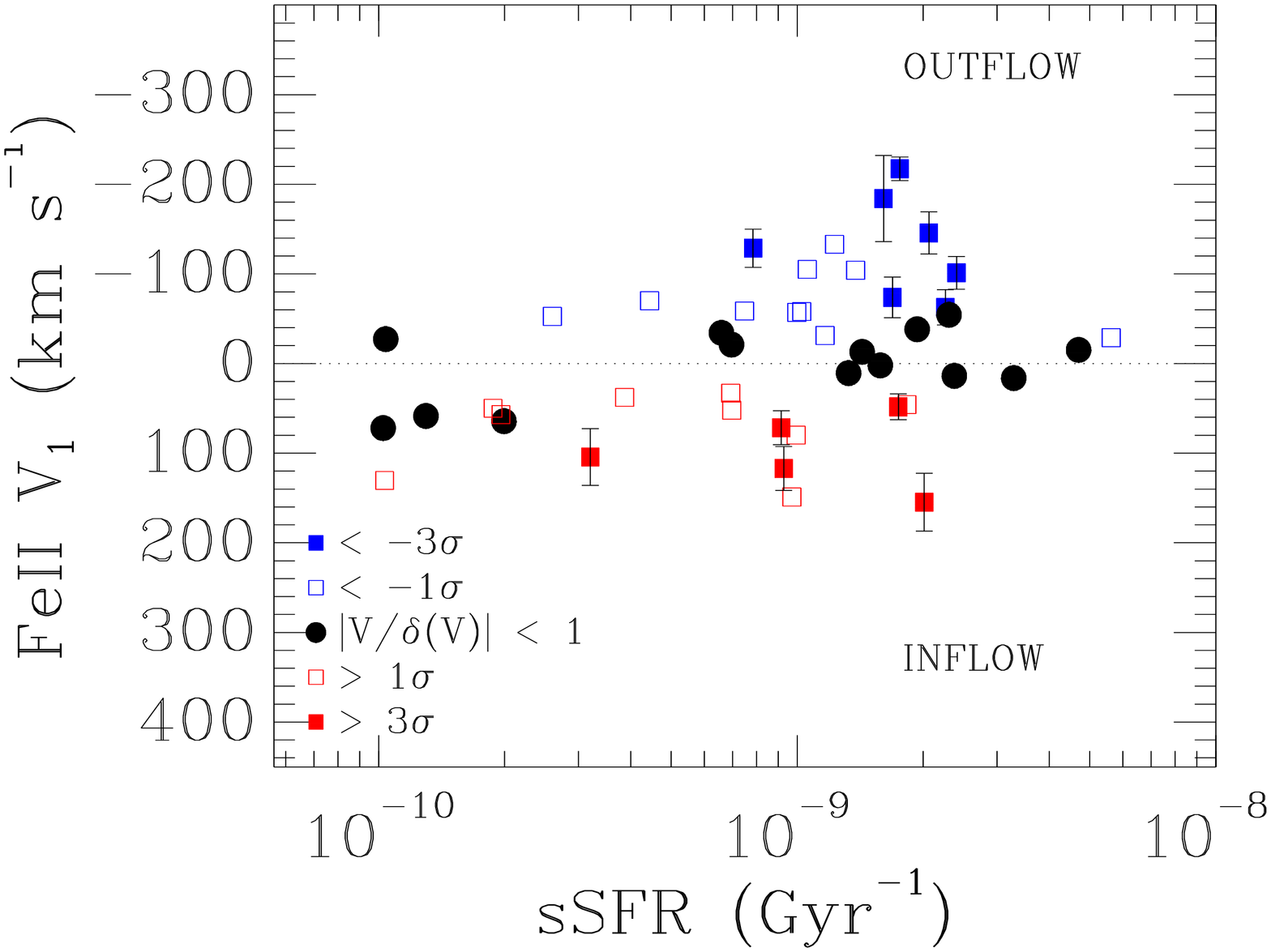}
                   \hfill}
 \hbox{\hfill \includegraphics[height=5.5cm,angle=-90,clip=true]{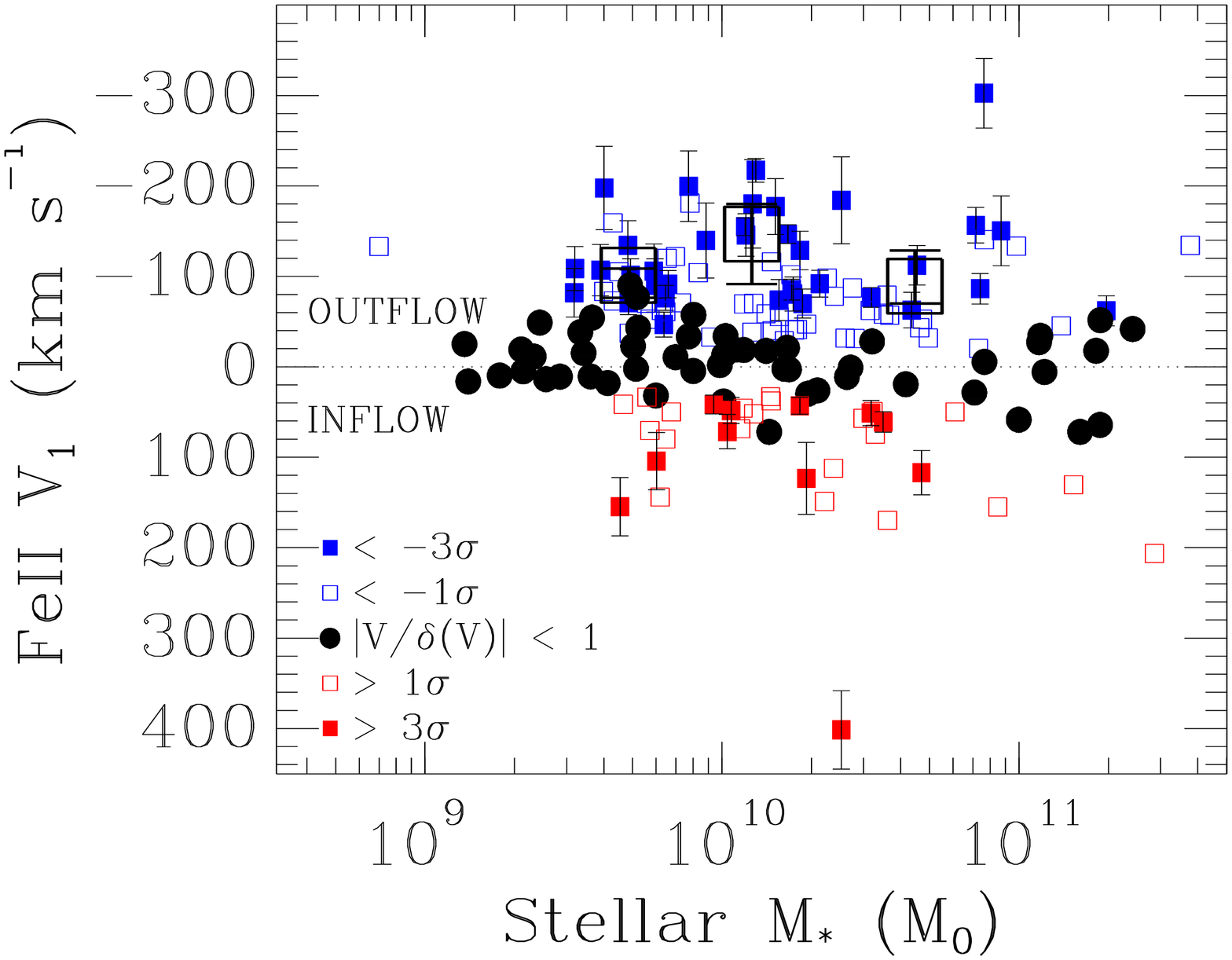}
               \includegraphics[height=5.5cm,angle=-90,clip=true]{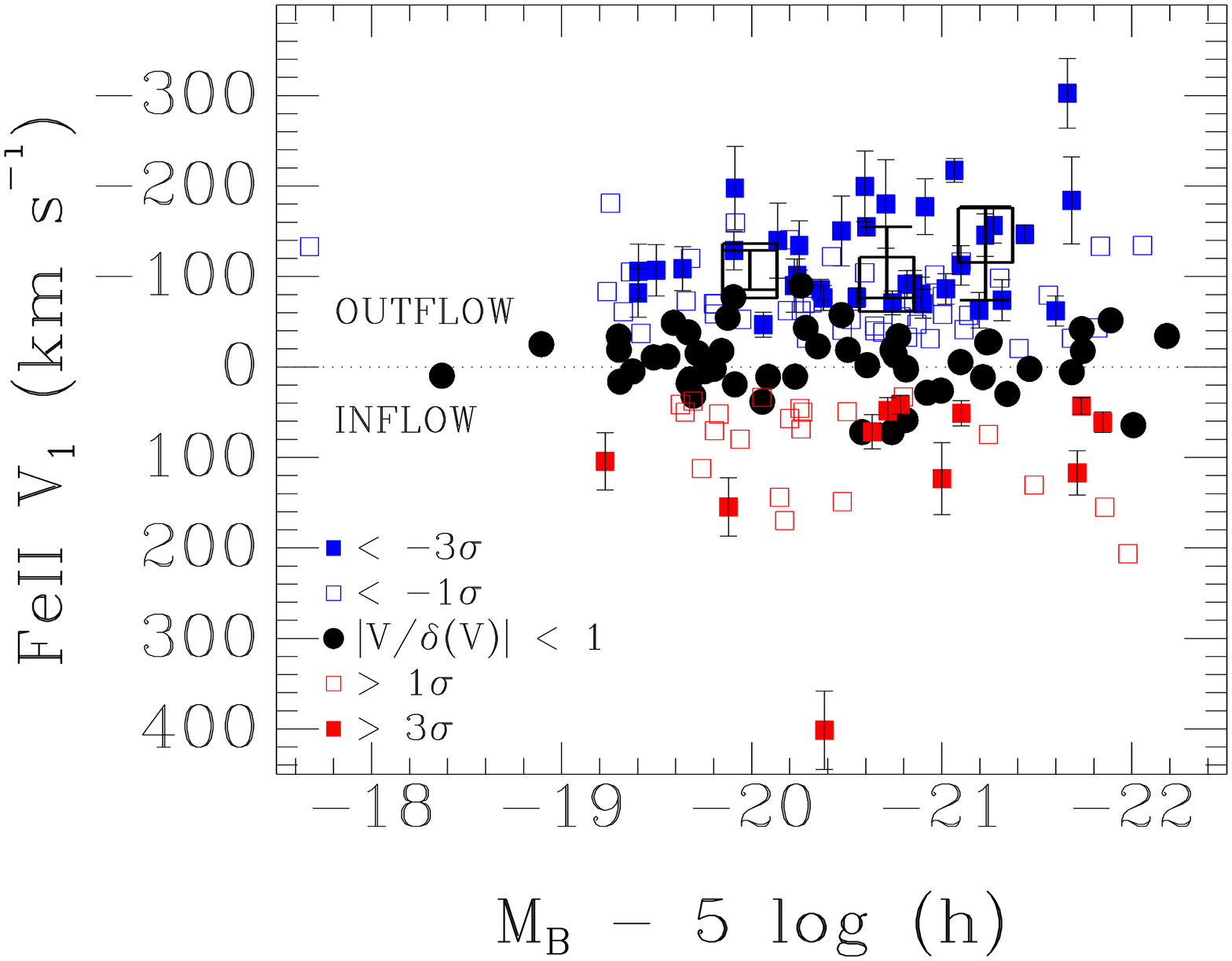} 
                \includegraphics[height=5.5cm,angle=-90,clip=true]{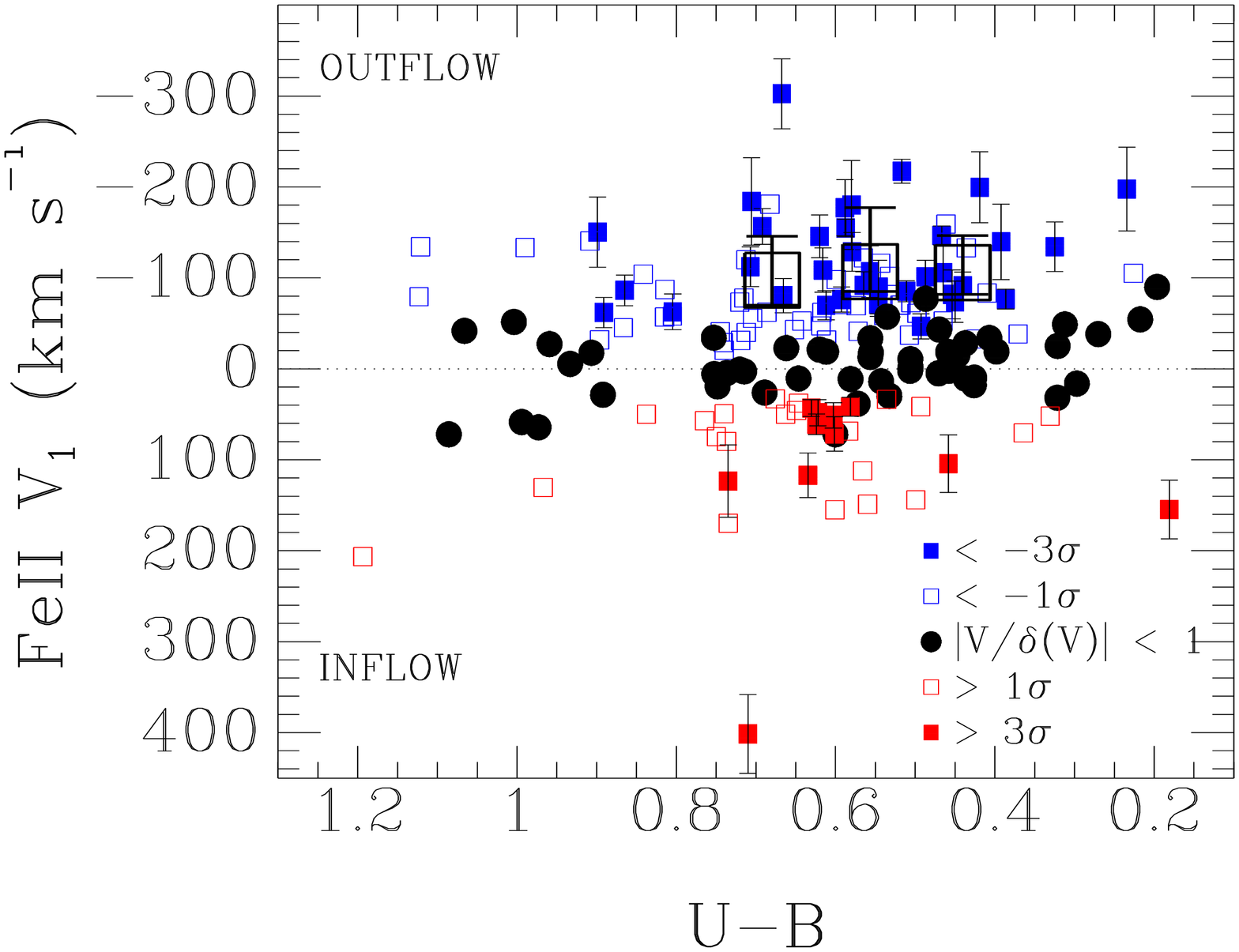}
                \hfill}
          \caption{\footnotesize
Range of Doppler shifts measured vs. galaxy properties.
The outflow velocities projected along our sightline show no correlation with 
stellar mass, SFR, specific SFR, $B$-band luminosity, or $U-B$ color.
{\it Top Left:}
Maximum blueshift of \mgII\ $ \lambda 2796$ absorption trough vs. stellar mass. 
Large symbols mark the median values and the $\pm34\% $ range.
{\it Bottom Left:}
Doppler shift of \feII\ absorption troughs vs. stellar mass.
Large symbols denote the median values of the spectra with significant ($> 3\sigma$)  blueshifts.
{\it Top Center:}
Doppler shift of \feII\ absorption troughs vs. star formation rate.
{\it Bottom Center:}
Doppler shift of \feII\ absorption troughs vs. $B$-band luminosity.
Large symbols denote the median values of the spectra with significant ($> 3\sigma$)  blueshifts.
{\it Top Right:}
Doppler shift of \feII\ absorption troughs vs. specific SFR.
{\it Bottom Right:}
Doppler shift of \feII\ absorption troughs vs. U-B color.
Large symbols denote the median values of the spectra with significant ($> 3\sigma$)  blueshifts.
}
 \label{fig:scaling} \end{figure*}

In their study of composite spectra, Weiner \et (2009) found that the Doppler
shift of the most blueshifted \mgII\ absorption increased with stellar mass 
to the power 0.17. They measured the Doppler shift $V_{10\%}$ at 90\% of the 
continuum intensity, which corresponds closely to our $V_{max}$ measurement 
for the median continuum S/N ratio of the LRIS spectra. The $V_{max}(\mgII)$ values measured for the 
LRIS sample, however, do not increase with $M_*$, as illustrated in \fig~\ref{fig:scaling}.
We found that $V_{max}(\mgII)$ does increase with $M_*$ when our 
analysis includes the K+A and green valley galaxies, which have high stellar masses
and older stars (contributing broad, stellar \mgII\ absorption). 
One reason our results may differ is that our \dlow\ spectra do not 
resolve the absorption troughs at blueshifts less than 435~km~s$^{-1}$, so  
we will overestimate the median velocity when a large fraction of spectra
have blue wings just below that resolution limit. To test this idea, we
identify the \dhigh\ spectra by open squares and see that these $V_{max}$ 
values become less blueshifted with increasing $M_*$ in \fig~\ref{fig:scaling}.
Furthermore, our measurements for the \feII\ centroid
velocity $V_1$, which are not affected much by the resolution, do not
vary strongly with stellar mass either in \fig~\ref{fig:scaling}. The
centroid velocities systematically underestimate the outflow velocity 
in more massive galaxies because the interstellar absorption at $V_{sys}$ is 
stronger; but when we estimated the velocity corrections with two-component fitting, 
the median blueshifts of the Doppler component did not reveal a significant trend with 
stellar mass. Spectral resolution is therefore not an obvious explanation for
the discrepant result, yet it remains unclear whether stellar absorption significantly
biases the Weiner \et (2009) composite spectrum for high stellar mass. 

\subsubsection{Do Outflows Escape?}
We follow Weiner \et (2009)
and use [OII] velocity dispersions, $\sigma_{[OII]}$ (Weiner \et 2006), to estimate
the depths of the gravitational potentials for galaxies in the LRIS 
sample.\footnote{To the extent that the line-of-sight 
           velocity dispersion 
           measured from the linewidth reflects the gravitational potential, the 3D velocity 
           dispersion will be approximately $\sqrt{3}$ times larger. Averaging over the 
           variation in the relation between the line-of-sight velocity dispersion,
           $\sigma_{[OII]}$, and the rotation velocity at different disk 
           inclinations (Rix \et 1997; Kobulnicky \& Gebhardt 2000; Weiner \et 2006), 
           Weiner \et (2009) argue that $V_c \sim 1.67 - 2 \sigma_{[OII]}$ with scatter 
           of $\sim 25\%$.}
The relationship between the local escape velocity and $\sigma_{[OII]}$ is sensitive
to the location of the absorbing gas along the sightline to each galaxy.
For a spherical outflow geometry (of any solid angle), much of the column density
of an outflow viewed down-the-barrel is necessarily contributed by the gas near the
launch radius due to the inverse-square dilution of the gas density. The most conservative 
assumption is therefore that the gas lies a few pressure scaleheights above the disk 
at radius $R = 1$~kpc. For an isothermal halo of radius 100~kpc, the escape velocity
at $R = 1 $~kpc is $V_{esc} = 6.1 \sigma_{[OII]}$. The median values of $V_{esc}$
computed this way are  281, 406, and 643\kms\ for the low, middle, and high stellar
mass tertiles; but the escape speeds are lower if the gas is further away. 
We detect absorption at $V_{max}$ values higher than these $V_{esc}$ estimates, 
see upper left panel of of \fig~\ref{fig:scaling}, and therefore support the 
conclusion of Weiner \et (2009) that outflowing gas detected in \mgII\ may not 
return to the galaxy. 

The absence of 
observed growth in $V_{max}$ with stellar mass implies that a higher fraction of
outflowing gas is 
recycled in more massive galaxies.  In light of the significant uncertainties inherent
in interpreting $V_{max}$ -- including projection, sensitivity to S/N ratio, 
dependence on spectra resolution, and possible
stellar contamination --  the results presented here are probably not the final word
on this scaling. In particular, we would not discount models that predict $V_{max} \propto 
V_{esc}$ (Murray \et 2005; Zhang \& Thompson 2010); their results are supported
by observations of small samples of local starbursts
observed at much higher resolution and S/N ratio (Martin 2005; 
Rupke \et 2005).

The \feII\ Doppler shifts and the $V_{max}(\mgII)$ values suggests that outflows 
reach the circumgalactic medium. For example, we have measured
\mgII\ absorption at blueshifts as high as 700\kms; and, these outflows would
coast to 70 kpc in 100~Myr, a short enough time period for the host galaxy to
remain blue even if the star formation rate declines due to the outflow. 
Bouch\'{e} \et (2006) found an anti-correlation between intervening absorber 
equivalent width and host halo mass. In contrast, we find the largest \mgII\ 
equivalent widths in the more massive galaxies. Hence, the strengths of intervening 
\mgII\ absorption and the total \mgII\ equivalent width in galaxy spectra scale differently 
with galaxy mass. This apparent contradiction may be resolved by appealing to two of our 
results: (1) from two-component fitting (see Section \ref{sec:measure_2comp}), the equivalent 
width of ISM absorption (at the systemic velocity) grows with increasing stellar mass and (2) 
emission filling is stronger in lower mass galaxies (see Section \ref{sec:composite_spectra}).
Since the resonance emission in halo sightlines is negligible, and many halo sightlines
intersect little ISM, the physical affects driving the equivalent width -- stellar mass
trend in galaxy spectra largely do not apply to intervening absorption.

Since our results indicate that outflows from $z \sim 1$ galaxies subtend a 
solid angle much less than $4\pi$, we expect the
intervening absorption from outflows to show a dependence on azimuthal angle,
specifically the location of the sightline relative to the gas disk and outflow axis. 
Simple dynamical arguments show
that the ``blowout'' will be perpendicular to the gaseous disk (De Young \& Heckman
1994), so a distingushing property of outflows is that they would be detected in sightlines 
passing near the minor axis of a galaxy. New evidence for a bimodal distribution of azimuthal 
angles among strong \mgII\ systems (Bouch\'{e} \et 2012; Kacprzak \et 2012)  supports our
conclusions that outflows (minor axis absorption) make a substantial contribution to the 
population of intervening absorbers. The average minor axis sightline within 60~kpc of a blue
galaxy (at $0.5 < z <  0.9$) shows more absorption than a typical sightline at the equivalent
impact parameter along the major axis (Bordoloi \et 2011). This excess absorption provides 
the best constraint to date on the distance to which the \mgII\ clouds survive in 
the galactic outflows.

\subsubsection{Scaling with SFR, sSFR, and $M_B$}

In Figure~\ref{fig:scaling}, the \feII\ centroid velocities are shown
against SFR, sSFR, absolute $B$-band magnitude, and color. The median velocities 
of both the $3\sigma$ outflows and the $1\sigma$ outflows show no systematic
trend with any of these galaxy properties. We examined similar
plots using the velocities of the Doppler component and found no correlations
with the Doppler component. In the SFR and sSFR plots, the errorbars on the median 
are quite large owing to the small number of galaxies in each bin. This large
uncertainty combined with the relatively small range in SFR ($\sim 1$ dex) could
hide any underlying correlation between $V_1$ and SFR.

\begin{figure}[t]
  \hbox{\hfill \includegraphics[height=9cm,angle=-90,clip=true]{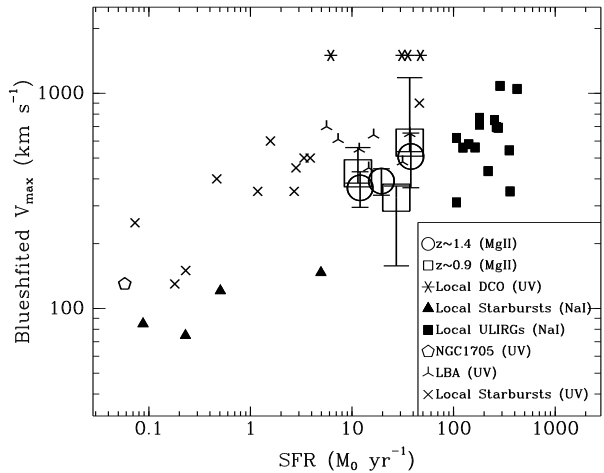} \hfill}
          \caption{\footnotesize
Maximum blueshift of low-ionization absorption vs. SFR. Large symbols show the
median values among the LRIS galaxies and the Weiner \et (2009) measurement of
composite spectra. The legend denotes the sample and the lines used to measure the maximum
blueshift of the absorption trough.
Measurements for local galaxies from 
Heckman \et 2011 (asterisk),
Schwartz \& Martin 2004 (filled triangle),
Martin 2005 (filled square), 
Grimes \et 2009 (crosses),
Heckman \et 2011 (trefoil), and
V\'{a}zquez \et 2004 (NGC 1705 with SFR from Marlowe \et 1997). 
Over the relatively small range in SFR covered by our LRIS sample, the median 
$V_{max}(\mgII)$ values measured among individual galaxies at $z \sim 1$ are 
comparable to the maximum blueshifts of resonance lines in local samples
of galaxies with the notable exception of the larger blueshifts among the
galaxies with a dominant central oject (DCO). Starbursts in dwarf galaxies
only show low-ionization absorption at significantly smaller Doppler shifts.
}
 \label{fig:vmax_sfr} \end{figure}

To increase the SFR baseline, we compare $V_{max}$ measurements to nearby dwarf
galaxies and extremely luminous starbursts in Fig.~\ref{fig:vmax_sfr}. First,
we note that our median $V_{max}(\mgII)$ measurements are consistent
with the Weiner \et (2009) measurements at slightly higher redshift. 
Second, when lower mass galaxies are included, the envelope describing the maximum
outflow velocity increases with SFR. Excluding the galaxies with dominant
central objects (DCOs), because their
high velocities may be related to AGN (Heckman \et 2011), the slope is consistent 
with the $V \propto SFR^{0.35}$ relation measured previously for local starbursts (Martin 2005).

The true scatter in the $V_{max} - SFR$ scaling may be even larger than it
appears in Fig.~\ref{fig:vmax_sfr}. Small Doppler shifts would go undetected
unless the spectral resolution is very high, and echelle resolution has only
been obtained for the nearby dwarf galaxies. Hence, the absence of data in 
the lower right corner of this plot is in part a selection effect. Moreover,
projection effects only make the measured Doppler shifts lower than the
outflow velocity, so the upper limit rather than the median is most interesting.
The low-redshift galaxies shown in these plots are all
starburst galaxies, and their SFR surface densities are
typically higher than the outflow threshold (Heckman \et 2003; Kornei \et 2012).
From this viewpoint, we argue that the absence of large Doppler shifts in 
dwarf galaxies -- i.e., the paucity of points in the upper left of \fig~\ref{fig:vmax_sfr} --
is the most important feature of this scaling relation. Considering the
multi-phase nature of winds, however, the apparent lack of high-velocity outflow in
dwarf galaxies may be limited to warm, low-ionization gas. Based on 
Figure 4 of Hopkins \et (2012), we suspect that most of the mass in a dwarf starburst
outflow may be in the hot phase where the flow is moving at more than a few 
hundred \kms.  If the highest velocity gas is in fact un-observable in low-ionization
lines, then outflow velocity may not increase appreciably with SFR as
argued, for example, based on the measured temperatures of the X-ray emitting
gas (Martin 1999).

\begin{figure}[t]
 \hbox{\hfill \includegraphics[height=9cm,angle=-90,trim= 0 0 0 0]{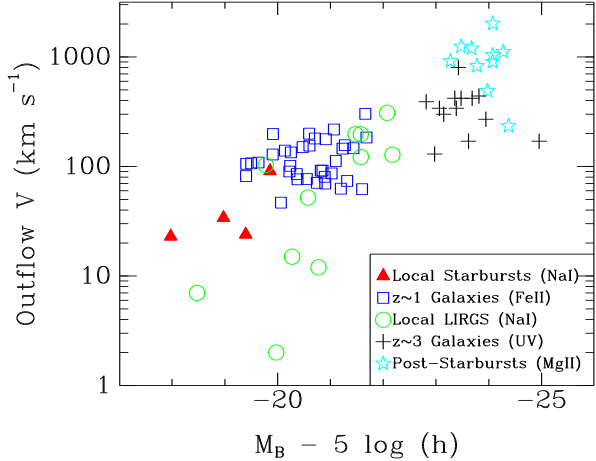} 
        \hfill}
          \caption{\footnotesize
Outflow velocity of individual galaxies vs. absolute $B$-band magnitude. The legend
identifies the samples and the transition(s) used to measure Doppler shift.
The $3\sigma$ 
outflows in the LRIS sample have velocities intermediate to local starburst galaxies 
(Schwartz \et 2004) and local luminous infrared galaxies (Heckman \et 2000). Local
post-starburst galaxies (Tremonti \et 2007) and high-redshift galaxies (Pettini
\et 2001) show larger outflow velocities but also have significantly larger
luminosities. 
}
\label{fig:all_v_mb} \end{figure}

Figure~\ref{fig:all_v_mb} compares our $V_1(\feII)$ measurements for the
outflow sample to other populations with
starburst-driven outflows.
Each of these samples includes spectra in which the resonance
absorption lines show no significant blueshift relative to the galaxy, but we 
plot only the outflow galaxies. Since the different surveys have different 
criterion for the minimum detectable Doppler shift, the absence of points in 
the lower right corner is not significant; but the maximum outflow speeds of
each population likely represent the ``down-the-barrel'' view of their outflows.
Taken together, these measurements show outflow speed increases
with $B$-band luminosity. Since $B$-band luminosity is correlated with SFR, this
result provides evidence for higher outflow velocities in galaxies with higher SFRs.

\section{GAS INFLOWS}  \label{sec:inflow}

In Section~\ref{sec:intro} we discused gas inflows in the context of
a popular model whereby the gas accretion rate regulates the galactic SFR 
(Dekel \& Birnboim 2006; Kere\~{s} \et 2009a; Kere\~{s} \et 2009b;
Bouch\'{e} \et 2010; Dav\'{e}, Finlator, \& Oppenheimer 2012).
Testing this theory requires observations 
of the gas inflows that reach galactic disks. Cosmological simulations agree
that the infalling gas which fuels star formation is {\it cold}, in the sense that
its temperature never exceeded $10^{5.5}$~K, and that these cold flows  gradually 
disappear at halo masses exceeding $10^{12}$\msun\ due to the formation of virial shocks 
which increase the gas cooling time (Kere\~{s} \et 2005; Dekel \& Birnboim 2006; Ocvirk, 
Pichon, \& Teyssier 2008; Kere\~{s} \et 2009; Brooks \et 2009; van de Voort \et 2011a).
The densest gas in these cold flows has a low covering factor but threads filaments 
of lower density gas with cross-sections much larger than a galaxy (Kimm \et 2010; 
Faucher-Gigu\`{e}re \et 2011; Fumagalli \et 2011). These inflowing streams carry
significant angular momentum that cause them to co-rotate with the galaxy
and form a warped, extended disk; hence cold flows may be recognized 
kinematically by a velocity offset in the same direction as the rotation of
the central disk (Stewart \et 2011a, 2011b). 

In contrast to the accretion rate onto {\it halos}, which is robustly predicted, 
the rate of gas accretion onto {\it galaxies} depends sensitively on feedback processes 
(van de Voort \et 2011b). Winds transport supernova ejecta into the circumgalactic medium 
(Martin, Kobulnicky, \& Heckman 2001), so these metals mark gas that was once in a 
galaxy. The mixing between these metals and cold flows is 
challenging to resolve in numerical simulations, so the metallicity of accreting gas near 
galaxies remains quite uncertain. At least in some models, however, recyling of the metals 
carried to large distances by winds substantially increases the metallicity of the infalling 
gas (Oppenheimer \et 2008; Dav\'{e} \et 2011b). In other simulations, in contrast, 
the cold flows are not even detectable in metal lines owing to their low metallicity
(Kimm \et 2011; Fumagalli \et 2011; Goerdt \et 2011). 

The obvious difference between galactic outflows and inflows is the kinematics
of the absorption-lines they imprint on galaxy spectra. Blueshifted (redshifted) 
lines in galaxy spectra necessarily mark outflowing (inflowing) gas because the 
absorbing material must lie on the near side of the galaxy. Near-UV spectra
are equally sensitive to redshifted and blueshifted \feII\ absorption. Because
blueshifts turn out to be more common, composite spectra miss the redshifted 
absorption, and spectra of individual galaxies turn out to be critical for
identifying inflows (Sato \et 2009; Rubin \et 2012). Inflow 
detections apparently require surveys of roughly a hundred or more galaxies. 
Both the measurements presented 
here (9 spectra out of 208) and recent work by Rubin \et (2012) identify
inflowing metals towards a few percent of $z \sim 1$ galaxies.

These infalling streams may not arise from cold flows; the metals may mark
gas recycled through galactic winds, tidally stripped gas, and/or material
condensing out of hot halo gas. To provide some basis for distinguishing
among these processes, in Section~\ref{sec:inflow_map} and Section~\ref{sec:inflow_properties} 
 we individually examine the spectra of $3\sigma$ inflow
galaxies and estimate physical properties of the inflowing gas.
In particular, we find evidence that the infalling gas
detected in front of the galaxies marks an inflowing stream which
subtends a larger solid angle than the galaxy on the sky.
We then return to the discussion of the source of the inflow
in Section~\ref{sec:inflow_discussion} focusing primarily 
on the cold inflow model due to recent interest in, and 
theoretical predictions for, this particular scenario.

\subsection{Inflow Confirmation} \label{sec:inflow_confirm}

We identify a net inflow of low-ionization gas by a statistically significant
redshift of the \feII\ absorption troughs, $V_1(\feII)   \ge\ 3 \sigma (V_1)$.
The uncertainty in this fitted centroid velocity does not take errors
in the galaxy redshift determination into account. We reviewed the redshift
determination and other sources of systematic error individually for the 11 
galaxies with $3\sigma$ inflows and identified possible systematic errors 
in two of the three cases described here.

All but one (22028473) of the inflow galaxies have strong emission lines in our LRIS
spectrum; and, as described in Section~\ref{sec:lris_observations}, we measure the
centroids of the emission lines to within $\pm 19$\kms. Among these inflow 
galaxies, the average difference between this LRIS redshift and the DEIMOS 
redshift from DEEP2 is $45$\kms\ with neither survey yielding systematically 
higher (or lower) redshifts. The magnitude of this discrepancy is consistent
with the internal error of $41$\kms\ reported by DEEP2 from multiple observations
of galaxies and presumably arises from velocity gradients across galaxies. 
For the inflow galaxy 22028473, however, the galaxy redshift comes from the [\ion{O}{2}] 
emission in the DEEP2 DEIMOS spectrum because our LRIS spectrum does not cover optical, 
nebular lines. The fitted Doppler shift of the \feII\ absorption, $41\pm10$\kms\ 
in Table~\ref{tab:inflow}, is comparable in magnitude to the systematic error in
the redshift determination.
Although correcting this error is equally likely to (a) double the inflow speed
or (b) negate the inflow detection, we drop this object from the inflow sample 
because it is not a significant ($> 3\sigma$) detection
when the error in the galaxy redshift is considered.

        Our adopted LRIS redshift for 22005270 exhibits one of the largest
        discrepancies from the DEEP2 redshift, 69\kms.  
        The velocity gradient between apertures 2 and 4 exceeds this discrepancy,
        so systematic differences in slit placement may well be the cause.
        The \Hb\ emission line in our spectrum of 22005270 appears very slightly 
        redshifted, but we attribute this shift to residuals from a sky line 
        93\kms redward of \Hb. We keep this object in the inflow sample.
        The LRIS redshift accurately determines the systemic velocity for
        the purposes of comparing the velocities of the galaxy and the
        resonance absorption. Even if we adopted the extremely conservative
        viewpoint that the redshift uncertainty was 60\kms,  the projected 
        infall velocity of $401\pm43$\kms\ still yields a $3\sigma$
        inflow detection.

We do remove 22036194 from the inflow sample because of the
systematic error introduced in our \feII\ series fit by intervening
absorption. Our 22036194 spectrum was recorded
      at a position angle of 25\deg, so the slit passes near the $z=0.913$
      galaxy 22035919 at an angular separation of
       7\farcs1. Intervening \mgII\ absorption from
      a foreground galaxy near this redshift appears slightly redward of
      the interstellar \feII\ $\lambda 2344 $ absorption trough in our 22036194 spectrum.
      The net, blended absorption trough is stronger and redder than the other 
      lines in the \feII\ series and therefore sysematically biased the fit.
      Rejection of the \feII\ $\lambda 2344 $ transition yields a fitted \feII\ absorption
      velocity consistent with the interstellar gas in 22036194. We note,
      however, that 22035919 is unlikely the source of the intervening
      absorption due to the strength of the line and the $ \sgreat\ 50$~kpc 
      impact parameter. We detect prominent emission lines (near $z \sim 0.831$),
      in fact, from another galaxy closer to the 22036194 sightline in our 
      2D spectrum and suggest this object is the source of the intervening
      \mgII\ absorption.

\begin{figure*}[t]
 \hbox{\hfill \includegraphics[height=26cm,angle=-90,trim=100 140 0 0]{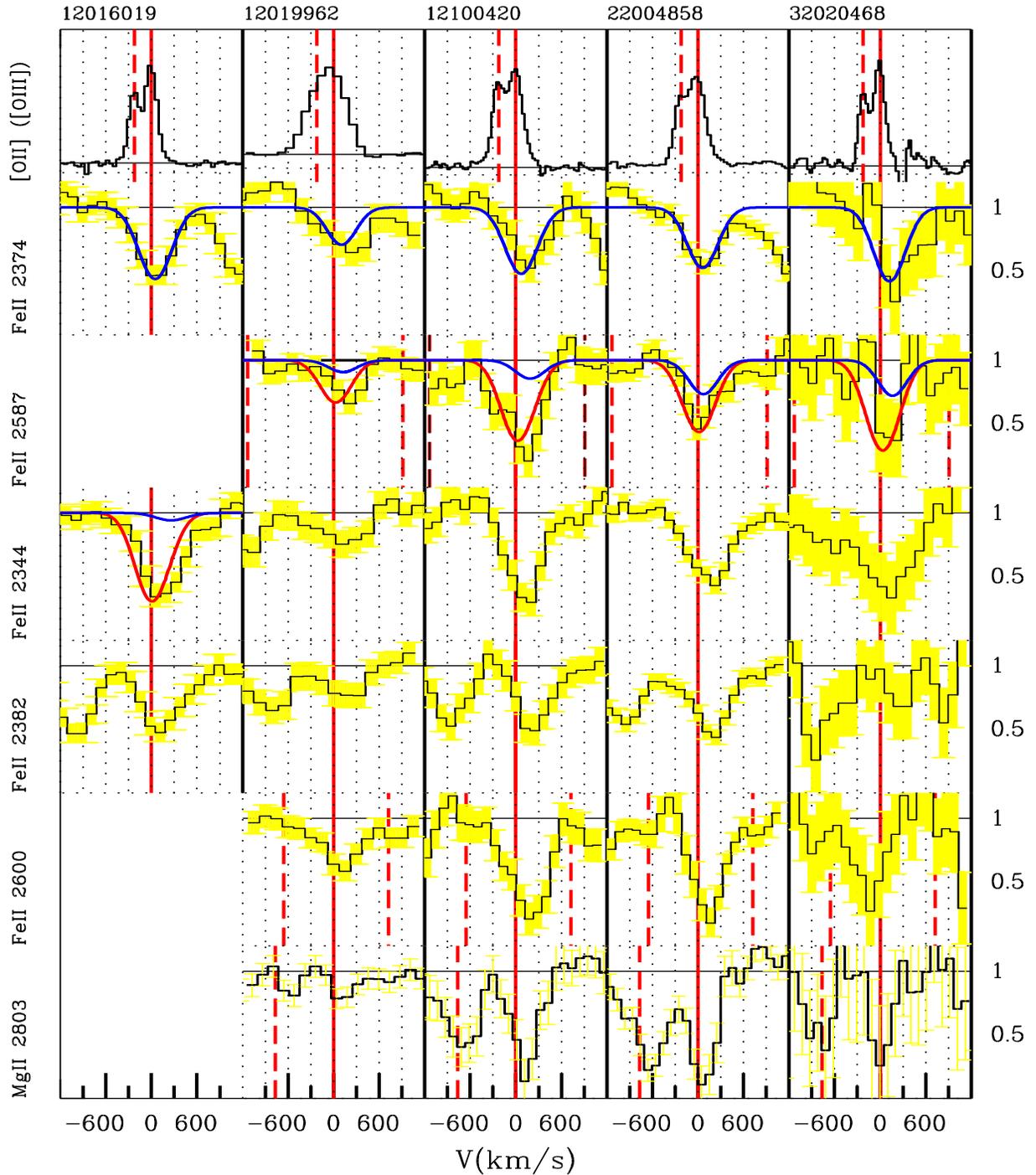}
                \hfill}
          \caption{\footnotesize
Spectra with resonance absorption lines (bottom 6 rows) significantly redshifted with 
respect to nebular emission lines (top row); see also Fig.~\ref{fig:inflows_4}. 
The solid (red) vertical line marks the systemic velocity. Dashed vertical lines 
denote the wavelengths of nearby transitions;  note the airglow line at -735\kms\ blueward 
of the \mgII\ $\lambda2796$ line in 12019996 which affects the wing of $\lambda2796$
blueward of -359\kms. 
The single-component fit to the \feII\ series is superposed on the \feII\ $\lambda2374$ profile;
and the two-component fit to the \feII\ series is plotted on the \feII\ 2587 profile (with
Doppler component shown by the dashed, blue line). 
These inflow galaxies were selected based on the significance of their \feII\ 
redshift, but their spectra typically show redshifted \mgII\ as well. Two-component fits
to the \feII\ absorption troughs yield slightly higher inflow velocities than the one-component
fits.
}
 \label{fig:inflows_x} \end{figure*}

\begin{figure*}[t]
 \hbox{\hfill \includegraphics[height=25cm,angle=-90,trim=100 160 0 0]{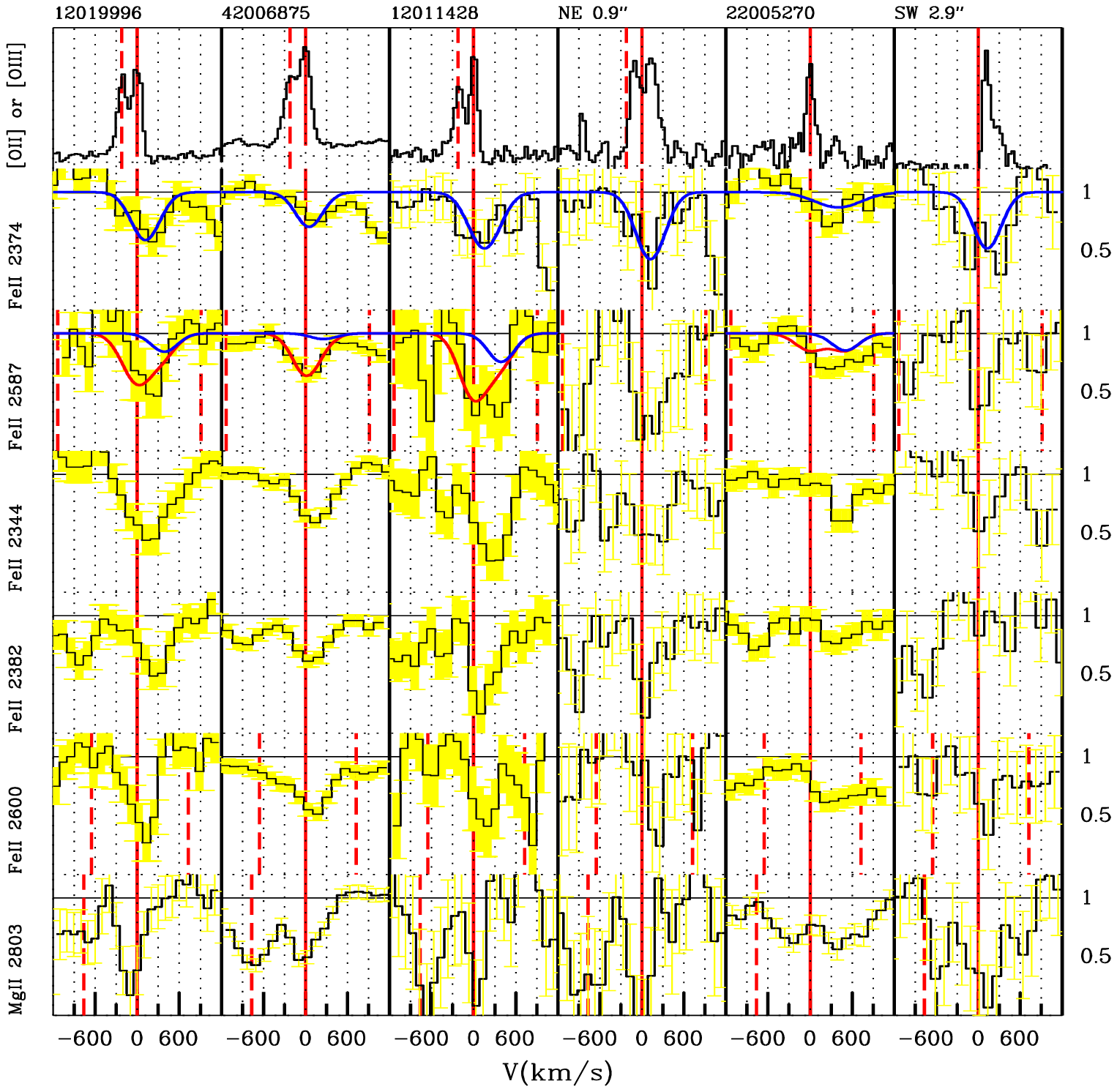}
                \hfill}
          \caption{\footnotesize
Spectra of the other galaxies with redshifted resonance absorption. Colored lines have the same
meanings as defined in the caption to Fig.~\ref{fig:inflows_x}. In addition to the {\it down-the-barrel}
spectra, i.e., a spectrum extracted from the aperture centered on a galaxy), spectra extracted off-center 
are shown for the two galaxies with spatially extended continuum emission. At locations
0\farcs85 northeast of 12011428 and 2\farcs94 SW of 22005270, the \feII\ Doppler shifts 
are $123\pm47$ and $129\pm62$\kms, respectively, consistent with the redshifts measured 
down-the-barrel (towards the centers) of those two galaxies. In contrast, the nebular
emission lines, which are by definition at zero velocity in the down-the-barrel
spectra, are prominently redshifted in the spectra extracted off center.
Off the center of each galaxy, the redshift of the nebular emission and the
\feII\ absorption are very similar, consistent with a common physical location along
the sightline.
}
 \label{fig:inflows_4} \end{figure*}

With 22028473 and 22036194 rejected, the $3\sigma$ inflow sample contains 9 galaxies.  
Figures~\ref{fig:inflows_x} and~\ref{fig:inflows_4} illustrate the $3\sigma$ redshifts 
of the \feII\ absorption troughs relative to the nebular emission. The joint fits to the
\feII\ series find a significant redshift because the absorption troughs of
multiple transitions show a net redshift. The redshifted absorption does not appear 
to be unique to the \feII\ lines; the \mgII\ 2803 absorption trough (row 7 of these figures) 
often has a net redshift as well. Table~\ref{tab:inflow} provides the measured properties of the 
redshifted absorption troughs.
Table~\ref{tab:properties} summarizes the quality of the individual spectra
and galaxy redshift, $B$-band luminosity, color, stellar mass, and (when
available) SFR.

\subsection{Spatial Extent of Infalling Gas} \label{sec:inflow_map}

\begin{figure*}[t]
\hbox{\hfill 
\includegraphics[height=18cm,angle=-90,clip=true]{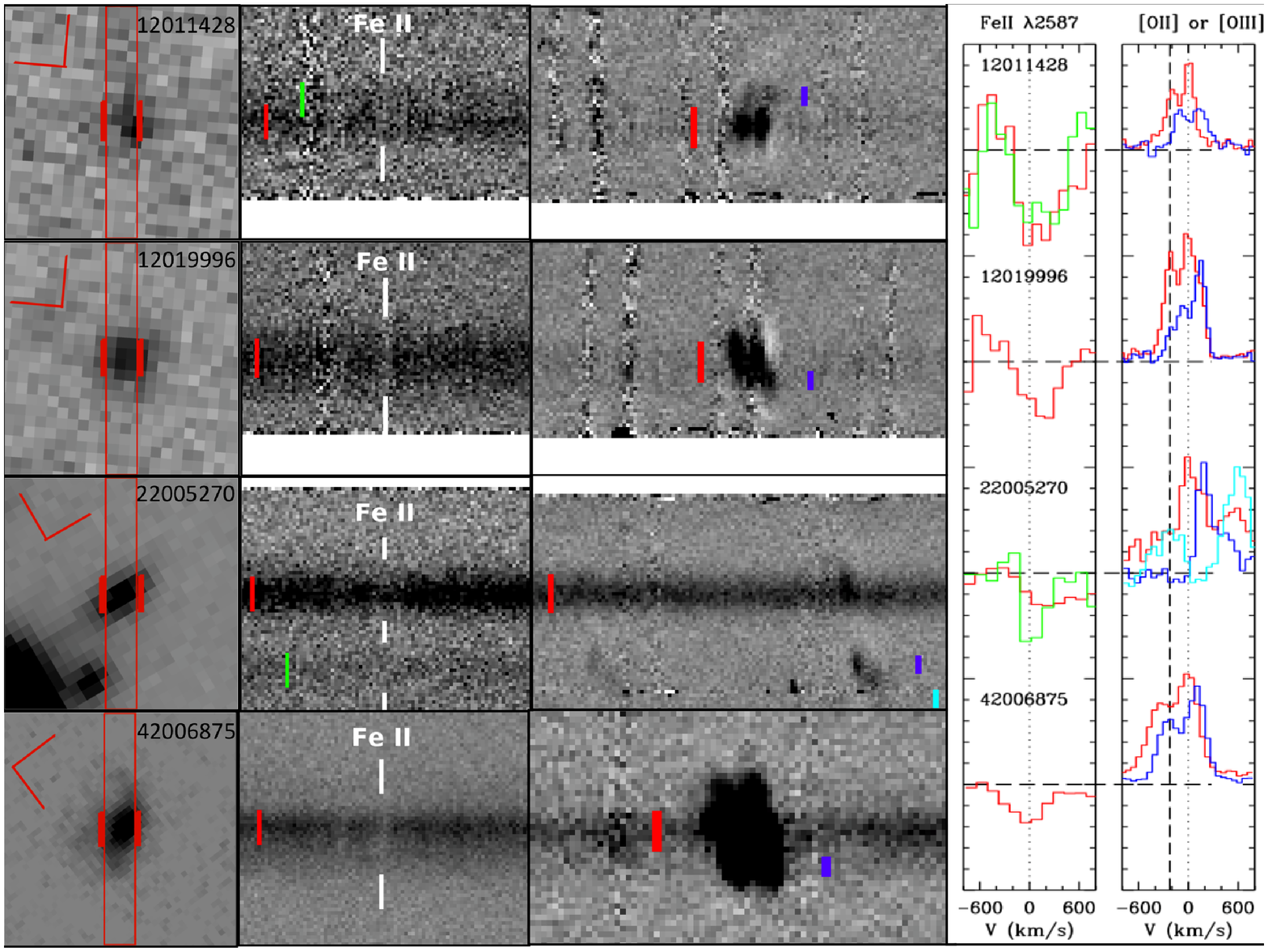}
                \hfill}
\caption{
Velocity gradients in the two-dimensional LRIS spectra of inflow galaxies.
{\it Column 1:} DEEP2 (CFHT, $R$-band) image with the 1\farcs2 by 9\farcs0 slitlet 
and central extraction aperture marked. The compass rose marks north and (90\deg\
counter-clockwise) east. The slit across 22005270 passes near a galaxy identified as 
22005066 in the DEEP2 photometric catalog. 
{\it Column 2:} Two-dimensional LRIS spectra near \feII\ $\lambda 2600 $ (white tic marks). 
Wavelength increases from left to right, and the width of the region shown
is approximately 183~\AA. We denote the central apertures and the apertures used 
for the off-center spectra shown in {\it Column 4} and \fig~\ref{fig:inflows_4}.
{\it Column 3:} Two-dimensional LRIS spectra near the [\ion{O}{2}] $\lambda \lambda
3726, 3729$ doublet (or, in the case of 22005270, near [\ion{O}{3}] $\lambda 5007$.
Wavelength increases from left to right, and the width of the region shown
is approximately 81~\AA. We mark the central apertures and the off-center 
apertures where we extracted the emission-line profiles shown in {\it Column 5}. 
The tilt of the emission lines reflects the rotation of dense gas in these galaxies;
and a satellite galaxy is clearly detected in the 22005270 spectrum.
{\it Column 4 and 5:} Extracted spectra zoomed in on the \feII\ $\lambda 2587$ and 
[\ion{O}{2}] (or [\ion{O}{3}]) lines.
In all 4 spectra, the gradient in the emission velocity along the slit is
as large as the redshift of the \feII\ absorption down-the-barrel (central
aperture). The infalling gas detected towards the center of each galaxy
has a redshift that matches the Doppler shift of either that galaxy's
outer disk or a satellite galaxy.
}
\label{fig:inflow_2d} \end{figure*}

We examined the two-dimensional spectra for velocity gradients along each slitlit.
For illustration, Figure~\ref{fig:inflow_2d} shows the positions of the slitlets
and selected regions of the two-dimensional spectra for the four galaxies with large velocity 
gradients.
The [\ion{O}{2}] emission lines are clearly tilted along the slits 
crossing 12011428, 12019996, and 42006875 as is the [\ion{O}{3}] $\lambda 5007$ emission  along
the 22005270 slit. We obtain additional insight about the inflows by investigating
whether the emission or absorption anywhere along the slit is at the same velocity as the
redshifted \feII\ absorption discovered in the galaxy spectra (i.e., extracted from apertures 
centered on the galaxy). We now consider the two-dimensional spectra of each of these
galaxies in further detail.

\paragraph{12011428:}
Along the slit that crosses 12011428, the line emission is not continuous.
To the north of 12011428, a second blob of [\ion{O}{2}] emission is detected 1\farcs14 
(9.6 kpc) away along the slit. The image shows a hint of faint, extended $R$-band emission 
in this direction. An [OII] emission-line spectrum extracted at this location exhibits a 
redshift of $125\pm15$\kms.  The emission redshift on this side of the galaxy is 
remarkable because resonance absorption is seen at a similar velocity, $154\pm32$\kms, 
in the galaxy spectrum (extracted from an aperture centered on the galaxy continuum). This 
coincidence in velocity space may indicate that the infalling stream crossing the center
of the galaxy reaches the galaxy near the star-forming knot 9.6~kpc away. Further evidence
supporting a structural connection between the redshifted emission and absorption is
found in the grism (blue) spectrum extracted on the northern side of the galaxy (aperture
marked in green in \fig~\ref{fig:inflow_2d}). The \feII\ resonance absorption remains redshifted 
in the off-center spectrum, and we find no spatial gradient in the absorption velocity across 
the 12011428 slit. The infalling gas apparently covers 12011428, and one side of the galactic
disk shares the redshift of this infalling stream.

\paragraph{22005270:}
Along the slit that crosses 22005270, the line emission is not continuous in the two-dimensional
spectrum shown in Figure~\ref{fig:inflow_2d}. Roughly 3\farcs5 southwest of 22005270, 
where the slit passes near galaxy 22005066, we detect a spatially extended blob 
of [\ion{O}{3}] emission (1\farcs6 or 12.2 kpc across). 
Further along this slit, we extracted another spectrum 4\farcs3 
(33 kpc) southwest of 22005270 because the [\ion{O}{3}] emission-line profile is double-peaked;
the stronger, longer wavelength maximum at 346\kms\ and a weaker component at $-158$\kms.
The large emission redshift near 22005066 
is remarkable because \feII\ resonance absorption is seen at a similar velocity in the spectrum
of 22005270. In fact, the spectrum of 22005270 shows the largest \feII\ redshift at
$401\pm43$\kms. Within the uncertainties, this absorption velocity is marginally consistent 
with the emission component at $346\pm19$\kms. Further evidence for a connection between the
redshifted [\ion{O}{3}] emission near 22005066 and the redshifted absorption in the 22005270 spectrum
is apparent upon re-inspection of the primary spectrum towards 22005270. In both  \fig~\ref{fig:inflows_4} 
and  \fig~\ref{fig:inflow_2d}, the bright [\ion{O}{3}] emission in the 22005270 spectrum falls
at zero velocity by definition, but a weak redshifted emission component is detected at 341\kms,
a redshift similar to the centroid velocity of the \feII\ absorption. We show the extracted blue 
spectrum where the slit passes near 22005066 in \fig~\ref{fig:inflows_4}.
The continuum S/N ratio in 
this spectrum is low relative to the blue spectrum of 22005270; but we see a net redshift of the 
\feII\ and \mgII\ absorption. These coincidences in velocity space indicate that the infalling gas 
seen in absorption towards 22005270 is part of a physical structure covering both 22005270 and 
22005066 and therefore at least 33~kpc across.

\paragraph{12019996:}
The tilt of the [\ion{O}{2}] emission in the 12019996 spectrogram
shows the usual signature of a rotating disk. The emission extends 
0\farcs68 (5.6 kpc) to the south where the Doppler shift reaches $102\pm10$\kms.
The velocity gradient between the central aperture and the receding side is 
nearly as large as the $117\pm25$\kms\ redshift of the \feII\ absorption in the 
central aperture. The infalling gas marked by the redshifted resonance absorption 
may therefore extend all the way across the 12019996 disk.

\paragraph{42006875:}
In the 42006875 spectrum, the tilt of the [\ion{O}{2}] emission 
shows the usual signature of a rotating disk. 
Across the 42006875 slit, the relative redshift of the emission reaches 68\kms\
1\farcs4 (10 kpc) to the northwest of the center of the galaxy. The velocity
gradient between the center of the galaxy and the receding side is
as large as the $51\pm14$\kms\ redshift of the \feII\ absorption in
the galaxy spectrum. 
The comparable redshifts of the spatially extended emission and the \feII\
absorption toward the center of the galaxy suggest to us 
that the infalling gas is part of a spatially extended structure, and 
this structure meets the disk roughly 10~kpc southeast of the galactic
center. The \mgII\ absorption trough exhibits a red wing in the $\lambda 2803$ 
consistent with the infalling structure contributing to the line profile.

How do the emission-line kinematics compare to the redshift of the \feII\ absorption in the 
galaxy spectrum in the other 5 spectra?
The two-dimensional spectra of 32020468 and 12100420 do not resolve the 
line emission spatially. The absence of spatially extended emission is not
surprising in light of their compact morphology in ground-based images, and 12100420
also looks compact in an HST image. 
Yet, as illustrated in the spectra shown in \fig~\ref{fig:inflows_x}, the \feII\ absorption 
is redshifted $123\pm40$\kms\ in 32020468 and $71\pm18$ in 12100420.
In 12100420, the \mgII\ absorption also displays a net redshift; but
in 32020468, the \mgII\ absorption is detected a low significance near the systemic 
velocity. In the remaining 3 galaxies -- 12016019, 12019962, and 22004858 --
the [\ion{O}{2}] emission along the slitlets is spatially resolved and very slightly 
tilted. Despite the small gradient in the velocity of the emission, however, the
spectra of these galaxies exhibit significantly redshifted resonance absorption. 
The \mgII\ absorption trough also exhibits a net redshift in 22004858. 
Hence, none of these 5 spectra show line emission at redshifts near the \feII\ absorption
redshift. It would be valuable to obtain spectra at other position angles
and determine whether or not these galaxies have any extended emission at the redshift
of the \feII\ absorption as in the case of the four objects in \fig~\ref{fig:inflow_2d}.

To summarize, the LRIS spectra cover nebular emission lines in all 9 inflow galaxies.
We have described the discovery of line emission near the redshift of the \feII\ resonance 
absorption in 4 of these: 12011428, 12019996, 22005270, and 42006875. Since the 
velocity of the spatially offset, redshifted emission is comparable to the Doppler
shift of the absorption measured directly towards the galaxy, the inflow may mark
a spatially extended infalling structure. To assess the probability of a random
coincidence in velocity, we also looked for emission features with blueshifts
similar to the blueshifted absorption. For 33 of the 35 spectra with $3\sigma$ \feII\ 
blueshifts, we could inspect the velocity gradient of the nebular emission along the slit.
We found four spectra with extended emission reaching the blueshift of the \feII\
absorption and one spectrum with emission significantly more blueshifted than the \feII\ 
emission. Among the spectra showing outflow detections, the fraction of coincidences between 
emission and absorption velocities (5 out of 33 spectra) is lower than the
fraction of inflow spectra showing extended emission near the velocity of the redshifted
absorption (4 matches among the 9 inflow galaxies). This result supports our suggestion of 
a physical association between extended disks and satellite galaxies (the source of the
spatially extended emission in the inflow spectra) and the redshifted \feII\ absorption 
component. We also note that the absence of nebular emission (anywhere along the slit)
at the velocity of the \feII\ blueshifts further strengthens our interpretation of
those blueshifts as evidence of outflowing gas.

\subsection{Properties of Infalling Gas} \label{sec:inflow_properties}

\subsubsection{Which Galaxies Have Infalling Metals?}

Nine of the 208 spectra show very significant ($3\sigma$) redshifts of the \feII\ absorption
relative to the host galaxy. The low fraction of unambiguous inflows appears to be consistent 
with the 6 inflows identified recently by Rubin \et (2012) among approximately 100 objects. Here,
we examine the properties of these 9 galaxies and their enriched infall. We cannot exclude a
higher inflow fraction at lower velocities and less significant equivalent widths; however,
the results presented for inflows in \fig~\ref{fig:fout} indicate not more than 20\% of
the spectra show inflows with Doppler shifts larger than the systematic error of $41$\kms\
in the determination of the systemic velocity.

In the SFR -- stellar mass plane shown in Figure~\ref{fig:mmUV_flow}, 
the masses of 4 of these 5 inflow galaxies fall in the high mass tertile. Given the
small number of objects, however, this excess is hardly significant and not supported
by the full sample in which 4 of the 9 inflow galaxies have stellar masses less 
than $\log M/\msun = 10.20$, the median stellar mass of the LRIS sample. 
We have only measured the SFRs for 5 of the galaxies with \feII\ redshifts 
(that lie in the AEGIS field). Three (two) of these 5 galaxies have specific SFRs 
lower (higher) than the average shown in in Figure~\ref{fig:mmUV_flow}, hence  
inflows do not depend strongly on specific SFR. We found 3/69, 2/68, and 4/71 galaxies with 
redshifted \feII\ absorption, respectively, in the low, middle, and high redshift 
tertiles of the LRIS sample. In spite of the limited sample size, we therefore conclude 
that inflow galaxies may sample the full range of SFR, specific SFR, stellar mass,
and redshift.

The galaxies with spectra showing redshifted \feII\ absorption do not 
have unusual properties in any way, aside from having infalling gas along the sightline
to them. Since the galaxies with detected inflows do not share some physical property, we
argue that most galaxies may have inflows and attribute the low incidence of
measured redshifts to the geometry of the infalling streams. Specifically,
we find 3-6\% sightlines towards  $z \sim 1$ galaxies show redshifted \feII, so 
the covering fraction of the inflows is quite small.

\subsubsection{Velocity}

In Table~\ref{tab:inflow}, we summarize the inflow speeds fitted to centroids of  the \feII\ 
absorption troughs. Except for the extreme case of 22005270, the fitted 
1-component velocities are less than 200\kms\ and therefore consistent with Rubin \et (2012).
The gas moves at speeds of material 
bound to galaxies. Evidence for high velocity gas can be seen in the absorption-line profiles 
shown in Figures~\ref{fig:inflows_x}, \ref{fig:inflows_4}, and \ref{fig:inflow_2d}.
 The largest inflow velocity is detected via the \mgII\ 
absorption trough in 22005270. In the red wings of the \feII\ line profiles,
absorption at velocities exceeding 600\kms\ is detected in 12016019, 12019962, 12100420, 
22004858, 22005270, and 42006875. Modeling the contribution of the ISM via 
two-component fitting increases the estimated inflow velocity. The median inflow velocity 
increases from 71\kms\ in the single-component fits to 191\kms\ in the two-component fits. 

\subsubsection{Inclination}

Rubin \et (2012) found that 5 of their 6 inflow galaxies have highly inclined ($i > 55\deg$), 
disk-like morphologies. They interpreted the high inclinations as evidence that rotation
dominates the kinematics of the infalling gas but pointed out a paradox -- i.e., that
the absorbing gas must be located on only the receding side of these galaxies.
Kornei \et (2012) show the morphologies of five of the LRIS inflow galaxies.
The axis ratio was not well measured for 12011428.
The measurements  for 12100420, 12019962, and 12019996 indicate $i < 55\deg$ however,
and only 12016019 is consistent with a large inclination of $\sim 61\deg$.  Hence,
it remains unclear whether high inclination is a necessary condition for the
detection of enriched infall.

\subsubsection{Infalling Mass}

The (median) equivalent widths of the \feII\ $\lambda2374$ and 2261 lines suggest the
total \feII\ column density, ISM plus inflow, typically lies in the range
$15.08 \le\ \log N(Fe^+) (\col) \le\ 15.85$. 
For purposes of illustration, we conservatively choose solar 
metallicity gas, no depletion of Fe onto grains, and an ionization fraction
of 100\% singly ionized iron. The total hydrogen column density is then
\begin{eqnarray}
\log N(H) (\col) \approx\ \nonumber \\
19.58 + (\log N(Fe^+) - 15.08) - \log ({\rm Z} / \zsun). 
\label{eqn:column_density} \end{eqnarray}
The implied column is $\log N(H) = 19.58$ for solar metallicity infall, but
lower metallicities are plausible and raise the inferred hydrogen column. 

The galactic ISM clearly produces some of the \feII\ equivalent width since our 
sightline directly intersects a galaxy. Although a true measurement of the inflowing 
column in ionized iron requires much higher spectral resolution, the potential impact 
of the ISM on the equivalent width of the inflowing component can be illustrated with 
two-component fits to the absorption trough as outlined in Section~\ref{sec:measure_2comp}.
For the Doppler component fitted to the \feII\ absorption, we integrated the 
equivalent widths of the $\lambda2374$ and $\lambda 2261$ profiles to obtain
the lower and upper bounds, respectively, on $N_{Dop}(Fe^+)$ in Table~\ref{tab:inflow}.
The median bounds provide a very rough estimate of typical inflow column 
$ 14.4  \le\ \log N_{Dop}(Fe^+) C_f \le\ 15.15$.  The total hydrogen column of 
the infalling column is therefore typically larger than
\begin{eqnarray}
\log N(H) (\col) \approx\ \nonumber \\
18.90 + (\log N(Fe^+) - 14.40) - \log ({\rm Z} / \zsun) .
\end{eqnarray}
Gas columns larger than those associated with galactic disks
and damped Lyman-$\alpha$ systems, $\log N_H (\col) \approx\ 20.3$, seem unlikely. Hence,
we expect the metallicity of the redshifted gas is greater than roughly $0.04 \zsun$.

\subsection{Comparison of Inflow Properties to Models} \label{sec:inflow_discussion}

Many properties of the infalling gas are consistent with the cold flow scenario.
The velocity components along our sightlines are largely consistent with virial
motion, the exception being the extremely large redshift towards 22005270 which
is associated with the satellite galaxy 22005066.  
The two-dimensional spectra provide spatial information about the inflow in 3
other objects, and in those the redshift of the infalling gas matches the projected
velocity on one side of a rotating, gas disk. This association could reflect a stream
crossing in front of each galaxy and connecting to the outer disk in the manner
predicted by Stewart \et (2011a, b). The low incidence of redshifted \feII\
absorption in spectra of $z \sim 1$ galaxies would also be expectd for cold flows.

To have any chance of detecting a cold flow in a metal line, the sightline must run 
along the filament (e.g., see Figure~2 in Kimm \et 2010), and the chance of such 
a favorable orientation is small. While differences in the radiative 
transfer calculations produce some variation in the average covering fraction 
among these models, galaxies with covering fractions of more than a few percent, 
as originally suggested by Dekel \et 2009a, are now thought to be quite rare (Kimm \et 2009). 
Numerical simulations suggest that only a few percent
of the sightlines passing within 100 kpc of a massive galaxy at $z \sim 2$ 
intersect dense ($\log N_H (\col) > 20.3$), infalling gas (Kimm \et 2010; 
Faucher-Gigu\`{e}re \et 2011; Fumagalli \et 2011). It must be acknowledged, however,
that these high columns represent the cool gas threading the filaments rather
than the smooth component of the cold streams, which dominates the cross section
of neutral hydrogen absorption below $\log N_{HI} (\col) = 18$ within $R_{vir}$ (Fumagalli \et 2011).
At $z < 2$, less than half the cross section in the range $19 < \log N_{HI}  (\col) < 20$,
and less at higher columns, is likely due to streams (Fumagalli \et 2011).

Theory predicts the disappearance of cold flows in halos more massive than
roughly $10^{12}$\msun.  The halo masses of many of the galaxies in the high
stellar mass tertile of the LRIS sample are likely above $10^{12}$\msun\
based on either their clustering properties (Coil \et 2008) or halo
abundance matching (Behroozi, Conroy, and Wechsler 2010). In the LRIS sample,
we did not find any change in the inflow fraction with increasing stellar mass;
however, this sample excludes red sequence galaxies. As can be seen by comparing the density
of red sequence and blue cloud galaxies in \fig~\ref{fig:sample}, the fraction of red galaxies
increases quite quickly among galaxies with stellar masses $\log M/ \msun > 10.3$, 
the typical stellar mass in a $10^{12}$\msun\ halo at $z \sim 1$ indicated by abundance
matching (Behroozi \et 2010).  Whether or not the fraction of inflow galaxies declines 
with mass in an unbiased sample cannot be determined from our data since we do not measure 
the inflow fraction towards red galaxies.  Likewise, the lack of any evolution 
with redshift might be limited to blue cloud galaxies which compose the LRIS sample.

According to some models (Kimm \et
2011; Fumagalli \et 2011), detection of the inflows via a metal line definitely implies
these are not cold accretion flows. From our perpsective, however, the metallicity
of the cold inflows near galaxies remains a highly uncertain quantity in the models
due to the unrealistic treatment of metal recycling via galactic winds. The ciculation
of metals might significantly enrich the cold flows before the infalling gas is
incorporated into the galactic disk.

In this recycling scenario, the distinction between cold flows and infalling relic outflows
may become blurred. Numerical simulations including outflows have suggested that
the covering factor of high column density (DLA-like) gas increases with halo
mass, due in large part to the kinematics of outflowing gas (Hong \et 2010; 
Faucher-Gigu\`{e}re \et 2011). Given the lack of a strong scaling relation
between outflow velocity and stellar mass, we expect a larger fraction of the outflowing
material to fall back onto the more massive galaxies. Hence, the recycling
of outflows should produce an increasing inflow column as stellar mass increases. 

The distinction between the increasing (decreasing) inflow column with stellar mass due to wind
recycling (cold flows) may provide a means to discriminate the primary origin of infalling 
metal-enriched gas. This proposal further emphasizes that inflow fraction measurements
should include red sequence galaxies. Although we expect only star-forming galaxies to drive
outflows, the galaxies may migrate across the color - magnitude diagram (due to the
cessation of star formation) before the bulk of the outflow slows down and turns
around (due to the gravitational attraction of the galaxy). The first inflow galaxies
identified appeared to support this scenario; their optical colors placed them on the red 
sequence, but their UV-optical colors indicated they were forming stars a few 100 Myr prior
(Sato \et 2009). A subsequent study of 13 K+A galaxies, however, discovered 2 inflow 
galaxies (Coil \et 2011) suggesting an inflow fraction similar to our result for
blue cloud galaxies.\footnote{
     The galaxies in our LRIS sample that show redshifted \feII\ absorption do not
     have the spectral signatures of K+A galaxies nor the the colors of green-valley
     galaxies.  They are normal blue cloud galaxies and are not post-starburst galaxies.}
Finally, mergers also bring gas into galaxies. The association of the galay 
22005066 with the redshifted \feII\ absorption in 22005270 may reflect this
scenario. Because 22005066 is the less luminous of the two galaxies,
we would expect a tidal stream to pull gas out of 22005066 onto 22005270.

In summary, cold flows, mergers, and wind recycling are all expected to
contribute to gas infall. The low incidence of infall suggests that whichever
process dominates, the covering fraction of the infalling streams is low.
Much larger samples of individual spectra will be needed to further distinguish
among these mechanisms on a statistical basis.

\section{SUMMARY \& IMPLICATIONS} \label{sec:conclusions}

Our understanding of galaxies will be greatly improved once we understood
what controls the flow of matter between galaxies and the circumgalactic
medium. Empirical constraints are essential in this regard because there are
many physical processes that may be important and these processes can
be nonlinear -- e.g., the cooling rate or the interplay between 
mechanical and radiative feedback.

In this paper, we used unusually deep spectroscopy of galaxies at $0.4 < z < 1.4$
to investigate which galaxies have low-ionization outflows. The blue sensitivity
of LRIS and collecting power of the Keck~I telescope made it possible to detect
NUV absorption lines in spectra of individual galaxies. These galaxies are
typical blue cloud galaxies and have higher SFRs than do most galaxies today.
We have fit a series of \feII\ absorption lines to describe the bulk flow
(to within $\pm40$\kms) and constrain the \feII\ column density (to within
an order of magnitude). These Doppler shifts have been compared to the most sensitive tracer
of high-velocity gas in the spectra, the \mgII\ $ \lambda \lambda 2796$, 2803 doublet, which
is frequently detected out to impact parameters of 70~kpc in galaxy halos.

We measure \feII\ blueshifts of $ -50, -100, {\rm ~and} -200$\kms\ or more in spectra
of, respectively, 45\%, 20\%, and 2.5\% of blue-cloud galaxies at $z \sim 1$.
This outflow fraction is roughly 3 times higher in galaxies with 
$ 21  < {\rm SFR~} (\msunyr) < 98$ than in galaxies with  $ 0.9 < {\rm SFR~} (\msunyr) < 10 $. 
Comparing the average spectra of the sample divided at the median SFR, the \feII\ 
absorption is significantly blueshifted ($-38 \pm 8$\kms) in the higher SFR, but not the 
lower SFR, composite spectrum. Among the individual spectra, however, the correlation 
between \feII\ Doppler shift and SFR, while in the direction of larger blueshifts
at higher SFR (Spearman rank-order correlation coefficient $r_S = -0.225$), 
deviates by only $1.6\sigma$ from the null hypothesis of no correlation. We find
the most significant outflows in spectra of individual galaxies with specific SFR 
larger than 0.8~Gyr$^{-1}$.  Galaxies with 
$1.6 < {\rm sSFR~} ({\rm ~Gyr}^{-1}) < 5.6$
are roughly 5 times more likely to show blueshifted \feII\ absorption than those with 
$0.07 < {\rm sSFR~} ({\rm ~Gyr}^{-1}) < 0.7$. 
These results for the outflow fraction and outflow velocity confirm the
expectation that the blueshifts of the \feII\ lines identify outflows powered by massive stars.
When combined with the measured cosmic SFR history, they also predict
an increase in the fraction of galaxies with blueshifted absorption towards higher redshift.
Among just blue galaxies (our LRIS sample), however, we do not detect a statistically
significant increase in the blueshifted fraction with redshift.

While we find no variation of \feII\ Doppler shift with stellar mass, blue
luminosity, or $U-B$ color, the \mgII\ equivalent width does increase towards higher
stellar mass, $B$-band luminosity, and redder color in both composite and individual
spectra. Fitting the ISM absorption explicitly shows that the increase in the \mgII\ 
equivalent width can be attributed in part to a stronger interstellar gas
component in the more massive galaxies. An associated increase in the dust mass
is consistent with the redder color of the more massive galaxies and reduces their 
emission in resonance lines. This reduction of emission filling in the spectra of massive 
galaxies (relative to lower mass galaxies) amplifies the positive correlation between the \mgII\ 
absorption equivalent width and stellar mass.

We have described gas flows over a range in stellar mass spanning the halo mass 
floor and mass ceiling required by equilibrium SFR models. As the mass on the red 
sequence grows with cosmic time, the fraction of massive ($\log M_*/\msun \sgreat\ 10.3$) 
galaxies having blue colors decreases. It follows that although we find no change in 
the outflow properties of blue cloud galaxies between redshift 0.4 and 1.4, the 
decline in the fraction of galaxies with blue galaxies may render outflows less
prevalent among massive galaxies at lower redshift.

We demonstrate that the fraction of spectra with blueshifted \feII\ absorption is much 
less than 100\%. While blueshifted absorption is found more often in spectra of
galaxies with high SFRs, particularly high specific SFRs, the blueshifted fraction 
is remarkably insensitive to galaxy properties overall. We conclude that the 
detection of blueshifted resonance absorption depends strongly on viewing angle
and the outflows must subtend solid angles much less than $4\pi$ steradians. This 
geometry is reminiscient of the bipolar outflows emanating from nearby starbursts
(Heckman \et 1990). Because outflows breakout of the ISM in the direction of 
least resistance, their collimation indicates the presence and prevalence of 
gas disks in blue galaxies at $z \sim 1$. The higher rotation speeds of these
disks in more massive galaxies may be the physical reason for the higher 
absorption equivalent widths observed at the systemic velocity in more massive galaxies.

Based on this geometrical consideration, we can describe the typical outflow properties.
The de-projected outflow velocity is measured directly when our sightline is
parallel to the outflow axis. If most of the galaxies have similar outflow properties,
then the largest blueshifts we measure, roughly 200\kms\ (line centroids)
and 500\kms\ (Doppler components), characterize the typical outflow
speed. Using bounds on the column density of outflowing \feII, the mass loss rates
in the warm phase are comparable to the average SFR for plausible assumptions 
about the launch radius, ionization fraction, and depletion. If the outflows persist 
at the observed rates for  $\sim 1$~ Gyr, then the low-ionization outfows would remove 
roughly $10^{10}$\msun\ of warm gas, a significant baryonic mass.

The most surprising aspect of this work was the discovery of 9 galaxies with 
robust redshifts of low-ionization metals lines. Since these galaxies do not have
unusual properties, inflows of low-ionization gas at velocities up to 100-200\kms\ 
(based on two-component fits) appear to be a common property of blue galaxies at $z \sim 1$.
We do not measure the hydrogen column density of these infalling streams directly; 
but for metallicities in the range of 1 to 0.1\zsun\  the streams would have total
hydrogen gas cloumns of $\log N(H) C_f  > 18.9$. At these columns, the inflows
subtend a small solid angle and escape detection in all but the largest surveys. We
find some evidence that the inflows are kinematically associated with satellite 
galaxies and extended disks, which if confirmed would help isolate the origin of
the infalling gas. The inflows may originate, for example, from tidal forces generated by 
galaxy interactions or multiphase cooling of hot halo gas. We emphasize that aside from 
their metallicity, the properties of these inflows are consistent with recent 
predictions for cold flows of accreting gas. Since future modeling of the mixing between
primordial inflow and recycled wind material may well increase the expected
metallicity of cold flows within a few tens of kpc of a galaxy, the detection of 
the infall in metal lines does not obviously rule out a cold flow origin

\acknowledgements
This research was supported by the National Science Foundation through grants
AST-0808161 and AST-1109288 (CLM) and CAREER award AST-1055081 (ALC), the David \& Lucile Packard Foundation (AES
and CLM), the Alfred P. Sloan Foundation (ALC), and a Dissertation Year Fellowship at UCLA (KK).  
We are grateful to the DEEP2 and AEGIS teams for providing both the
galaxy sample and ancillary data on galaxy properties. We thank  Donald Marolf, Andrey 
Kravtzov, Norman Murray, Dawn Erb, and Nicolas Bouch\'{e} for discussions that improved this
work. The Aspen Center for Physics provided a stimulating environment for completing much of the writing; and this research was partially 
supported by the National Science Foundation under Grant No. NSF PHY05-51164.
We also wish to recognize and acknowledge the highly significant
cultural role that the summit of Mauna Kea
has always had within the indigenous Hawaiian community. It is a
privilege to be given the opportunity to conduct observations from
this mountain.

{\it Facilities:} \facility{Keck}

\begin{appendix}
\section{Outflow Fraction Calculation} \label{sec:fout_calc}

As an example of how to calculate the outflow fraction, or more precisely
the fraction of spectra with blueshifted resonance absorption,
from data, we will use the results of
our single-component fitting to the \feII\ series of absorption lines.
For each spectrum, the fitted Doppler shift, $V_1$, and its
 uncertainty, $\delta V_1$,  provide the best estimate of
the distribution of Doppler shifts (i.e., its mean and standard deviation)
which would be obtained after many observations. Equivalently, in the special 
case of a symmetric probability distribution, we can describe the probability 
of measuring a Doppler shift $V_1$ when the true velocity is equal to or past
$V_{crit}$.

Let $P_i(V \le\ V_{crit})$ represent the probability that the absorption lines in 
a particular spectrum have a Doppler shift bluer than $ V_{crit}$. 
The probability, $P_i(V \le\ V_{crit})$, 
is the fractional area of this distribution with Doppler shifts $V \le\ V_{crit}$. 
Since this distribution is well described by a Gaussian function of 
width $\sigma \equiv \delta V_1$ (Section~\ref{sec:spectral_features}), we simply
compute $P_i(V \le\ V_{crit})$ from an expression closely related to the 
normal error integral.

The expression for the entire integral depends on the relation of 
$V_{crit}$ and $V$. We adopt a sign convention that outflow velocities are
negative.
When a spectrum has a low Doppler shift relative to the outflow threshold,
$V_{crit} \le\ V_{1} $, the chance that the true value of the Doppler
shift is faster than $V_{crit}$ is small and given by
\begin{eqnarray}
P_i(V \le\ V_{crit}|V_1,\delta V_1) = 0.5 - Q.
\end{eqnarray}
When the fitted Doppler shift is bluer than the outflow threshold,
$V_{crit} \ge\ V_{1}$, the chance that the true Doppler shift is faster than
$V_{crit}$ is much more significant; and the total integral is
\begin{eqnarray}
P_i(V \le\ V_{crit}|V_1,\delta V_1) = 0.5 + Q.
\end{eqnarray}
The function 
\begin{eqnarray}
Q(t) = \frac{1}{\sqrt{2\pi}} \int_0^t e^{-z^2/2} dz = \frac{1}{2} erf(t).
\end{eqnarray}
describes the integral over a portion of a Gaussian distribution, 
where we have defined the dimensionless variable $t$ as
\begin{eqnarray}
t \equiv |V_{crit} - V_1| / \delta V_1.
\end{eqnarray}

For each discrete value of the threshold velocity, we compute the outflow 
fraction by adding the outflow probabilities $P_i(V \le\ V_{crit})$ 
of all the spectra. We normalize this sum by the number of spectra, $N_{spec}$
to obtain the outflow fraction
\begin{eqnarray}
P_{out}(V_{crit}) = N_{spec}^{-1} \Sigma_{i=1}^{N_{spec}} P_i(V \le\ V_{crit}).
\end{eqnarray}
The probability of finding a Doppler shift redder than $V_{crit}$ 
is $1 - P_{out}$. For positive values of $V_{crit}$,  the fraction
of spectra with $V > V_{crit}$ obtained from $1 - P_{out}$ describes
the frequency of redshifted velocities in the sample. 

Error bars are calculated directly from the probability distribution.
For a sample of $N_{spec}$ objects, we compute the probability of finding
$0, 1, 2, ..., N_{spec} - 1, N_{spec}$ objects with $V \le\ V_{crit}$.
     In practice, our code uses a subroutine that starts with the first 
     spectrum in the list, calculates the probability of finding 0 or 1 objects
     with $V_1 \le\ V_{crit}$, and then proceeds to calculate probabilities
     for larger samples in an iterative fashion advancing through lists of
      $2, 3, 4, ...$ objects until the probabilities are obtained for the
     full list of $N_{spec}$ objects.
The sum of the first $i+1$ values in this sequence yields the cumulative
probability that the number of outflows with $V_1 \le\ V_{crit}$ is less than or equal to
$N_i$, where $i = 0$ to $N_{spec}$. The boundaries of the 68.27\% confidence 
interval are found by interpolating between the $N_i$ values to estimate
the number of spectra where the cumulative probability that $V_1 \le\ V_{crit}$
is 0.1587 (lower bound) and 0.8413 (upper bound).

\end{appendix}

\clearpage

\begin{turnpage}
 \input table1.tex
  \end{turnpage}

\clearpage
\input table2.tex

 \clearpage

\begin{turnpage}
 \input table3.tex

  \end{turnpage}

\clearpage

\begin{turnpage}
 \input table4.tex

  \end{turnpage}

 \input table5.tex

\clearpage

\input table6.tex

\end{document}

%% file: mymac.tex
%  Sept. 1993
%  CLM
%  Use ``\input mymac.tex'' in a tex file to use these definitions.
%
%\renewcommand{\deg}{\mbox{$^{\circ}$}}
%
\def \dlow {\mbox{$400 {\rm ~lines~mm}^{-1}$}}
\def \dhigh {\mbox{$600 {\rm ~lines~mm}^{-1}$}}
% PAPER SPECIFIC
% EVAN S.
\newcommand{\be}{\begin{equation}} \newcommand{\ba}{\begin{eqnarray}}
\newcommand{\ee}{\end{equation}} \newcommand{\ea}{\end{eqnarray}}
\def\etal{{\it et al.\thinspace}}
\def\-{{\em{---}}}
\def \mA {\mbox{${\rm m \AA} $} }
\def \rr {\mbox{${\rm RR}$} }
\def \rarb {\mbox{${\rm R_AR_B}$} }
\def \rara {\mbox{${\rm R_AR_A}$} }
\def \dd {\mbox{${\rm DD}$} }
\def \dada {\mbox{${\rm D_AD_A}$} }
\def \dadb {\mbox{${\rm D_AD_B}$} }
\def \dr {\mbox{${\rm DR}$} }
\def \darb {\mbox{${\rm D_AR_B}$} }
\def \dara {\mbox{${\rm D_AR_A}$} }
\def \dbra {\mbox{${\rm D_BR_A}$} }
\def \hMpc      {h^{-1}{\rm\ Mpc}}
\def \hkpc      {h^{-1}{\rm\ kpc}}
\def \h         {\hbox{$\, h$} }
\def \hinv      {\hbox{$\, h^{-1}$} }
\def \hinvseven    {\hbox{$\, h_{70}^{-1}$} }
\def\ewr{\mbox {EW$_r$}}
\def\ewo{\mbox {EW$_o$}}
\def\H7{\mbox {$h_{0.7}$}}
\def\naI{\mbox {\ion{Na}{1}}}
\def\mgI{\mbox {\ion{Mg}{1}}}
\def\feI{\mbox {\ion{Fe}{1}}}
\def\oVI{\mbox {\ion{O}{6}}}
\def\znII{\mbox {\sc Zn~II~}}
\def\crII{\mbox {\sc Cr~II~}}
\def\alI{\mbox {\sc Al~I~}}
\def\alII{\mbox {\sc Al~II~}}
\def\alIII{\mbox {\sc Al~III~}}
\def\mgII{\mbox {\ion{Mg}{2}}}
\def\mnII{\mbox {\ion{Mn}{2}}}
\def\niII{\mbox {\ion{Ni}{2}}}
\def\feII{\mbox {\ion{Fe}{2}}}
\def\feIII{\mbox {\ion{Fe}{3}}}
\def\cIV{\mbox {\ion{C}{4}}}
\def\sV{\mbox {\ion{S}{5}}}
\def\siIV{\mbox {\ion{Si}{4}}}
\def\siII{\mbox {\ion{Si}{2}}}
\def\siI{\mbox {\ion{Si}{1}}}
\def\cII{\mbox {\ion{C}{2}}}
\def\cIII{\mbox {\ion{C}{3}}}
\def\llambda{\mbox {$\lambda$}}
\def\mstar{\mbox {$M_*$}}
\def\hlen{\mbox {$h_{0.7}^{-1}$}}
\def\lstarlya{\mbox {$L^*_{Ly\alpha}$}}
\def\IZw18{I~Zw~18}
\def\m82{M82}
\def\Ab{Abell~}
\def\gi{\mbox {\rm g-i}}
\def\ug{\mbox {\rm u-g}}
\def\br{\mbox {\rm b-r}}
\def\eqn{equation}
\def\vesc{\mbox {$v_{\rm esc}$}}
% HE PAPER
\def\heha{\mbox {He~I~$\lambda 5876$ / H$\alpha$}}
\def\xhe{\mbox {$\chi({\rm He}) / \chi({\rm H})$} }
\def\heii{\mbox {${\rm He}^+$}}
\def\he{\mbox {\rm He}}
\def\hii{\mbox {${\rm H}^+$}}
\def\h{\mbox {\rm H}}
\def\mab{\mbox {$\rm m_{AB}$}}
\def\ssp{\baselineskip=13pt plus 1pt minus 1pt}
\def\tsp{\baselineskip=5pt plus 1pt minus 1pt}
%
% ASTRO SYMBOLS (revised to work in/out mathmode).
%
\def\deg{\mbox {$^{\circ}$}}
\def\msun{\mbox {${\rm ~M_\odot}$}}
\def\zsun{\mbox {${\rm ~Z_{\odot}}$}}
\def\lsun{\mbox {${~\rm L_\odot}$}}
\def\msunyr{\mbox {$~{\rm M_\odot}$~yr$^{-1}$}}
\def\angs{\mbox {~\AA}}
\def\lya{\mbox {Ly$\alpha$~}}
\def\Ha{\mbox {H$\alpha$~}}
\def\Hb{\mbox {H$\beta$~}}
\def\Hg{\mbox {H$\gamma$~}}
\def\tion{\mbox {$T_{\rm ion}$~}}
\def\ch{\mbox {$\bigtriangleup$}}
\def\grad{\mbox {$\bigtriangledown$}}
\def\lstar{\mbox {$L^*$}}
\def\line{\mbox {~$\lambda$}}
\def\lines{\mbox {~$\lambda\lambda$~}}
\def\h0{\mbox {~H$_0$}}
\def\q0{\mbox {~q$_0$}}
%
% **** LINE RATIOS ****
%
\def\auroral{[OIII]~$\lambda4363$~}
\def\auroral{[OIII]~$\lambda4363$~}
\def\ohsun{\mbox {(O/H)$_{\odot}$~}}

\def\o3hb{[OIII]$\lambda5007$~/~H$\beta$~}
\def\O1ha{[OI]$\lambda6300$~/~H$\alpha$~}
\def\Ru{[OII]$\lambda\lambda3727$~/~[OIII]$\lambda5007$~}
\def\s2ha{[SII]$\lambda\lambda6717,31$~/~H$\alpha$~}
\def\2z2{HeII~$\lambda4686$~}
\def\z7{[NII]~$\lambda6583$ }
\def\N2{[NII]~$\lambda6583$~/~H$\alpha$~}
\def\16z2{[SII]~$\lambda\lambda6717, 6731$ }
\def\HgI{HgI~$\lambda4358$~}
\def\Sdensity{[SII]~$\lambda6717 / \lambda6731$}
\def\Temp{[OIII]~$\lambda\lambda4959 + 5007 ~{\rm to}~ \lambda4363$~}
%
% **** end ****
%
\def\n{NGC~}
\def\asec{\ifmmode {'' }\else $''~$\fi}  % arc sec
\def\amin{\ifmmode {' }\else $'~$\fi}    % arc min
\def\arcsper{\ifmmode \rlap.{'' }\else $\rlap{.}'' $\fi} % '' %Arcsec period
\def\arcmper{\ifmmode \rlap.{' }\else $\rlap{.}' $\fi} % '  %Arcmin period
\def\sles{\lower2pt\hbox{$\buildrel {\scriptstyle <}
   \over {\scriptstyle\sim}$}} % approximately less than
\def\sgreat{\lower2pt\hbox{$\buildrel {\scriptstyle >}
    \over {\scriptstyle\sim}$}} % approximately greater than
%These are smaller.
%\def\gapp{$_>\atop{^\sim}$}  % approximately greater than
%\def\lapp{$_<\atop{^\sim}$}  % approximately less than
\def\gapp{\mbox {$_>\atop{^\sim}$}}  % approximately greater than
\def\lapp{\mbox {$_<\atop{^\sim}$}}  % approximately less than
%
% UNITS
\def\kms{~km~s$^{-1}$~}
\def\ergsec{~ergs~s$^{-1}$~}
\def\sb{~ergs~s$^{-1}$~cm$^{-2}$~arcsec$^{-2}$}
\def\flux{~ergs~s$^{-1}$~cm$^{-2}$}
\def\flam{~ergs~s$^{-1}$~cm$^{-2}$ \AA$^{-1}$}
\def\cm3{~cm$^{-3}$}
\def\col{\mbox {~cm$^{-2}$}}
\def\mpc3{~Mpc$^{3}$}
\def\mpc-3{~Mpc$^{-3}$}
\def\rate{~sec$~{-1}$}
\def\um{~${\mu}$m~}
% ABBREVIATIONS
\def\fig{{Figure}}
\def\figs{{Figures}}
\def\tbl{{Table}~}
\def\sec{{Sec.}~}
\def\x{{X-ray}~}
\def\xs{{X-rays}~}
\def\X{{X-Ray}~}

%
% REFERENCES
\def\et{{\rm et\thinspace al.}\ }   % et al.
\def\ets{{\rm et\thinspace al.'s}\ }   % et al.'s
\def\reff{\par\noindent\parskip=1pt\hangindent=3pc\hangafter=1}
\def\apj{ApJ}
\def\apjs{ApJS}
\def\pasp{PASP}
\def\aj{AJ}
\def\mn{MNRAS}
\def\nat{Nature}
\def\aa{A\&A}
\def\aasup{A\&AS}
\def\baas{BAAS}
\def\annrev{ARA\&R}
\def\aar{A\&AR}
\def\pasj{PASJ}
%

% USE THIS FOR TABLE REFERENCES
%
\def\beginrefs{
         {\normalsize}
         {\noindent}
         \small
        \baselineskip=11pt
        \parindent=0pt
        \frenchspacing
        \parskip=1pt plus 1pt
%        \interlinepenalty=1000\tolerance=400
        \everypar={\hangindent=0.42in}}

%% file: table1.tex
\begin{deluxetable}{lrrrcclccl}
\tablecaption{Keck/LRIS Observations}
\tabletypesize{\tiny}
\tablewidth{0pt}
\tablehead{
\colhead{Field} &
\colhead{R.A.} &
\colhead{Decl.} &
\colhead{Mask PA}   &
\colhead{Grism / Grating} &
 \colhead{Dichroic} &
\colhead{Conditions} &
\colhead{Exposure Time} &
\colhead{Number of} &
\colhead{Observing Run}  \\
\colhead{}      &
\colhead{(J2000.0)}  &
\colhead{(J2000.0)} &
\colhead{($\deg$)}  &
\colhead{(l~mm$^{-1}$)\tablenotemark{a}}  &
\colhead{}         &
\colhead{Cloud Cover \& Seeing(\asec)} &
  \colhead{LRIS-B/LRIS-R (s)\tablenotemark{b}}                      &
  \colhead{Exposures (B/R)}                      &
 \colhead{Dates}      }
\startdata
msc42\_1           & 02 28 56.8 & +00 33 39   & 120.0    & 400 / 831 &  d680 & Clear to Mostly Cloudy \& 0.9-1.6 & 11340 / 11100 & 6/6 & 2007 Oct 6-7  \\
msc42\_5           & 02 30 35.6 & +00 28 17   & 143.0    & 400 / 831 &  d680 & Mostly Cloudy to Light Cirrus \& 0.7-0.9 & 26883 / 25400 & 15/14 & 2008 Sep 28-29 \\
msc12\_d           & 14 17 09.8 & +52 31 18   & 165.0    & 400 / 831 &  d680 & Clear to Mostly Cloudy \& 0.6-2.0& 18500 / 18000 & 10/10& 2008 June 5-6  \\
msc12\_8           & 14 18 19.4 & +52 34 33   & 5.0      & 400 / 831 &  d680 & Clear to Partly Cloudy \& 0.6-2.0 & 28800 / 26940 & 15/31& 2009 June 18-21\tablenotemark{c}  \\
msc12\_ee          & 14 18 46.8 & +52 38 59   & 170.0    & 600 / 600 &  d560 & Clear to Partly Cloudy \& 0.6-2.0 &11520 / 10800  & 6/12 & 2009 June 18-21\tablenotemark{c}  \\
msc22\_bb          & 16 51 09.2 & +34 58 15   & 25.0     & 600 / 600 &  d560 & Clear to Mostly Cloudy \& 0.6-2.0 &18400 / 17950  & 9/9 & 2008 June 5-6 \\
msc22\_6           & 16 51 28.8 & +34 48 09   & 60.0     & 400 / 831 &  d680 & Clear to Partly Cloudy \& 0.6-2.0 &30760 / 23100  & 16/26 & 2009 June 18-21\tablenotemark{c}  \\
msc32\_aa          & 23 29 19.5 & +00 07 38   & 82.0     & 600 / 600 &  d560 & Mostly Cloudy to Light Cirrus \& 0.7-0.9 &11070 / 10800 & 6/6 & 2008 Sep 28-29 \\
msc32\_1           & 23 30 18.9 & +00 12 53   & 80.0     & 400 / 831 &  d560 & Clear to Light Cirrus to Mostly Cloudy &25810 / 24182  & 14/14 &2007 Oct 6-7 \& 2008 Sep 28-29 \\
\enddata
\tablenotetext{a}{After binning by 2 pixels in the dispersion direction, \dlow\ and \dhigh\
                  blue spectra have 2.18~\AA~pix$^{-1}$ and 1.26~\AA~pix$^{-1}$, respectively.
                  The unbinned 831~mm$^{-1}$ and 600~mm$^{-1}$ red spectra have dispersions of
                  0.93~\AA~pix$^{-1}$ and 1.28~\AA~pix$^{-1}$, respectively, for all runs except
                  2009 June.
                  }
\tablenotetext{b}{The typical exposure time per frame was 1800~s.}
\tablenotetext{c}{The red spectra from 2009 June were obtained with a new detector.
                  After binning by 2 pixels in the dispersion direction,
                  the 831~mm$^{-1}$ and 600~mm$^{-1}$ red spectra have dispersions
                  of 1.16~\AA~pix$^{-1}$ and 1.60~\AA~pix$^{-1}$, respectively.
                  Shorter exposure times were used for individual frames due to high cosmic ray rates. 
                  }
\label{tab:observations}  
\end{deluxetable}
\normalsize

%% file: table2.tex
\begin{deluxetable}{lcrclllll}
\tabletypesize{\footnotesize}
\tablewidth{0pt}
\tablecaption{Properties of Galaxies with Net \feII\ Doppler Shifts}
\tablehead{
\colhead{Object } &
\colhead{Grism }     &
\colhead{SNR } &
\colhead{Redshift} &
\colhead{$z$ } &
\colhead{M$_B -5\log h$ } &
\colhead{U-B } &
\colhead{$\log M_*$ } &
\colhead{SFR } \\
\colhead{ } &
\colhead{($l~mm^{-1}$) }     &
\colhead{ } &
\colhead{Source} &
\colhead{ } &
\colhead{(AB mag) } &
\colhead{ } &
\colhead{(\msun) } &
\colhead{(\msunyr) } \\
\colhead{(1)     } &
\colhead{ (2)     }     &
\colhead{   (3)   } &
\colhead{    (4)                        } &
\colhead{    (5)       } &
\colhead{    (6)    } &
\colhead{ (7)            } &
\colhead{ (8)           } &
\colhead{   (9)         } 
}
\startdata
12008197 & 400     &  12.2  &   L    & 0.9802& -21.19& 0.804& 10.63$^{\dagger}$      &98      \\
12008550 & 400     &   9.86  &   L   & 1.3024& -21.22& 0.620& 10.08$^{\dagger}$     & 25     \\
12011836 & 400     &   9.38  &   L   & 0.9270& -19.90& 0.579& 10.26$^{\dagger}$     & 14     \\
12012777 & 400     &  11.68 &   L    & 1.2742& -21.06& 0.516& 10.11$^{\dagger}$     & 23     \\
12013242 & 400     &   8.83  &   L   & 1.2867& -21.31& 0.450& 10.19                  & 26     \\
12011428$^*$ & 400 &   4.6     &   L & 1.2840& -19.87& 0.180&  9.65              & 9     \\
12015177 & 600     &   5.3   &   L   & 0.9860& -20.23& 0.487&  9.69                  & 12     \\
12016019$^*$ & 400 &  11.57   &   L  & 1.0846& -20.71& 0.622& 10.03$^{\dagger}$  & 19     \\
12019542 & 600     &   3.73  &   D   & 1.2784& -21.68& 0.705& 10.40$^{\dagger}$     & 40     \\
12019962$^*$ & 400 &  10.1    &   L  & 0.6444& -19.22& 0.458&  9.77$^{\dagger}$  & 2      \\
12019996$^*$ & 400 &  7.36    &   L  & 1.2812& -21.71& 0.634& 10.67$^{\dagger}$  & 43     \\
12100420$^*$ & 400 &  8.60    &   L  & 1.1995& -20.63& 0.601& 10.01$^{\dagger}$  & 10     \\
22004858$^*$ & 400 & 15.00   &   L   & 1.2686& -21.84& 0.624& 10.54              & \nodata   \\
22005216 & 400     &  9.27  &   L    & 0.9130& -19.49& 0.556&  9.59                  & \nodata   \\
22005270$^*$ & 400 & 10.57   &   L   & 0.8308& -20.38& 0.710& 10.40              & \nodata   \\
22006172 & 400     & 10.65 &   L     & 0.8328& -21.10& 0.707& 10.65                  & \nodata   \\
22006207 & 400     &  9.42  &   L    & 1.2709& -20.59& 0.419&  9.88                  & \nodata   \\
22029066 & 600     & 10.45 &   L     & 0.7852& -21.59& 0.891& 11.29$^{\dagger}$     & \nodata   \\
22029224 & 600     &  8.66  &   L    & 0.8603& -20.90& 0.611& 10.27                  & \nodata   \\
22036854 & 600     &  7.29  &   L    & 0.9373& -20.85& 0.564& 10.32                  & \nodata   \\
22036912 & 600     &  6.66  &   L    & 0.8099& -21.02& 0.864& 10.86                  & \nodata   \\
32010773 & 600     &  9.23  &   L    & 0.8039& -19.91& 0.234&  9.60$^{\dagger}$     & \nodata   \\
32011098 & 600     &  5.01  &   L    & 0.9561& -20.22& 0.546&  9.77$^{\dagger}$     & \nodata   \\
32011099 & 600     &  6.51  &   L    & 0.8832& -20.06& 0.492&  9.80                  & \nodata   \\
32011192 & 600     &  9.51  &   L    & 0.8478& -20.60& 0.587& 10.07$^{\dagger}$     & \nodata   \\
32011682 & 600     &  8.79  &   L    & 0.8359& -20.36& 0.511& 10.23$^{\dagger}$     & \nodata   \\
32016857 & 600     &  8.32  &   L    & 0.9391& -20.81& 0.440&  9.81$^{\dagger}$     & \nodata   \\
32017112 & 600     &  5.85  &   L    & 1.0084& -20.89& 0.665& 10.23$^{\dagger}$     & \nodata   \\
32020468$^*$ & 400 &  3.16    &   L  & 1.2355& -21.00& 0.734& 10.28$^{\dagger}$  & \nodata   \\
32022156 & 600     &  4.60  &   L    & 1.0434& -20.70& 0.580& 10.10$^{\dagger}$     & \nodata   \\
42006781 & 400     & 10.98 &   D     & 1.2859& -20.74& 0.546&  9.68$^{\dagger}$     & \nodata   \\
42006915 & 400     & 25.09 &   L     & 0.8945& -20.55& 0.386& 10.50$^{\dagger}$     & \nodata   \\
42014101 & 400     &  8.39  &   L    & 0.7494& -20.47& 0.899& 10.94$^{\dagger}$     & \nodata   \\
42014154 & 400     &  9.75  &   L    & 0.8427& -19.40& 0.465&  9.77$^{\dagger}$     & \nodata   \\
42014585 & 400     &  8.17  &   D    & 1.2705& -20.91& 0.588& 10.18                  & \nodata   \\
42014618 & 400     & 17.64 &   L     & 1.0130& -20.37& 0.592&  9.81                  & \nodata   \\
42014718 & 400     &  9.21  &   L    & 1.1904& -21.27& 0.692& 10.85$^{\dagger}$     & \nodata   \\
42014732 & 400     & 16.01 &   L     & 0.7502& -19.40& 0.453&  9.50$^{\dagger}$     & \nodata   \\
42021266 & 400     & 11.66 &   L     & 0.9765& -19.63& 0.616&  9.50                  & \nodata   \\
42022173 & 400     &  8.15  &   D    & 1.3112& -20.25& 0.324&  9.68                  & \nodata   \\
42022307 & 400     & 20.72 &   D     & 1.2595& -21.43& 0.466& 10.22                  & \nodata   \\
42026243 & 400     &  3.30  &   L    & 1.3456& -21.66& 0.667& 10.88$^{\dagger}$     & \nodata   \\
42033991 & 400     &  6.48  &   L    & 0.8714& -20.13& 0.392&  9.94$^{\dagger}$     & \nodata   \\
42006875$^*$ & 400 & 19.18   &   L   & 0.8696& -21.10& 0.602& 10.50$^{\dagger}$  & \nodata   \\
\enddata
\tablecomments{
(1) - Object identification from DEEP2. The \feII\ absorption in these spectra
is Doppler shifted (relative to the galaxy) at a significance level greater than
$3\sigma$. Inflows (redshifts) are denoted by $^*$; the objects 22028473 and
22036194 have been dropped per the discussion in \S~5.1. 
(2) - Blue grism where \dlow\ and \dhigh\ imply FWHM of 435\kms\ and 282\kms, respectively.
(3) - Continuum SNR between 2400 and 2500 \AA.
(4) - Source of redshift. The flag L/D indicates (L)RIS (the spectra presented here)
      or (D)EIMOS, DEEP2 redshift, respectively. 
(5) - Redshift.
(6) - DEEP2 absolute magnitude (Willmer \et 2006).
(7) - DEEP2 color (Willmer \et 2006).
(8) - Stellar mass from SED fitting (Bundy \et 2006).
Galaxies with an SED fit that includes K-band photometry (most certain) are
denoted by $^{\dagger}$; the other fits are based on BRI photometry.
(9) - Star formation rate derived from dust-corrected UV continuum
        luminosity and converted to a Chabrier initial
        mass function.
}
\label{tab:properties}
\end{deluxetable}

%% file: table3.tex
\begin{deluxetable}{llllllllllll}
\tabletypesize{\scriptsize}
\tablewidth{0pt}
\tablecaption{Outflow Properties}
\tablehead{
\colhead{Object} &
\colhead{W(2374)} &       
\colhead{W(2261)} &
\colhead{ $\log [N(Fe^{+}) C_f]$                   } &
\colhead{  $V_{1}$           } &
\colhead{ $b_{ISM}$ } &
\colhead{ $V_{DOP}$              } &
\colhead{ $\log [N_{ISM}(Fe^{+}) C_f]$ } &
\colhead{ $\log [N_{DOP}(Fe^{+}) C_f]$} &
\colhead{ $V_{max}(\lambda2374)$  } &
\colhead{ $V_{max}(\lambda2796)$  } \\
\colhead{}         &
\colhead{    (\AA)       } &
\colhead{  (\AA)  } &
\colhead{  (\col)                       } &
\colhead{  (\kms)      } &
\colhead{  (\kms)   } &
\colhead{ (\kms)         } &
\colhead{   (\col)          } &
\colhead{  (\col)         } &
\colhead{      (\kms)    } &
\colhead{      (\kms)  } \\
\colhead{(1)     } &
\colhead{ (2)     }     &
\colhead{   (3)   } &
\colhead{    (4)                        } &
\colhead{    (5)       } &
\colhead{    (6)    } &
\colhead{ (7)            } &
\colhead{ (8)           } &
\colhead{   (9)         } &
\colhead{     (10)          } &
\colhead{   (11)          } 
}
\startdata
12008197    & $1.21 \pm 0.23 $ &    0.21  & 14.89 - 15.75  & $-62 \pm 19$ &    126&  $-93 \pm   43  $ & $14.80 -15.34$  & $14.58 -15.93$    &      \nodata         &     $ -466 \pm 120 $ \\
12008550    & $1.05 \pm 0.19 $ &    0.17  & 14.83 - 15.66  & $-145 \pm 23$ &    77&  $-322 \pm  135 $ & $14.69 -15.27$  & $14.32 -15.79$    &     $ -547 \pm 93 $ \\
12011836    & $2.14 \pm 0.32 $ &    0.27  & 15.14 - 15.86  & $-128 \pm 21$ &   114&  $-465 \pm  142 $ & $15.00 -15.65$  & $14.42 -15.01$      &      $-697 \pm 117 $ &     $ -667 \pm 94 $ \\
12012777    & $2.13 \pm 0.19 $ &    0.18  & 15.13 - 15.68  & $-217 \pm 13$ &    95&  $-443 \pm  70  $ & $15.07 -15.89$  & $14.79 -15.59$      &      $-704 \pm 122 $ &     $ -909 \pm 146 $ \\
12013242    & $1.14 \pm 0.23 $ &    0.24  & 14.86 - 15.81  & $-73 \pm 22$ &    121&  $-97 \pm   45  $ & $14.89 -15.48$  & $14.78 -15.81$      &      \nodata         &     \nodata \\
12015177    & $2.34 \pm 0.32 $ &    0.33  & 15.17 - 15.94  & $-101 \pm 18$ &    91&  $-106 \pm  49  $ & $14.95 -15.64$  & $14.72 -15.77$      &      $-418 \pm 51$   &     \nodata \\
12019542    & $0.72 \pm 0.30 $ &    0.39  & 14.66 - 16.02  &$-183 \pm 60^{4}$ & 69&  $-311 \pm  273 $ & $14.51 -15.04$  & $14.51 -15.04$     &       \nodata         &     \nodata \\
22005216    & $0.98 \pm 0.26 $ &    0.25  & 14.80 - 15.82  & $-106 \pm 28$ &    59&  $-450 \pm  146 $ & $14.90 -15.76$  & $14.20 -14.66$     &      \nodata         &     $ -504 \pm 56 $ \\
22006172    & $1.57 \pm 0.28 $ &    0.27  & 15.00 - 15.86  & $-112 \pm 21$ &   118&  $-95 \pm   47 $  & $14.84 -15.41$  & $14.80 -15.80$       &      $-588 \pm 178 $ &     $ -514 \pm 128 $ \\
22006207    & $0.76 \pm 0.15 $ &    0.16  & 14.69 - 15.63  & $-199 \pm 39$ &    76&  $-34 \pm   35  $ & $14.57 -15.11$  & $14.68 -16.06$      &      $-487 \pm 84 $  &     $ -693 \pm 94 $ \\
22029066    & $1.85 \pm 0.20 $ &    0.27  & 15.07 - 15.86  & $-62 \pm 16$ &    152&  $-209 \pm  46  $ & $14.93 -15.49$  & $>14.05 $       &      $-396 \pm 115 $ &     $ -919 \pm 117 $   \\
22029224    & $1.09 \pm 0.19 $ &    0.22  & 14.84 - 15.77  & $-69 \pm 16$ &    117&  $-146 \pm  53  $ & $14.81 -15.37$  & $14.37 -14.94$       &      \nodata         &     $ -450 \pm 39 $ \\
22036854    & $2.08 \pm 0.24 $ &    0.24  & 15.12 - 15.81  & $-91 \pm 14$ &     58&  $-121 \pm  37  $ & $14.91 -15.81$  & $14.55 -15.30$       &      $-444 \pm 46 $  &     \nodata \\
22036912    & $0.84 \pm 0.21 $ &    0.34  & 14.73 - 15.96  & $-86 \pm 16$ &    167&  $-238 \pm  87  $ & $14.80 -15.29$  & $14.14 -14.49$       &      $-297 \pm 53 $  &     $ -653 \pm 58 $ \\
32010773    & $0.63 \pm 0.17 $ &    0.19  & 14.60 - 15.70  & $-197 \pm 46$ &    45&  $-189 \pm  126 $ & $14.45 -15.03$  & $14.13 -14.60$      &      $-511 \pm 95 $  &     $ -703 \pm 43 $ \\
32011098    & $0.59 \pm 0.31 $ &    0.33  & 14.58 - 15.94  & $-89 \pm 29$ &     87&  $-78 \pm   101 $ & $14.54 -15.04$  & $14.56 -15.57$       &      \nodata         &     \nodata \\
32011099    & $1.79 \pm 0.25 $ &    0.23  & 15.06 - 15.79  & $-46 \pm 13$ &     72&  $-97 \pm   24  $ & $15.06 -16.12$  & $14.65 -15.30$       &      $-356 \pm 69 $  &     $ -463 \pm 105 $ \\
32011192    & $1.85 \pm 0.19 $ &    0.22  & 15.07 - 15.77  & $-154 \pm 9$ &     97&  $-154 \pm  38  $ & $14.96 -15.65$  & $14.74 -15.30$       &      $-527 \pm 86 $  &     $ -739 \pm 74 $ \\
32011682    & $1.30 \pm 0.20 $ &    0.23  & 14.92 - 15.79  & $-85 \pm 11$ &     96&  $-246 \pm  60  $ & $14.90 -15.55$  & $14.18 -14.68$       &      $-317 \pm 71 $  &     $ -521 \pm 62 $ \\
32016857    & $0.88 \pm 0.18 $ &    0.20  & 14.75 - 15.73  & $-91 \pm 15$ &     97&  $-227 \pm  82  $ & $14.75 -15.32$  & $14.42 -15.53$       &      $-297 \pm 84 $  &     \nodata \\
32017112    & $1.25 \pm 0.25 $ &    0.30  & 14.90 - 15.90  & $-80 \pm 18$ &     99&  $-255 \pm  106 $ & $14.80 -15.38$  & $14.24 -14.88$      &      \nodata         &     \nodata \\
32022156    & $0.95 \pm 0.34 $ &    0.43  & 14.78 - 16.06  & $-180 \pm 48$ &    74&  $-318 \pm  182 $ & $14.82 -15.49$  & $14.41 -14.98$      &      $-355 \pm 42 $  &     $ -436 \pm 27 $ \\
42006781    & $2.07 \pm 0.17 $ &    0.15  & 15.12 - 15.60  & $-70 \pm 13$ &     59&  $-143 \pm  28  $ & $15.00 -16.11$  & $14.06 -14.42$       &      $-565 \pm 77 $  &     $ -495 \pm 68 $ \\
42006915    & $1.29 \pm 0.10 $ &    0.097 & 14.92 - 15.41  & $-76 \pm 10$ &    206&  $-322 \pm  75  $ & $14.67 -15.08$  & $13.91 -14.03$       &      $-621 \pm 117 $ &     $ -576 \pm 91 $ \\
42014101    & $1.08 \pm 0.35 $ &    0.32  & 14.84 - 15.93  & $-150 \pm 38$ &   107&  $-281 \pm  251 $ & $14.65 -15.17$  & $14.48 -15.29$      &      $-514 \pm 68 $  &     $ -480 \pm 66 $ \\
42014154    & $0.84 \pm 0.22 $ &    0.25  & 14.73 - 15.82  & $-105 \pm 30$ &    63&  $-315 \pm 163  $ & $14.77 -15.46$  & $13.93 -14.42$      &      \nodata         &     \nodata \\
42014585    & $1.63 \pm 0.27 $ &    0.27  & 15.02 - 15.86  & $-177 \pm 30$ &    61&  $-297 \pm  85  $ & $14.96 -15.94$  & $14.55 -15.20$      &      $-579 \pm 155 $ &     $ -456 \pm 47 $ \\
42014618    & $1.29 \pm 0.14 $ &    0.15  & 14.92 - 15.60  & $-76 \pm 13$ &    121&  $-115 \pm  33  $ & $14.83 -15.39$  & $14.43 -15.67$       &      $-548 \pm 107 $ &     \nodata  \\
42014718    & $2.06 \pm 0.31 $ &    0.30  & 15.12 - 15.90  & $-156 \pm 19$ &   199&  $-219 \pm  77  $ & $14.96 -15.49$  & $14.84 -15.90$      &      $-712 \pm 216 $ &     $ -651 \pm 128 $ \\
42014732    & $0.91 \pm 0.17 $ &    0.16  & 14.76 - 15.63  & $-81 \pm 26$ &     79&  $-52 \pm   43  $ & $14.59 -15.13$  & $14.47 -15.33$      &      $-450 \pm 82 $  &     \nodata  \\
42021266    & $1.06 \pm 0.20 $ &    0.19  & 14.83 - 15.70  & $-108 \pm 24$ &    76&  $-388 \pm  152 $ & $14.75 -15.37$  & $14.28 -15.26$      &      $-590 \pm 163 $ &     \nodata  \\
42022173    & $0.81 \pm 0.21 $ &    0.23  & 14.71 - 15.79  & $-134 \pm 27$ &   106&  $-125 \pm  85  $ & $14.67 -15.20$  & $14.70 -15.85$      &      \nodata         &     \nodata  \\
42022307    & $1.15 \pm 0.09 $ &    0.10  & 14.87 - 15.43  & $-146 \pm 10$ &   165&  $-381 \pm  79  $ & $14.71 -15.18$  & $14.20 -14.66$      &      $-554 \pm 62 $  &     $ -529 \pm 50 $ \\
42026243    &\nodata           &    0.70  & \nodata - 16.27  & $-302 \pm 38$ &  93&  $-476 \pm  201 $ & $15.17 -16.22$  & $15.03 -16.29$      &      \nodata         &     $ -843 \pm 135 $ \\
42033991    & $1.39 \pm 0.46 $ &    0.40  & 14.95 - 16.03  & $-139 \pm 41$ &   132&  $-369 \pm  319 $ & $14.69 -15.19$  & $14.52 -15.36$      &      $-466 \pm 136 $ &     \nodata \\
\enddata
\tablecomments{
(1) DEEP2 survey identification.
(2) Rest-frame equivalent width of \feII\ 2374 absorption from direct integration of the trough. Error estimated from bootstrap resampling.
(3) Upper limit ($1\sigma$) on \feII\ 2261 equivalent width. Computed from an assumed linewidth
    that is twice the FWHM of the line response function and the continuum S/N ratio.
(4) Limits on \feII\ column density for a fiducial covering factor of $C_f = 1$. For optically thin absorption, the linear portion
    of the curve-of-growth gives $N(\feII) = 6.40 \times 10^{14}\col W_{2374}$, which we adopt as a lower bound.
    We derive an upper bound,     $N(\feII) = 9.06 \times 10^{15}\col W_{2261}$, by taking
    the $3\sigma$ upper limit on the \feII\ 2261 equivalent width.
(5) Doppler shift of the single velocity component model fitted to \feII\ absorption as described in Section~\ref{sec:measure_fe2}.
(6) Doppler parameter of the systemic component in the two-component fits to
    the \feII\ series.  We adopt characteristic velocities of gas in the galaxy as probed
    by cooling radiation from \ion{H}{2} regions. Specifically, we require the FWHM of
    the absorption optical depth profile match that of [OII] measured previously
    by Weiner \et (2006), which gives $b_{sys} = \sqrt{2} \sigma ([OII]) $.
(7) Doppler shift of the {\it Doppler component} fitted to \feII\ lines in two-component model described in Section~\ref{sec:measure_2comp}.
(8) Limits on \feII\ column density of ISM component in two-component fits. The model line profile for \feII\ 2261 and 2374 were
integrated and the above formulae, from comment \#4, applied.
(9) Limits on \feII\ column density of Doppler component in two-component fits. The model line profile for \feII\ 2261 and 2374 were
integrated and the above formulae, from comment \#4, applied.
(10) Velocity of maximum \feII\ 2374 absorption for $I + \delta(I) = 1$.
(11) Velocity of maximum \mgII\ 2796 absorption for $I + \delta(I) = 1$.
}
\label{tab:outflow}
\end{deluxetable}

%% file: table4.tex
\begin{deluxetable}{lllllllllll}
\tabletypesize{\scriptsize}
\tablewidth{0pt}
\tablecaption{Inflow Properties}
\tablehead{
\colhead{Object} &
\colhead{W(2374)} &       
\colhead{W(2260)} &
\colhead{ $\log [N(Fe^{+}) C_f]$                  } &
\colhead{  $V_{1}$           } &
\colhead{ $b_{ISM}$ } &
\colhead{ $V_{DOP}$              } &
\colhead{ $\log [N_{ISM}(Fe^{+}) C_f]$ } &
\colhead{ $\log [N_{DOP}(Fe^{+}) C_f]$ } &
\colhead{ $V_{max}(\lambda2374)$  } &
\colhead{ $V_{max}(\lambda2382)$  } \\
\colhead{}         &
\colhead{    (\AA)       } &
\colhead{  (\AA)  } &
\colhead{  (\col)                       } &
\colhead{  (\kms)      } &
\colhead{  (\kms)   } &
\colhead{ (\kms)         } &
\colhead{   (\col)          } &
\colhead{  (\col)         } &
\colhead{      (km/s)    } &
\colhead{      (km/s)  } \\
\colhead{(1)     } &
\colhead{ (2)     }     &
\colhead{   (3)   } &
\colhead{    (4)                        } &
\colhead{    (5)       } &
\colhead{    (6)    } &
\colhead{ (7)            } &
\colhead{ (8)           } &
\colhead{   (9)         } &
\colhead{     (10)          } &
\colhead{   (11)          } 
}
\startdata
12011428& $2.14 \pm 0.48$ & 0.44     &  15.1         - 16.0          & $ 154 \pm  32$ &   86 &  $392 \pm 124 $      & 15.14 -16.19  & 14.71 -15.49            &  $416 \pm 0.9$  &  $523 \pm  0.1$  \\
12016019  & $2.44 \pm 0.20$ & 0.19     &  15.1         - 15.7          & $ 48 \pm  14$   &   97 &  $258 \pm  40 $   & 15.17 -16.15  & $>13.58$                &  $655 \pm 34$   &  $723 \pm  0.6$  \\
12019962& $1.88 \pm 0.34$ & 0.32     &  15.0         - 15.9          & $ 104 \pm  31$ &  113 &  $122 \pm  97 $      & 14.68 -15.20  & 13.86 -15.15            &  $621 \pm 9$    &  $472 \pm  1$    \\
12019996  & $1.69 \pm 0.31$ & 0.30     &  15.0         - 15.9          & $ 117 \pm  24$  &  158 &  $386 \pm  138 $  & 14.92 -15.47  & 14.56 -15.57          &  $510 \pm 3$    &  $577 \pm  5$    \\
12100420  & $1.86 \pm 0.28$ & 0.26     &  15.0         - 15.8          & $ 71 \pm  18$   &  109 &  $190 \pm  45 $   & 15.14 -15.98  & 14.37 -14.82            &  $539 \pm 1$    &  $641 \pm  1$    \\
22004858  & $1.94 \pm 0.14$ & 0.14     &  15.0         - 15.6          & $ 61 \pm  11$   &  151 &  $71 \pm  18 $    & 15.03 -15.63  & 14.76 -15.53           &  \nodata        & $737 \pm  2$     \\
22005270& $0.98 \pm 0.23$ & 0.25     &  14.8         - 15.8          & $ 401 \pm  43$ &  145 &  $505 \pm  588 $     & 14.39 -14.71  & 14.37 -14.81           &  $546 \pm 2$    &  $724 \pm  9$    \\
32020468  & $2.36 \pm 0.69$ & 0.62     &  15.1         - 16.2          & $ 123 \pm  39$  &   95 &  $164 \pm  79 $   & 15.21 -16.33  & 14.82 -15.97          &  $465 \pm 1$    &  $311 \pm  3$    \\
42006875  & $1.01 \pm 0.12$ & 0.12     &  14.8         - 15.5          & $ 51 \pm  14$   &   62 &  $256 \pm  50 $   & 14.87 -15.66  & 13.86 -14.21          &  \nodata        & $663 \pm  3$     \\
\enddata
\tablecomments{
(1) DEEP2 survey identification.
(2) Rest-frame equivalent width of \feII\ 2374 absorption from direct integration of the trough. Error estimated from bootstrap resampling.
(3) Upper limit ($1\sigma$) on \feII\ 2261 equivalent width. Computed from an assumed linewidth
    that is twice the FWHM of the line response function and the continuum S/N ratio.
(4) Limits on \feII\ column density for a fiducial covering factor of $C_f = 1$. For optically thin absorption, the linear portion
    of the curve-of-growth gives $N(\feII) = 6.40 \times 10^{14}\col W_{2374}$, which we adopt as a lower bound.
    We derive an upper bound,     $N(\feII) = 9.06 \times 10^{15}\col W_{2261}$, from
    the $3\sigma$ upper limit on the \feII\ 2261 equivalent width.
(5) Doppler shift of the single velocity component model fitted to \feII\ absorption as described in Section~\ref{sec:measure_fe2}.
(6) Doppler parameter of the systemic component in the two-component fits to
    the \feII\ series.  We adopt characteristic velocities of gas in the galaxy as probed
    by cooling radiation from \ion{H}{2} regions. Specifically, we require the FWHM of
    the absorption optical depth profile match that of [OII] measured previously
    by Weiner \et (2006), which gives $b_{sys} = \sqrt{2} \sigma ([OII]) $.
(7) Doppler shift of the {\it Doppler component} fitted to \feII\ lines in two-component model described in Section~\ref{sec:measure_2comp}.
(8) Limits on \feII\ column density of ISM component in two-component fits. The model line profile for \feII\ 2261 and 2374 were
integrated and the above formulae, from comment \#4, applied.
(9) Limits on \feII\ column density of Doppler component in two-component fits. The model line profile for \feII\ 2261 and 2374 were
integrated and the above formulae, from comment \#4, applied.
(10) Velocity of maximum \feII\ 2374 absorption for $I + \delta(I) = 1$.
(11) Velocity of maximum \mgII\ 2796 absorption for $I + \delta(I) = 1$.
}
\label{tab:inflow}
\end{deluxetable}

%% file: table5.tex
\begin{deluxetable}{lclll}  
\tablewidth{0pt}
\tablecaption{Properties of Composite Spectra}
\tablehead{
\colhead{Feature} &
\colhead{$V_1$ (\feII)} &
\colhead{EW($\lambda 2374$) } &
\colhead{$V_{max}$(\mgII)} &       
\colhead{EW($\lambda2796$)} \\       
\colhead{Spectrum}         &
\colhead{    (\kms)       } &
\colhead{    (\AA)       } &
\colhead{    (\kms)       } &
\colhead{    (\AA)       } \\
\colhead{(1)     } &
\colhead{ (2)     }     &
\colhead{   (3)   } &
\colhead{    (4)                        } &
\colhead{    (5)       } 
}
\startdata
Lowest Mass ($8.85 \le\ \log (M_*/\msun) < 9.85$ )     & $-28 \pm 8 $ & 1.26  & $ -649 \pm 86 $   & $0.61 \pm 0.01  $  \\
Intermediate Mass ($9.85 \le\ \log(M_*/\msun)< 10.35$) & $-32 \pm 6 $ & 1.47  & $ -770 \pm 103 $  & $1.40 \pm 0.03  $  \\
Highest Mass ($10.35 \le \log(M_*/\msun) \le 11.29$)   & $-23 \pm 8 $ & 1.59  & $ -713 \pm 119 $  & $1.63 \pm 0.05  $  \\     
Bluest ($0.189 \le\ U-B < 0.519$)                      & $-32 \pm 8 $ & 1.31  & $ -693 \pm 86 $   & $0.72 \pm 0.02  $  \\     
Bluer ($0.519 \le\ U-B < 0.662$)                       & $-27 \pm 6 $ & 1.48  & $ -649 \pm 91 $   & $1.17 \pm 0.02  $  \\
Blue ($0.662 \le\ U-B \le\ 0.907$)                     & $-26 \pm 7 $ & 1.52  & $ -888 \pm 173 $  & $1.85 \pm 0.07  $  \\
Lowest Luminosity ($-17.67 \ge\ M_B -5\log h > -20.03$)         & $-23 \pm 9 $ & 1.19  & $ -666 \pm 88 $   & $0.72 \pm 0.02  $  \\     
Intermediate Luminosity ($-20.03 \ge\ M_B-5\log h > -20.79$)   & $-18 \pm 7 $ & 1.56  & $ -667 \pm 88 $   & $1.08 \pm 0.01  $  \\
Highest Luminosity ($-20.79 \ge\ M_B-5\log h > -22.19$)        & $-39 \pm 6 $ & 1.65  & $ -782 \pm 92 $   & $1.79 \pm 0.03  $  \\
Lowest SFR ($0.86 \le\ SFR(\msunyr) < 12.24$)          & $26 \pm 10$  & 1.48  & $ -629 \pm 104$   & $0.81 \pm 0.04 $    \\     
Highest SFR ($12.24 \le\ SFR(\msunyr) < 97.81$)        & $-38 \pm 8$  & 1.47  & $ -755 \pm 91 $   & $1.88 \pm 0.02 $   \\
Lowest sSFR ($1.00 < sSFR^{-1}(Gyr) \le\ 15.08$)       & $19 \pm 10$  & 1.55  & $ -779 \pm 136$   & $1.59 \pm 0.06 $    \\     
Highest sSFR ($0.178 < sSFR^{-1}(Gyr) \le 1.00$)       & $-24 \pm 9$  & 1.41  & $ -610 \pm 81 $   & $0.92 \pm 0.02 $    \\
$3\sigma$ FeII Outflow  &$-119 \pm 6 $ & 1.47  & $ -901 \pm 99 $   & $1.59 \pm 0.02 $   \\
$3\sigma$ FeII Inflow   & $95 \pm 8$   & 1.89  & $ -559 \pm 113$   & $1.25 \pm 0.07 $    \\
\enddata
\tablecomments{
Col. 1 -- Subsample averaged together to form composite spectrum.
Col. 2 -- Doppler shift of the \feII\ absorption measured as described in Section~\ref{sec:measure_fe2}.
Col. 3 -- Equivalent width of the \feII\ $\lambda 2374$ absorption trough.
Col. 4 -- Maximum blueshift, $V_{max}$, of \mgII\ $\lambda 2796$ absorption measured as described in 
Section~\ref{sec:vmax}.
Col. 5 -- Equivalent width of the \mgII\ $\lambda 2796$ trough from 0 to $V_{max}$.
}
\label{tab:composite_spectra} \end{deluxetable}

%% file: table6.tex
\begin{deluxetable}{lccccc}
\tablewidth{0pt}
\tablecaption{Correlations Among Properties of Individual Galaxies}
\tablehead{
\colhead{Quantity}  &   
\colhead{$V_1(\feII)$}  &   
\colhead{$W_{\lambda2374}$}       &    
\colhead{$W_{\lambda2796}(V<0)$}     &  
\colhead{$V_{max}$}        &   
\colhead{$W_{MgII}$}}
\startdata
$M_*$       &   0.099  (1.21)\tablenotemark{a}   &     0.139  (1.70)\tablenotemark{a}   &    0.320  (1.87)\tablenotemark{c} & -0.305  (1.78)\tablenotemark{c}   &   0.493 (2.9)\tablenotemark{c}    \\ % 0.377  (2.20)   \\
$M_B -5 \log h$ &   0.048  (0.58)\tablenotemark{a} &  -0.133  (1.62)\tablenotemark{a}   &   -0.338  (1.97)\tablenotemark{c} &  0.097  (0.56)\tablenotemark{c}   &  -0.495 (2.9)\tablenotemark{c}    \\ % -0.250  (1.46)    \\
$U-B$       &   0.123  (1.50)\tablenotemark{a}   &     0.116  (1.42)\tablenotemark{a}   &    0.443  (2.58)\tablenotemark{c} & -0.205  (1.20)\tablenotemark{c}   &   0.685 (4.0)\tablenotemark{c}    \\ % 0.462  (2.69)                      \\
$z$         &  -0.054  (0.66)\tablenotemark{a}   &     0.180  (2.20)\tablenotemark{a}   &    0.066  (0.38)\tablenotemark{c} &  0.143  (0.84)\tablenotemark{c}   &  0.401 (2.34)\tablenotemark{c}    \\ % -0.139  (0.81)                      \\
SNR         &   0.021  (0.26)\tablenotemark{a}   &    -0.065  (0.80)\tablenotemark{a}   &   -0.165  (0.96)\tablenotemark{c} & -0.059  (0.34)\tablenotemark{c}   &  -0.050 (0.29)\tablenotemark{c}   \\ % -0.148  (0.86)                      \\
$V_1(\feII)$ &    \nodata          &     0.221  (2.70)\tablenotemark{a}  &   -0.1421 (0.83)\tablenotemark{c} &  0.160  (0.93)\tablenotemark{c}   &                     0.159 (0.92)\tablenotemark{c}    \\ % -0.020  (0.11)                      \\
$W_{\lambda2374}$ &   0.221  (2.70)\tablenotemark{a}   &     \nodata           &    0.451  (2.63)\tablenotemark{c} & -0.033  (0.19)\tablenotemark{c}   &                0.410 (2.39)\tablenotemark{c}    \\ % -0.129  (0.75)                      \\
$V_{max}$(\mgII) &   0.160  ($0.93$)\tablenotemark{c} &    -0.033  (0.19)\tablenotemark{c}   &   -0.451  (2.63)\tablenotemark{c} &   \nodata          &                -0.302 (1.76)\tablenotemark{c}   \\ %-0.332  (1.94)                      \\
$W_{\lambda2796}(V<0)$ &  -0.142  ($0.83$)\tablenotemark{c} &     0.451  (2.63)\tablenotemark{c}   &    \nodata         &  -0.451  (2.63)\tablenotemark{c}  &          0.717 (4.2)\tablenotemark{c}         \\ %0.774  (4.51)                      \\
SFR       &  -0.225  (1.40)\tablenotemark{b} &    -0.005  ($0.03$)\tablenotemark{b} &    0.086  ($0.19$)\tablenotemark{d} &  0.257  ($0.57$)\tablenotemark{d} &   0.714 (1.60)\tablenotemark{d}     \\ % 0.143  (0.32)                      \\
sSFR      &  -0.257  (1.61)\tablenotemark{b} &    -0.079  ($0.49$)\tablenotemark{b} &   -0.257  ($0.58$)\tablenotemark{d} &  0.086  ($0.19$)\tablenotemark{d} &   0.314 (0.70)\tablenotemark{d}     \\ % -0.314  (0.70)      \\
\enddata
\tablecomments{
We computed the Spearman
rank-order correlation coefficient, $r_S$, between the continuum
S/N ratio near \feII\  and various galactic properties. To determine
whether a significant correlation was present, we examined the variance
in the sum-squared difference of ranks, $D$, which is approximately
normally distributed as described in chapter~14 of Press \et
(1992). The value in parentheses is
the number of standard deviations from the null hypothesis (no correlation
between the two quantities).
}
\tablenotetext{a}{Sample of 150 blue-cloud galaxies with measured FeII Doppler shift $V_1$.}
\tablenotetext{b}{Sample of 40 blue-clouds galaxies with  measured FeII Doppler shift $V_1$ and extinction-corrected, ultraviolet SFR.}
\tablenotetext{c}{Sample of 35 high S/N ratio spectra with measured maximum \mgII\ blueshift $V_{max}$.}
\tablenotetext{d}{Sample of 6 high S/N ratio spectra with measured $V_{max}(\mgII)$ and extinction-corrected, ultraviolet SFR.}
\label{tab:xc_measure}\end{deluxetable}

%% file: ms.bbl
\begin{references}

Aird, J. \et 2012, \apj, 746, 90

Arav, N. \et 2005, \apj, 620, 665

Behroozi, P. S., Conroy, C., \& Wechsler, R. H. 2010, \apj, 717, 379

Bell, E. F. \et 2003, \apjs, 149, 289

Bell, E. F. \et 2005, \apj, 625, 23

Bouch\'{e}, N. \et 2006,  \mn, 371, 495 % MgII - halo mass anti-correlation

Bouch\'{e}, N. \et 2010, \apj, 718, 10001

Brooks, A. M.  \et 2009, \apj, 694, 396

Bundy, K. \et 2006, \apj,  651, 120

Bregman, J. N. 1980, \apj, 236, 577

Chabrier, G. 2003, PASP, 115, 763

Chen, Y.-M. \et 2010, \aj, 140, 445 %Absorption-line Probes of the Prevalence and Properties of Outflows in Present-day Star-forming Galaxies

Chevalier, R. A. \& Clegg, A. W. 1985, Nature, 317, 44

Churchill, C. \et 2003, \aj, 125, 98

Coil, A. L. \et 2008, \apj, 672, 153 % Color and L dependence of clustering

Coil, A. L. \et 2011, \apj, 743, 46  % arXiv:1104.0681

Cooper, J. \et 2008, \apj, 674, 157

Cooper, J. \et 2009, \apj, 703, 330

Cresci, G. \et 2009, \apj, 697, 115

Croton, D. J. \et 2006, \mn, 365, 11 %The many lives of active galactic nuclei: cooling flows, black holes and the luminosities and colours of galax


Dav\'{e}, R., Oppenheimer, B. D., \& Finlator, K. 2011a, \mn, 415, 11 %Galaxy evolution in cosmological simulations with outflows - I. Stellar masses and star formation rates

Dav\'{e}, R., Finlator, K., \& Oppenheimer, B. D.  2011b, \mn, 416, 1354  %Galaxy evolution in cosmological simulations with outflows - II. Metallicities and gas fractions

Dav\'{e}, R. Finlator, K., \& Oppenheimer, B. D. 2012, \mn, 421, 98 % An analytic model for the evolution of the stellar, gas and metal content of galaxies

Davis, M. \et 2003, SPIE, 4834, 161

Davis, M. \et 2007, \apj, 660, L1 % AEGIS

De Young, D. S., \& Heckman, T. M. 1994, \apj, 431, 598

Dekel, A. \et Birnboim, Y.  2006, \mn, 368, 2 % Galaxy bimodality due to cold flows and shock heating

Dekel, A. \et 2009a, Nature, 457, 451 % Cold streams in early massive hot haloes as the main mode of galaxy formation

Dekel, A. \et 2009b, \apj, 703, 785  % Formation of Massive Galaxies at High Redshift: Cold Streams, Clumpy Disks, and Compact Spheroids

Elbaz, D. \et 2007, \aa, 468, 33

Ellison, S., et al. 2000, AJ, 120, 1175

Erb, D. K. \et 2006, \apj, 644, 813

Erb, D. K. \et 2012, arXiv:1209.4903

Faber, S. M. 2007, \apj, 665, 265

Faucher-Gigu\`{e}re, C.-A. \& Kere\~{s}, D. 2011, \mn, 412, L118


Fumagalli, M. \et 2011, \mn, 418, 1796  % preprint (arXiv:1103.2130)

Fujita, A. \et 2009, \apj, 698, 693

Goerdt, T. \et 2012, arXiv:1205.2021  % Detectability of cold streams into high-z galaxies by absorption lines

Granato, G. L. \et 2004, \apj, 600, 580

Heckman, T. M. \et 1990, \apjs, 

Heckman, T. M. \et 2000, \apjs, 129, 493

Heckman, T. M. 2002, ASP Conference Series 254: Extragalactic Gas
at Low Redshift, 292

Hong, S. \et 2010, arXiv:1008.4242


Hopkins, A. M. \& Beacom, J. F. 2006, \apj 651, 142 %On the Normalization of the Cosmic Star Formation History

Hopkins, P. F., Quataert, E., \& Murray, N. 2012, \mn, 421, 3522

Hopkins, P. F. \et 2007, \apj, 659, 976

Houck, J. C. \& Bregman, J. N. 1990, \apj, 352, 506

Jones, T., Stark, D. P., \& Ellis, R. S. 2012, \apj, 751, 51

Kauffmann, G., et al. 2003, MNRAS, 346, 1055  % see also -. 2003b, MNRAS, 341, 33

Kennicutt, R. C. 1998, \araa, 36, 189.

Kere\~{s}, D \et 2005, \mn, 363, 2

Kere\~{s}, D \et 2009a, \mn, 396, 2332

Kere\~{s}, D \& Hernquist, L. 2009b, \mn, 396, 2332

Kimm, T. \et 2011, \mn, 413, 51

Kobulnicky, H. A. \& Gebhardt, K. 2000, \aj, 119, 1608


Lilly, S. J. \et 1996, \apj, 460, L1

Lu, L. 1991, ApJ, 379, 99

Madau, P. \et 1996, \mn, 283, 1388

Marchesini, D. \et 2009, \apj, 701, 1765

Marinacci, F. \et 2011, \mn, 415, 1534

Marlowe, A. \et 1997, \apjs, 112, 285 %The Taxonomy of Blue Amorphous Galaxies. I. H alpha and UBVI Data

Martin, C. L. 1999, \apj, 513, 156

Martin, C. L. 2005, \apj, 621, 227

Martin, C. L. Kobulnicky, H. A. \& Heckman, T. M. 2002, \apj, 574, 663

Martin, C. L. \et 2010, \apj, 721, 174

Martin, C. L. \et 2009, \apj, 721, 174

McGaugh, S. S. \et 2010, \apj, 708, 14 %The Baryon Content of Cosmic Structures

Mendez, A. J. \et 2011, \apj, 736, 110

Meyer, D. M. \& York, D. G. 1987, ApJ, 315, 5

Morton, D. C. 2003, \apjs, 149, 205

Mostek, N. \et 2012, \apj, 746, 124 %arXiv:1108.2503 % Calibrating the Star Formation Rate at z=1 from Optical Data

Nestor, D. \et 2006, \apj, 643, 75  % .... break in MgII DF

Newman, J. A. \et 2012, arXiv:1203.3192

Noeske, K. G. 2007a, \apj, 660, L43

Noeske, K. G. 2007b, \apj, 660, L47

Ocvirk, P., Pichon, C., Teyssier, R. 2008, \mn, 390, 1326

Oke, J.B., \et  1995, PASP, 107, 375 

Oppenheimer, B. \et 2008, \apj, 679, 1574

Oppeheimer, B. \et 2010, \mn, 406, 2325

Oppenheimer, B. \et 2012, \apj, 420, 829

Pettini, M. \et 2000, \apj, 528, 96

Pettini, M. \et 2001, \apj, 554, 981

Phillips, A. C., Miller, J., Cowley, D., Wallace, V. 2006, SPIE, 6269, 56 

Prochaska, J. X., Kasen, D., \& Rubin, K. 2011, \apj, 734, 24 % Simple Models of Metal-line Absorption and Emission from Cool Gas Outflows

Press, W. H., Teukolsky, S. A., Vetterling, W. T., \& Flannery, B. P. 1992, {\it Numerical
Recipes}, Press Syndicate of the Cambridge University (New York, NY)

Rix, S. A. \et 2004, \apj, 615, 98

Rix, H.-W. \et 1997, \mn, 285, 229

Rubin, K. H. R. \et 2012, \apj, 747, 26 % The Direct Detection of Cool, Metal-enriched Gas Accretion onto Galaxies at z ~ 0.5

Rubin, K. H. R. \et 2011a, \apj, 728, 55 % Low-ionization Line Emission from a Starburst Galaxy: A New Probe of a Galactic-scale Outflow

Rubin, K. H. R. \et 2010b, \apj, 719, 1503 % The Persistence of Cool Galactic Winds in High Stellar Mass Galaxies between z ~ 1.4 and ~1

Rubin, K. H. R. \et 2010a, \apj, 712, 574  % FeII 2328

Rupke, D. \et 2005, \apjs, 160, 115


Savage, B. D. \& Sembach, K. R. 1996, \araa, 34, 279

Scannapieco, E. \& Oh, S. P. 2004, \apj, 608, 62

Schaye, Y., et al. 2003, ApJ, 596, 768

Schawinski, K. \et 2007, \mn, 382, 1415 %Observational evidence for AGN feedback in early-type galaxies

Schwartz, C. M. \& Martin, C. L. 2004, \apj, 610, 201

Schwartz, C. M. \et 2006, \apj, 646, 858

Shapley, A. E. \et 2003, \apj, 588, 65

Shapley, A. E. \et 2006, \apj, 651, 688



Simcoe, R. A. 2011, \apj, 738, 159

Simcoe, R. A. \et 2006, \apj, 637, 648 %Observations of Chemically Enriched QSO Absorbers near z~2.3 Galaxies: Galaxy Formation Feedback Signatures in the Intergalactic Medium

Somerville, R. S. \et 2008, \mn, 391, 481

Songaila, A., \& Cowie, L. L. 1996, AJ, 112, 335

Spitzer, L. Jr., 1978, Physical Processes in the Interstellar Medium, John Wiley \& Sons, Inc., (New York, NY)

Springel, V., Di Matteo, T., \& Hernquist, L. 2005, \mn, 361, 776

Steidel, C. S. \et 2003, \apj, 592, 728

Steidel, C. S. \et 2004, \apj, 604, 534

Steidel, C. S. \et 2010, \apj, 717, 289

Stewart, K. R. et 2011b, \apj, 738, 39 % preprint (arXiv:1103.4388) Orbiting Circumgalactic Gas as a Signature of Cosmological Accretion

Stewart, K. R. et 2011a, \apj, 735, 1  % Observing the End of Cold Flow Accretion Using Halo Absorption Systems

Tinker, J. L. \& Chen, H.-W. 2008, \apj, 679, 1218 % On The Halo Occupation of Dark Baryo

Tremonti, C. \et 2004, \apj, 613, 898

Tremonti, C. \et 2007, \apj, 663, 77    % The Discovery of 1000 km s-1 Outflows in Massive Poststarburst Galaxies at z=0.6

Weiner, B. J. \et 2006, \apj, 653, 1027

Weiner, B. J. \et 2009, \apj, 692, 187

Willmer, C. N. 2006 \apj, 647, 853

van de Voort, F. \et 2011a, \mn, 414, 2458  % rates of modes of gas accretion onto galaxies/halos

van de Voort, F. \et 2011b, \mn, 415, 2782  % drop in cosmic SFR caused by a change in the mode of gas accretion

Yan, R. \et 2009, \mn, 398, 735

Zhang, D. \& Thompson, T. A. 2010, arXiv:1005.4691

\end{references}
